\documentclass[11pt,a4paper]{article}
\usepackage[utf8]{inputenc}
\pdfoutput=1  
\usepackage{epsfig}
\usepackage{graphicx,psfrag}
\usepackage{multirow}
\usepackage{subfig}
\usepackage{slashed}
\usepackage{caption}
\usepackage{url}
\usepackage{braket}
\usepackage{jheppub}
\usepackage{enumerate}
\usepackage{wasysym} 
\usepackage{mathrsfs} 
\usepackage{amsfonts} 
\usepackage{amsbsy} 
\usepackage{amscd}
\usepackage{color}
\usepackage{changepage}
\usepackage{floatrow}
\usepackage{makecell}
\usepackage{tikz}
\usepackage[normalem]{ulem}
\usetikzlibrary{snakes}

\makeatletter

\def\eeq{\end{equation}}
\def\beq{\begin{equation}}

\newcommand{\Rmnum}[1]{\expandafter\@slowromancap\romannumeral #1@}

\def\mdm{m_{\rm DM}}
\def\mdma{m_{\rm DM_1}}
\def\mdmb{m_{\rm DM_2}}
\def\mhd{m_{\rm HD}}

\newcommand{\bea} {\begin{eqnarray}}
\newcommand{\eea} {\end{eqnarray}}

\newcommand{\gsim}{\raisebox{-0.13cm}{~\shortstack{$>$ \\[-0.07cm]
      $\sim$}}~}
\newcommand{\lsim}{\raisebox{-0.13cm}{~\shortstack{$<$ \\[-0.07cm]
      $\sim$}}~}
\makeatother

\title{Distinguishing two dark matter component particles at $e^+e^-$ colliders}
\author[a]{Subhaditya Bhattacharya,}
  \affiliation[a]{Department of Physics, Indian Institute of Technology Guwahati,
North Guwahati, Assam - 781039, India} 
\author[b]{Purusottam Ghosh,}
  \affiliation[b]{School Of Physical Sciences, Indian Association for the Cultivation of Science,
2A and 2B, Raja S.C. Mullick Road, Kolkata 700032, India} 
\author[a]{Jayita Lahiri,}
\author[c]{Biswarup Mukhopadhyaya} 
  \affiliation[c]{Department of Physical Sciences, Indian Institute of Science Education and Research Kolkata, Mohanpur - 741246, India} 
\emailAdd{subhab@iitg.ac.in}
\emailAdd{pghoshiitg@gmail.com}
\emailAdd{jayitalahiri@rnd.iitg.ac.in}
\emailAdd{biswarup@iiserkol.ac.in}
\abstract{
 We investigate ways of identifying two kinds of dark matter (DM) component particles at high-energy colliders. The strategy is to notice and distinguish double-peaks(humps) in the missing energy/transverse energy distribution. The relative advantage of looking for {\em missing energy} is pointed out, in view of the fact that the longitudinal component of the momentum imbalance becomes an added input. It thus turns out that an electron-positron collider is better suited for discovering  a two-component DM scenario, so long as both of the components are kinematically accessible. This and a number of associated conclusions are established, using for illustration a scenario including a scalar and a spin-1/2 particle. We also formulate a set of measurable quantities which quantify the distinguishability of the two humps, defined in terms of double-Gaussian fits to the missing energy distribution. The efficacy of these variables in various regions of the parameter space is discussed, using the aforesaid model as illustration.}

\DeclareUnicodeCharacter{2212}{-}
\begin{document}
\maketitle
\section{Introduction}
\label{sec:intro}

Evidence for dark matter (DM) has accumulated from different astrophysical observations like rotation curves of galaxies \cite{Rubin:1970zza,Zwicky:1937zza}, 
gravitational lensing effects around bullet clusters \cite{Hayashi:2006kw}, and cosmological observations like the anisotropy of cosmic microwave background radiation 
(CMBR) \cite{Hu:2001bc} in WMAP \cite{Hinshaw:2012aka,Spergel:2006hy} or PLANCK \cite{Planck:2018vyg} data. Observations further suggest that DM constitutes a 
large portion $(\simeq 24\%)$ of the energy budget of the universe, often expressed in terms of relic density $\Omega h^2\simeq 0.12$ \cite{Planck:2018vyg}, where 
$\Omega={\rho}/{\rho_c}$ represents the cosmological density with $\rho$ being the DM density, $\rho_c$ the critical density and $h$ represents Hubble expansion rate 
in units of 100 km/s/Mpc. However, direct evidence of DM in reproducible terrestrial observations is yet to be found.
We are thus still unable to confirm whether DM consists of elementary particles of the weakly or feebly interacting
types. All one can say with certainty is that neutrinos cannot be the dominant components of DM, and thus
physics beyond the standard model (SM) have to be there if particle DM exists.

Two major classes of ideas, both of which can account for correct relic density, are often discussed. The first category consists of 
weakly interacting massive particle (WIMP) scenarios where the DM particles were in thermal and chemical equilibrium in early universe, 
and have frozen out  when their annihilation rate dropped below the Hubble expansion rate \cite{Bertone:2004pz,Roszkowski:2017nbc,Kolb:1990vq}. 
In the second category, one can have feebly interacting massive particles (FIMP)\cite{Hall:2009bx} which do not thermalise with
the cosmic bath, and are presumably produced from the decay or scattering of some massive particles in thermal bath. 
We shall be concerned here with WIMPs, since they are the likeliest one to be detected in collider experiments which constitute the theme of this paper.

It is of course possible to have more than one DM components simultaneously, and this is the possibility we are concerned with. 
While many of the existing multicomponent DM studies are in the context of WIMPs \cite{Aoki:2012ub,Liu:2011aa,Cao:2007fy,Bhattacharya:2013hva,Esch:2014jpa,Karam:2016rsz,Ahmed:2017dbb,Poulin:2018kap,Aoki:2018gjf,YaserAyazi:2018lrv,Aoki:2017eqn,Biswas:2013nn,Bhattacharya:2016ysw,Bhattacharya:2017fid,Barman:2018esi,Bhattacharya:2018cgx,Bhattacharya:2019fgs,Borah:2019aeq,Chakraborti:2018lso,Chakraborti:2018aae,Bhattacharya:2018cqu,Yaguna:2021rds,Belanger:2021lwd,VanLoi:2021dzv,Yaguna:2021vhb,DiazSaez:2021pfw,Chakrabarty:2021kmr,Nam:2020twn,Betancur:2020fdl,Nanda:2019nqy,Bhattacharya:2019tqq,Elahi:2019jeo,Herrero-Garcia:2018lga,Das:2022oyx}, scenarios with more than one DM types have   
also been studied \cite{Bhattacharya:2021rwh,DuttaBanik:2016jzv,Choi:2021yps}. 
Direct search experiments \cite{XENON:2018voc,PandaX-4T:2021bab} might probe two component WIMP frameworks via observation 
of a kink in the recoil energy spectrum \cite{Herrero-Garcia:2017vrl,Herrero-Garcia:2018qnz}. We devote the present
discussion to the collider detectability and distinguishability of two DM components, both of whom are of the WIMP type. 
However, studies on collider searches are relatively fewer \cite{Hernandez-Sanchez:2020aop,Konar:2009qr}\footnote{Counting 
number of DMs simultaneously produced in cascade from the end point of the spectrum is studied in \cite{Agashe:2010tu,Giudice:2011ib}.}. 
Here we develop some criteria for the discrimination of two peaks in missing energy distributions, in an illustrative scenario 
where the DM components are pair-produced in cascades, along with the same kinds of visible particles.


More specifically, we consider cases where  the two DM components belong to two separate
`dark sectors'. The initial hard scattering pair-produces members of either sector, and each of
these members initiate a decay chain culminating in the DM candidate of either kind. The visible particles
produced alongside happen in our examples to be multileptons.  Our purpose is to maximise the visibility
of the two peaks, via discriminants based on their heights, separation and spreads.

We also emphasize that the main principle(s), on which our suggested method of analysis is based,
do not depend on the model used here for illustration. Certain features of the model are course best suited
for substantial production of the two kinds of  DM particles, and the decay chains that occur here
simplify and facilitate what we wish to demonstrate. More complicated avenues of DM production are
of course within our horizon, but the points we make here serve as leitmotifs in any analysis. 

We establish further that an electron-positron collider is in most cases better suited for thus 
discerning the two DM components, as compared to hadronic machines. The main reason, as we shall show, 
is that in electron-positron collisions, the full kinematic information, especially that on longitudinal
components of momenta (including missing momenta) can be utilised. In addition, it helps the suppression
of standard model (SM) backgrounds. The option of using polarised beams, too, can be of advantage. 

On the whole, the essential points developed and established in this study are:

\begin{itemize}
\item A two-component WIMP scenario which leads to the same final state via two different kinds of cascade, can lead to double-peaks (bumps) in kinematic 
distributions such as missing energy.
\item An $e^+ e^-$ machine having the requisite kinematic reach is advantageous in this respect, since (a) it makes use of the longitudinal components of missing momenta, 
and (b) beam polarization can reduce backgrounds.
\item A set of measurable quantities, formulated by us for this specific purpose, are useful in making the double-peaking behaviour prominent.
\end{itemize}

%

The paper is organised as follows. We first discuss WIMP signal at colliders in section \ref{sec:WIMPsignal} including 
some general aspects of kinematics in both single-and two-component DM scenarios. In section \ref{sec:model} 
we discuss the model chosen for illustration, followed by the selection of benchmark 
points in section \ref{sec:bp}. Section \ref{sec:signal} contains a discussion of the signal from two-component dark sector vis-a-vis 
SM backgrounds, while in section \ref{sec:analysis} we analyse the predicted results in detail.  Some proposed criteria for distinguishing 
the peaks are exemplified in section \ref{sec:distinction}. We summarise and conclude in section \ref{summary}. 
In Appendix~\ref{app1}-\ref{app5}, we include some details that are omitted in the main text.

\section{WIMP signal at colliders}
\label{sec:WIMPsignal}
WIMPS are produced at colliders either via electroweak hard scattering processes or in weak decays
of other particles.  But no component of
currently designed collider detectors is equipped to register their presence. Their smoking gun signatures,
therefore, result from energy/momentum imbalance in the final state, rising above SM backgrounds
as well as the imbalance due to mis measurement\footnote{Such signals may be contrasted with some others, mostly 
associated with FIMPs, where only the particle in bath responsible for DM production can be produced at collider, leading to a 
disappearing/long-lived charge track or displaced vertex \cite{Belanger:2018sti,Alimena:2019zri,Banerjee:2018uut} as an indirect signal of DM.}. Such imbalance can be 
quantified in terms of the following kinematic variables:

\begin{itemize}
\item {\it Missing Transverse Energy or MET ($\slashed{E}_T$)},  defined as:
\bea
\slashed{E}_T = -\sqrt{(\sum_{\ell,j,\gamma} p_x)^2+(\sum_{\ell,j,\gamma} p_y)^2};
\label{eq:MET}
\eea
where the sum runs over all visible objects that include leptons ($\ell$), photons ($\gamma$), jets ($j$), and also unclustered components. 

\item{\it Missing Energy or ME ($\slashed{E}$)}   with respect to the centre-of-mass (CM) energy ($\sqrt{s}$),
defined as:  
\bea
\slashed{E}=\sqrt{s}-\sum_{\ell,j,\gamma} E_{\rm vis}\,;
\label{eq:ME}
\eea
where the sum runs over visible objects like $\ell, j,\gamma$ and unclustered components.

\item {\it Missing Mass or MM ($\slashed{M}$)}, defined as:
\bea
\slashed{M}^2=\left(\sum_ip_{i}-\sum_f p_f\right)^2 \,,
\label{eq:MM}
\eea
\noindent which requires the knowledge of initial state four momenta ($p_i$) and final state ones ($p_f$), where $f$ runs over all the 
visible particles. For mono-photon process $e^+ e^- \to \chi \bar\chi \gamma$, where $\chi(\bar\chi)$ are DM, 
$\slashed{M}^2=s-2\sqrt{s} E_\gamma$. Here, $E_\gamma$ is the energy of the outgoing photon.
\end{itemize}


$\slashed{E}$ and $\slashed{M}$ are measurable in $e^+e^-$ (or $\mu^+\mu^-$) machines while
hadron colliders can only measure $\slashed{E_T}$. The scalar sum of transverse momentum, sometimes 
referred as {\it Effective mass} ($H_T$)  is another variable of interest for hadron colliders. 
$\slashed{E}$ or $\slashed{E_T}$ are reconstructible from the energies and momenta of visible particles, 
against which the DM particle(s) recoil.  The resulting signals can be

\medskip

\noindent
$\bullet$ {\bf mono-X + $\slashed{E}~(\slashed{E_T}$)}, where X is a jet, a photon, a weak boson or a Higgs, or, \\
$\bullet$ {\bf $n$-leptons + $m$-jets + $p$ photons + $\slashed{E}~(\slashed{E_T}$)} 


\medskip


The mono-X signature usually arises when two DM particles are produced directly via either a portal to dark sector~\cite{Liew:2016oon,Kahlhoefer:2017dnp,Boveia:2018yeb} or effective operators \cite{Abercrombie:2015wmb,Abdallah:2015ter,Barman:2021hhg}. 
The kinematic observables at our disposal in such a case are the four-momentum of particle X and $\slashed{E_T}/\slashed{E}/\slashed{M}$. 
Among them, as we shall show from a general argument below, $\slashed{E}$ (whenever measurable) 
is likely to contain the best usable information for discriminating between two unequal-mass DM particles,
both of which are  produced at a collider. The longitudinal component of the momentum imbalance makes
the all-important difference.


The second class of final states, namely, multi-jet/multi-lepton channels usually arises when a pair of heavy particles (usually members of the 
dark sector, described here as `Heavier Dark Sector Particles' (HDSP)) are produced. Each of them further decays into a DM together with jets 
and leptons in the final state. Obviously such an event topology, with a greater multitude of visible particles, will have richer kinematics. For example, 
the kinematics is governed here by both the HDSP and DM masses, a phenomenon which we discuss below
in detail. The resulting features of the final state play an important role in differentiating the contributions from different DM components in event distributions.

Our aim here is to investigate the correlation between $\slashed{E_T} ,~ \slashed{E}~\rm{or}~\slashed{M}$ with DM mass and the 
mass of the HDSP, in cases where two kinds of DM particles are produced via HDSP cascades. Peaks/bumps in the distributions can be treated as  tell-tale 
signature of multi-component DM framework, when the different bumps are distinguishable. We systematically 
address the issues to suggest some distinction criteria. But before we discuss 
some relevant features of kinematic observables for both one-and two-component WIMP scenarios.

\subsection{Aspects of kinematics}
\label{sec:kinematics}


We focus on multilepton/multijet final states in a scenario where the DM particles are produced at the end of decay chains in $e^-e^+$ collisions. 
Parameters that play decisive roles in shaping $\slashed{E_T},~ \slashed{E}~\rm{or}~\slashed{M}$ distributions are, 
(a) the mass of the DM particle ($\mdm$), and (b) its mass difference with the HDSP ($\Delta m$). 


\subsubsection{Single-component DM}
\label{single-component}

Let us first consider a single DM with mass $\mdm$ and an HDSP with $\mhd=\mdm + \Delta m$, which decays to DM with 
one/or more massless SM fermions. For simplicity, we assume the HDSPs are pair-produced nearly at rest\footnote{Arises when $\sqrt{s}\simeq 2\mhd$, 
as assumed mostly in the rest of the analysis for $e^+e^-$ collisions.}, 
in which case, the lab frame in an $e^+ e^-$ machine can be identified, approximately at least, with the rest frame of the HDSP. 

\begin{figure}[!hptb]
	$$
	\subfloat[]{\includegraphics[width=7cm,height=6cm]{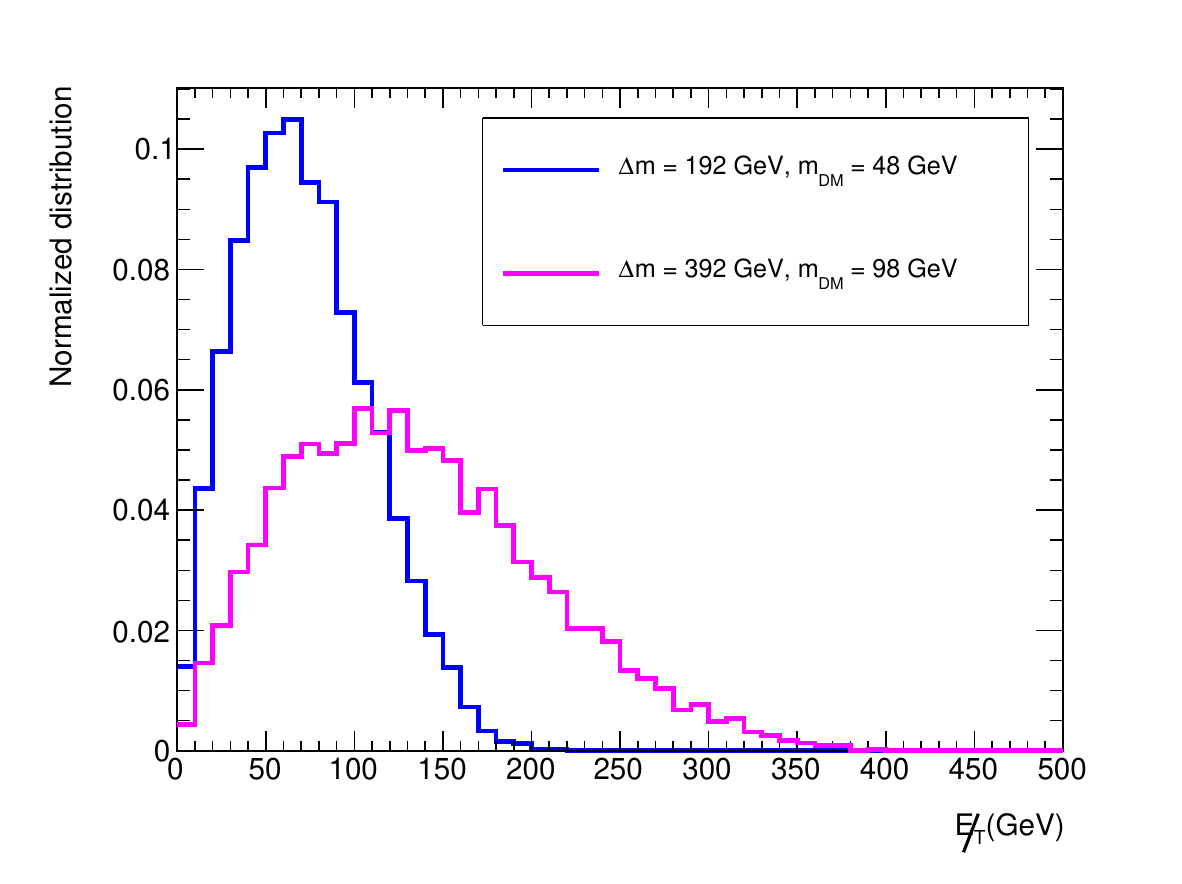}}
        \subfloat[]{\includegraphics[width=7cm,height=6cm]{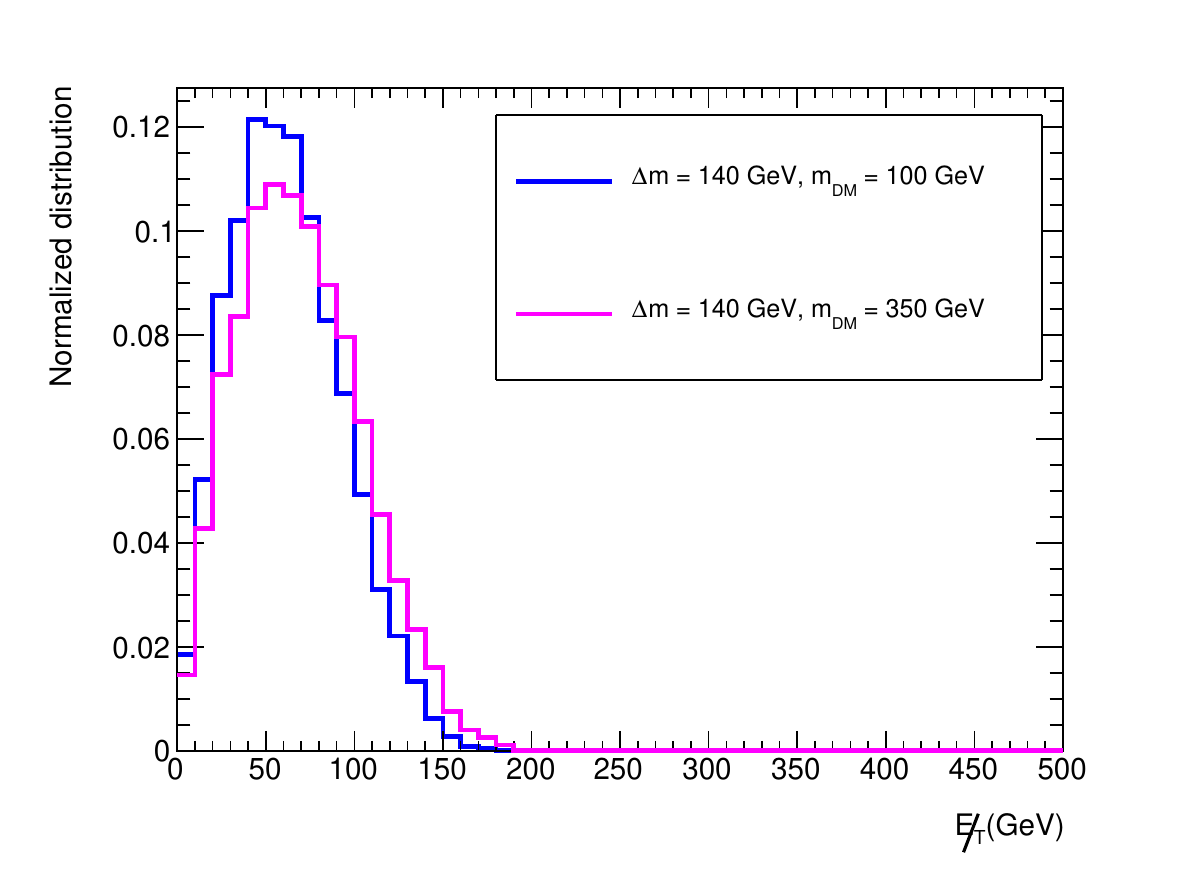}}
        $$
	\caption{Normalized $\slashed{E_T}$ distribution at $e^+e^-$ collider with (a) fixed $r$, different $\Delta m$, and (b) with fixed $\Delta m$, different $\mdm$ 
	 (see figure insets for details). We consider pair production of the singly charged scalar HDSP
	of an inert scalar doublet, which further decays to DM in association with on/off shell $W^\pm$ bosons at 
	$\sqrt s=500 (1000)$ GeV, as necessary to pair produce the HDSPs on-shell.}
	\label{deltam_ratio_comparison}
\end{figure}

The maximum $\slashed{E_T}$ is obtained when both DMs are 
moving in the same direction. Then (see Appendix~\ref{app1} for details),
\begin{equation}
\slashed{E_T}^{max} \simeq \Delta m \left( 1+ r \right)\,;
\label{eqn:MET-max}
\end{equation}
where $r = \frac{\mdm}{\mhd}$ is the ratio of the DM and HDSP masses. 
Clearly $\slashed{E_T}^{max}$ depends on both $\Delta m$ and $\mdm$. In order to verify that
we plot $\slashed{E_T}$ distribution in Fig.~\ref{deltam_ratio_comparison} for inert scalar doublet model 
(to be discussed in detail in Sec.~\ref{sec:model})\footnote{The shape and end point of the distribution depends on the model only when 
intermediate states differ.}, where singly-charged scalar HDSPs are pair-produced, 
followed by decays to DM with on/off-shell $W^\pm$ bosons. The CM energy is chosen $\sqrt s=500 (1000)$ GeV, 
as necessary to pair produce the HDSPs on-shell. In the plot on the left panel, we fix $r =5$ and choose different $\Delta m=\{48,98\}$ GeV, while on the right plot
we fix $\Delta m=140$ GeV and choose different values of $\mdm=\{100,350\}$ GeV. From the left plot, we see that the end-point shifts significantly towards 
higher value of $\slashed{E_T}$ with the increase of $\Delta m$\footnote{The end point of the blue $\slashed{E_T}$ curve 
($\{\Delta m,\mdm\}=\{192,48\}$ GeV) in Fig.~\ref{deltam_ratio_comparison} (a) appears at 200 GeV, whereas the estimated end point is at 
192 GeV using Eqn.~\eqref{eqn:MET-max}.}. More importantly, the peak of the $\slashed{E_T}$ distribution also shifts to higher values 
with larger $\Delta m$. The plot on the right-hand side show a relatively weaker dependence on $\mdm$. 
This establishes that $\slashed{E_T}$ distribution has a more pronounced dependence on $\Delta m$ than on $\mdm$.

\begin{figure}[!hptb]
	$$
	\subfloat[]{\includegraphics[width=7.5cm,height=6.5cm]{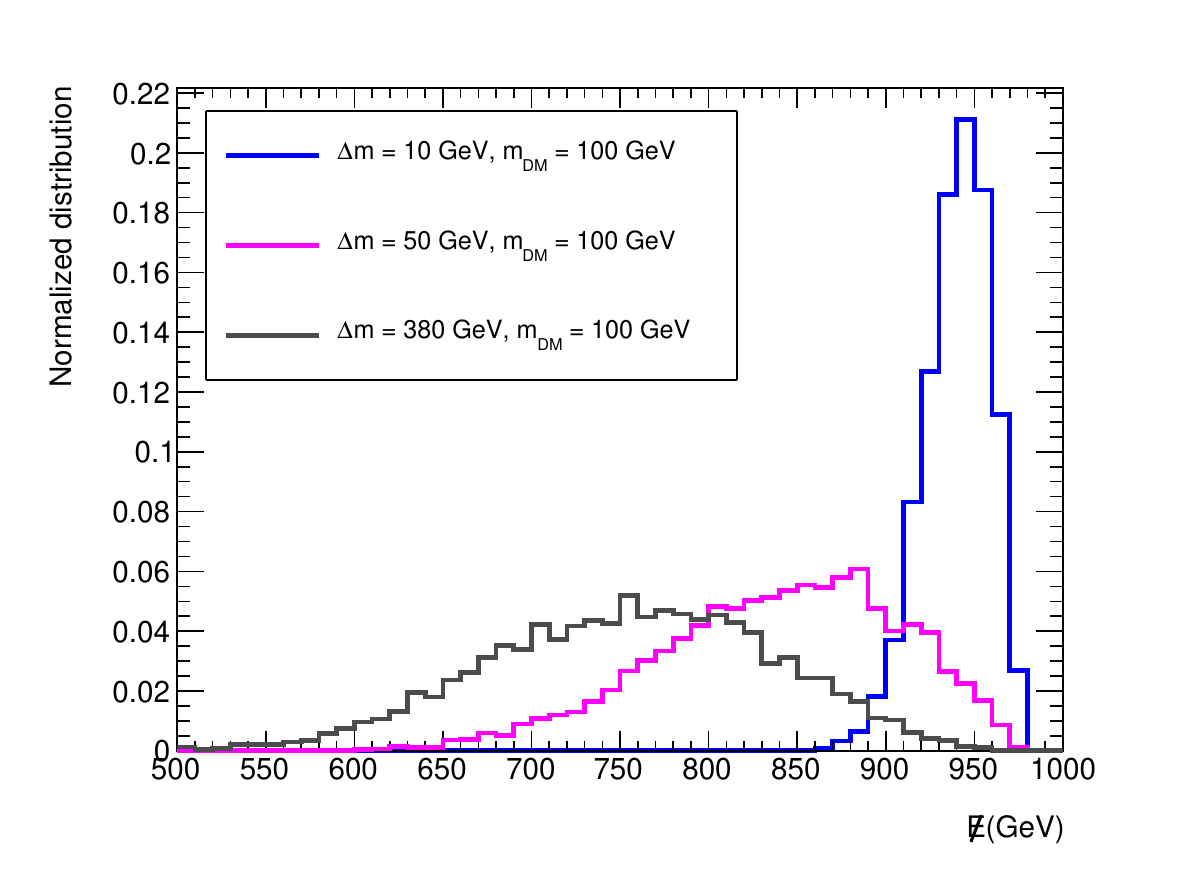}} 
        \subfloat[]{\includegraphics[width=7.5cm,height=6.5cm]{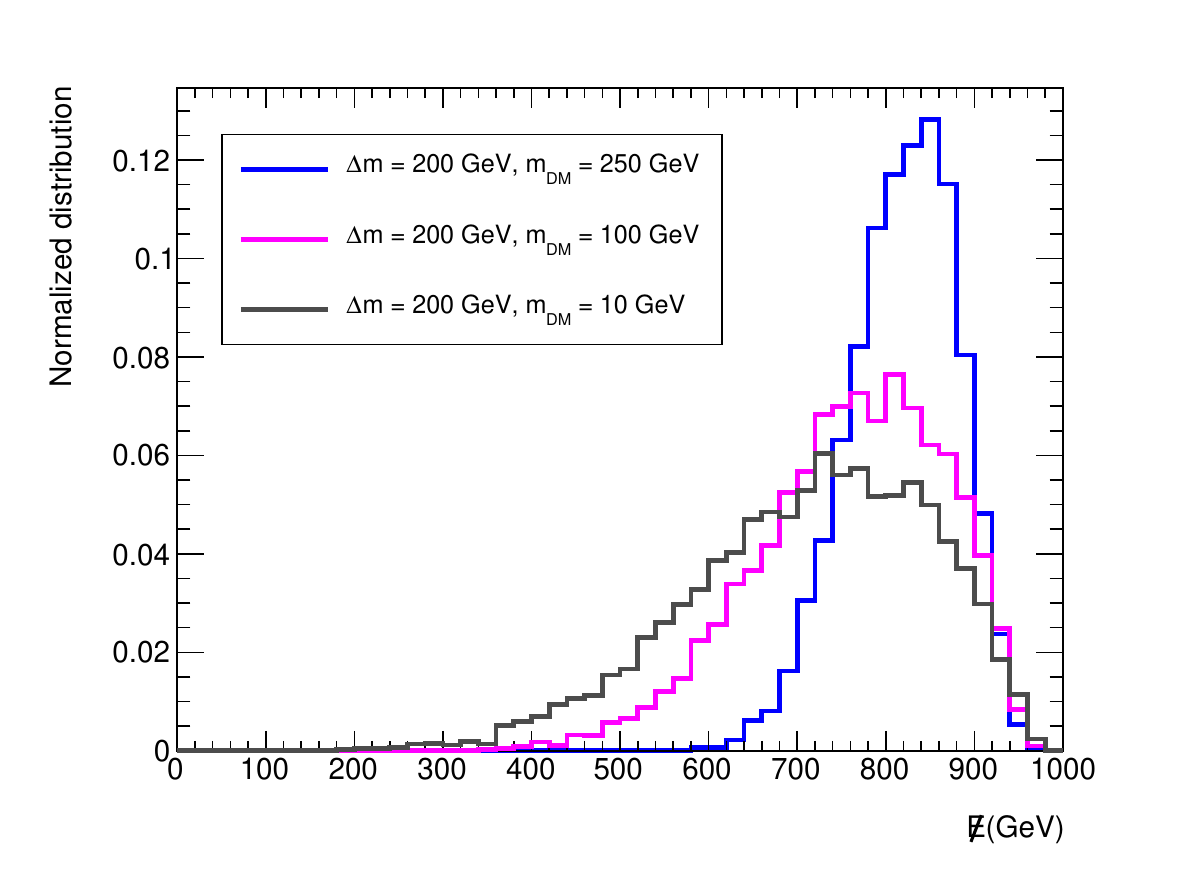}}
        $$
	\caption{Normalised $\slashed{E}$ distribution at $e^+e^-$ collider with (a) fixed DM mass, different $\Delta m$ and with (b) fixed $\Delta m$, 
	 different DM mass (see figure insets for details). We consider pair production of singly charged scalar HDSP of 
	an inert scalar doublet, which further decays to DM in association with on/off shell $W^\pm$ bosons at $\sqrt s=1000$ GeV.}
	\label{deltam_comparison}
\end{figure}


We examine $\slashed{E}$ distribution next and compare with $\slashed{E_T}$ distribution. 
Following, ${E}_{\rm DM} = \sqrt{{|\vec{p}|^2_{\text{ DM}}} + \mdm^2}$, 
an event with DM pair cascading from HDSP pair production yields,
 \begin{equation}
\slashed{E} = \sqrt{{|\vec{p}_1|^2_{\text{ DM}}} + \mdm^2}+\sqrt{{|\vec{p}_2|^2_{\text{ DM}}} + \mdm^2}\,.
\label{eq4}
\end{equation}

In order to illustrate the more noticeable presence of $\mdm$ in $\slashed{E}$ distribution, 
as evinced from Eq.~\ref{eq4}~\footnote{We note here that $|\vec{p}_1|_{\text{DM}}$ or $|\vec{p}_2|_{\text{DM}}$ in Eq.~\ref{eq4} are not observable quantities. 
Here they have been used purely for the demonstration purposes, to understand the behaviour of the actual observable quantities, i.e. $\slashed{E_T}$ and $\slashed{E}$.}, 
we present such distributions for different values of 
$\Delta m$  in Fig.~\ref{deltam_comparison}(a), where $\mdm$ is kept fixed.  
Fig.~\ref{deltam_comparison}(b) shows similar plots with the roles of $\Delta m$ and $\mdm$ reversed.
The $\slashed{E}$ spectrum peak shifts to the left side with larger $\Delta m$. 
In Fig.~\ref{deltam_comparison}(b), too, the $\slashed{E}$
distribution gets sharper with the peaks shifting to larger values with larger $\mdm$. 
On the whole, the $\slashed{E}$ distribution, if available, is comparably sensitive to both $\mdm$ and $\Delta m$, 
while $\slashed{E_T}$ is sensitive primarily to $\Delta m$. We thus conclude that an $e^+ e^-$ collider is more
effective in distinguishing DM components with different masses, as compared to a hadronic machine,
provided that both DM components are within the kinematic reach of such a collider. 
$\slashed{M}$ distribution turns rather similar to $\slashed{E}$ distributions, and
does not offer much advantage in our context. $\slashed{M}$ distributions and their comparison with $\slashed{E}$ 
are provided in Appendix~\ref{app1}.




\subsubsection{Two-component DM}
\label{two-component}
We now extend the study to the two-component scenario.
While the details of a two component DM model will be taken up in Section \ref{sec:model}, here we focus on 
a simple situation having two scalar DM particles with masses $\mdma,\mdmb$ and mass splitting with the corresponding HDSP as 
${\Delta m}_1~{\rm and}~ {\Delta m}_2$ respectively\footnote{We reiterate that kinematic features as described here,
do not depend on the details of the theoretical framework, especially when each HDSPs are produced with very little boost.}.
As previous, pair production of singly charged HDSPs and subsequent decay to the respective DM components with leptonically 
decaying on/off-shell $W^\pm$ is considered. Subsequent $\slashed{E}$ and $\slashed{E_T}$ distributions for different choices of 
$\mdm$ and $\Delta m$ are studied:


\begin{itemize}

\item ${\Delta m}_1 \approx {\Delta m}_2$ and $\mdma<\mdmb$

In this case, the $\slashed{E_T}$ distribution for both dark sectors is expected to merge since ${\Delta m}_1 \approx {\Delta m}_2$, whereas 
the $\slashed{E}$ distributions are expected to show a double peak, owing to the dependence of $\slashed{E}$ on the DM mass with $\mdma<\mdmb$. 
We can see this feature in Fig.~\ref{comparison1}. 
We assume the production cross-sections for the HDSP pairs to be equal for both the cases\footnote{
Relative cross-sections of the DM components indeed play a vital role, which will be discussed in context of a specific model. For the current discussion, 
the two HDSP pair production cross-sections have been taken in a specific ratio to make the peaks discernible.}. 



\begin{figure}[!hptb]
	\centering
	\subfloat[]{\includegraphics[width=7.5cm,height=6.5cm]{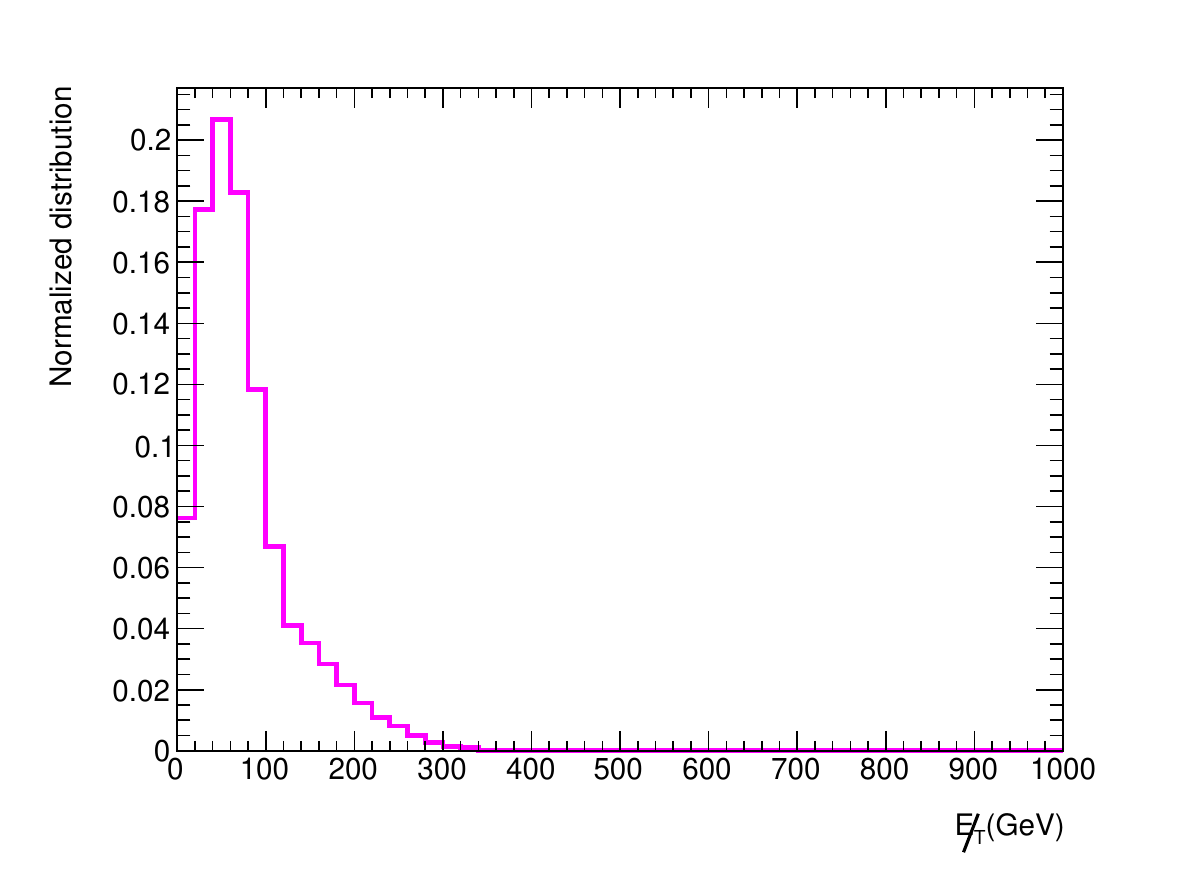}} 
        \subfloat[]{\includegraphics[width=7.5cm,height=6.5cm]{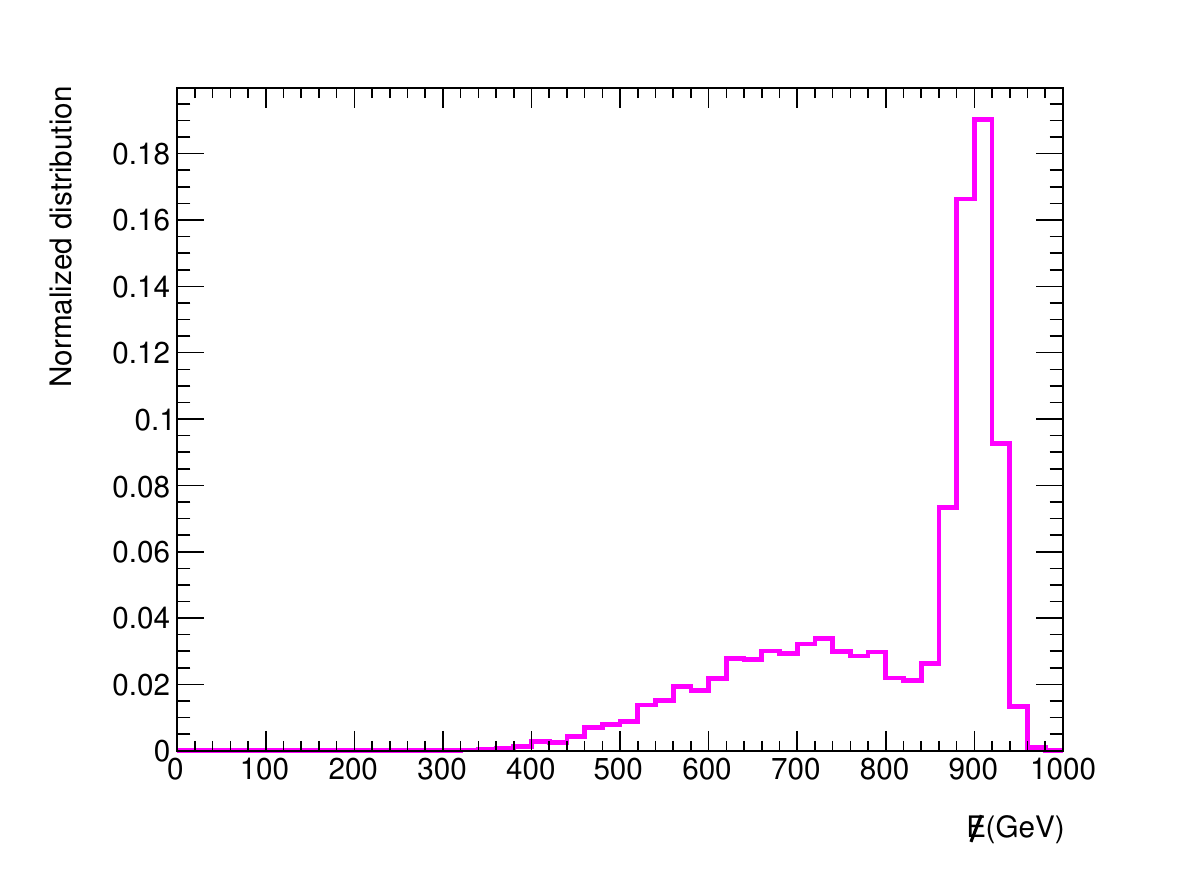}}
	\caption{Normalized (a) $\slashed{E_T}$ and (b) $\slashed{E}$ distribution for two component scalar DM scenario 
	with $\{\mdma, \mdmb, {\Delta m}_1, {\Delta m}_2\}=\{50,390,100,100\}$ GeV where HDSP pair production and subsequent 
	 decay chain is chosen as in Fig.~\ref{deltam_comparison} (see text). The production cross-sections for both the HDSP pairs are assumed equal at $\sqrt s=1000$ GeV.}
	\label{comparison1}
\end{figure}

\item ${\Delta m}_1 < {\Delta m}_2$ and $\mdma\approx\mdmb$ 

We consider next a scenario characterized by different mass splittings but similar DM masses. As an example, 
we chose $\mdma=\mdmb= 100$ GeV, ${\Delta m}_1 = 10$ GeV and ${\Delta m}_2 = 380$ GeV. 
Corresponding $\slashed{E_T}$ and $\slashed{E}$ distributions are shown in Fig.~\ref{comparison3}. The production cross-section for the lighter 
HDSP pair is assumed to be 50\% of that of the heavier HDSP. Here the $\slashed{E_T}$ distribution shows a double-peak nature unlike the previous case,
(see Fig.~\ref{comparison3}(a)), while $\slashed{E}$ distribution produces a better distinction where the second peak is even 
more prominent and well separated, see Fig.~\ref{comparison3}(b).

\begin{figure}[!hptb]
	\centering
	\subfloat[]{\includegraphics[width=7.5cm,height=6.5cm]{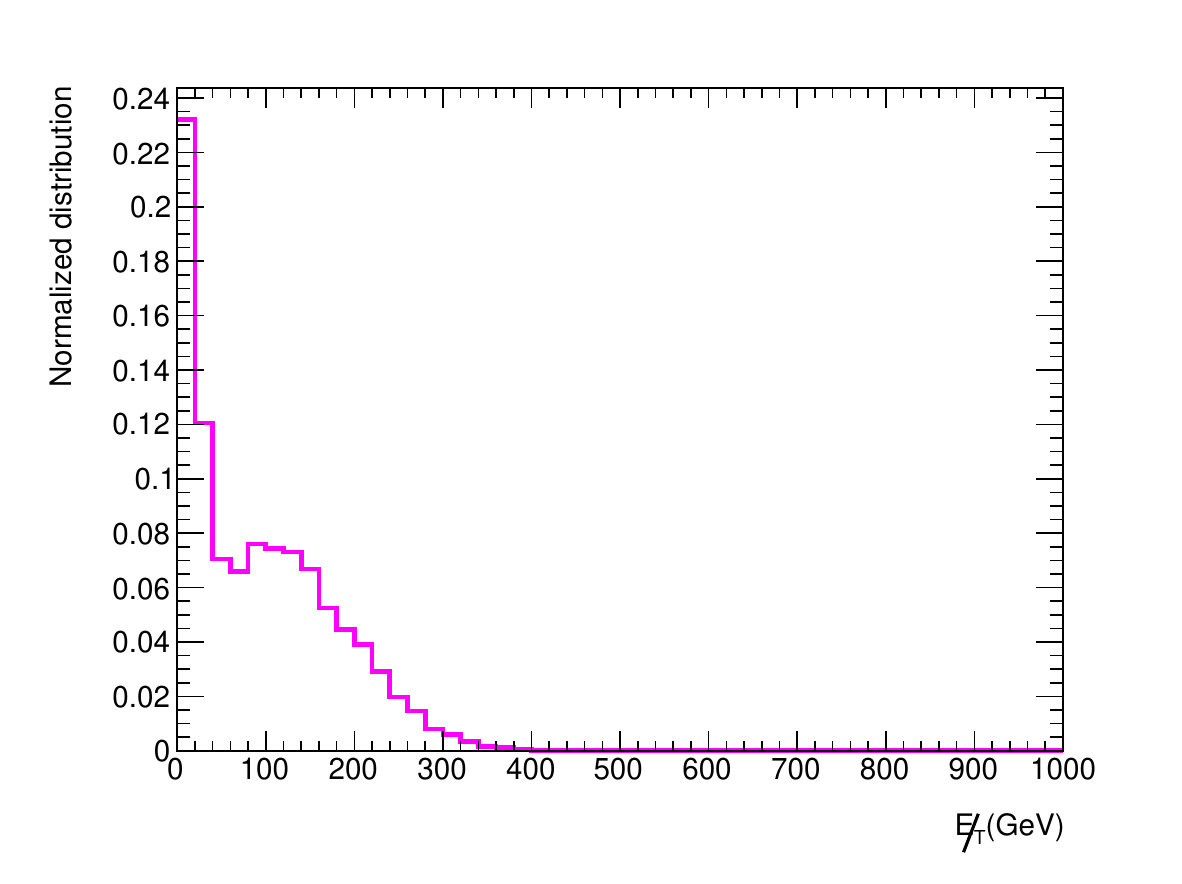}} 
        \subfloat[]{\includegraphics[width=7.5cm,height=6.5cm]{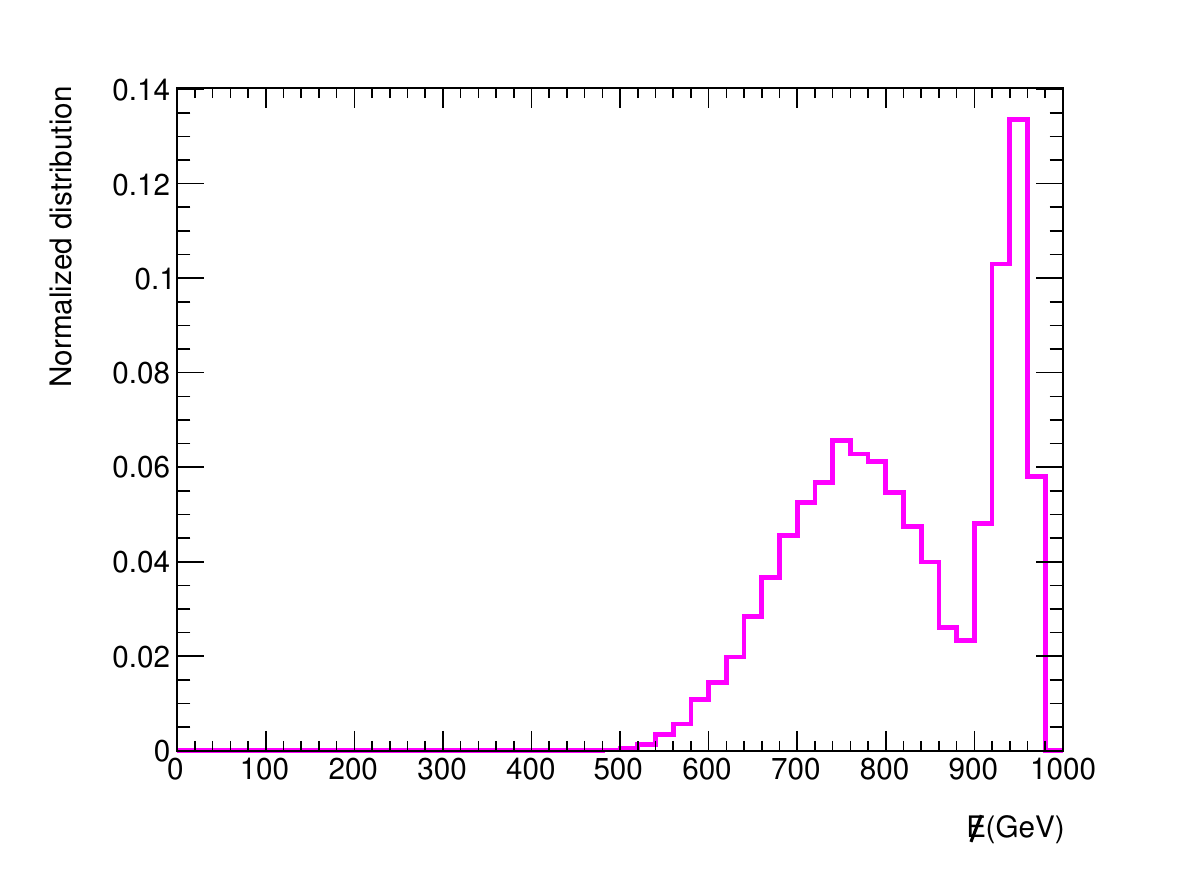}}
	\caption{ Same as Fig.~\ref{comparison1}, with $\{\mdma, \mdmb, {\Delta m}_1, {\Delta m}_2\}=\{100,100,10,380\}$ GeV. 
	The production cross-section of the lighter HDSP pair is assumed half of that for the heavier HDSP pair at $\sqrt s=1000$ GeV.}
	\label{comparison3}
\end{figure}

\item ${\Delta m}_1 < {\Delta m}_2$ and $\mdma<\mdmb$ 
 
In Fig.~\ref{comparison2} we depict a situation where both DM mass and splitting with the corresponding HDSPs are different, following a hierarchy 
${\Delta m}_1 < {\Delta m}_2$ and $\mdma<\mdmb$ (see caption for model inputs). 
We see that the $\slashed{E_T}$ distributions (on the left) for each DM component are almost overlapping and thus show a single peak. 
On the other hand, $\slashed{E}$ distribution (on the right) shows the presence of two peaks coming from two different dark sectors.
Thus, the difference in $\Delta m$ may not show up in $\slashed{E_T}$ in such cases, while $\slashed{E}$ still highlights it.

\begin{figure}[!hptb]
	\centering
	\subfloat[]{\includegraphics[width=7.5cm,height=6.5cm]{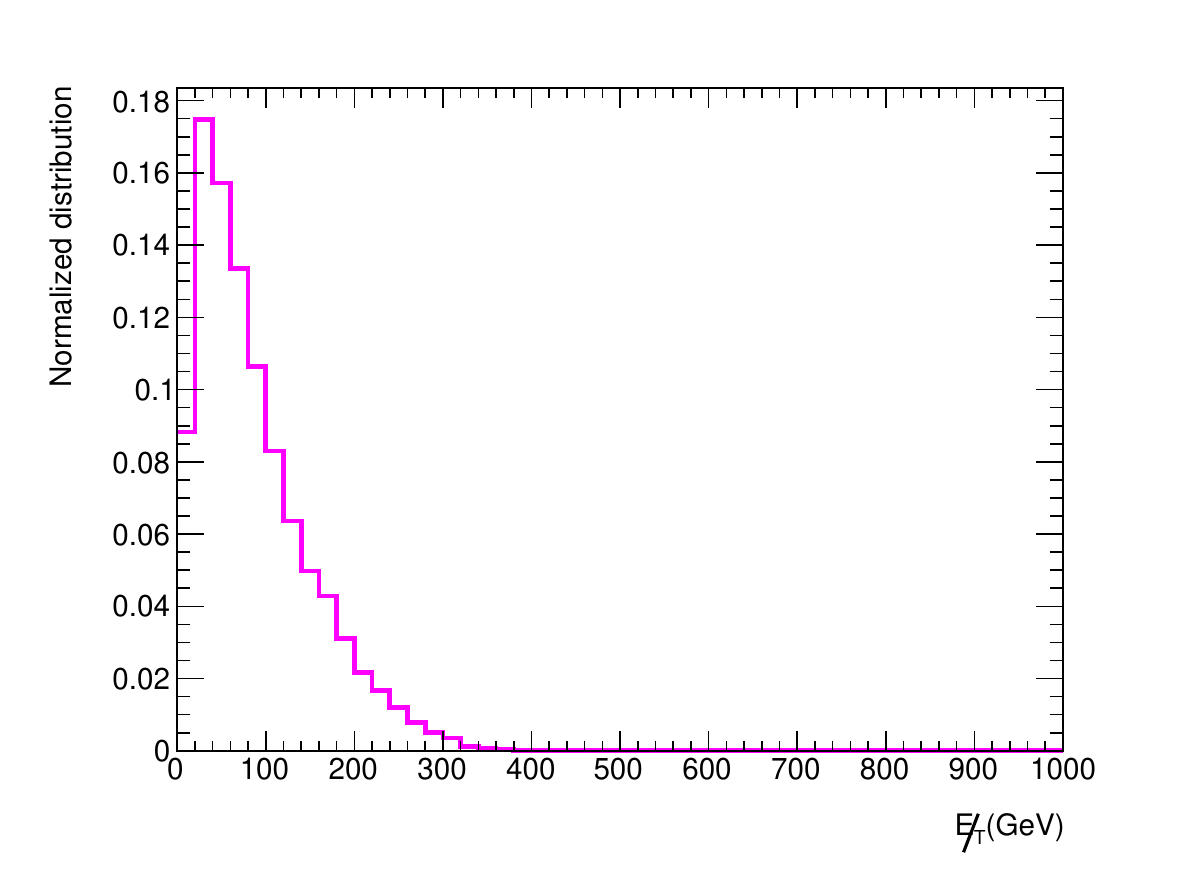}} 
        \subfloat[]{\includegraphics[width=7.5cm,height=6.5cm]{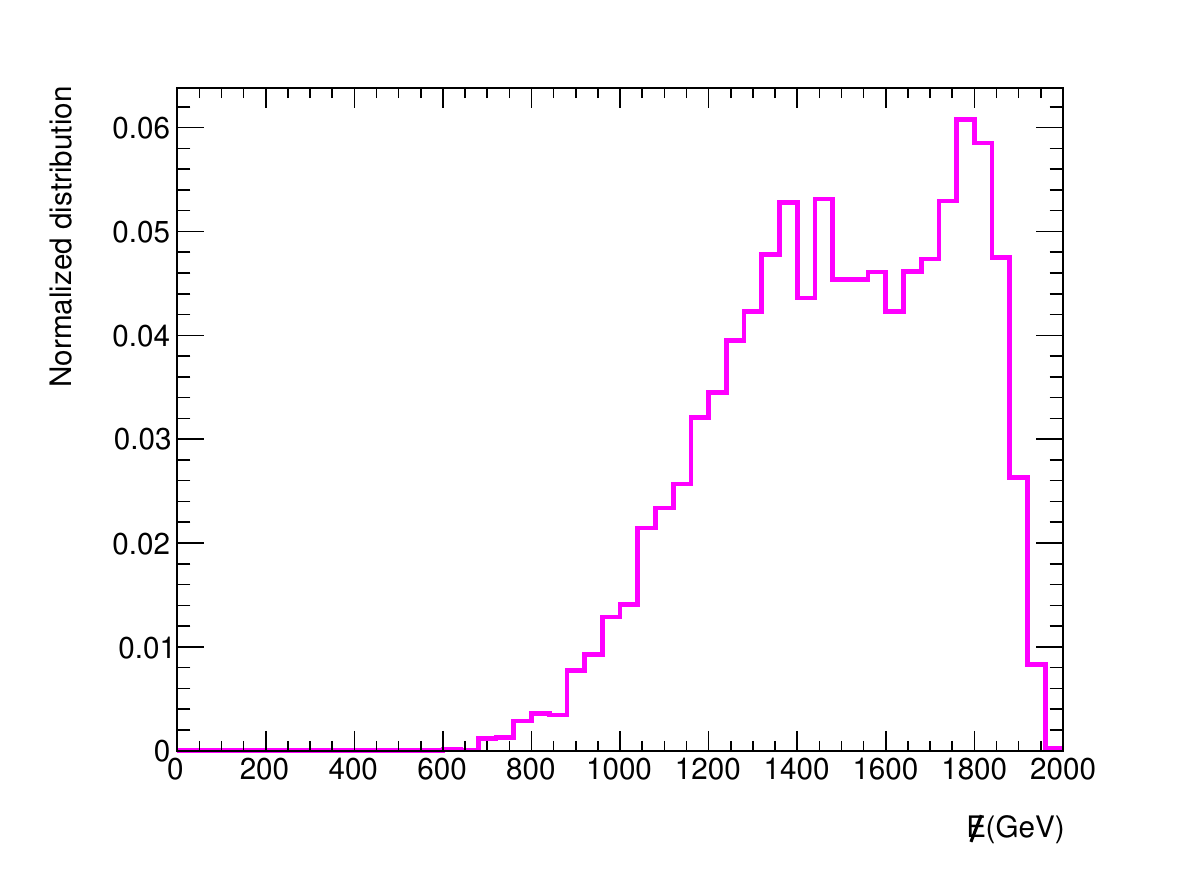}}
	\caption{Same as Fig.~\ref{comparison1}, with $\{\mdma, \mdmb, {\Delta m}_1, {\Delta m}_2\}=\{50,700,100,200\}$ GeV. 
	The production cross-sections for the heavier HDSP is assumed to be one quarter of that of the lighter ones at $\sqrt s=2000$ GeV.}
	\label{comparison2}
\end{figure}

\item ${\Delta m}_1 < {\Delta m}_2$ and $\mdma>\mdmb$

Finally we explore the possibility where the hierarchy in the DM masses is opposite to that of the mass splittings with the corresponding HDSPs 
as shown in Fig.~\ref{comparison4}. Again $\slashed{E}$ distribution (Fig.~\ref{comparison4} 
(b)) shows a clear two-peak behaviour while the $\slashed{E_T}$ (Fig.~\ref{comparison4} (a)) distribution shows a mere distortion. 
Note that the peak on the right side in Fig.~\ref{comparison4} (b) corresponds to $\mdma$, the higher DM mass.


\begin{figure}[!hptb]
	\centering
	\subfloat[]{\includegraphics[width=7.5cm,height=6.5cm]{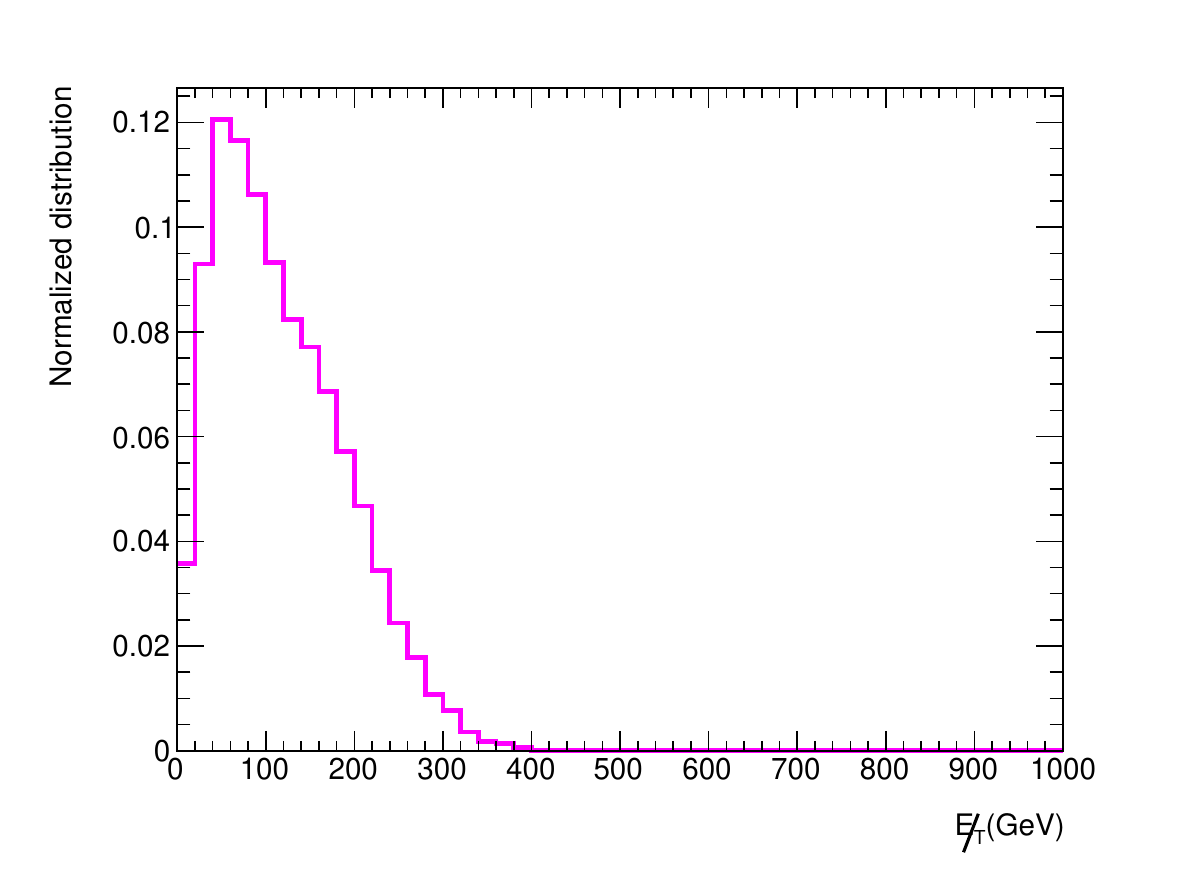}} 
        \subfloat[]{\includegraphics[width=7.5cm,height=6.5cm]{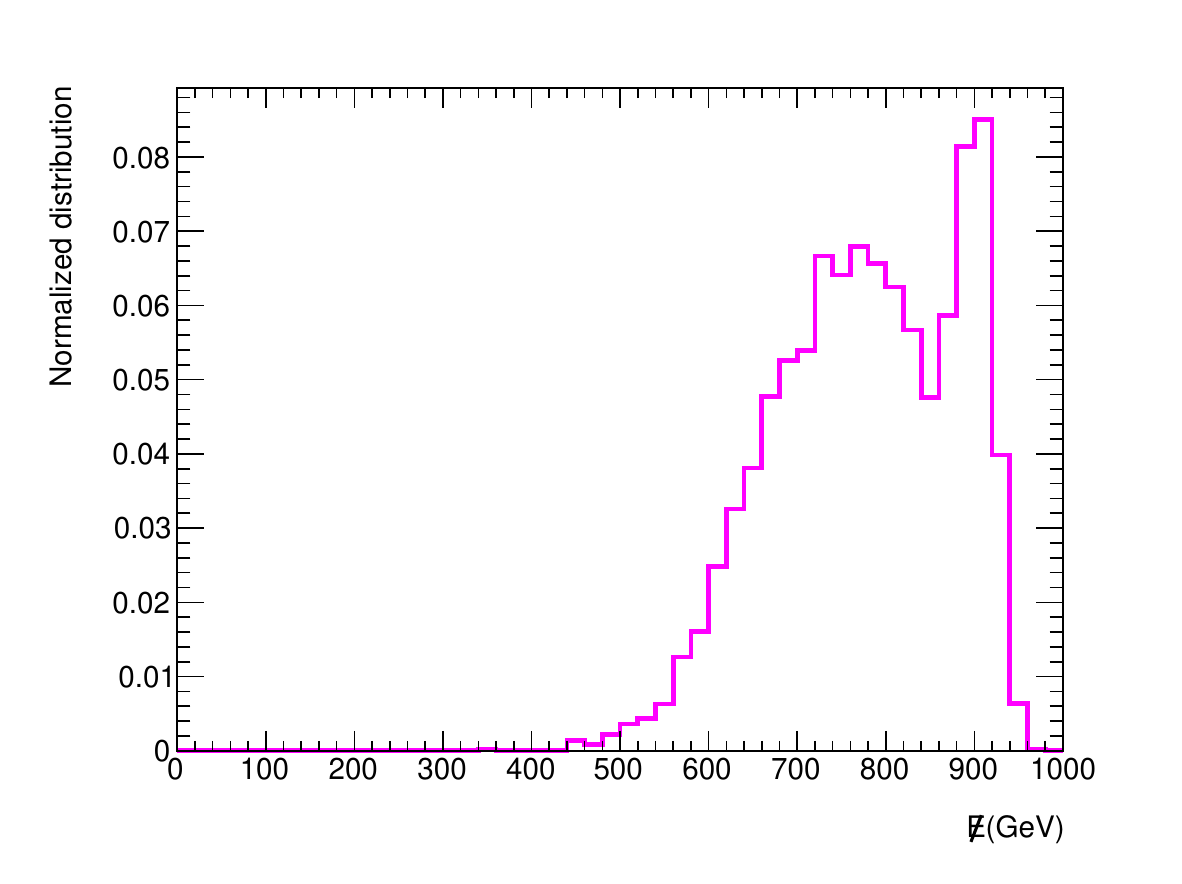}}
	\caption{Same as Fig.~\ref{comparison1}, with $\{\mdma, \mdmb, {\Delta m}_1, {\Delta m}_2\}=\{390,50,100,400\}$ GeV. 
	The production cross-sections for the heavier HDSP is taken to be one-quarter of that of the lighter ones at $\sqrt s=1000$ GeV.}
	\label{comparison4}
\end{figure}
\end{itemize}

\noindent

It is clear from the preceding discussion that multipartite DM scenario can show up in terms of double peak in $\slashed{E}$ and/or $\slashed{E_T}$ distributions
\footnote{We note here that the above possible kinematic conditions are exhaustive in a two component set up when the two dark sectors are identical.}. 
However, $\slashed{E}$ is more useful as it is more sensitive to $\mdm$ than $\slashed{E_T}$, in both the limits, $\mdm >> \Delta m$ and 
$\Delta m >> \mdm$. Complete knowledge of the total centre-of-mass energy of each collision and the very possibility of having $\slashed{E}$ 
distribution can therefore be advantageous for $e^+e^-$ collider over LHC where one has to resort to $\slashed{E_T}$, so long as DM particle 
productions are sufficiently copious and kinematically allowed. 

So far we have discussed mainly the relative positions of the peaks. The larger is the splitting between $\mdma$ and $\mdmb$, or 
${\Delta m}_1$ and ${\Delta m}_2$, the better is the separation between the two peaks in $\slashed{E}$ distribution and easier it is to hint for two component DM.
However, another important factor is the relative heights of the peaks, which are determined by the cross-section of the process. 
Two peaks of comparable size are more amenable to distinction, provided that their separation is adequate. Assuming for simplicity that the HDSPs 
decay into the DM candidate via a single channel with 100\% branching ratio, we can estimate the size of the peaks from the pair production 
cross-section of the respective HDSPs. Now, HDSP production depends on its mass and thus on the parameters $\mdm$ and $\Delta m$ 
(following $\mhd=\mdm + \Delta m$), both of which are also crucially involved in segregating the peaks in resultant $\slashed{E}$ 
(or $\slashed{E_T}$) distribution. The cross-section depends further on the model; whether the HDSPs
are scalar, fermion or vector boson and what mediators contribute to the production process. We will refer to these features quantitatively when 
we encounter the model. We must note here that HDSP production cross-section diminishes with larger $\mdm$ and $\Delta m$, so 
whichever component is heavier contributes with a smaller cross-section at least kinematically. So, the suitable scenario for discriminating DM components
occurs when we do not lose cross-section significantly due to kinematics. This is better addressed if one dark sector is 
fermionic, while the other is scalar, as the fermion production cross-section is larger than that of a scalar
with similar mass\footnote{Having both DM components scalar or fermion can also provide such distinctive $\slashed{E}$ ($\slashed{E_T}$) 
distribution, but with a more fine tuned choices of $\Delta m$ and $\mdm$.}. We therefore consider HDSPs in the form of singly charged scalars ($\phi^{\pm}$) 
and singly charged fermions ($\psi^{\pm}$) and their subsequent decays to respective DM components at $e^+e^-$ collider. 
We shall examine in specific how they can yield distinguishable peaks in $\slashed{E}$ ($\slashed{E_T}$) distributions 
with comparable cross-sections after satisfying cosmological constraints, in context of an illustrative 
model, which we elaborate next. 




\section{A model with two DM components }
\label{sec:model}
We illustrate our numerical results in terms of a model which consists of two dark sectors having two stable DM candidates: 
$(i)$ Scalar DM (SDM), as the lightest neutral state of an inert scalar doublet
$\Phi^T_i=\left(\begin{matrix} \phi^+ && \frac{\phi^0 + i A^0}{\sqrt{2}} \end{matrix}\right)$ and $(ii)$ Fermion DM (FDM), the lightest neutral state that arises from an admixture 
of a vectorlike fermion doublet $\Psi^T=\left(\begin{matrix} \psi && \psi^- \end{matrix}\right)$ and a right handed (RH) fermion singlet $\chi_R$. 
Stability of both DM components is ensured by additional ${\mathcal{Z}_2 \otimes \mathcal{Z}^\prime_2 }$ symmetry, where the inert scalar 
doublet transforms under $\mathcal{Z}_2$, while the fermion fields transform under a different discrete 
symmetry, namely,  $\mathcal{Z}^\prime_2 $. The quantum numbers of additional 
fields are shown in the Table \ref{tab:tab1}. The minimal renormalizable Lagrangian for this model then reads, 
\bea
\mathcal{L} \supset \mathcal{L}^{\rm SDM}+\mathcal{L}^{\rm FDM}.
\eea


\begin{table}[h]
\resizebox{\linewidth}{!}{
 \begin{tabular}{|c|c|c|c|}
\hline \multicolumn{2}{|c}{Fields}&  \multicolumn{1}{|c|}{ $\underbrace{ SU(3)_C \otimes SU(2)_L \otimes U(1)_Y}$ $\otimes \underbrace{ \mathcal{Z}_2 \otimes \mathcal{Z}^\prime_2 }$} \\ \hline
\multirow{2}{*}
{SDM} & $\Phi = \left(\begin{matrix}
 \phi^+ \\  \frac{1}{\sqrt{2}}(\phi^0+iA^0) 
\end{matrix}\right)$ & ~~1 ~~~~~~~~~~~2~~~~~~~~~~1~~~~~~~~~~~-~~~~~+  \\
\hline
{FDM} &  $\Psi_{L,R}=\left(\begin{matrix} \psi \\ \psi^- \end{matrix}\right)_{L,R}$&  ~~1 ~~~~~~~~~~~2~~~~~~~~~~-1~~~~~~~~~~+~~~~~- \\ [0.5em] \cline{2-3}
     & $\chi_{_R}$ & ~~1 ~~~~~~~~~~~1~~~~~~~~~~0~~~~~~~~~~+~~~~~- \\
\hline
\end{tabular}
}
\caption{Charge assignment of the BSM fields under the gauge group $\mathcal{G} \equiv \mathcal{G}_{\rm SM} \otimes \mathcal{G}_{\rm DM}$ in the toy scenario. Here $\mathcal{G}_{\rm SM}\equiv SU(3)_C \otimes SU(2)_L \otimes U(1)_Y$ and $\mathcal{G}_{\rm DM} \equiv \mathcal{Z}_2 \otimes \mathcal{Z}^\prime_2 $.  }
    \label{tab:tab1}
\end{table}

The Lagrangian for the SDM sector, having inert scalar doublet $\Phi$ can be written as : 
\bea
\mathcal{L}^{\rm SDM}&=& \Big|\Big(\partial^\mu-ig_2 \frac{\sigma^a}{2} {W^{a}}^\mu-ig_1 \frac{Y} {2} B^\mu\Big){\Phi}\Big|^2 - V(\Phi,H) \,; \nonumber \\ 
V(\Phi,H) &=&\mu_\Phi^2(\Phi^\dagger \Phi)+\lambda_\Phi (\Phi^\dagger \Phi)^2
 +\lambda_1 (H^\dagger H)(\Phi^\dagger \Phi) +\lambda_2 (H^\dagger \Phi)(\Phi^\dagger H) +\frac{\lambda_3}{2}[(H^\dagger \Phi)^2+h.c.]~ \,.\nonumber \\
\eea
Note that we assume $\mu_\Phi^2>0$ which keep vacuum extension value (vev) $\langle \Phi \rangle=0$ and ensure 
$\mathcal{Z}_2$ remains unbroken. $H$ represents SM Higgs isodoublet. After Electroweak Symmetry Breaking (EWSB), 
the SM Higgs acquires non zero vev, $\langle H \rangle=v ~(246 ~{\rm GeV})$ and generates physical states in the form of CP even state $\phi^0$ 
(mass $m_{\phi^0}$), CP-odd state $A^0$ (mass $m_{A^0}$) and charged scalar states $\phi^\pm$ (mass $m_{\phi^\pm}$) with adequate mass 
splitting between them. Using minimization conditions of the scalar potential $V(H,\Phi)$ along different field directions and the definitions 
of physical states, one can obtain the following relations between the mass eigenvalues and parameters involved in the scalar potential (for details, see \cite{Dolle:2009fn,Bhattacharya:2019fgs}):
\bea
&\mu_H^2 ~=~ \frac{m_h^2}{2} , ~~~~ \mu_\Phi^2~=~m_{\phi^0}^2-\lambda_L v^2 ,~~\lambda_1~=~2\lambda_L-\frac{2}{v^2}(m_{\phi^0}^2-m_{\phi^\pm}^2)~, \nonumber \\
& \lambda_2~=~\frac{1}{v^2}(m_{\phi^0}^2+m_{A^0}^2-2 m_{\phi^\pm}^2)~~{\rm{and}} ~ \lambda_3~=~\frac{1}{v^2}(m_{\phi^0}^2-m_{A^0}^2)~~;
\label{eq:mass-coupling}
\eea
where $\lambda_L=\frac{1}{2}(\lambda_1+\lambda_2+\lambda_3)$ and $m_h$ represents the mass  of SM-like neutral
scalar found at LHC with $m_h=125.09$ GeV. In our analysis we consider the mass hierarchy: $ m_{\phi^0} < m_{A^0}, m_{\phi^\pm}$ and hence 
the lightest neutral state $\phi^0$ with mass $m_{\phi^0}$ is a viable DM candidate\footnote{Note here that other neutral state, $A^0$ can also be 
a viable DM candidate with $ m_{A^0} < m_{\phi^0} , m_{\phi^\pm}$, by adjusting model parameters.}. 
The parameters relevant for DM phenomenology from SDM sector are given by:
\bea
\{\mdma,~\Delta m_1,~\lambda_L \}.
\eea
Here $\mdma=m_{\phi^0},~\Delta m_1=m_{\phi^\pm}-m_{\phi^0}\simeq  m_{A^0}-m_{\phi^0}$, assuming
$m_{\phi^\pm}\simeq m_{A^0}$, which does not alter subsequent DM phenomenology or collider analysis significantly.


Let us now turn to FDM sector. The minimal renormalizable Lagrangian for FDM having one vector-like doublet 
($\Psi$) and one right-handed singlet ($\chi_R$) reads \cite{Dutta:2020xwn}: 
\bea
\mathcal{L}^{\rm FDM} &=& \overline{\Psi}_{L(R)}~[i\gamma^{\mu}(\partial_{\mu} - i g_2 \frac{\sigma^a}{2}W_{\mu}^a - i g_1\frac{Y'}{2}B_{\mu})]~\Psi_{L(R)}
+\overline{\chi_{_R}}~(i\gamma^\mu \partial_{\mu})~\chi_{_R} \nonumber \\
&-& m_{\psi} \overline{\Psi} \Psi- \Big(\frac{1}{2} m_\chi \overline{\chi_{_R}} (\chi_{_R})^c + h.c \Big)  - \frac{Y}{\sqrt2} \Big(~\overline{\Psi_L}~\widetilde{H}{\chi_R}+~\overline{\Psi_R}~\widetilde{H}{\chi_R}^c+h.c \Big) \,;
\eea
where $\Psi_{L(R)}=P_{L(R)}\Psi;~P_{L/R}=\frac{1}{2}(1\mp\gamma_5)$.
After EWSB, the neutral component of the doublet $\Psi$ mixes with the neutral 
singlet $\chi_R$ via the Yukawa coupling $Y$.
The resulting mass matrix in the basis of $ ({\psi_R}^c, \psi_L, {\chi_{_R}}^c)^T$ is given by: 

\begin{equation}
\mathcal{M}=
\left(
\begin{array}{ccc}
0 & m_{\psi} &\frac{Y v}{2}\\
m_{\psi} &0 &\frac{Y v}{2}\\
\frac{Y v}{2} &\frac{Y v}{2} & m_\chi\\
\end{array}
\right)\, ,
\end{equation}
where $v=\langle H\rangle~ (246~{\rm GeV})$ is the vacuum expectation value (vev) of the SM Higgs. 
Upon diagonalisation by a unitary matrix, the mass eigenvalues of the physical states are: 
\bea
m_{\psi_{_1}} & =& m_\chi \cos^2\theta + m_{\psi} \sin^2\theta - \frac{Y v}{\sqrt{2}}\sin2\theta , \nonumber \\
m_{\psi_{_2}} & =& m_{\psi}, \\
m_{\psi_{_3}} & =&  m_{\psi} \cos^2\theta + m_\chi \sin^2\theta + \frac{Y v}{\sqrt{2}}\sin2\theta, \nonumber \\
m_{\psi^\pm} &=& m_{\psi} \nonumber ~;
\eea
where $\theta$ denotes singlet-doublet mixing angle; in terms of model parameters, it is given by:
\begin{equation}
\tan 2\theta= \frac{Y v}{\sqrt{2}(m_{\psi}-m_\chi)} \,.
\end{equation}

%
%

For $Y v/\sqrt{2} <  m_\chi \ll m_{\psi}$, $m_{\psi_{_3}}>m_{\psi_{_2}} (\equiv m_{\psi^\pm}) > m_{\psi_{_1}}$ makes $\psi_{_1}$ the DM. 
Note here all the neutral states, $\psi_{1,2,3}$ are Majorana fermions. The phenomenology of FDM mainly depends on the
following independent parameters:
\bea
\{\mdmb,~\Delta m_2,~\sin\theta \}\,;
\eea
where $\mdmb=m_{\psi_1}, \Delta m_2= m_{\psi^\pm}-m_{\psi_1} =  m_{\psi_2}-m_{\psi_1}$. In terms of these parameters,
\bea
Y &=& \frac{\sqrt{2}~ \Delta m_2~ \sin2\theta}{v} \,, \nonumber \\
m_{\psi} &=& m_{\psi_1} \sin^2\theta + m_{\psi_3} \cos^2\theta \,, \\
m_{\chi} &=& m_{\psi_1} \cos^2\theta + m_{\psi_3} \sin^2\theta .\nonumber 
\eea
The dependence of $Y$ on $ \Delta m_2$ as well as $\theta$ plays an important role in the consequent 
FDM phenomenology. We note further that the two dark sectors have no renormalizable interaction term\footnote{The interaction terms appear
at the dimension-five level, such as $(\overline{\Psi_L} \Psi_R)(\Phi^\dagger \Phi),~(\overline{\chi_{_R}} (\chi_{_R})^c)(\Phi^\dagger \Phi)$ with hermitian conjugates 
and at higher dimensions.}, so that DM-DM conversions occur mainly via Higgs portal coupling.                                      
\subsection{Constraints on the model parameters}
\label{sec:constraints}
The parameters of the model are subject to the following constraints, which we abide by:

\begin{itemize}
\item A stable vacuum at the electroweak scale \cite{Kannike:2012pe,Chakrabortty:2013mha} is ensured by,
   \begin{align}
     \lambda_H,\,\lambda_\Phi > 0\, ;~
     \lambda_1 + 2\sqrt{\lambda_H\lambda_\Phi} > 0\, ;~
     \lambda_1 + \lambda_2 - |\lambda_3| + 2\sqrt{\lambda_H\lambda_\Phi} > 0\, .
   \end{align}
\item Perturbativity at the electroweak scale requires, 
\begin{align}
 |\lambda_H| < {4 \pi}~;~~|\lambda_\Phi| < {4 \pi}~;~
 |\lambda_{1,2,3} | < {4 \pi}~~{\rm and}~~ |Y| < \sqrt{4\pi}~.
\end{align}
\item Limits on the masses from non-observation of the BSM particles at collider search, the most significant of those here are $m_{\chi^{\pm}} > 102.7$ GeV~\cite{LEP} and 
$m_{\phi^{\pm}} > 70$ GeV~\cite{Pierce:2007ut}.
\item The requirement that the two DM components $\phi^0$ and $\psi_1$ saturate the relic-density according to the Planck data~\cite{Planck:2018vyg} at 
$2\sigma$ level, 
\bea
\Omega_{\rm DM} h^2= \Omega_{\rm DM_1} h^2 + \Omega_{\rm DM_2} h^2=\Omega_{\phi^0} h^2+\Omega_{\psi_1} h^2 \equiv 0.12 \pm 0.001.
\label{eq:relic}
\eea
\item Limit on spin independent DM-nucleon cross-section from non-observation of WIMP in direct search experiments, the latest being XENON-1T,
at $2\sigma$ level \cite{XENON:2018voc,PandaX-4T:2021bab} provides
\bea
\sigma^{SI}_{\rm{min}}\lesssim 10^{-47}~{\rm{cm}^2}.
\eea
\end{itemize} 

We focus mainly on the DM constraints and demonstrate the allowed parameter space of the model, 
which in turn provides some characteristic benchmark points that is taken up for subsequent collider analysis. 

\bigskip
\bigskip
\noindent
{\bf Dark matter constraints }
\medskip
\noindent
The most important constraint comes from the DM relic density, where individual densities 
($\Omega_{\rm DM_{i}} (i=1,2)$) add to the total observed relic as dictated by PLANCK data \cite{Planck:2018vyg}. 
We may further recall that relic (over/under) abundance of a particular WIMP is inversely proportional to the
thermal average of its annihilation cross-section to visible sector particles as well as to other DMs, 
\bea
\Omega_{\rm DM_{i}} \sim \frac{1}{\langle \sigma v \rangle_i}\,.
\eea

Now, one may assume 
that the DM number density of any component around the earth
is proportional to the share of that component in the total
relic density. In such a situation, the upper limit on the scattering cross-section of the 
$i$th DM component ($\sigma_i^{\rm{eff}}$) with the detector nucleon for direct search 
is related to $\sigma_i$, the cross-section if the corresponding 
component were the sole WIMP DM constituent, by the fraction of relic density 
with which it is present in a two component set up, following \cite{Bhattacharya:2016ysw},

\begin{equation}
\sigma_i^{\rm{eff}} = {\frac{\Omega_{\rm{DM_i}}}{\Omega_{\rm{DM}}}}\sigma_i\,.
\label{DD-eff}
\end{equation}      

Thus the direct search limit on a DM component
is different from the case where it is the only WIMP candidate.
Since any WIMP is subject to both the limits from relic density and direct search, 
it is useful to note two more points. First, the tension between the limit on the 
cross-section in direct searches and the relic density limit is relaxed in scenarios 
where the dark sector permits co-annihilation, as occurs for both the SDM and FDM 
components of our illustrative model. Secondly, the right-chiral fermion $\chi_R$ in FDM 
has no Z-coupling (having $Q = Y = 0$), its mixing
with $\psi_0$ suppresses the Z-induced contribution to the elastic
scattering and relieves the pressure from direct search limits. Annihilation, 
co-annihilation and conversion processes for both DM components of the model are mentioned 
in Appendix~\ref{app2}. 

In order to find the consistent DM parameter space, we implement the model in 
{\tt Feynrules} \cite{Degrande:2011ua} and in {\tt MicrOmegas} \cite{Belanger:2018ccd} 
to perform a numerical scan on the relevant parameters in the following range:
 
\begin{align}
30 ~\leq m_{\phi^0} \leq 600~{\rm GeV},~1 ~\leq \Delta m_1 \leq 500~{\rm GeV},~0.01 ~\leq \lambda_L \leq 0.1,
\nonumber\\  
30 \leq~ m_{\psi_1} \leq 600~ \rm GeV,~1 ~\leq \Delta m_2 \leq 500~{\rm GeV},~0.01 ~\leq \sin\theta \leq 0.3 ~.
\label{eq:par}
\end{align}

\begin{figure}[!hptb]
	\centering
	\subfloat[]{\includegraphics[width=7.5cm,height=6cm]{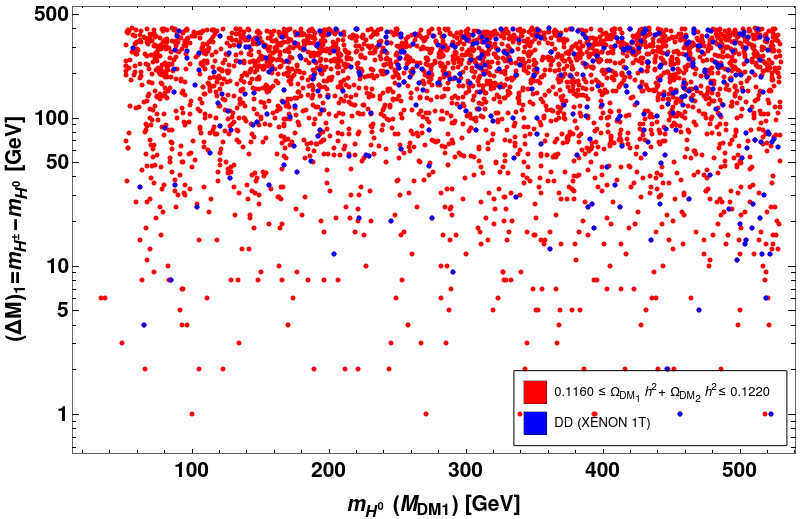}}
	\subfloat[]{\includegraphics[width=7.5cm,height=6cm]{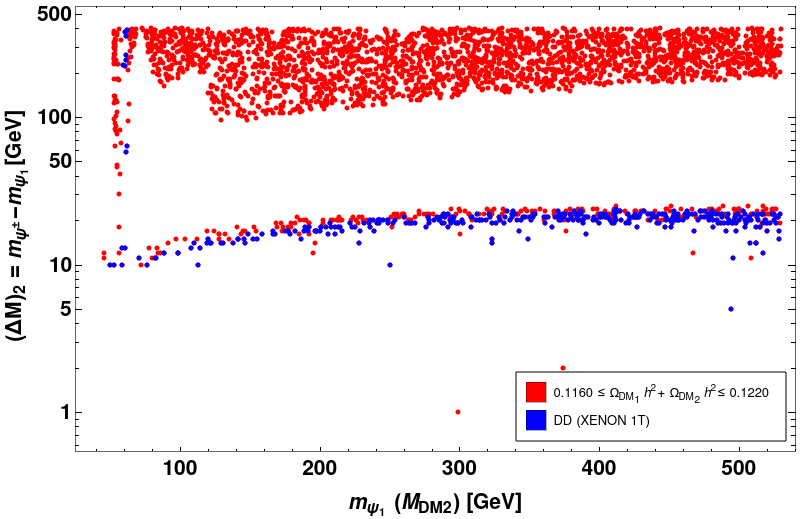}}
	\caption{Relic and direct search outcome for (a) scalar DM and (b) fermion DM. The red points satisfy relic density requirement and 
	the blue points satisfy direct detection bound (XENON 1T) as well as relic density constraints.}
	\label{fig:fig4}
\end{figure}

\noindent
We present in Fig.~\ref{fig:fig4}, the allowed region of the model in the planes of (a) $\mdma-\Delta m_1$ and (b) $\mdmb-\Delta m_2$. 
Red points satisfy relic density and the blue points satisfy additionally the direct search bound from XENON1T. 
It is seen that for the scalar DM, relic (under)abundance and direct detection constraints are both satisfied for almost the entire range of 
$\mdma, ~\Delta m_1$ scanned here. This is achieved mainly due to two reasons: one, the gauge mediated depletion processes (see Appendix~\ref{app2} 
for details) ensure significant annihilation cross-section, while suitable choice of the parameter $\lambda_L$, which governs the interaction between 
the DM and SM Higgs, ensures the consistency with direct detection bound. For FDM case (see Fig.~\ref{fig:fig4}(b)), both relic density and direct 
search constraints are satisfied consistently for $\Delta m_2 \lesssim 20$ GeV. This is due to co-annihilation processes contributing 
significantly to relic density with smaller $\Delta m_2$, which do not contribute to DM elastic scattering for direct search. We also note that 
for $m_{\psi_1} \simeq \frac{m_h}{2}$, in the Higgs resonance region, one can only simultaneously satisfy both relic (under) abundance and 
direct detection constraints even for larger $\Delta m_2$, where the co annihilation contribution is compensated by  resonant enhancement. 
We further note that there are two white regions in Fig.~\ref{fig:fig4}(b), where the relic density constraints can not be satisfied. 
For $\Delta m_2 \simeq 20-100$ GeV, the co-annihilation contribution turns drastically low, subject to the range of $\sin\theta$ used for the scan, 
so that this region provides relic over abundance. For smaller mass splitting, $\Delta m_2 \lesssim 10$ GeV, the co-annihilation contribution
turns so large that it yields under abundance. On the other hand, for $\Delta m_2 \gtrsim 100$ GeV, 
the Yukawa coupling ($Y\sim \Delta m_2$) turns large enough to provide required under abundance for FDM to make up for SDM contribution. 
Also note that the allowed parameter space in this two component model most often has a dominant FDM contribution over 
the scalar one, simply because satisfying relic under abundance together with direct search constraint for FDM relies heavily on 
co-annihilation contribution given a minimal $\sin\theta$, limiting the minimum relic density that FDM can achieve.

\begin{figure}[!hptb]
	\centering
	\includegraphics[width=7.5cm,height=6cm]{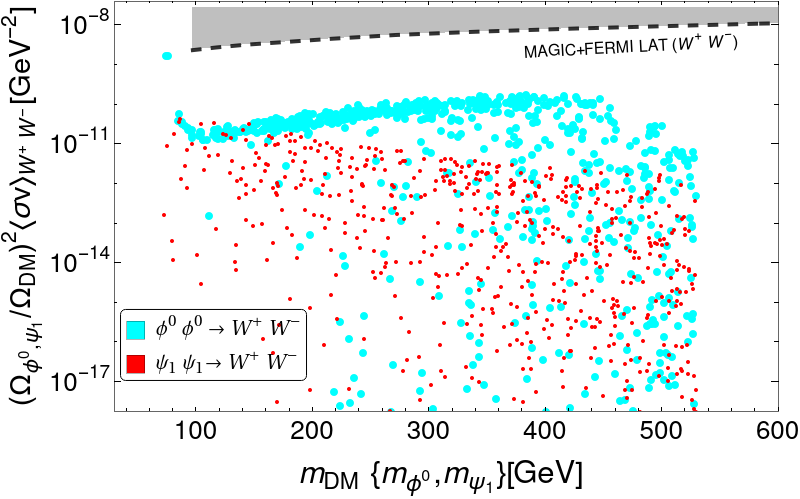}
	\caption{Regions allowed by indirect detection constraints from $W^+W^-$ mode for scalar and fermionic DM. Their individual contributions have been scaled by their respective number density.}
	\label{fig:id}
\end{figure}

In Fig.~\ref{fig:id}, we show the regions allowed by indirect detection constraints. Due to gauge interaction, the major annihilation channels for 
both SDM and FDM occurs into $W^+W^-$ final state. Therefore, the most stringent upper limit comes from this particular channel in MAGIC~\cite{MAGIC:2016xys} plus 
Fermi-LAT data~\cite{Fermi-LAT:2015att}. Similar to direct search, while calculating the annihilation cross-section of each DM into $W^+W^-$ final state, we have 
scaled their individual contribution consistently with the corresponding DM densities to the total relic density. One can see that the entire range of DM parameters 
scanned here satisfy the upper limit from indirect searches. 

\section{Selection of benchmark points}
\label{sec:bp}
\begin{figure}[!hptb]
	\centering
	\subfloat[]{\includegraphics[width=7.5cm,height=6cm]{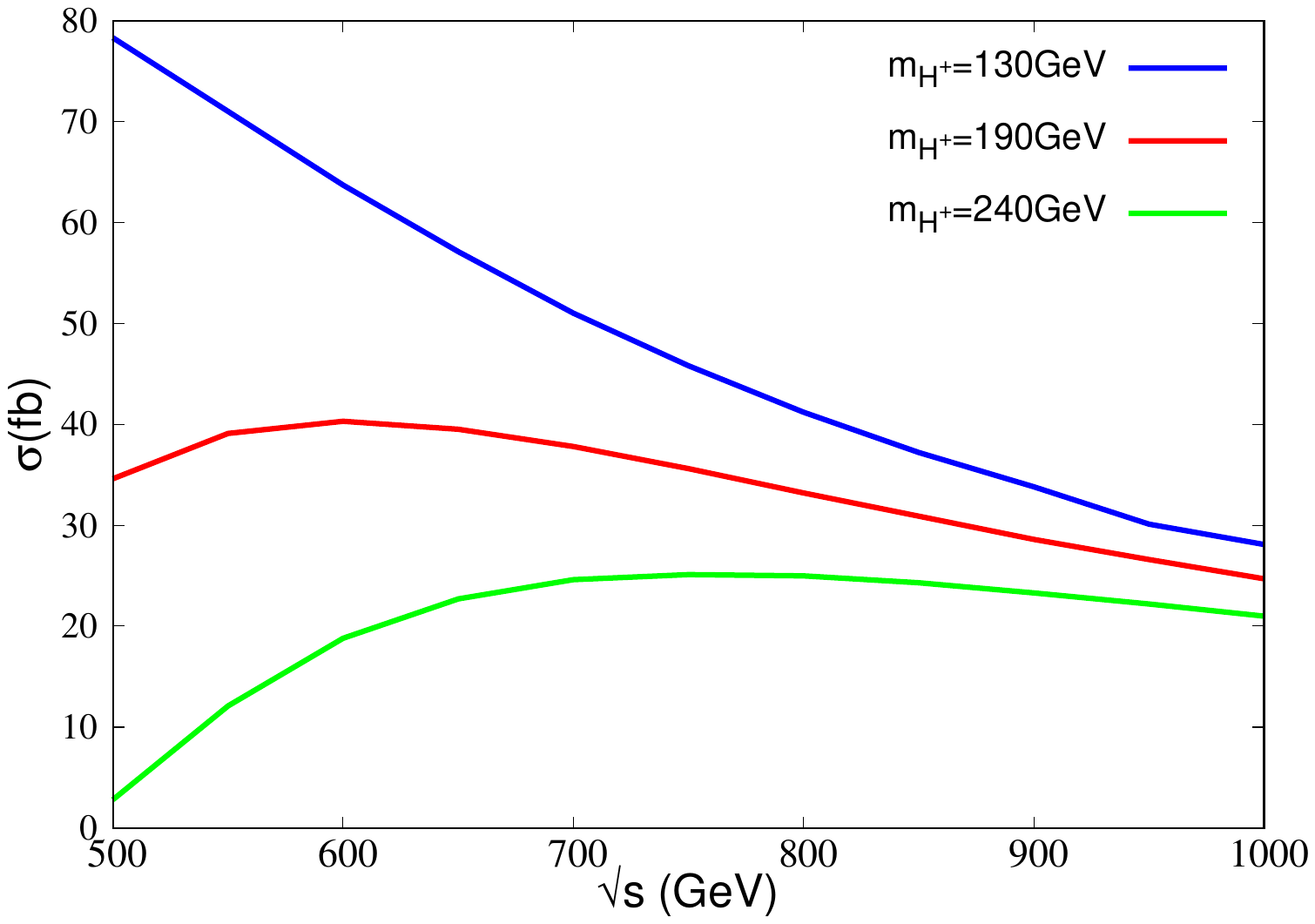}}
	\subfloat[]{\includegraphics[width=7.5cm,height=6cm]{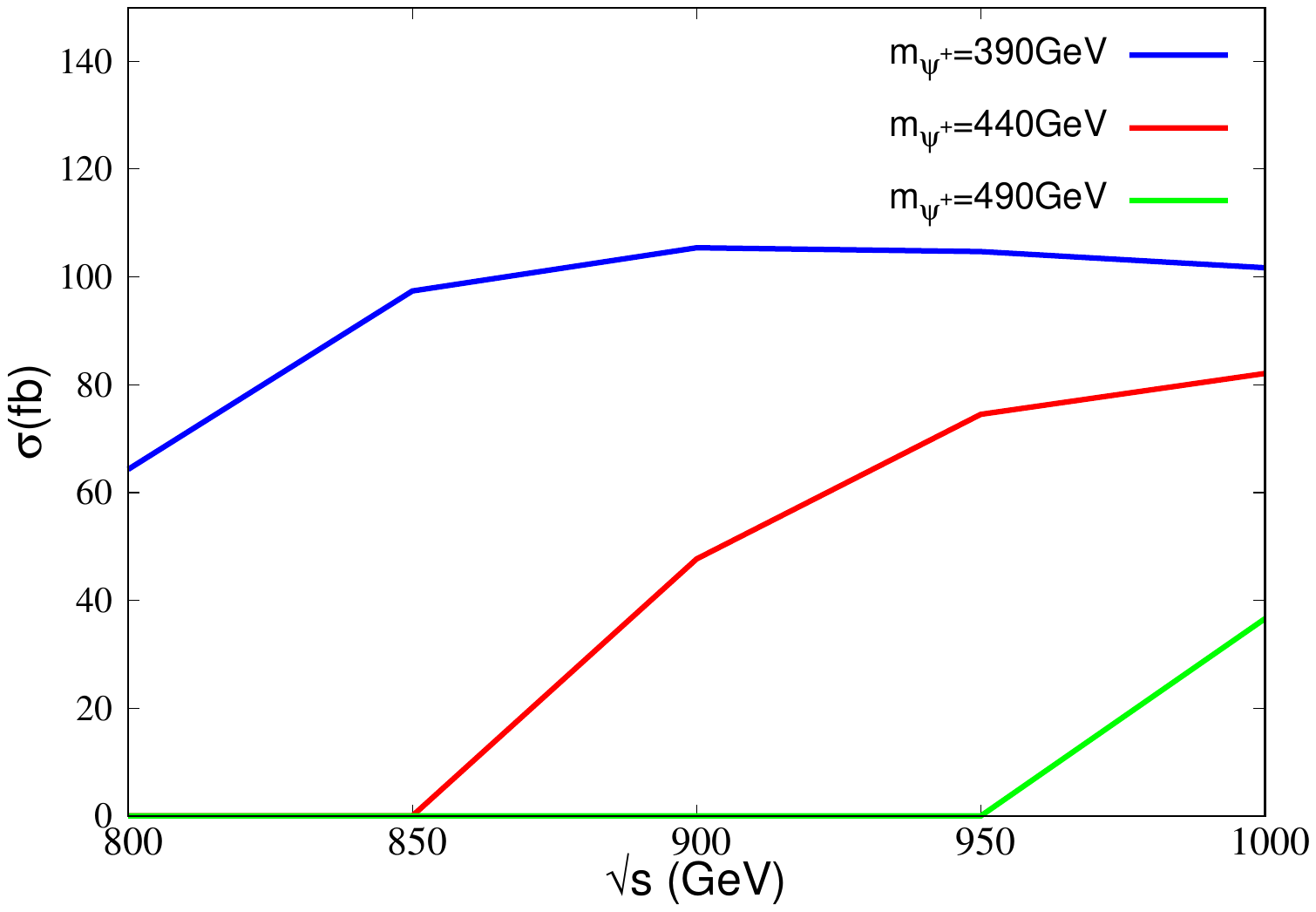}}
	\caption{Production cross-section of HDSP pair for (a) scalar $\phi^+\phi^-$ and (b) fermion $\psi^+\psi^-$ as a function of $\sqrt{s}$ for 
	different choices of HDSP masses at ILC. We use unpolarized initial beams. All the relevant couplings for the production of HDSP's are given in Eqs.~\ref{eq:vertex1}-\ref{eq:vertex4}.}
	\label{crosssec}
\end{figure}

Having discussed all the constraints on the model and carefully examining various possibilities for distinguishing two 
peaks in kinematic distributions for two-component DM, we are in a position to choose a few benchmark points, 
suited for the likely centre-of-mass energies of $e^+e^-$ colliders planned for the near future, including the International Linear 
Collider (ILC). Relevant kinematic regions (allusions to some have already been discussed) where the peaks can be distinguished are as follows:
\begin{itemize}
\item {\bf Region I}:  ~$\mdma > \mdmb$  and $\Delta m_1 > \Delta m_2$~,
\item {\bf Region II}:  ~$\mdma < \mdmb$  and $\Delta m_1 < \Delta m_2$~,
\item {\bf Region III}:  ~$\mdma < \mdmb$  and $\Delta m_1 > \Delta m_2$~,
 \item {\bf Region IV}:  ~ $\mdma > \mdmb$  and $\Delta m_1 < \Delta m_2$~.
\end{itemize}
Recall that subscript 1 above refers to SDM sector and 2 to FDM sector. 
However, distinguishability of the peaks at collider crucially depend on the relative cross-sections ($\sigma$) 
of the two dark sector particles, which makes some of these kinematic regions more favourable than the others. 
We therefore make a quantitative comparison between the scalar and fermionic HDSP pair production cross-sections
 in Fig.~\ref{crosssec}, where (a) $\sigma_{\phi^{+}\phi^{-}}$ and (b) $\sigma_{\psi^{+}\psi^{-}}$ are plotted 
as a function of $\sqrt{s}$ for different choices of HDSP masses using unpolarised initial beams \footnote{
We will discuss the effect of polarization on the signal cross-section shortly.}. In Fig.~\ref{crosssec}, the range of $\sqrt{s}$ is varied between 500 GeV to 1 TeV as per the technical design report 
of ILC~\cite{Adolphsen:2013kya,Bambade:2019fyw,Zarnecki:2020ics}. 

We recall here that $\psi^+$ and $\psi^-$ have vector-like couplings to both $\gamma$ and $Z$ as follows,

\beq
\psi^+\psi^-Z:- ~i e_0 \cot(2\theta_W) \gamma^\mu~,
\label{eq:vertex1}
\eeq

\beq
\psi^+\psi^-\gamma:- ~ie_0\gamma^\mu \,.
\label{eq:vertex2}
\eeq

\noindent
The charged scalar couplings with $\gamma$ and $Z$ are momentum-dependent and are given by,

\beq
\phi^+\phi^-Z:- ~i e_0 \cot(2\theta_W) (p_1^\mu-p_2^\mu)~,
\label{eq:vertex3}
\eeq

\beq
\phi^+\phi^-\gamma:- ~ie_0 (p_1^\mu-p_2^\mu) \,,
\label{eq:vertex4}
\eeq

\noindent
where $p_1^\mu$ and $p_2^\mu$ are the four-momenta of the outgoing charged scalars.



Scalar HDSP has smaller production cross-section compared to the fermionic ones for the same mass, 
because of the absence of polarization sum in the former, as is also clear from Fig.~\ref{crosssec}. We would then ideally use for illustration, a lighter $\phi^{\pm}$ and heavier $\psi^{\pm}$ 
in order for them to have comparable production rates. From this viewpoint, Region I proves to be underwhelming while Region II appears favourable. 
However, with large mass splitting in the FDM sector ($\Delta m_1 < \Delta m_2$ as in Region II), we need to restrict to the 
Higgs resonance region with $m_{\psi_1} \approx \frac{m_h}{2} \sim 60$ GeV (see Fig. \ref{fig:fig4}(b)) to address DM constraints. 
Therefore, in Region II, one requires even lighter scalar DM with relatively large $\Delta m_1$, so that cross-sections 
for SDM is comparable to that of FDM. This leads to a broad peak in $\slashed{E}$ from the scalar sector which eventually merges 
partially with the peak coming from the fermion sector.  

We will actually have large accessible regions 
consistent with $\mdma>\mdmb$ in Region IV.  As small $\Delta m$ would imply a narrow $\slashed{E}$ distribution, $\Delta m_1 < \Delta m_2$ 
provides a narrow distribution in the scalar dark sector (with smaller cross-section), and a wider one for fermions (with larger cross-section), 
a much needed feature for the separation of two peaks. In Region III with $\mdma< \mdmb$, and $\Delta m_1 > \Delta m_2$, 
the FDM is almost degenerate with HDSP to address DM constraints, while the SDM has a larger mass gap from the 
corresponding HDSP, which is possible only with $m_{\phi^0} \sim  \frac{m_h}{2}$ to account for DM constraints. 
In this case, we will then end up with a narrow distribution in the FDM sector, and a wider one for the SDM sector. 
The relative cross-section of the scalar and fermion DM production makes this scenario hard to distinguish. Therefore, Region IV turns 
most favourable in the context of the model at hand for probing two-component DM signature at ILC with reasonable $\sqrt{s}$. The discussion 
above also elucidates how DM constraints together with production cross-section can favor a specific region over the others.



In Table~\ref{tab:dm}, we identify four benchmark points from Region IV, out of which BP1 and BP2 can be probed with $\sqrt{s} = $ 1 TeV, while BP3 and BP4 can be 
probed with $\sqrt{s} = 500$ GeV. The scalar sector is kept fixed at $\mdma=100$ GeV, $\Delta m_1=10$ GeV and $\lambda_L=0.01$. As argued before, 
FDM is chosen around the Higgs resonance for cosmological constraints, so that the essential difference amongst the benchmark points lies in the values of  
$\Delta m_2$ and $\sin\theta$ in FDM sector. We also present the individual DM relic densities and spin-independent direct detection cross-section for all the benchmarks together with 
invisible branching ratio of Higgs for each cases to show that they satisfy all the limits. The relic density for SDM is minuscule $\Omega_{\phi^0}h^2 \approx 0.0022$, 
while the effective spin independent direct search cross-section (following Eq.~\ref{DD-eff}) $\sigma_{\phi^0}^{\rm eff} \approx 10^{-46}$ cm$^2$ is well within the limit. 
FDM shares the major relic density with $\Omega_{\psi_1} h^2\approx 0.11$ with $\sigma_{\psi_1}^{\rm eff} \approx 10^{-47}$ cm$^2$. The relative contribution of SDM to observed relic can be increased via increasing SDM mass. However, that will come at a cost of reduced production cross-section of SDM and consequently, poor distinguishability of the two peaks.

%
%
%


\begin{table}

{\scriptsize{ \begin{tabular}{|c|c|c|c|c|c|c|c|}
\hline
\makecell{BPs } & \makecell{SDM sector \\ $\{m_{\phi^0},~\Delta m_1,~\lambda_L\}$ } & \makecell{FDM sector \\ $\{m_{\psi_1},~\Delta m_2,\sin\theta\}$ } & $\Omega_{\phi^0} h^2$    &  $\Omega_{\psi_1} h^2$  & $\sigma_{\phi^0}^{\rm eff}$ (cm$^{2}$) &  $\sigma_{\psi_1}^{\rm eff}$ (cm$^{2}$) & \rm{BR}($H_{\rm inv}$)$\%$\\  \hline \hline 

BP1& $100,~ 10, ~0.01$ & $60.5,~ 370, ~0.022$ & $0.00221$ & $0.1195$ & $3.45 \times 10^{-46}$ & $2.03 \times 10^{-47}$ & $0.25$ \\ \hline
BP2& $100,~ 10, ~0.01$ & $58.91,~ 285,~ 0.032$ & $0.00221$ & $0.10962$ & $3.45 \times 10^{-46}$ & $5.38 \times 10^{-47}$ & $1.60$ \\ \hline    
BP3& $100,~ 10, ~0.01$ & $58.87,~ 176,~ 0.04$ & $0.00221$ & $0.11941$ & $3.45 \times 10^{-46}$ & $5.00 \times 10^{-47}$ & $1.50$ \\ \hline
BP4& $100,~ 10, ~0.01$ & $58.48,~ 190,~ 0.042$ & $0.00221$ & $0.1114$ & $3.45 \times 10^{-46}$ & $7.01 \times 10^{-47}$ & $2.4$ \\ \hline
\hline
\end{tabular}}}

\caption{Benchmark points of the model; contribution to relic density, spin-independent direct detection cross-section as well as that of invisible Higgs decay 
branching ratios of the DM components $\phi^0$ and $\psi_1$ are mentioned.}
    \label{tab:dm}
\end{table}




\section{Signal and background}
\label{sec:signal}
\begin{figure}[!hptb]
	\centering
	\subfloat[]{\includegraphics[width=7.0cm,height=6.5cm]{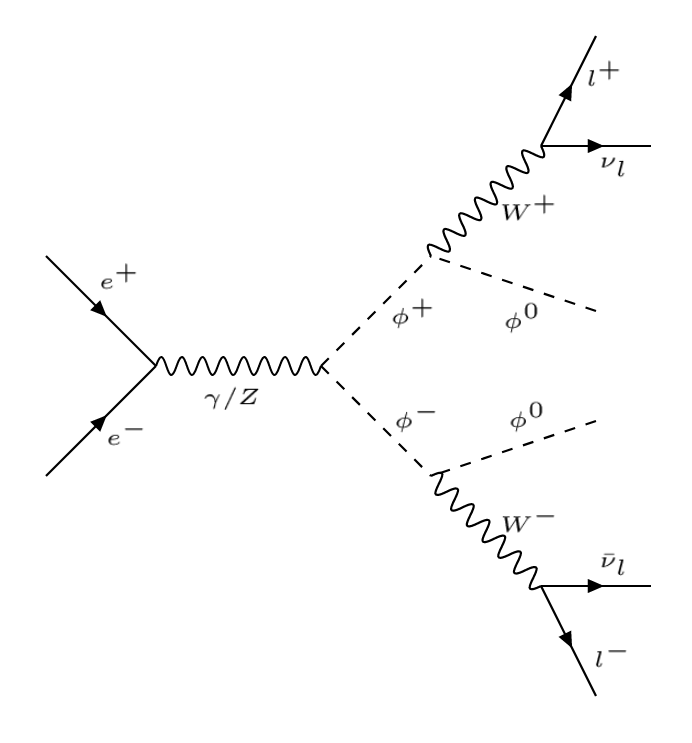}} 
        \subfloat[]{\includegraphics[width=7.0cm,height=6.5cm]{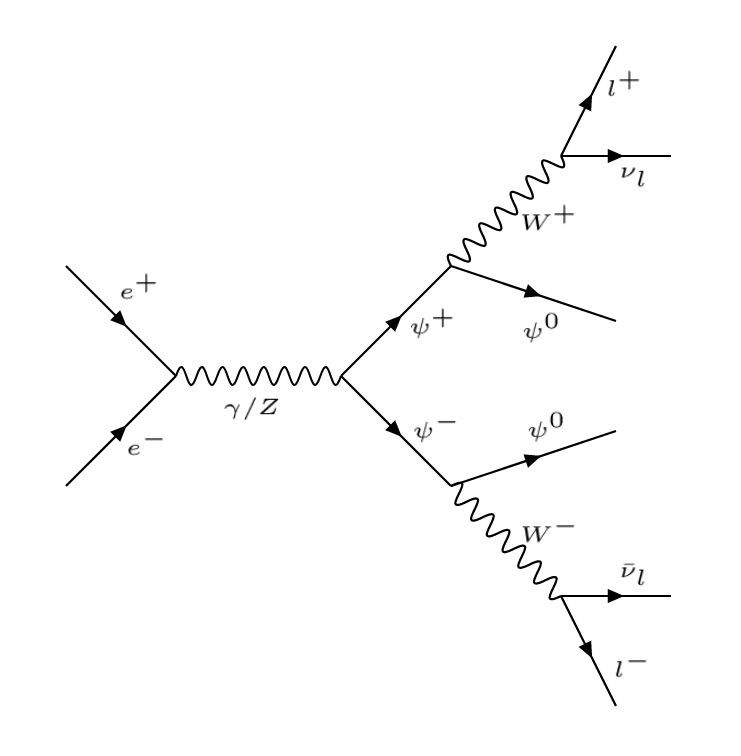}} \\
        \subfloat[]{\includegraphics[width=7.5cm,height=6.5cm]{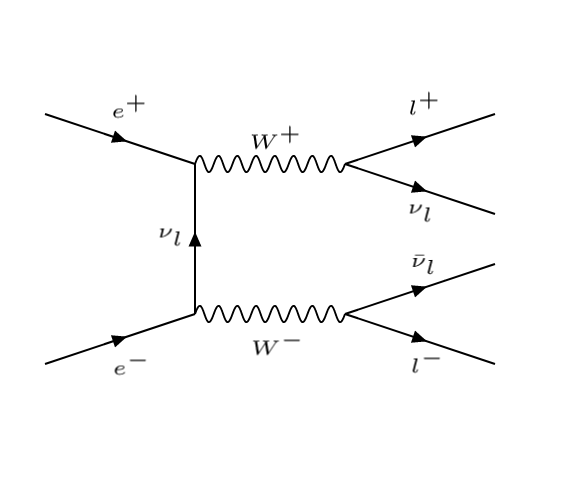}}
        \subfloat[]{\includegraphics[width=7.5cm,height=6.5cm]{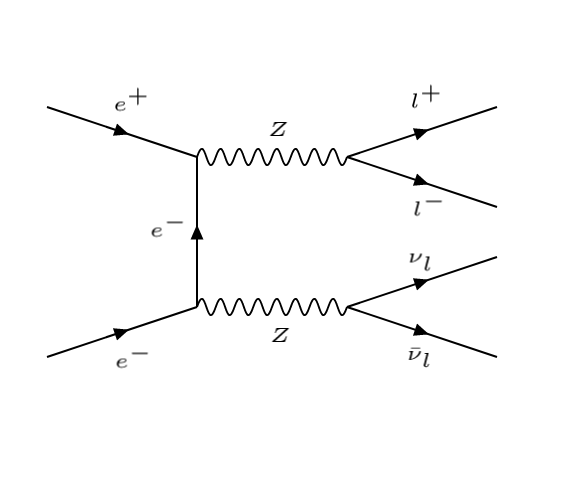}}
	\caption{Dominant process contributing to $\ell^+ \ell^- + \slashed{E}$ at ILC from the two component dark sector signal (top) and 
	$WW$ and $ZZ$ background (bottom).}
	\label{diagrams}
\end{figure}

The final state of our interest is $\ell^+ \ell^- + \slashed{E}$. We demand two isolated opposite-sign dileptons (OSD) with both $e,\mu$ and $p_T > 10$ GeV for each of them.
The isolation criterion for the leptons demands ${\frac{|p^{\ell}_T|}{\Sigma^i |p^i_T|}} < 0.12$, where the summation in 
the denominator is over all the particles within $\Delta R = 0.5$ around each lepton.
Events for the signal and backgrounds have been generated using Madgraph@MCNLO~\cite{Alwall:2014hca}. The detector simulation has been taken care 
of by Delphes-3.4.1~\cite{deFavereau:2013fsa} with ILD specifications. 

The following HDSP production and subsequent decay chains involving NP as per the model contribute dominantly to this final state. 

\bigskip

$e^{+} e^{-} \rightarrow \phi^{+}\phi^{-}, \phi^{+} \rightarrow W^{+(*)} \phi^0, W^{+(*)} \rightarrow \ell^+ \nu_{\ell}, \phi^{-} \rightarrow W^{-(*)} \phi^0, W^{-(*)} \rightarrow \ell^- \bar\nu_{\ell}$ \\

$e^{+} e^{-} \rightarrow \psi^{+}\psi^{-}, \psi^{+} \rightarrow W^{+(*)} \psi_1, W^{+(*)} \rightarrow \ell^+ \nu_{\ell}, \psi^{-} \rightarrow W^{-(*)} \psi_1, W^{-(*)} \rightarrow \ell^- \bar\nu_{\ell}$ \\

The SM background processes that will lead to same final state are as follows: \\

$e^+ e^- \rightarrow W^+ W^-, W^+ \rightarrow \ell^+ \ell_{\nu}, W^- \rightarrow \ell^- \bar{\nu_\ell}$ \\

$e^+ e^- \rightarrow W^+ W^- Z, W^+ \rightarrow \ell^+ {\nu_\ell}, W^- \rightarrow \ell^- \bar{\nu_\ell}, Z \rightarrow \nu_{\ell} \bar\nu_{\ell}$ \\

$e^+ e^- \rightarrow Z Z, Z \rightarrow \ell^+ \ell^-, Z \rightarrow \nu_{\ell} \bar\nu_{\ell}$ \\

In Fig.~\ref{diagrams}, we show all the major Feynman diagrams contributing to the final state 
$\ell^+ \ell^- + \slashed{E}$ from the DM signal and the dominant SM background processes.
Out of the SM contributions, $e^+ e^- \rightarrow W^+ W^-$ has the largest production cross-section. 
The most efficient way to reduce this background is the use of beam polarization, which we discuss next.
 $e^+e^- \rightarrow ZZ$ constitutes another copious background, while $e^+ e^- \rightarrow W^+W^-Z$ has much lesser 
 production cross-section. It should be noted that a large background contribution comes from non-resonant 
 production of $e^+e^-\rightarrow \nu\bar\nu Z$, where the $\nu\bar\nu$ pair is not coming from $Z$. This process primarily 
 involves $t$-channel $W$ mediated diagrams. The cross-section at $\sqrt{s}=1$ TeV is 50\% of $WW$ background and at 
 $\sqrt{s}=500$ GeV is of similar magnitude to that of $ZZ$ production. However, a strong cut on the invariant 
 mass of the lepton pair, (demanding it to be outside $Z$ mass window) will considerably reduce this background.

\subsection{Effect of beam polarization}

\begin{figure}[!hptb]
	\centering
	\subfloat[]{\includegraphics[width=0.5\linewidth]{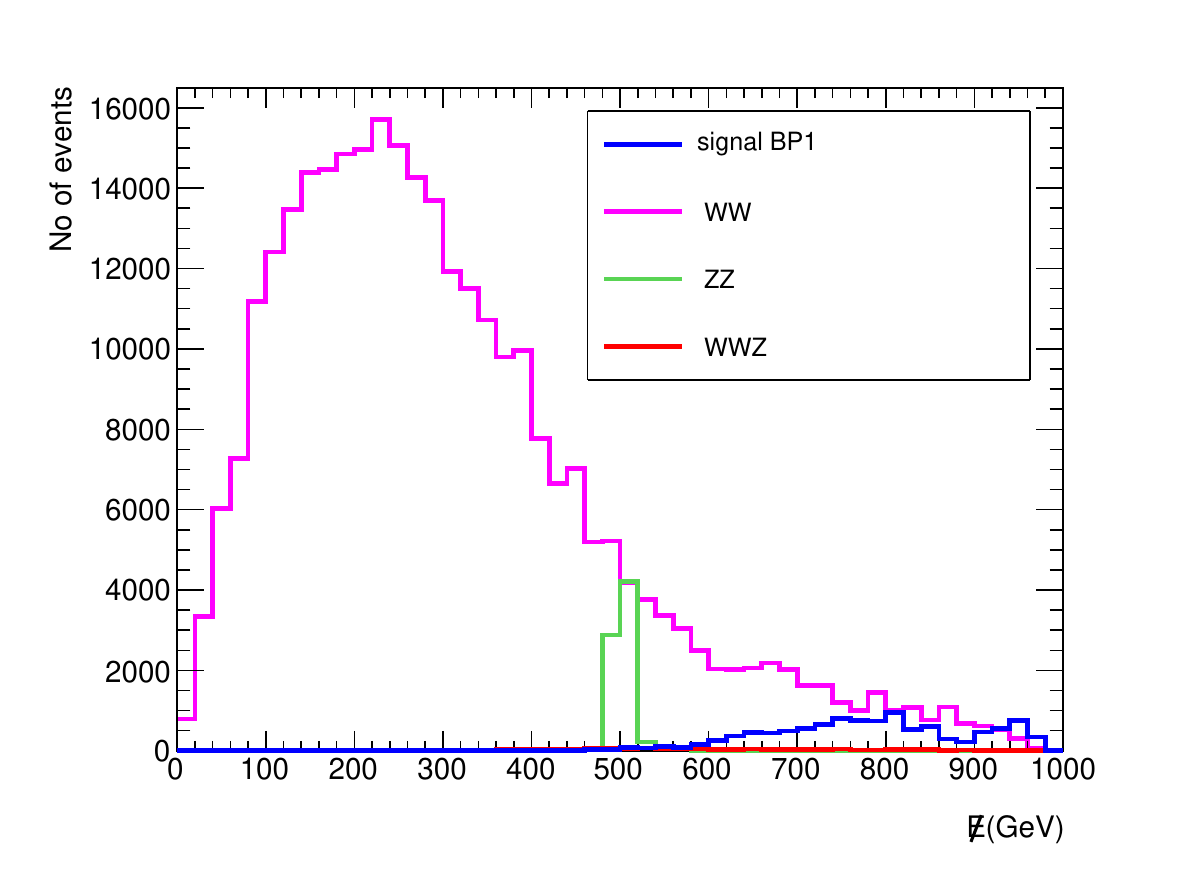}\label{bp1pol1}} 
	\subfloat[]{\includegraphics[width=0.5\linewidth]{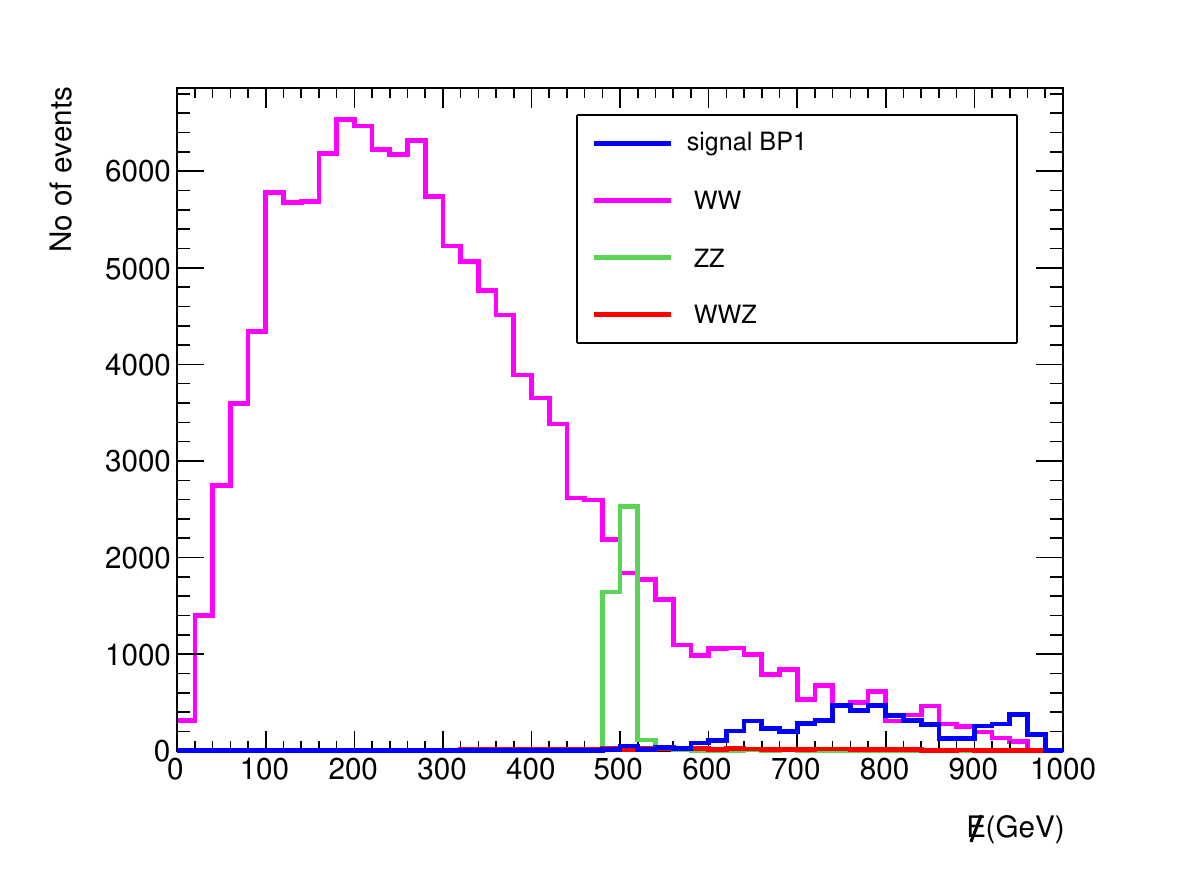}\label{bp1plo2}} \\
        \subfloat[]{\includegraphics[width=0.5\linewidth]{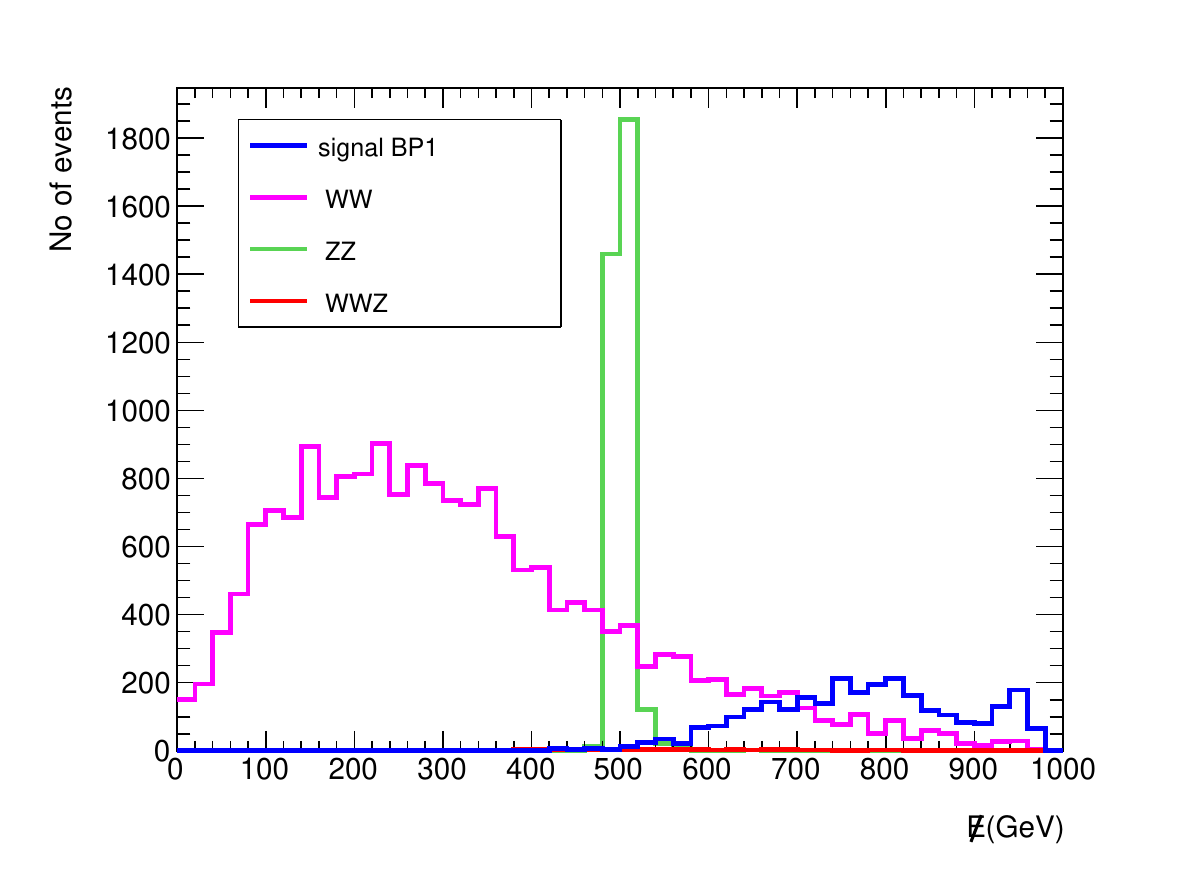}\label{bp1pol3}} 
	\caption{Unnormalized $\ell^+ \ell^-$ event distribution for $\slashed{E}$ at ILC with $\sqrt{s}$ = 1 TeV and $\mathcal{L}=$ 1000 ${\rm fb}^{-1}$ for BP1 along with 
	$WW$ and $ZZ$ backgrounds for various degrees of beam polarisation, (a) P1, (b) P2 and (c) P3 (see caption of Table ~\ref{table:BP} and text).}
	\label{dist1}
\end{figure}

$e^+ e^- \rightarrow W^+ W^-$ receives dominant contribution via $t$-channel neutrino exchange (see Fig.~\ref{diagrams}), 
involving left-handed electrons and right-handed positrons. This process can thus be 
suppressed with maximally right polarized $e^-$ beam and maximally left polarised $e^+$ beam. 
At ILC, the electron polarization is proposed to be maximum 80\% whereas, the positron 
polarization is expected to be 30\% with a possible upgradation to 60\%~\cite{Behnke:2013lya}. In Fig.~\ref{dist1}, 
we compare $\slashed{E}$ distributions for BP1 (at $\sqrt s=$ 1TeV) with dominant SM backgrounds for 
various degrees of beam polarization, defined by P1 $\equiv$ $\{P_{e^{-}}: -0.8, P_{e^{+}}: +0.3\}$, P2 $\equiv \{P_{e^{-}}: 0, P_{e^{+}}:0\}$ 
and P3 $\equiv \{P_{e^{-}}: +0.8, P_{e^{+}}: -0.3\}$, where $+(-)$ denotes right(left) polarisation.
We clearly see the effect of polarization in reducing the dominant $WW$ background. One can see that the initial 
beam polarization affects the signal cross-section too. However, the effect is much milder compared to the background. 
A similar observation can be made for BP3 at $\sqrt s=$ 500 GeV. We note further that the sub-dominant $e^+e^- \rightarrow ZZ$, 
gets major contribution from lepton mediated $t$-channel diagrams (see Fig.~\ref{diagrams}), thus receives little suppression from the 
beam polarization of the above type. The non-resonant $\nu\bar\nu Z$ background, on the other hand, receives substantial suppression from the particular beam polarization configuration that we consider. The cross-sections for DM signal at the benchmark points are mentioned in Table \ref{table:BP}, 
while those from SM background is mentioned in Table~\ref{bkg}. It is evident that $S/B$, where $S$ refers to signal and $B$ refers to 
background events, will be the largest for polarization configuration P3. We further see that $WWZ$ background plays negligible role 
in the analysis, due to its dismal cross-section. Therefore, we omit the distributions for $WWZ$ background in all the forthcoming discussions.

\begin{table}
{\scriptsize{\begin{tabular}{| c |c | c c c c c c c c c | } 
\hline
\hline
\multicolumn{2}{|c|}{Benchmarks} &
\multicolumn{9}{|c|}{Collider cross-section (fb)} \\
\cline{3-11}
\multicolumn{2}{|c|}{}  &
\multicolumn{3}{|c|}{$\sigma_{\rm total}$(OSD)} &
\multicolumn{3}{|c|}{$\sigma_{\phi^+\phi^-}$(OSD)} &
\multicolumn{3}{|c|}{$\sigma_{\psi^+\psi^-}$(OSD)}\\
\cline{1-11}
$\sqrt{s}$ & Points & 
 \multicolumn{1}{|c|}{P1} & \multicolumn{1}{|c|}{P2} & \multicolumn{1}{|c|}{P3} & \multicolumn{1}{|c|}{P1} & \multicolumn{1}{|c|}{P2} & \multicolumn{1}{|c|}{P3} & \multicolumn{1}{|c|}{P1} & \multicolumn{1}{|c|}{P2} & \multicolumn{1}{|c|}{P3} \\ 
\hline
\multirow{2}*{{1000}} 
& BP1 & \multicolumn{1}{|c|}{232(10.8)} & \multicolumn{1}{|c|}{115(5.5)} & \multicolumn{1}{|c|}{58.5(2.75)} & \multicolumn{1}{|c|}{57.4(2.9)} & \multicolumn{1}{|c|}{28.9(1.5)} &  \multicolumn{1}{|c|}{14.5(0.75)} & \multicolumn{1}{|c|}{173(8.4)} & \multicolumn{1}{|c|}{83.0(4.0)} & \multicolumn{1}{|c|}{44.0(2.0)}  \\
\cline{3-11}
& BP2 & \multicolumn{1}{|c|}{276(13.4)} & \multicolumn{1}{|c|}{141(6.6)} & \multicolumn{1}{|c|}{70.0(3.3)} & \multicolumn{1}{|c|}{57.4(2.9)} & \multicolumn{1}{|c|}{28.9(1.5)} & \multicolumn{1}{|c|}{14.5(0.75)} &\multicolumn{1}{|c|}{218(10.4)} & \multicolumn{1}{|c|}{111(5.3)} & \multicolumn{1}{|c|}{55.5(2.7)} \\
\hline
\multirow{2}*{{500}} 
& BP3 & \multicolumn{1}{|c|}{686(33.0)} & \multicolumn{1}{|c|}{339(15.9)} & \multicolumn{1}{|c|}{168.1(7.8)} & \multicolumn{1}{|c|}{180(8.9)}  & \multicolumn{1}{|c|}{90.3(4.5)} & \multicolumn{1}{|c|}{44.3(2.3)} & \multicolumn{1}{|c|}{494(22.2)} & \multicolumn{1}{|c|}{253(11.3)} & \multicolumn{1}{|c|}{123.8(5.5)}  \\
\cline{3-11}
& BP4 & \multicolumn{1}{|c|}{345(16.7)} & \multicolumn{1}{|c|}{170(8.4)}  & \multicolumn{1}{|c|}{83.5(3.9)} & \multicolumn{1}{|c|}{180(8.9)} & \multicolumn{1}{|c|}{90.3(4.5)} & \multicolumn{1}{|c|}{44.3(2.3)} &\multicolumn{1}{|c|}{171.4(7.4)} & \multicolumn{1}{|c|}{82.4(3.9)} & \multicolumn{1}{|c|}{39.2(1.9)} \\
\hline
\hline
\end{tabular}}}
\caption{Signal cross-sections for HDSP pair production (OSD final state) at ILC. Total cross-section ($\sigma_{\rm tolal}$), as well as individual contributions from SDM
($\sigma_{\phi^+\phi^-}$) and FDM ($\sigma_{\psi^+\psi^-}$) are mentioned. Three choices of beam polarisation are used: P1 $\equiv$ $\{P_{e^{-}}: -0.8, P_{e^{+}}: +0.3\}$, P2 $\equiv \{P_{e^{-}}: 0, P_{e^{+}}:0\}$ and P3 $\equiv \{P_{e^{-}}: +0.8, P_{e^{+}}: -0.3\}$. CM energy ($\sqrt s$) is in the units of GeV.}
\label{table:BP}
\end{table}

\begin{center}
\begin{table}
\begin{tabular}{| c c | c  c  c| } 
\hline
\hline
\multicolumn{2}{|c|}{Backgrounds} &
\multicolumn{3}{|c|}{Cross-section(fb)} \\
\cline{1-5}
\multicolumn{1}{|c|}{$\sqrt{s}$} &
\multicolumn{1}{|c|}{Processes} & \multicolumn{1}{|c|}{P1} &
\multicolumn{1}{|c|}{P2} &
 \multicolumn{1}{|c|}{P3}\\
\cline{1-5}
\cline{1-2}
\multicolumn{1}{|c|}{\multirow{2}*{{1 TeV}} }
& \multicolumn{1}{|c|}{$WW$}& \multicolumn{1}{|c|}{296} & \multicolumn{1}{|c|}{128} &  \multicolumn{1}{|c|}{18.3}\\
\cline{2-5}
& \multicolumn{1}{|c|}{$ZZ$} & \multicolumn{1}{|c|}{7.5} & \multicolumn{1}{|c|}{4.4} &  \multicolumn{1}{|c|}{3.5}\\
\cline{2-5}
& \multicolumn{1}{|c|}{$WWZ$} & \multicolumn{1}{|c|}{1.2} & \multicolumn{1}{|c|}{0.5} &  \multicolumn{1}{|c|}{0.08}\\
\hline
\multicolumn{1}{|c|}{\multirow{2}*{{500 GeV}} }
& \multicolumn{1}{|c|}{$WW$} & \multicolumn{1}{|c|}{802} & \multicolumn{1}{|c|}{342} & \multicolumn{1}{|c|}{51}\\
\cline{2-5}
& \multicolumn{1}{|c|}{$ZZ$} & \multicolumn{1}{|c|}{21} & \multicolumn{1}{|c|}{12} &  \multicolumn{1}{|c|}{9.6}\\
\cline{2-5}
& \multicolumn{1}{|c|}{$WWZ$} & \multicolumn{1}{|c|}{0.8} & \multicolumn{1}{|c|}{0.37} &  \multicolumn{1}{|c|}{0.06}\\
\hline
\hline
\end{tabular}
\caption{Production cross-sections for $W^+(\ell^+\nu)W^-(\ell^-\bar\nu)$, $Z(\ell^+\ell^-)Z(\nu\bar\nu)$ and $W^+(\ell^+\nu)W^-(\ell^-\bar\nu)Z(\nu\bar\nu)$ 
background at $\sqrt{s}$ = 1 TeV and 500 GeV for various polarization combinations P1, P2 and P3 (see caption of Table \ref{table:BP}).
 }
\label{bkg}
\end{table}
\end{center}

\section{Analysis with selected benchmark points}
\label{sec:analysis}
\subsection{Analysis at $\sqrt{s}=$ 1 TeV}

\begin{figure}[!hptb]
	\centering
	\subfloat[]{\includegraphics[width=7cm,height=6cm]{missET_rl_1tev.pdf}}
	\subfloat[]{\includegraphics[width=7cm,height=6cm]{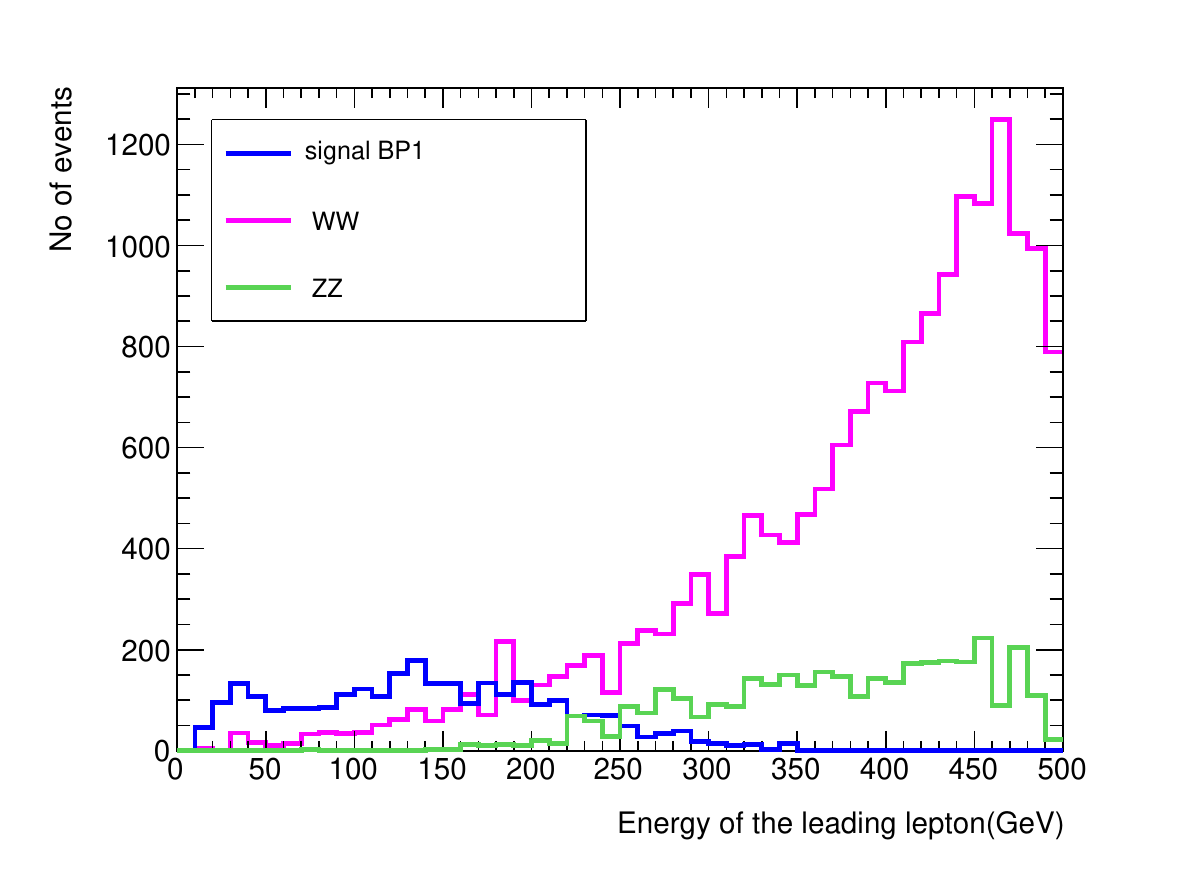}} \\
        \subfloat[]{\includegraphics[width=7cm,height=6cm]{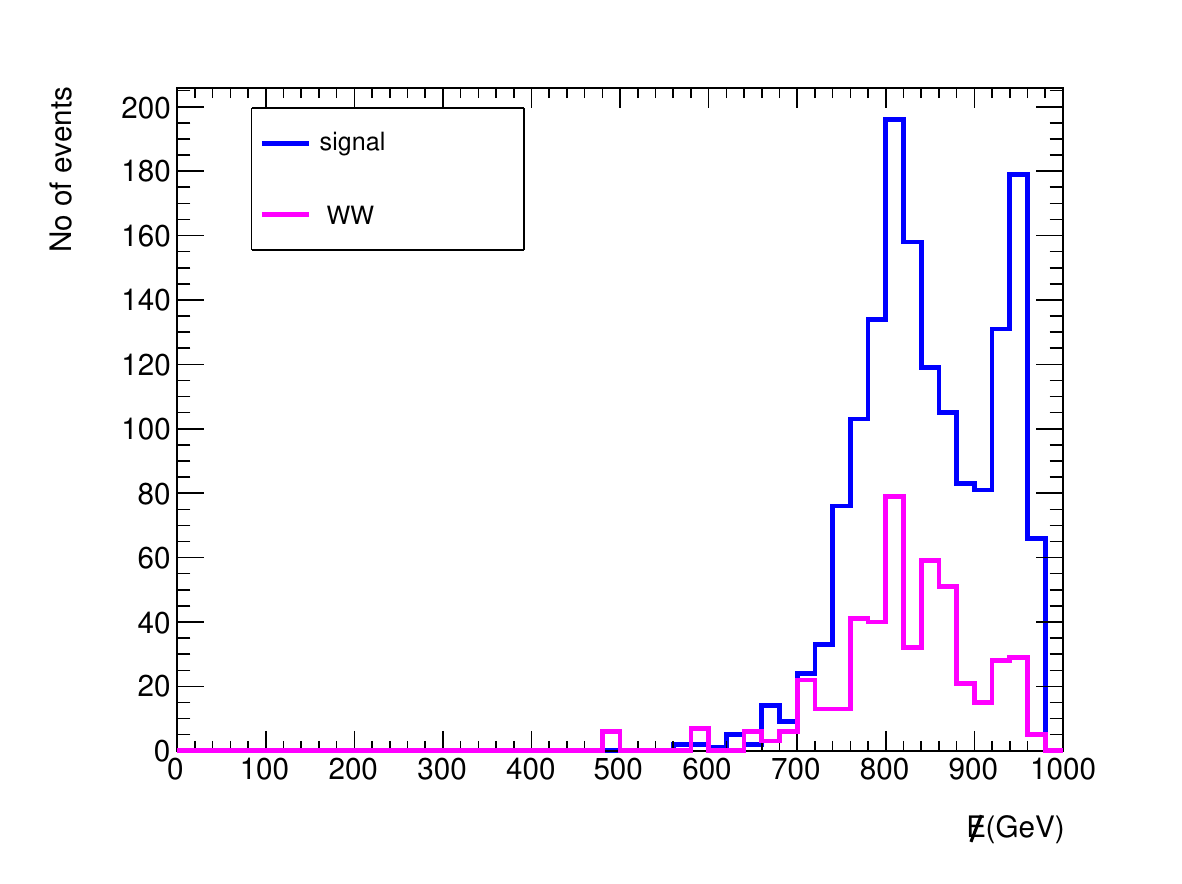}}
        \subfloat[]{\includegraphics[width=7cm,height=6cm]{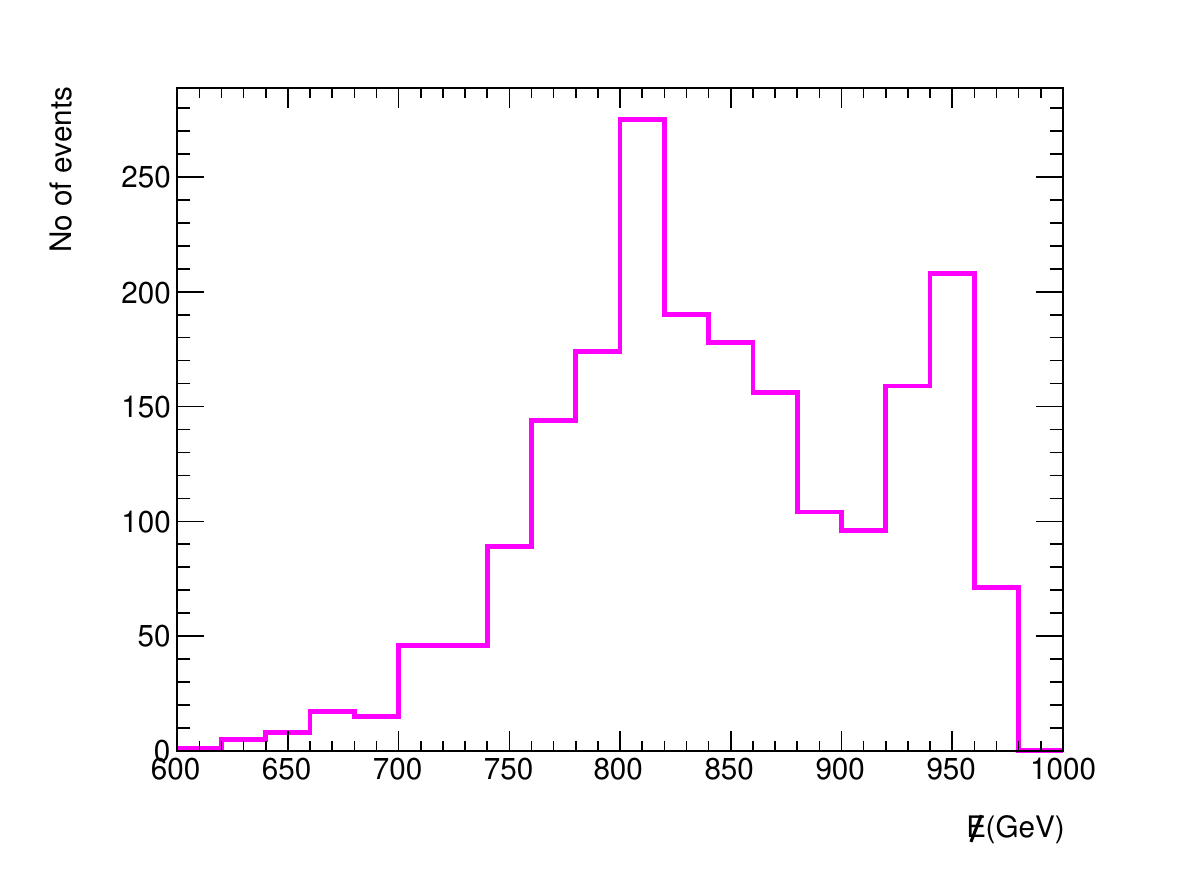}} 
	\caption{Unnormalized distribution of (a) missing energy, (b) energy of the leading lepton, (c) missing energy after applying the cut on the 
	energy of leading lepton $<$ 150 GeV for DM signal and SM backgrounds, (d) missing energy distribution of signal from BP1 plus background after 
	applying the cut $E_{\ell_1} < 150$ GeV at ILC with $\sqrt{s}=$ 1 TeV, beam polarisation P3 (see text) at $\mathcal{L}=$1000 ${\rm fb}^{-1}$. }
	\label{bp1}
\end{figure}

\begin{figure}[!hptb]
	\centering
	\subfloat[]{\includegraphics[width=7cm,height=6cm]{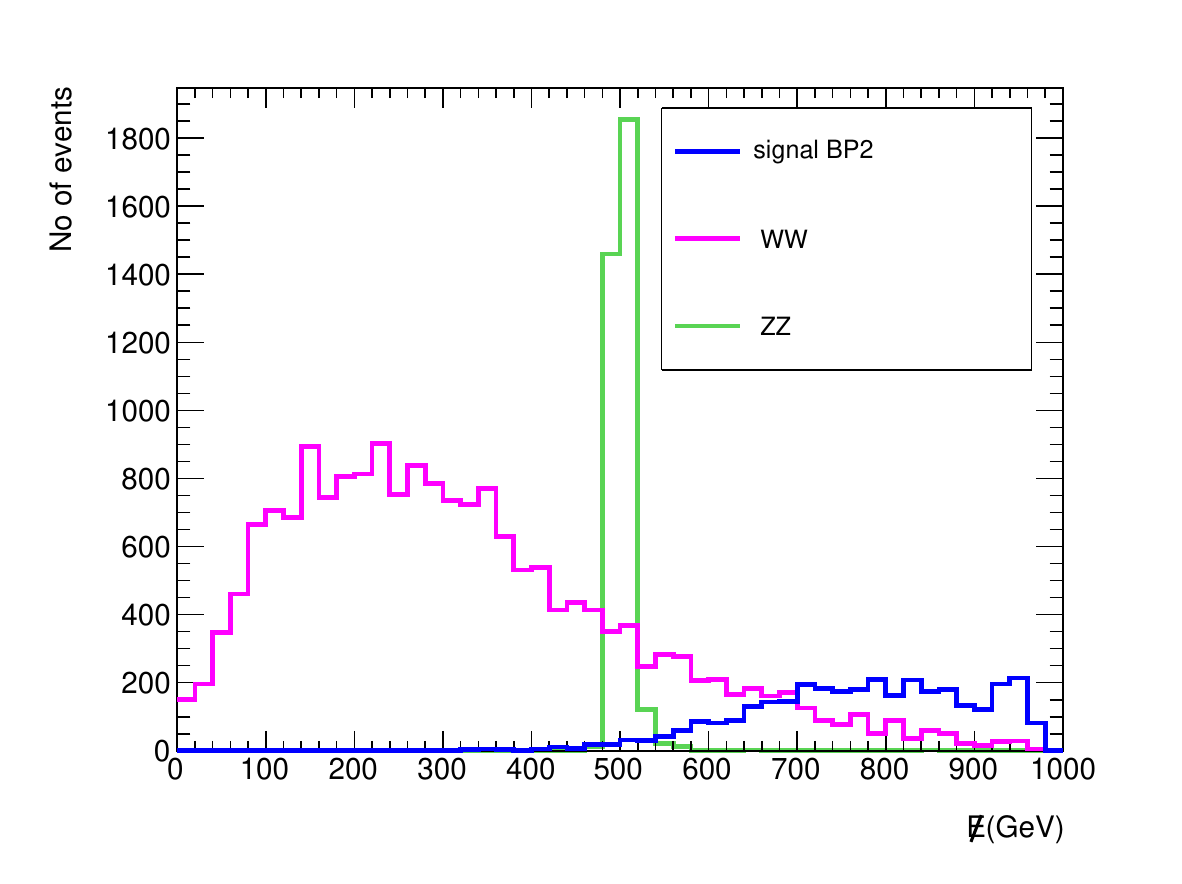}}
	\subfloat[]{\includegraphics[width=7cm,height=6cm]{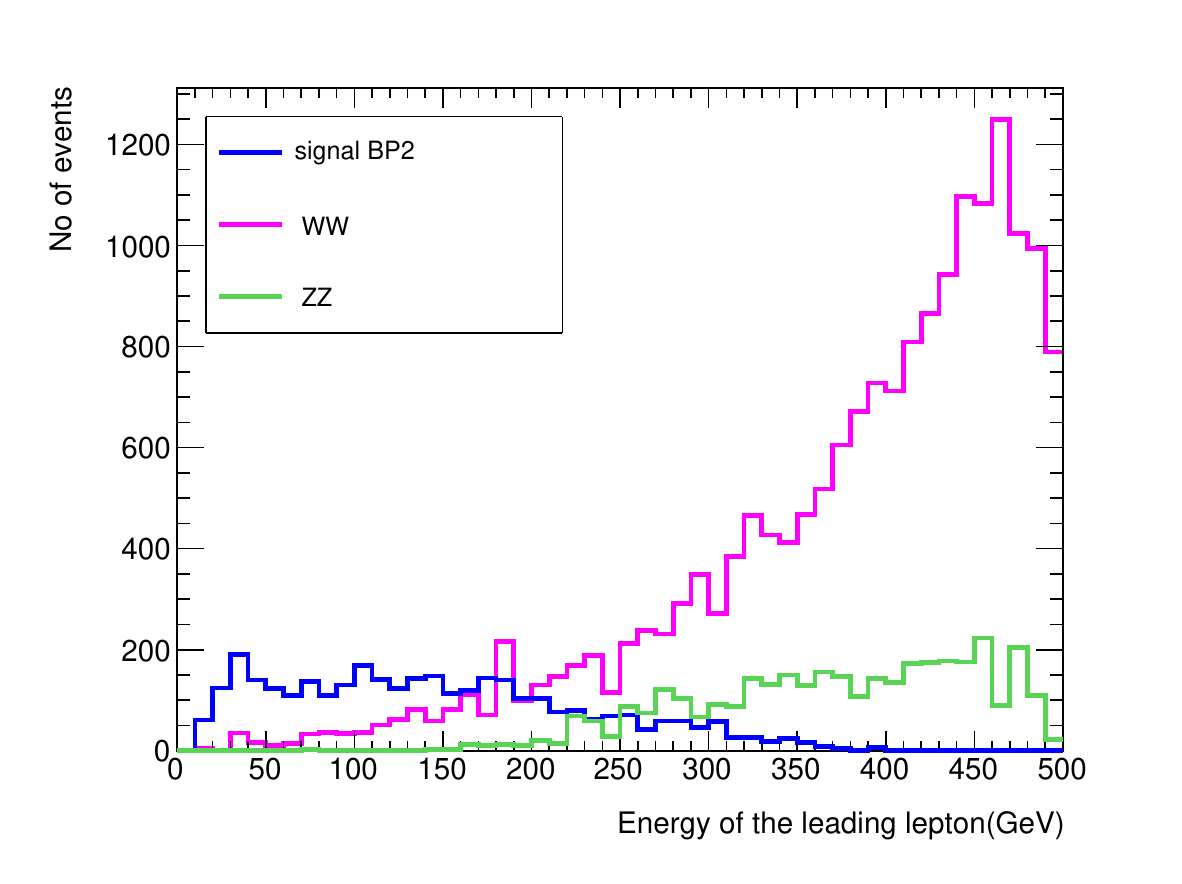}} \\
        \subfloat[]{\includegraphics[width=7cm,height=6cm]{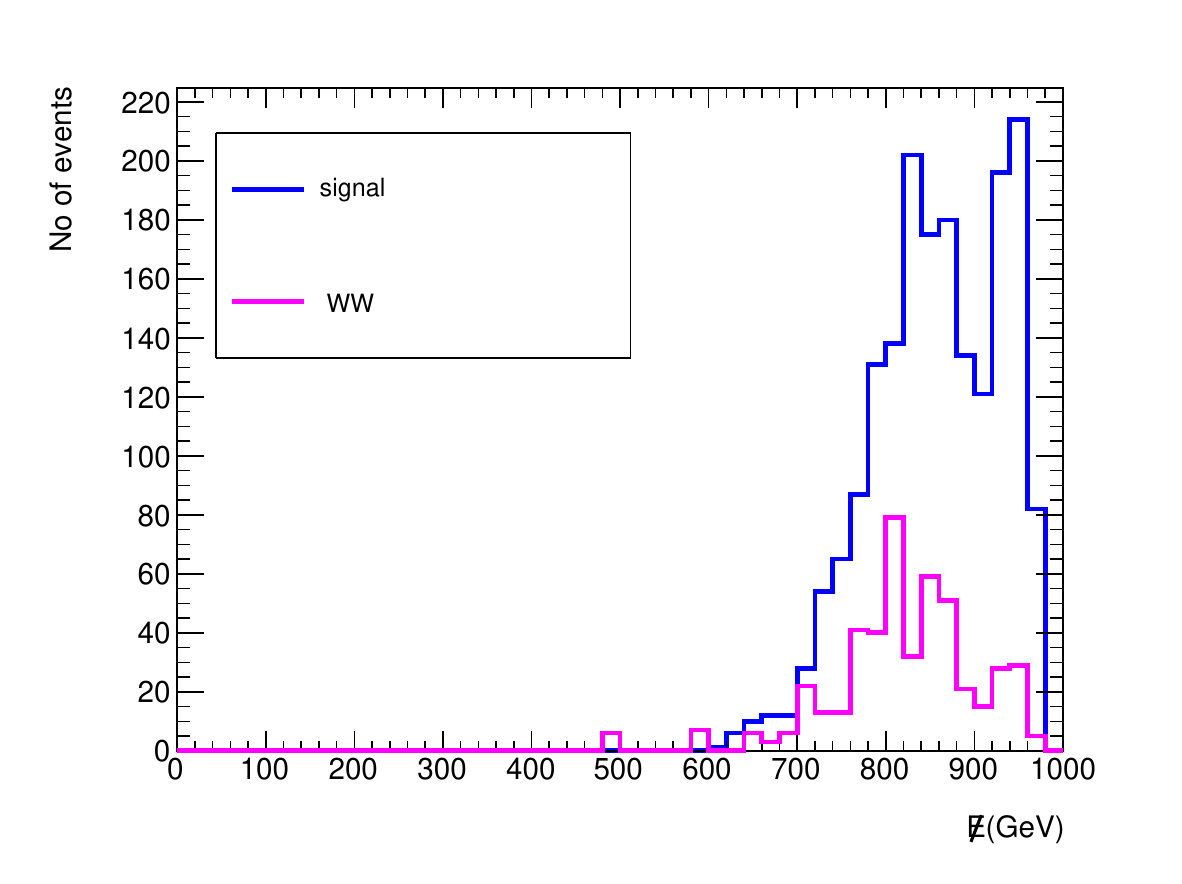}}
        \subfloat[]{\includegraphics[width=7cm,height=6cm]{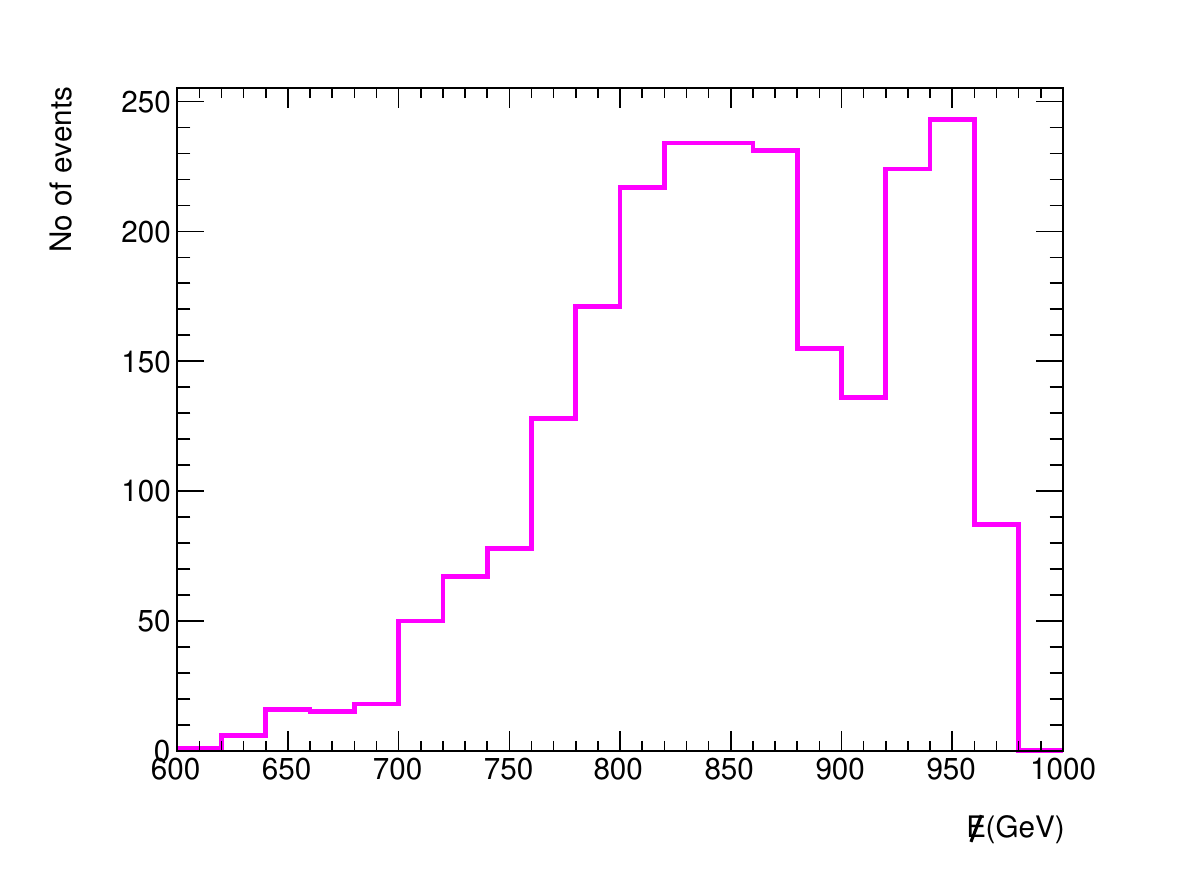}} 
	\caption{Same as Fig.~\ref{bp1} but for BP2.}
	\label{bp2}
\end{figure}

We take up first BP1 and BP2, which, due to their higher masses, can only be probed 
at $\sqrt{s} = 1$ TeV. In Fig.~\ref{bp1} (a), we show the corresponding $\slashed{E}$ distribution for signal (BP1) and dominant backgrounds at 
$\mathcal{L}=$1000 ${\rm fb}^{-1}$ of integrated luminosity with beam polarisation P3 $ \equiv \{P_{e^{-}}: 0.8, P_{e^{+}}: -0.3\}$. 
As one can clearly see, the $\slashed{E}$ distribution for the signal peaks at a higher value compared to that for the SM backgrounds. 
It is observed that, in case of $W^+W^-$ background, the leptons carry most 
of the energy and therefore, the $\slashed{E}$ peaks at a lower value. In case of $ZZ$ background, the energy is shared comparably between the two $Z$ bosons. 
Therefore $\slashed{E}$ peaks at $\frac{\sqrt{s}}{2}$ i.e. at 500 GeV. The signal distribution on the other hand, has two peaks. The peak at lower $\slashed{E}$ corresponds 
to FDM and the higher one arises from SDM, as $m_{\psi_1}<m_{\phi_0}$.
 The scalar sector being almost degenerate shows much narrower peak compared to the fermionic case, where $\Delta m_2$ is comparatively large and 
 the distribution is flatter. In Fig.~\ref{bp1} (b), we plot the energy of the leading lepton. As the total energy of collision is fixed at ILC, 
 the energy distribution of the leading lepton shows a sharp complementarity with the $\slashed{E}$. It peaks at a higher value for the backgrounds and at 
 a lower value for signal. Lepton energy distribution also retains the double-peak behaviour like $\slashed{E}$ distribution. Next we apply a cut on the energy 
 of the leading lepton and as a result, $\slashed{E}$ distribution of the $W^+W^-$ background becomes significantly softer (see Fig.~\ref{bp1}(c)). 
 The resulting position and size of the peak of the $W^+W^-$ background distribution is pretty sensitive to the lepton energy cut applied. 
 For example, in this case when we apply $E_{\ell_1} < 150$ GeV, 
 the peak from $W^+W^-$ distribution coincides with the first peak of the signal, and the distinction between the resulting first and second peak become 
 even more prominent. The extent at which the value of this cut affects our analysis will be discussed shortly. It may be noted that the contribution from $ZZ$ 
 background becomes negligible after applying the leading lepton energy cut and therefore is not shown explicitly in Fig.~\ref{bp1}(c). We further 
 show in Fig.~\ref{bp1}(d) the final signal plus background distribution after applying the aforementioned cut. We can see that the two peaks can 
 be distinguished rather well, even in the presence of background events.

We perform the same analysis for BP2 and resulting distributions are shown in Fig.~\ref{bp2}. The separation between two signal peaks is slightly worse here 
pertaining to smaller $\Delta m_2$ compared to BP1. However, after applying the leading lepton energy cut, 
the first peak, as well as $WW$ background gets diminished and the heights of the two peaks in the signal plus background $\slashed{E}$ distribution become 
comparable. We further note here that, in this case after applying the cut $E_{\ell_1} < 150$ GeV, the background distribution does not exactly coincide with the first peak 
of the signal (Fig.~\ref{bp2}(c)) and therefore, the first peak of the signal plus background distribution (Fig.~\ref{bp2}(d)) is flatter compared to BP1 as in Fig.~\ref{bp1}.

\subsection{Analysis at $\sqrt{s}=$ 500 GeV}

\begin{figure}[!hptb]
	\centering
	\subfloat[]{\includegraphics[width=7cm,height=6cm]{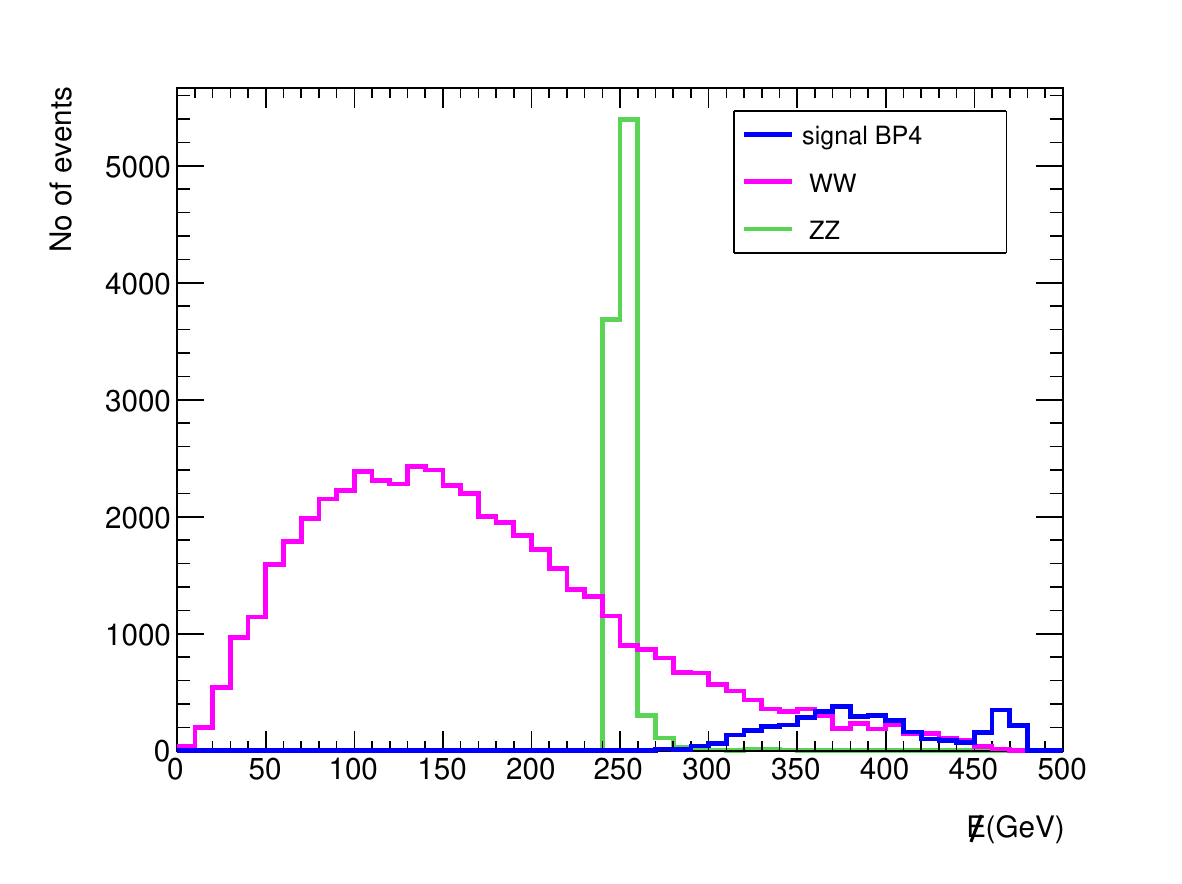}}
	\subfloat[]{\includegraphics[width=7cm,height=6cm]{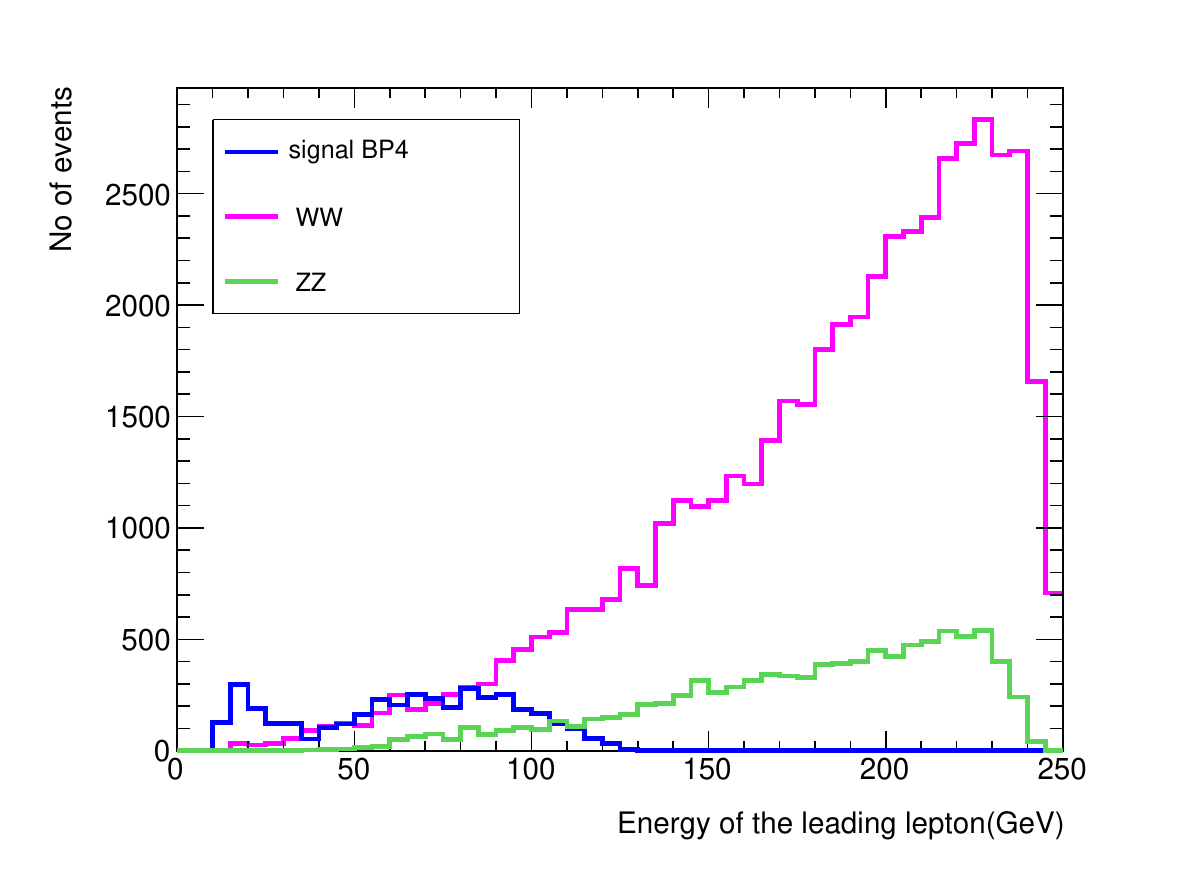}} \\
        \subfloat[]{\includegraphics[width=7cm,height=6cm]{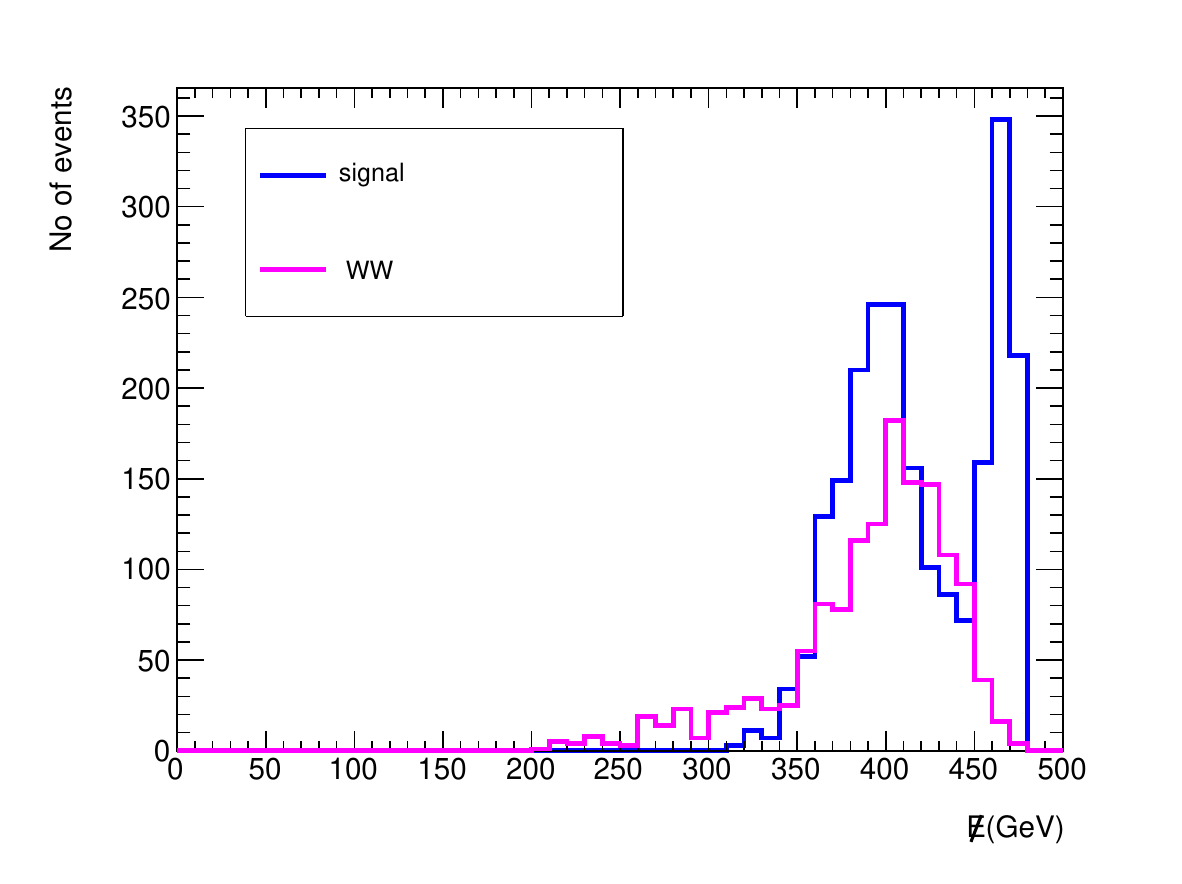}}
        \subfloat[]{\includegraphics[width=7cm,height=6cm]{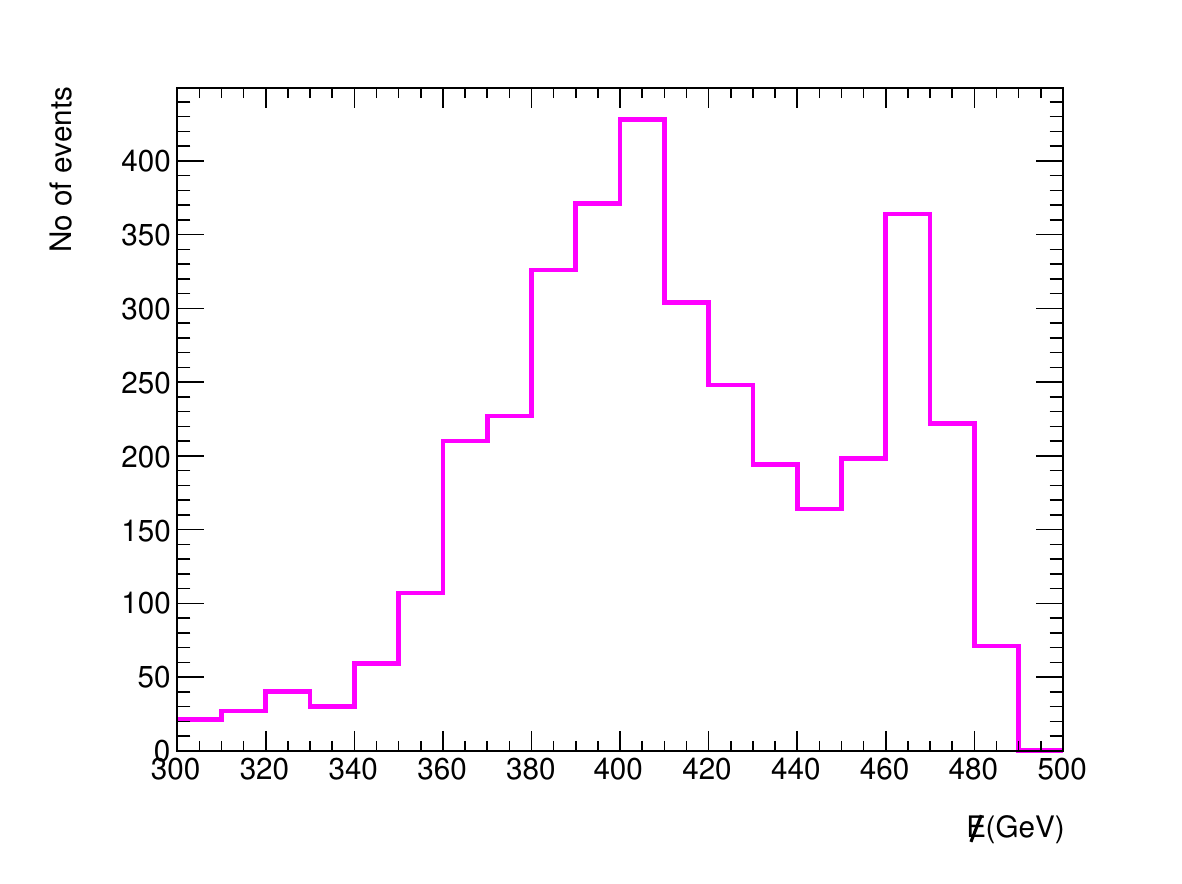}} 
	\caption{Same as Fig.~\ref{bp1}, but for BP4 at $\sqrt{s}=$ 500 GeV with $E_{\ell_1}<$ 75 GeV.}
	\label{bp4}
\end{figure}

For BP3 and BP4, it is possible to produce HDSPs on-shell at $\sqrt{s}=500$ GeV. It is evident that with smaller charged fermion mass, the size of the first peak in 
$\slashed{E}$ distribution is bigger in both cases compared to the second. This reduces the visibility of the double hump behaviour. The lepton energy cut is modified to 
$E_{\ell_1} < 75$ GeV for both these cases, which reduces the area under first peak. Remaining $WW$ background events adds to the first peak, resulting a
double peak in $\slashed{E}$ distribution for both the benchmark points. The case for BP4 is only shown in Fig.~\ref{bp4}; the organisation remains same as 
in Fig.~\ref{bp1}. The limitations of kinematical regions like Region III to see a distinguishable double peak in presence of SM background is discussed in 
Appendix \ref{app3}.

\subsection{Impact of lepton energy cuts}

\begin{figure}[!hptb]
	\centering
	\subfloat[]{\includegraphics[width=7cm,height=6cm]{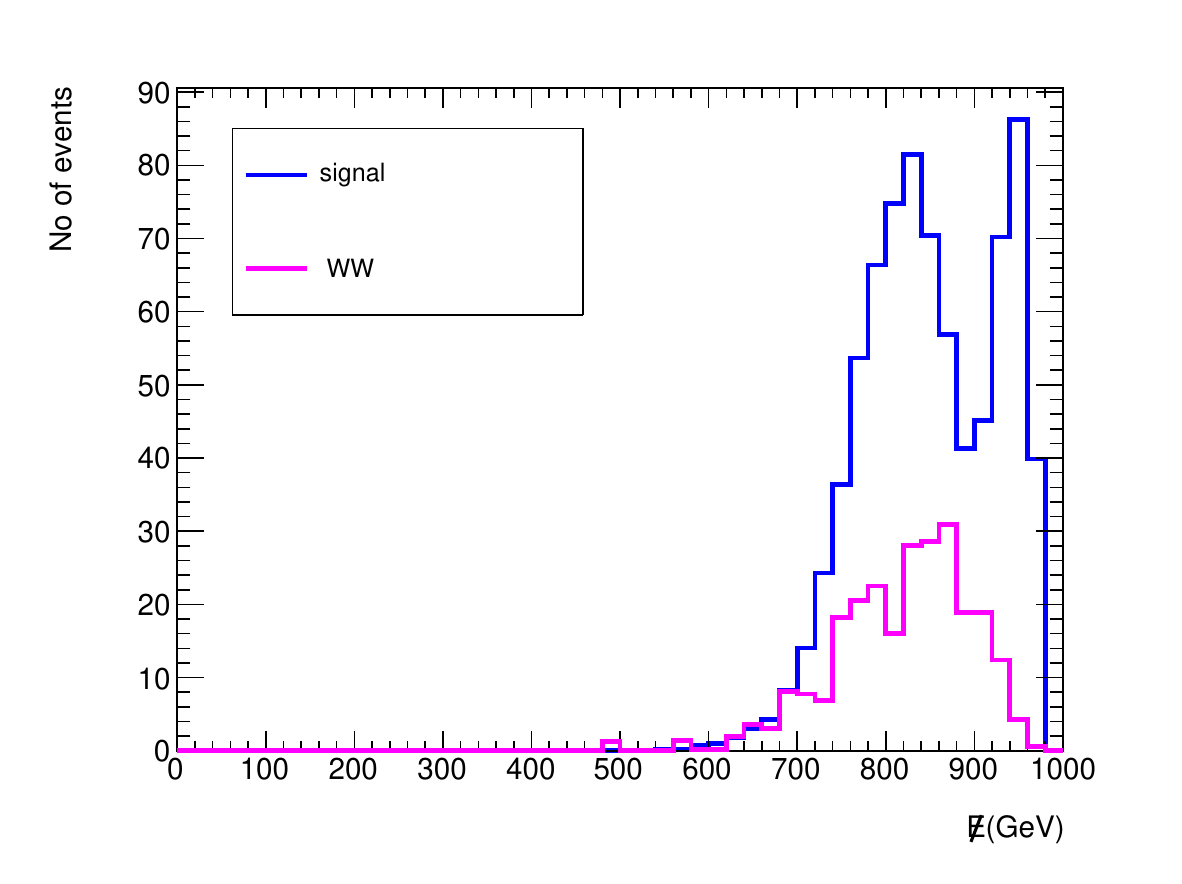}} 
        \subfloat[]{\includegraphics[width=7cm,height=6cm]{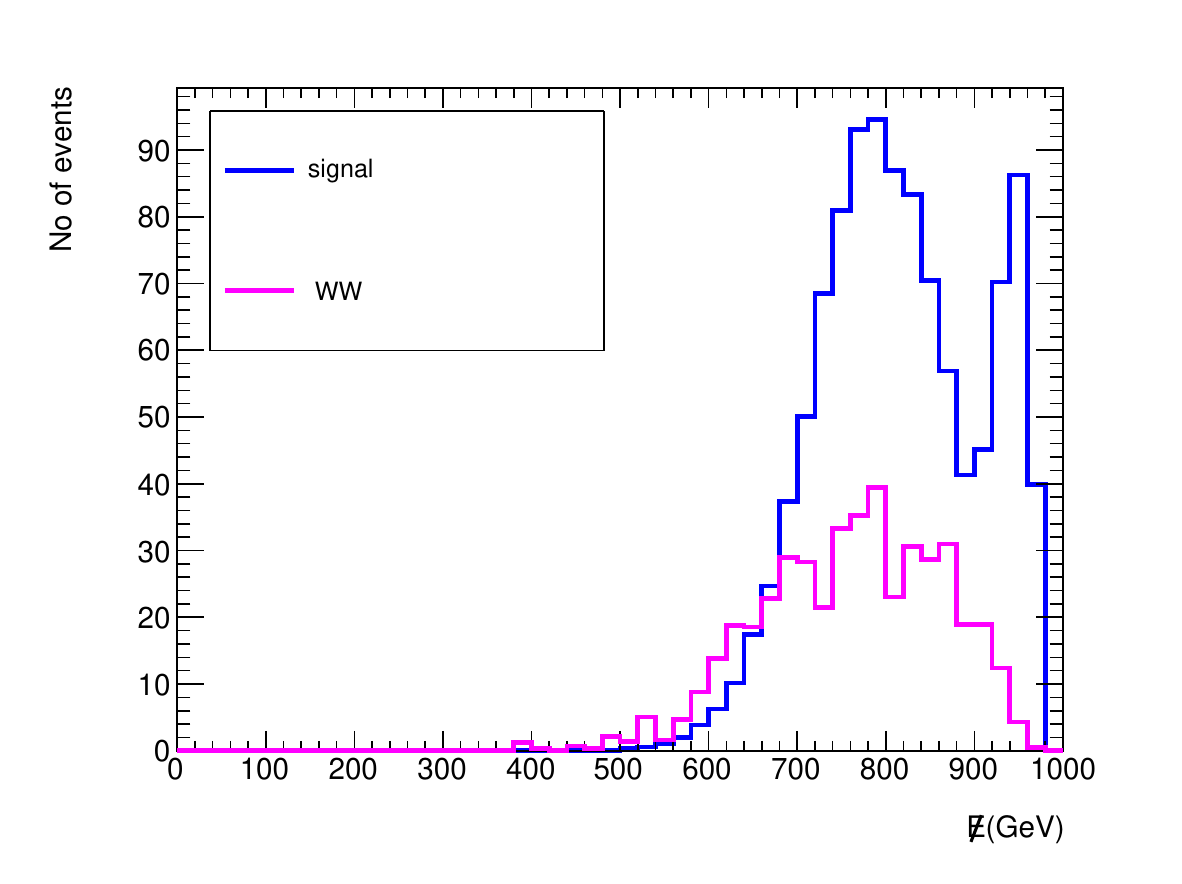}}
	\caption{Signal (BP1) and SM background distribution after applying cuts (a) $E_{\ell_1}< 150$ GeV and (b) $E_{\ell_1} < 200$ GeV with $\sqrt{s}=$ 1 
	TeV for polarisation P3 (see text) at ILC.}
	\label{cut_comparison}
\end{figure}

As pointed out earlier, at the $e^+ e^-$ collider, the lepton energy distribution shows a complementarity with $\slashed{E}$ distribution. 
A cut on the energy of the leading lepton suppresses the background and also helps us 
retain the Gaussian nature of the resulting distribution. However, the choice of the cut is crucial. As the total energy of collision 
in each event is fixed, higher lepton energy naturally corresponds to smaller $\slashed{E}$. Therefore, an upper cut on the lepton energy 
immediately indicates a lower cut-off in $\slashed{E}$ distribution. Given that $\slashed{E}$ distribution peaks at a much lower value for $WW$
compared to the DM signal, an upper cut on lepton energy distribution eliminates the low $\slashed{E}$ region. Harder lepton energy cut pushes 
the $\slashed{E}$ distribution from $WW$ events towards larger $\slashed{E}$. Naturally, the first peak on lower $\slashed{E}$ from the signal is also
affected by this cut. But, the second peak remains unaffected as long as the lepton energy cut is not too hard. Therefore optimising lepton energy cut 
so that $WW$ distribution coincides with the first peak of the signal elucidates the double hump behaviour. This is shown in Fig.~\ref{cut_comparison}, 
where we compare $\slashed{E}$ distribution after applying (a) $E_{\ell_1}<$ 150 GeV and (b) $E_{\ell_1}<$ 200 GeV. 

\subsection{Signal significance}
Before going into the quantitative discussion on distinguishability of the two DM peaks, we briefly examine the discovery potential of $\ell^+\ell^- + \slashed{E}$ signal 
at ILC by calculating signal significance (${\cal S}$) defined as follows:

\begin{equation}
{\cal S} = \sqrt{2 [(S+B) \text{Log}(1+\frac{S}{B}) - S]}\,;
\label{significance}
\end{equation}

\noindent
where $S$ and $B$ are the signal and background events surviving after all the analysis cuts are applied. We use the following selection cuts, apart from the basic ones: 
\begin{itemize}
\item  {\bf $\sqrt{s} =1$ TeV}: (i) Cut on the energy of the leading lepton $E_{\ell_1}<150$ GeV, (ii) $\slashed{E} > 600$ GeV, 
(iii) Invariant mass cut on the lepton pair, $60<m_{\ell\ell} < 120$ GeV.
\item {\bf $\sqrt{s} =500$ GeV}: (i) Cut on the energy of the leading lepton $E_{\ell_1}<75$ GeV, (ii) $\slashed{E} > 300$ GeV, (iii) Invariant mass cut on the lepton pair, 
$70<m_{\ell\ell} < 110$ GeV.
\end{itemize} 


\noindent
We remind the reader that the strong invariant mass cut of the lepton pair has been applied in order to reduce the non-resonant $\nu\bar\nu Z$ background, as discussed earlier.
We present ${\cal S}$ for all the benchmarks for various polarization combinations of initial beams in Table~\ref{significanceall} for $\mathcal{L}$=100 ${\rm fb}^{-1}$. 
We note here that for the particular polarization combination P1, one can achieve the maximum ${\cal S}$. However, with P2 and P3 too, it is possible to achieve 
${\cal S} \gsim 8\sigma$ for all the benchmark points. Therefore, all the aforementioned polarization combinations can ensure a significant discovery potential of our 
purported signal.  

In addition to ${\cal S}$, we present $S/B$ in Table~\ref{significanceall}, which gives us an estimate of the amount of signal purity or background contamination 
in the final kinematical distribution. To ensure that the two-peak signature is actually coming from the signal pertaining to two different DM particles, one should demand a large 
$S/B$. We see that $S/B$ maximizes for the polarization combination P3, which justifies our choice of polarization of initial beams as chosen in the rest of the analysis. 
We would like to point out that along with ${\cal S}$ and $S/B$, the distinguishability of two peaks also depend on the absolute number of observed events, which is proportional
to the integrated luminosity $\int L dt$. We use abbreviation $\int L dt \equiv\mathcal{L}$ to denote the same in the rest of the 
analysis and discuss the effect of $\mathcal{L}$ on the distinguishability of the two peaks in the next section.

\begin{table}
{\scriptsize{\begin{tabular}{| c | c c c c c c  | } 
\hline
\hline
\multicolumn{1}{|c|}{Benchmarks} &
\multicolumn{3}{|c|}{$S/B$} &
\multicolumn{3}{|c|}{${\cal S}$}\\
\cline{2-7}
 & 
 \multicolumn{1}{|c|}{P1} & \multicolumn{1}{|c|}{P2} & \multicolumn{1}{|c|}{P3} & \multicolumn{1}{|c|}{P1} & \multicolumn{1}{|c|}{P2} & \multicolumn{1}{|c|}{P3} \\
\cline{1-7}
BP1 & \multicolumn{1}{|c|}{1.07} & \multicolumn{1}{|c|}{0.91} & \multicolumn{1}{|c|}{3.7} & \multicolumn{1}{|c|}{14.5} & \multicolumn{1}{|c|}{9.7} &  \multicolumn{1}{|c|}{11.3} \\
\hline
BP2 & \multicolumn{1}{|c|}{1.2} & \multicolumn{1}{|c|}{1.05} & \multicolumn{1}{|c|}{4.2} & \multicolumn{1}{|c|}{16.2} & \multicolumn{1}{|c|}{11.0} &  \multicolumn{1}{|c|}{12.4} \\
\hline
BP3 & \multicolumn{1}{|c|}{1.05} & \multicolumn{1}{|c|}{1.13} & \multicolumn{1}{|c|}{3.4} & \multicolumn{1}{|c|}{22.0} & \multicolumn{1}{|c|}{16.1} &  \multicolumn{1}{|c|}{17.4} \\
\hline
BP4 & \multicolumn{1}{|c|}{0.4} & \multicolumn{1}{|c|}{0.5} & \multicolumn{1}{|c|}{1.5} & \multicolumn{1}{|c|}{10.7} & \multicolumn{1}{|c|}{8.0} &  \multicolumn{1}{|c|}{8.5} \\
\hline
\hline
\end{tabular}}}
\caption{$S/B$ and signal significance (${\cal S}$) for different benchmark points for various beam polarizations, P1, P2, P3 (see text). 
${\cal S}$ has been calculated for $\mathcal{L}$ = 100 ${\rm fb}^{-1}$.}
\label{significanceall}
\end{table}

\section{Distinction criteria for two peaks in $\slashed{E}$ spectrum}
\label{sec:distinction}
\begin{figure}[!hptb]
	\centering
	\subfloat[]{\includegraphics[width=6.8cm,height=5.5cm]{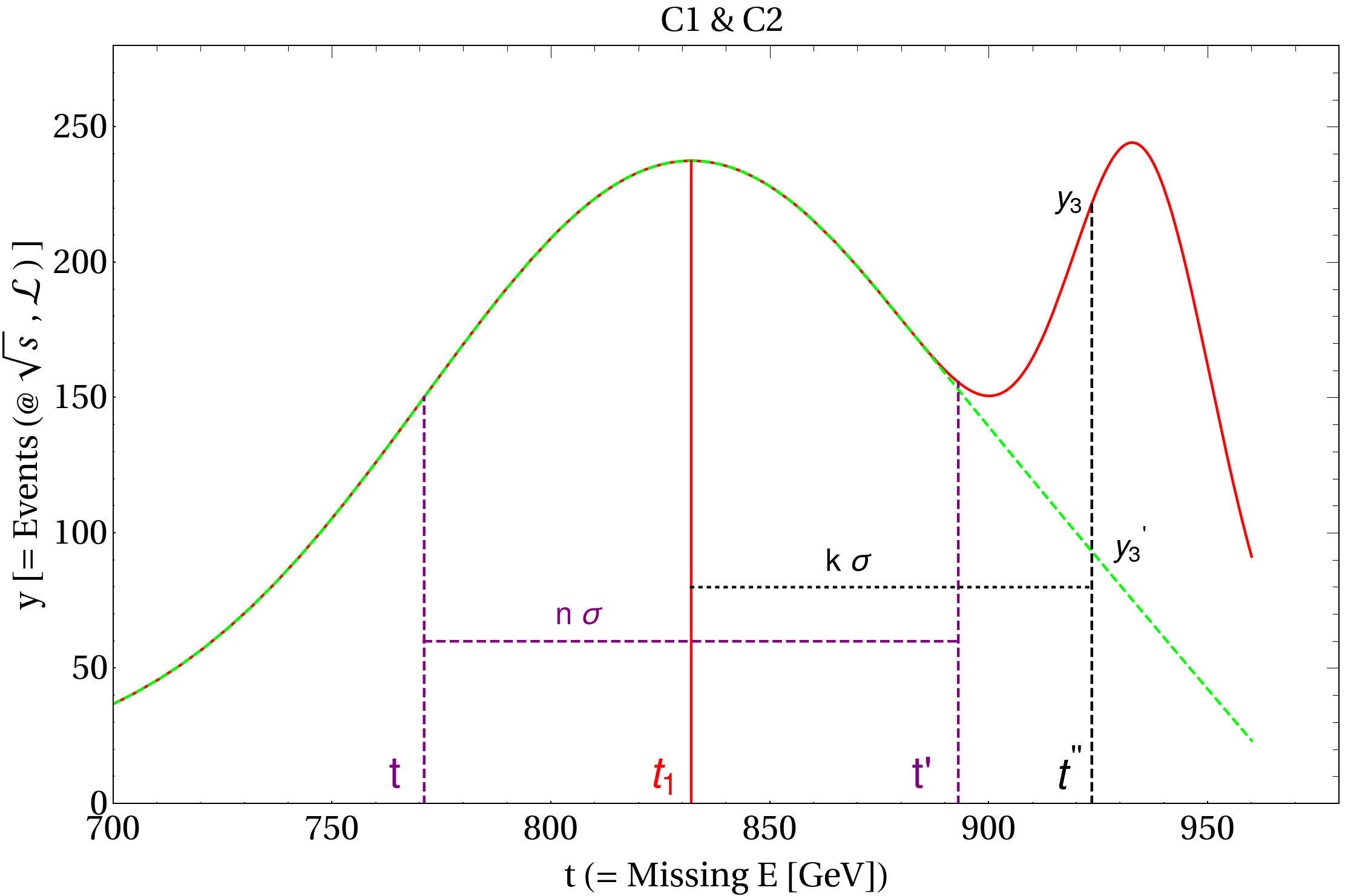}}~~
	\subfloat[]{\includegraphics[width=6.8cm,height=5.5cm]{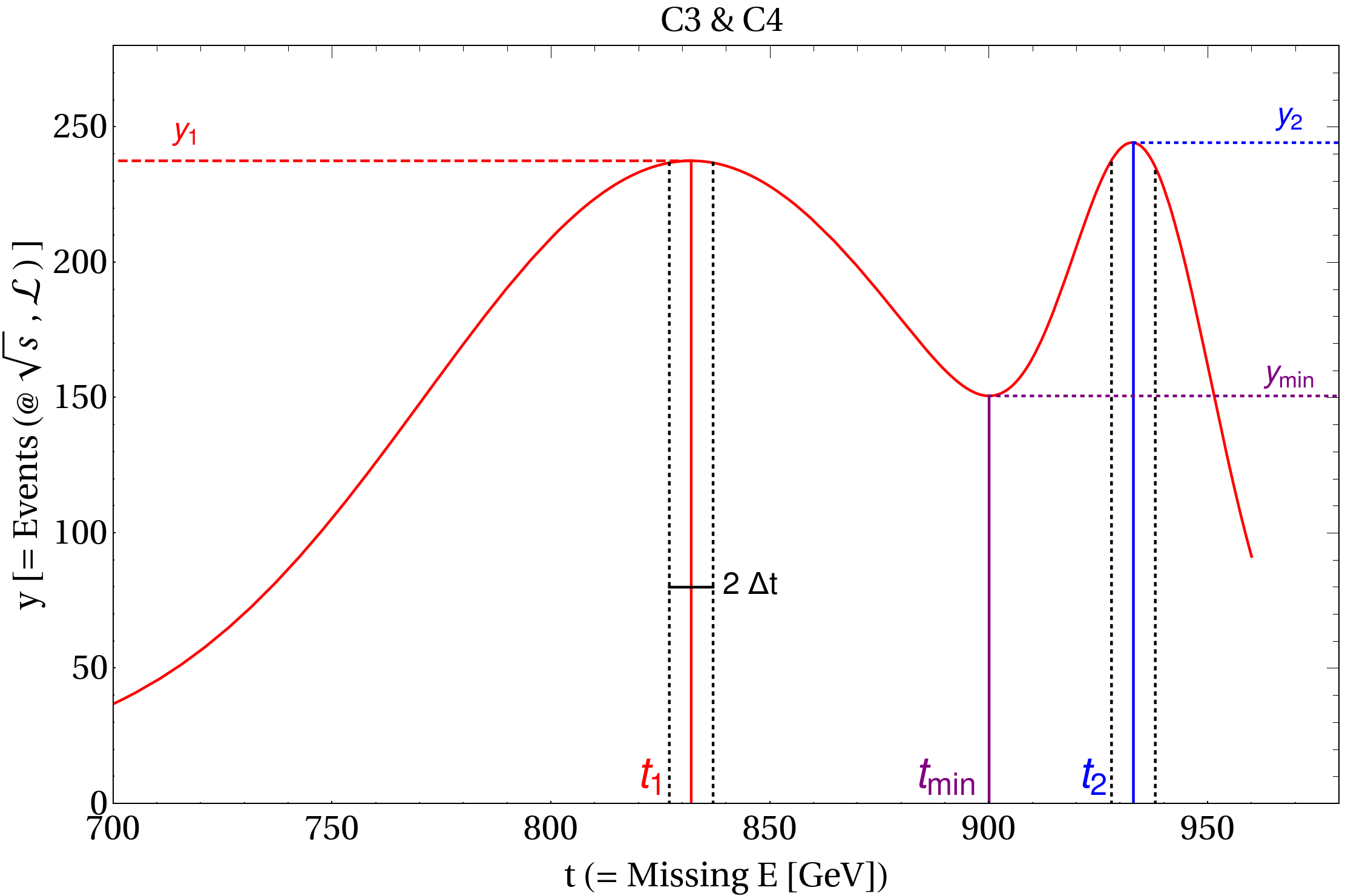}}
	\caption{ A schematic representation of a typical $\slashed{E}$ distribution in $\frac{d\sigma}{d\slashed{E}}(\equiv y)~\rm{vs}~\slashed{E}(\equiv t)$ 
	plane with two peaks when fitted with Gaussian function as in Eq.~\ref{eq:gf} is shown. Left: quantities required for conditions C1 and C2 are defined 
	(see text for details); Right: quantities required for conditions C3 and C4 are defined.}
 \label{fig:cartoon2}
\end{figure}

As demonstrated above, two different dark sectors having DM masses $\mdma$, $\mdmb$, and mass differences $\Delta m_1$ and $\Delta m_2$ with the HDSP, 
yielding same collider signal, will provide peaks at different values in $\slashed{E}$ distribution. These peaks partially overlap when the full signal produced from both 
dark sector is analysed. It is also shown that when the difference between masses and/or splitting becomes large, the two peaks in distorted $\slashed{E}$ distribution is more 
prominent. Such $\slashed{E}$ distributions of the signal with reduced SM background can therefore be fitted into a two-peak asymmetric Gaussian distribution as a function 
of any variable $x$ as:

 \bea\label{eq:gf}
 G(x)&=& G_1(x)+G_2(x) + \cal{B} \nonumber\\
 &=& A_1~ e^{-\frac{(x-\mu_1)^2}{2\sigma_1^2}}+A_2 ~ e^{-\frac{(x-\mu_2)^2}{2\sigma_2^2}} + \cal{B} ~.
 \eea
Here $G_1(x)$ is the Gaussian function corresponding to the first peak with amplitude $A_1$, mean $\mu_1$, and standard deviation $\sigma_1$; while $G_2(x)$ is 
the function containing the corresponding quantities for the second DM peak. 
The constant (or slowly varying) parameter $\cal{B}$ is further introduced to account 
for various theoretical uncertainties as well as those due to $\slashed{E}$ mis-measurement etc\footnote{In the results presented here, we have treated $\cal{B}$ as a 
constant function of $\slashed{E}$ for simplicity. However, we have checked that peaks do not change appreciably on fitting $\cal{B}$ as a polynomial 
function upto second degree. Differences, if any, are noticeable mainly in regions away from both peaks.}.  A schematic of such a function is shown in 
Fig. \ref{fig:cartoon2}. Here onwards we also introduce a simplified notation to denote $\slashed{E}$ by $t$ and $\frac{d\sigma}{d\slashed{E}}$ by $y$.

\noindent
Let us now assume that we can identify two peaks at $\slashed{E}$ values $t_1 (\equiv\mu_1)$ and $t_2 (\equiv\mu_2)$ as shown in Fig. \ref{fig:cartoon2}. 
The number of events at those peaks are denoted by $y_1=y(t_1)$ and $y_2=y(t_2)$ respectively; which are nothing but the area under the curve in a small 
interval around the peak as indicated in the right hand side figure. The minima between two peaks is identified as $t_{\rm min}$ and the 
corresponding event rate along y-axis is denoted by $y_{\rm min}$. With this preliminaries, we are now ready to set up the criteria for 
distinguishing the two peaks in $\slashed{E}$ spectra. Each such criterion must address either or both of the following questions: 
(a) How to elicit the prominence of the second (read smaller) peak relative to the first (bigger) peak, and (b) How to resolve best the 
separation between the two peaks.

\begin{itemize}
 \item {\bf C1}: The first condition examines how much the presence of second peak distorts the symmetry of the distribution about the first peak. 
 This will require us to compare the number of events within n$\sigma$ ($n \geq 1$) range of the first peak on both sides. Let us assume, $t$ (on left) and 
 $t^\prime$ (on right) are the two positions which are n$\sigma$ ($n \geq 1$) away from the first peak $t_1$ as shown in Fig.~\ref{fig:cartoon2}(a). 
 Assume number of events within n$\sigma$ on both sides of the first peak as:
 \bea
 \Delta N_1=\int_t^{t_1} y dt, ~~\Delta N_2=\int_{t_1}^{t^{'}} y dt \,.
 \eea
 We define  
 \bea
 R_{C1} = \frac{|\Delta N_2-\Delta N_1|}{\sqrt{\Delta N_1}}\,.
\label{defc1}
 \eea
 Then if
\bea
 R_{C1} > 2\,,
\label{c1}
 \eea

 we can stipulate that the peaks are resolved to 2$\sigma$ significance or larger. Note that the denominator in Equation~\eqref{c1} represents the fluctuation of the 
 distribution corresponding to the first peak. The criterion C1, defined in terms of integrated quantities, is useful when the two peaks are not visually 
 prominent but a distortion to the $\slashed{E}$ spectrum occurs. It is also worth registering that this criterion has its limitation when the second peak is 
 much smaller than the first. 

 \item {\bf C2}: The second condition addresses how high does the second peak go.
Let us assume $y_3= y(t^{''})$ is the number of events at $t^{''}$ which is assumed to be $k\sigma$ ($k \geq 1$) away from the first peak $t_1$ on the right side and 
$y_3^\prime= y^\prime(t^{''})$ is the corresponding number of events in the absence of second peak upon a Gaussian fit (Fig.~\ref{fig:cartoon2}(a)). 
Then we define $R_{C2}$ in terms of the difference between these two numbers as follows:
 \bea
R_{C2} = \frac{y(t^{''}) -y^\prime(t^{''})}{\sqrt{y^\prime(t^{''})}} \equiv \frac{y_3 -y_3^{'}}{\sqrt{y_3^{'}}}\,.
 \label{c2}
\eea

If $R_{C2} > 2$, we will be able to say that the fluctuation is more than 2$\sigma$ and can be termed as a second peak. Note that C2 offers a 
 criterion at the differential level, in contrast with C1, which comes at an integral level.
 
 \item {\bf C3}: Another possibility to check how significant the second peak is to compare the number of events within a close window ($\Delta t \sim$ 5 GeV) 
 around two peaks as shown in Fig.~\ref{fig:cartoon2}(b). One may take a ratio between the difference and sum of the following quantities as: 
 \bea
 R_{C3} &=& \frac{\int_{t_1-\Delta t}^{t_1+\Delta t} y dt-\int_{t_2-\Delta t}^{t_2+\Delta t} y dt}{\int_{t_1-\Delta t}^{t_1+\Delta t} y dt+\int_{t_2-\Delta t}^{t_2+\Delta t} y dt} ~~
 \underrightarrow{\{\Delta t \to 0\}}~~ \frac{y_1-y_2}{y_1+y_2}.
\label{c3}
 \eea
This essentially compares the number of events about the two maxima and thus the smaller is $R_{C3}$, the more significant is the second peak.
 
  \item {\bf C4}: This condition addresses how significant is the second peak with respect to the minimum in-between.
 Obviously if there exists a second peak (maxima) along with the first, there should be a local minimum between them ($t_1 < t_{\rm min} < t_2$). Then the significance 
 of the second peak with respect to the local minimum can be obtained as follows in terms of a quantity $R_{C4}$: 
 \bea
 R_{C4} = \frac{y(t_{2}) -y(t_{\rm min})}{\sqrt{y(t_{\rm min})}} \equiv \frac{y_2 -y_{\rm min}}{\sqrt{y_{\rm min}}}\,.
\label{c4}
\eea
$R_{C4} > 2$ indicates that the second peak rises more than 2$\sigma$ with respect to the fluctuation of the intermediate minimum. This too is a criterion at the differential level.

 \end{itemize}
 
It is obvious from the discussion above that the criteria to distinguish two peaks significantly depend
on statistics/luminosity and the parameters $n(k)$. We will explore the effect of these factors explicitly for our chosen 
BP's, in the context of the collider signal chosen for the analysis.


\begin{figure}[htb!]
 
   
   \subfloat[]{\includegraphics[width=6.8cm,height=5.5cm]{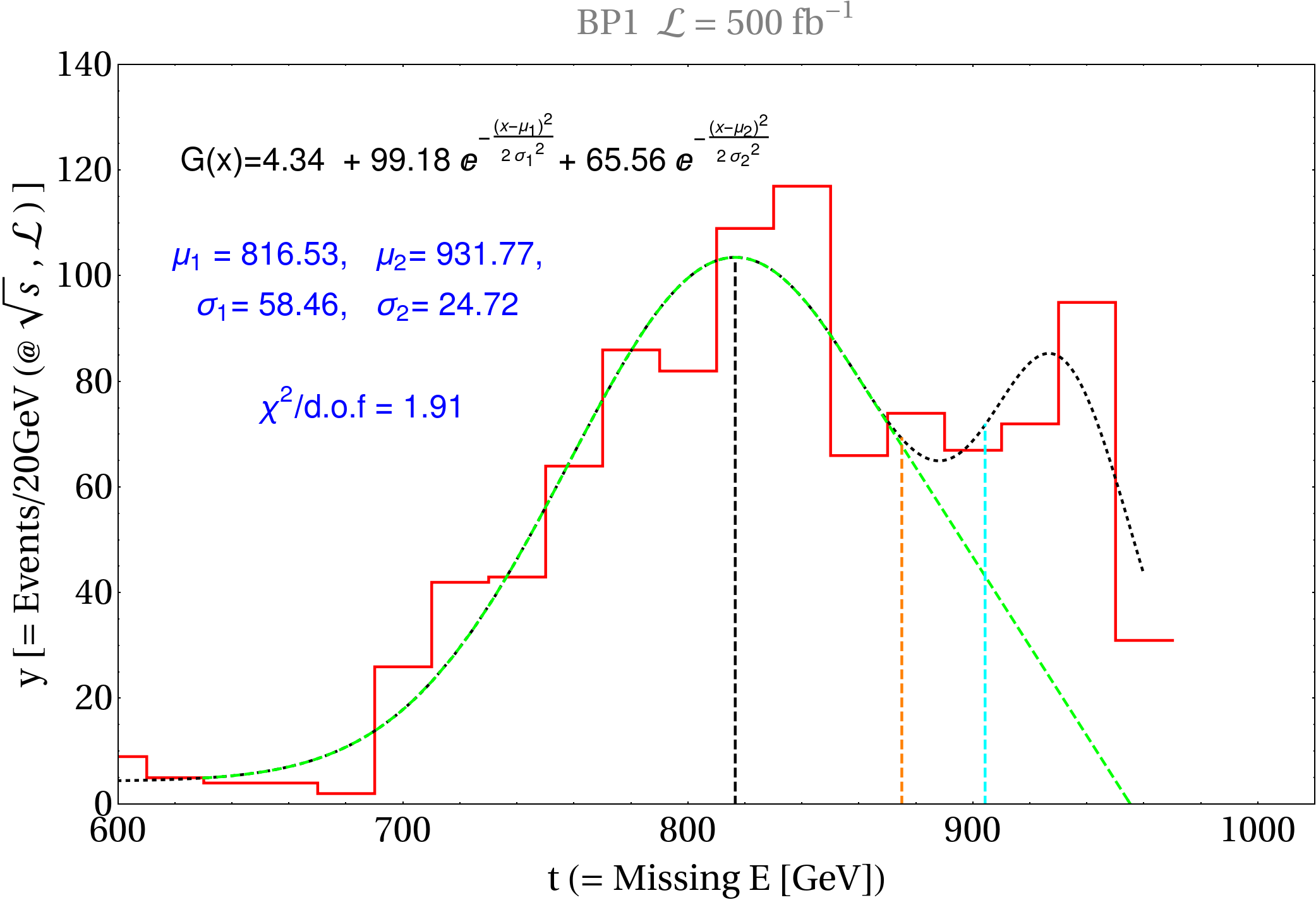}}~
   \subfloat[]{\includegraphics[width=6.8cm,height=5.5cm]{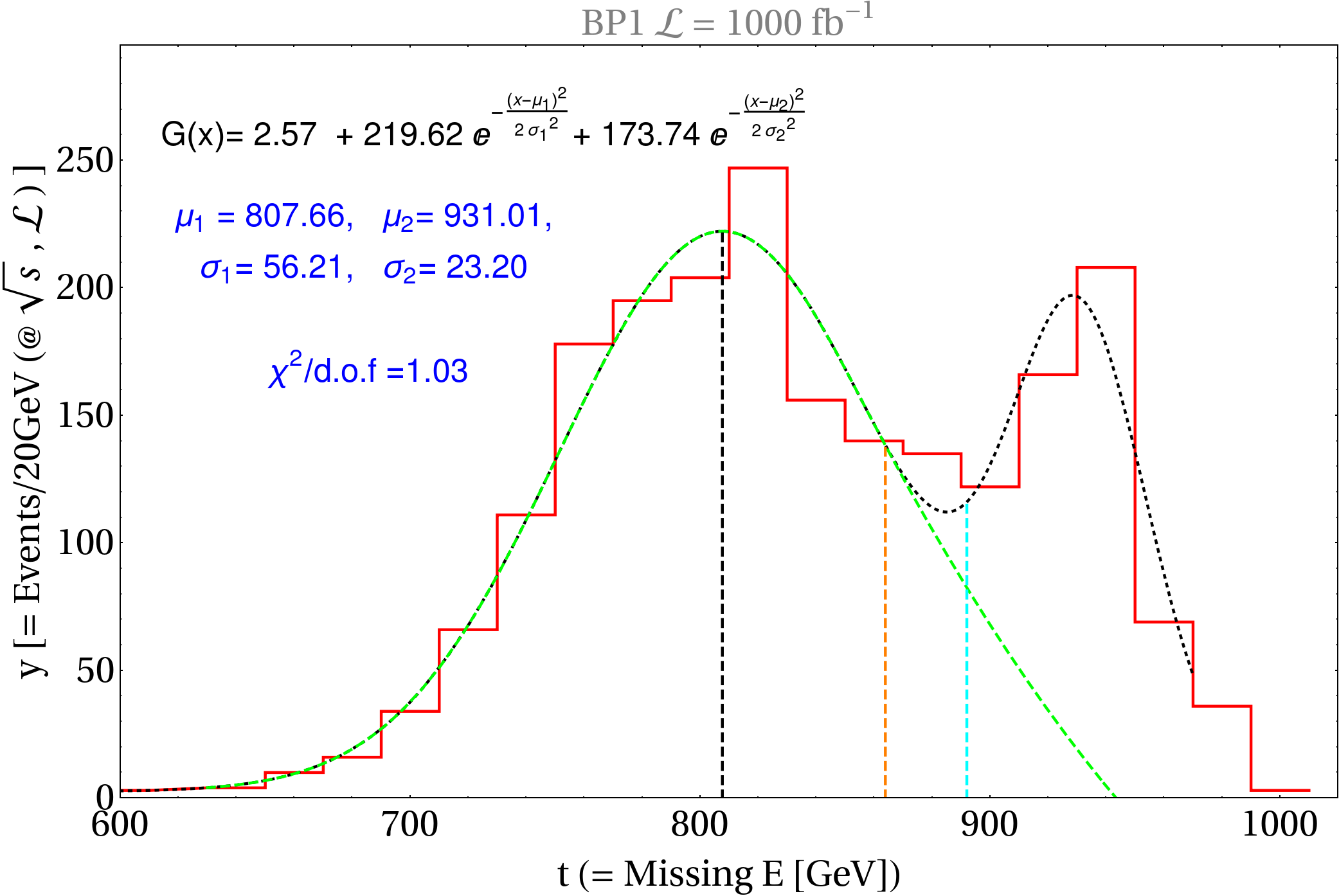}}\\
   \subfloat[]{\includegraphics[width=6.8cm,height=5.5cm]{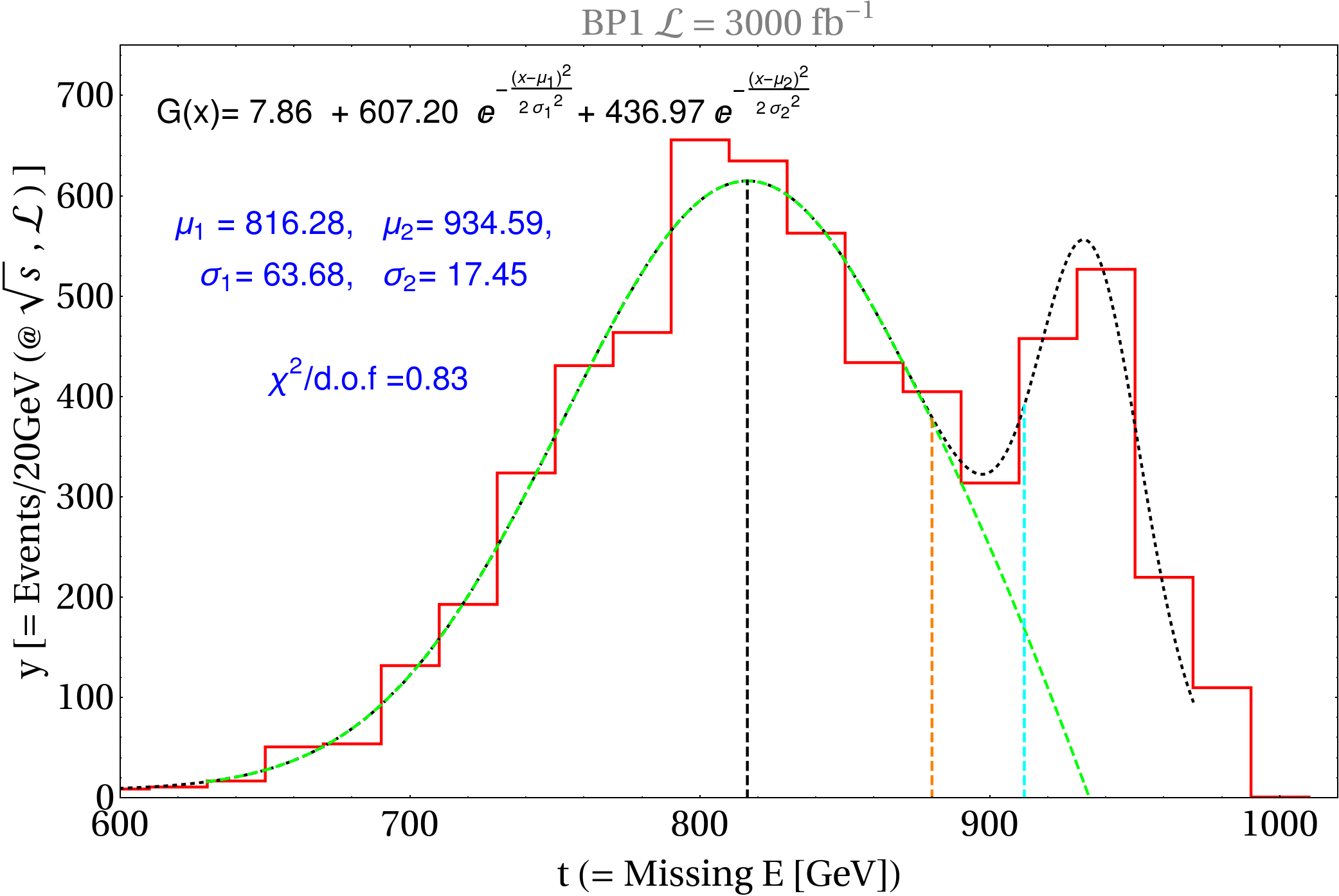}}~
   \subfloat[]{\includegraphics[width=6.8cm,height=5.5cm]{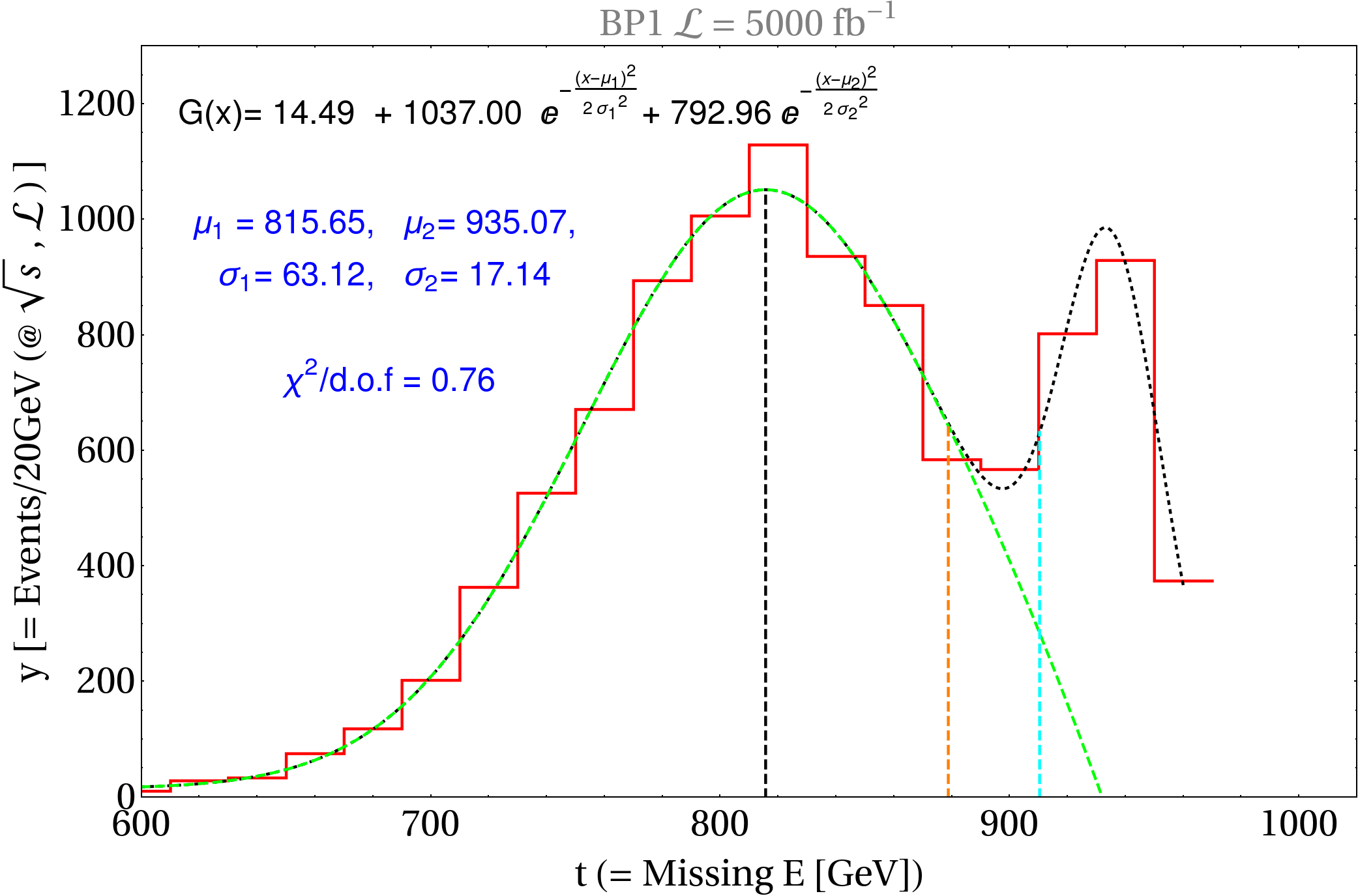}}~
   
   
  \caption{Gaussian Fitting of $\slashed{E}$ distribution for signal (BP1) plus SM background for different $\mathcal{L}$. 
  The polarization of initial beams are chosen to be P3.}
  \label{scbp1}
 \end{figure}
 
   \begin{figure}[htb!]
    \subfloat[]{\includegraphics[width=6.8cm,height=5.5cm]{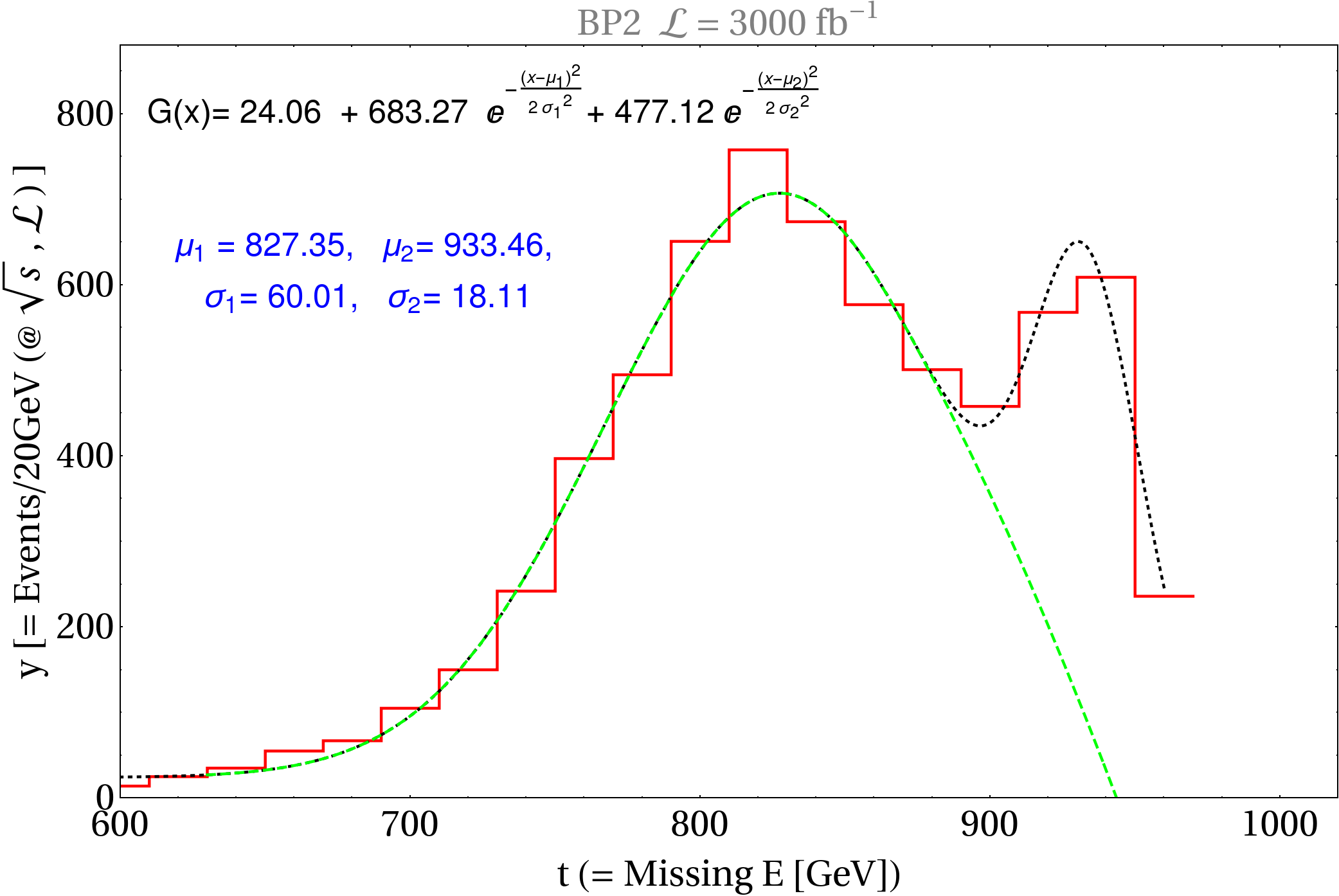}}~
   \subfloat[]{\includegraphics[width=6.8cm,height=5.5cm]{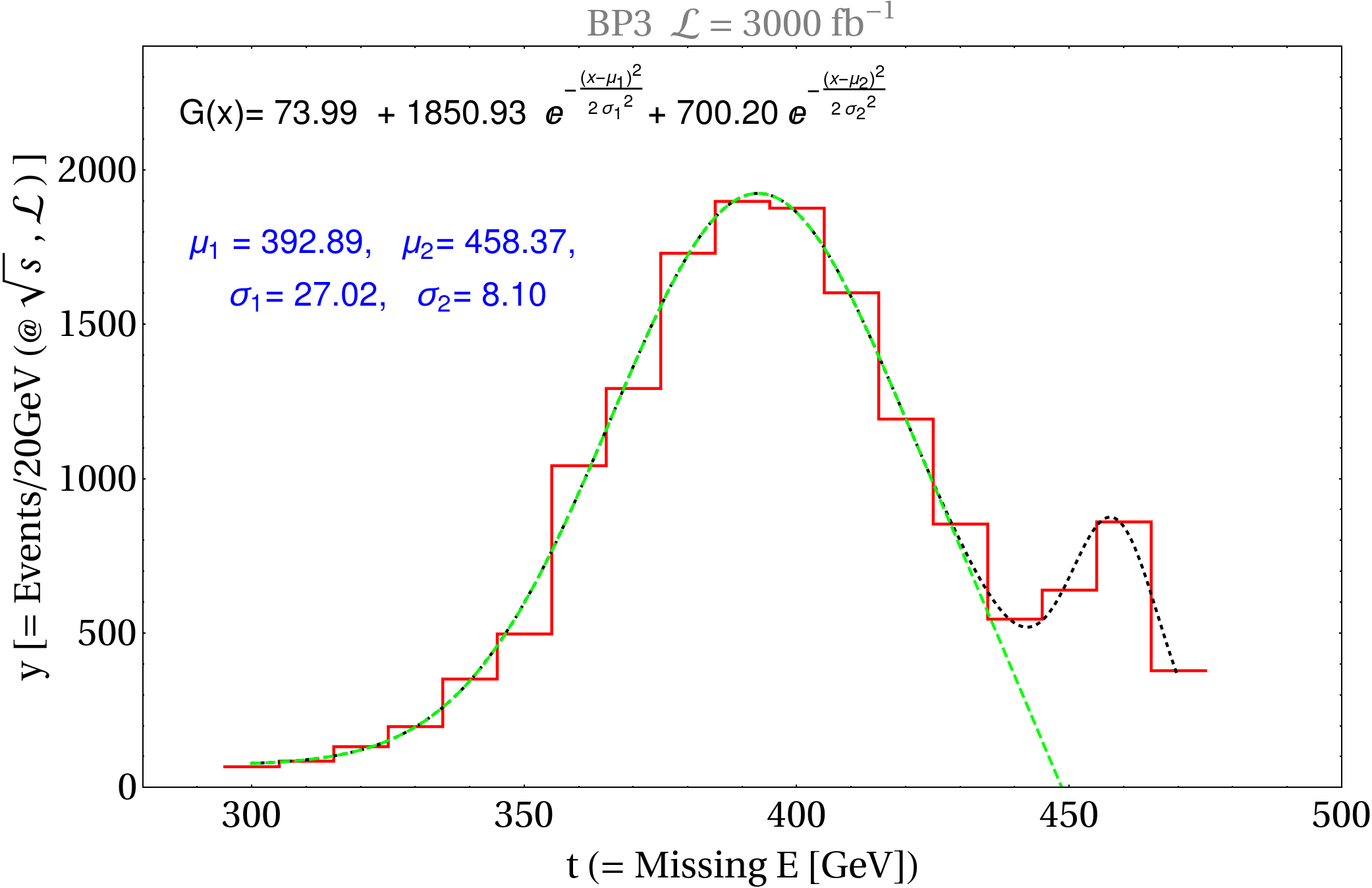}}\\
    \subfloat[]{\includegraphics[width=6.8cm,height=5.5cm]{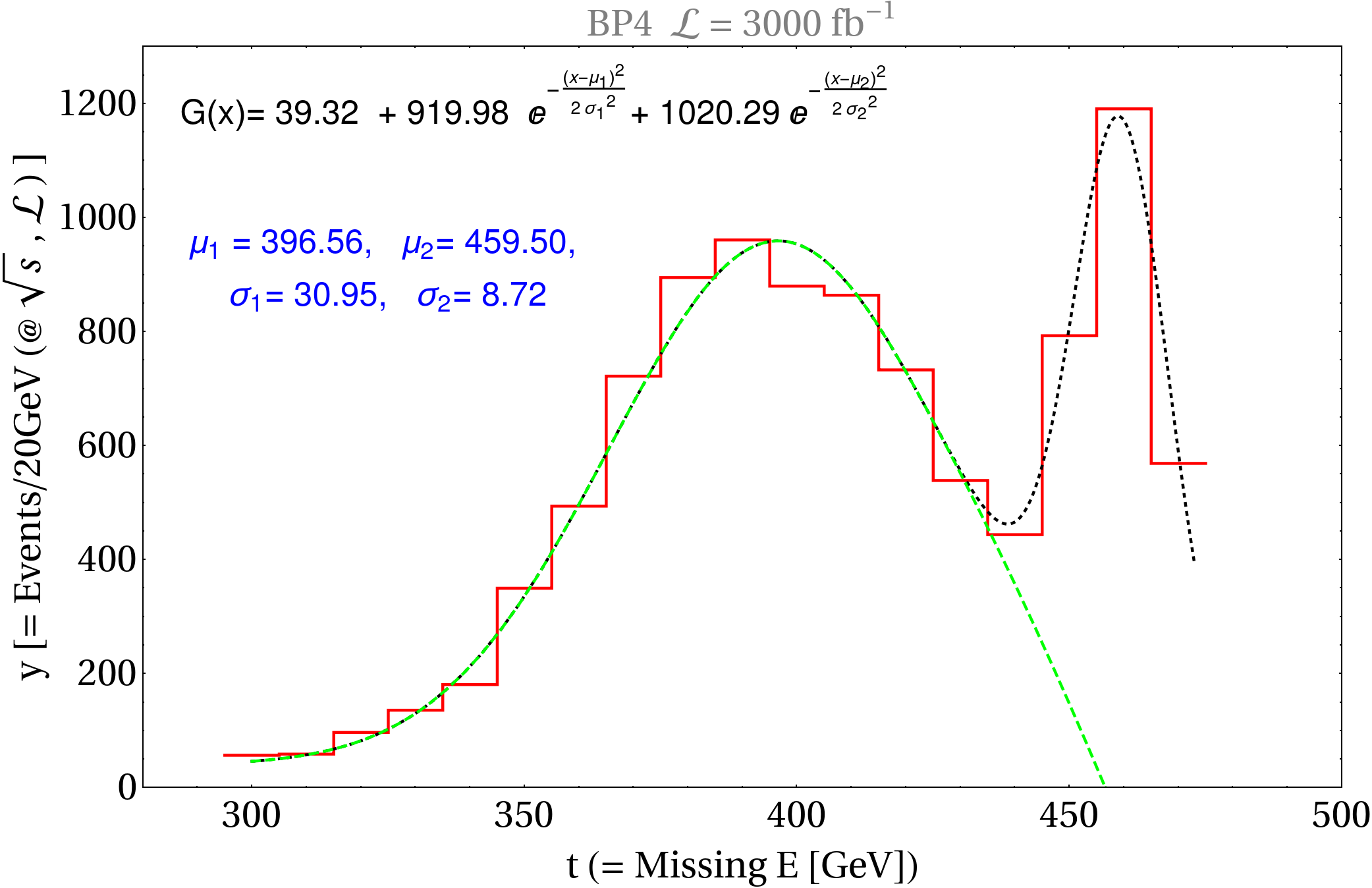}}
    
  \caption{Same as Fig.~\ref{scbp1} for BP2 (a), BP3 (b) and BP4 (c) at $\mathcal{L}=3000~{\rm fb}^{-1}$.}
  \label{bp2bp3bp4}
 \end{figure}

%
%
%
 
 \begin{figure}[htb!]
 
  \subfloat[]{\includegraphics[width=6.8cm,height=5.5cm]{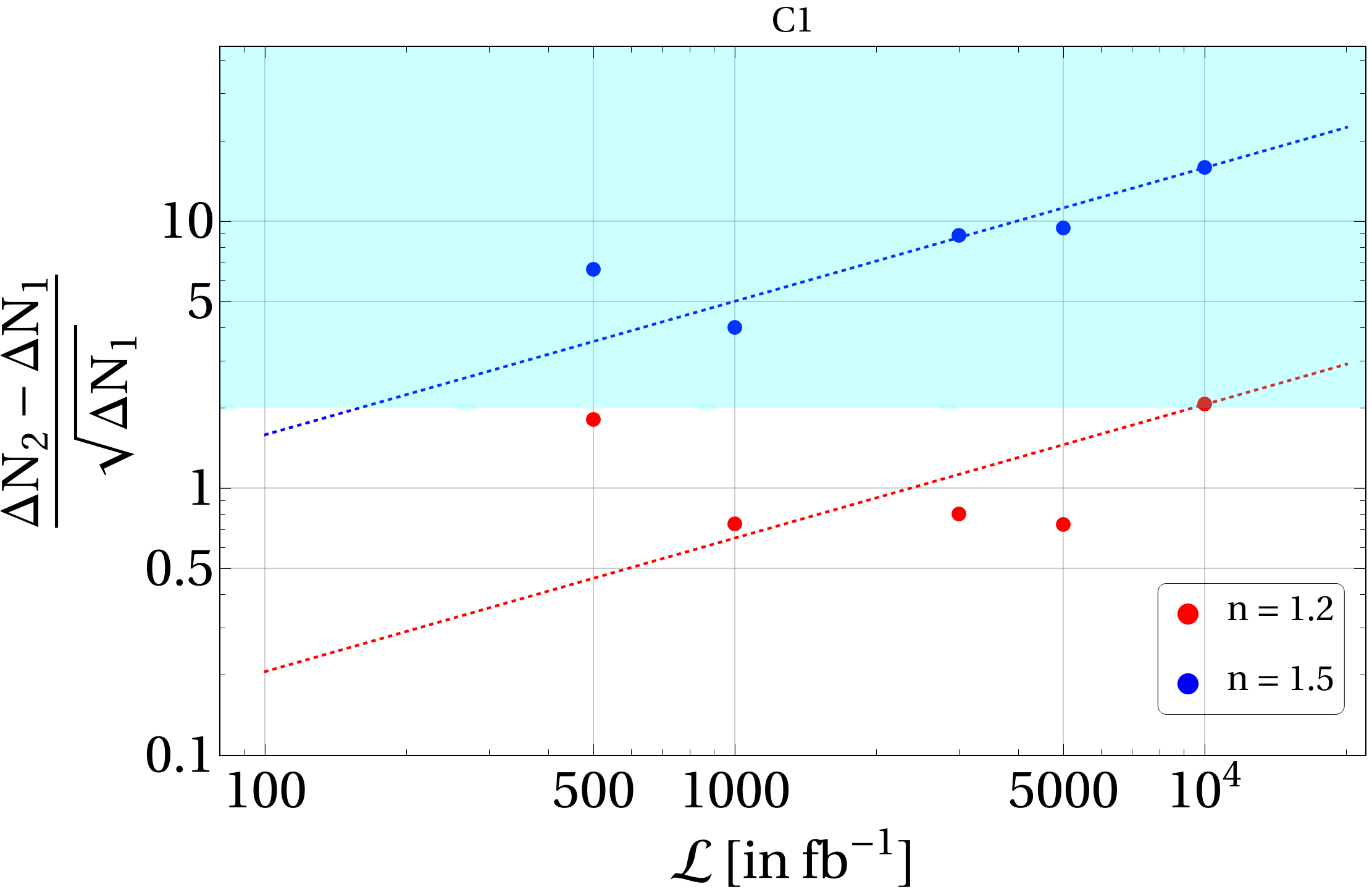}}~
    \subfloat[]{\includegraphics[width=6.8cm,height=5.5cm]{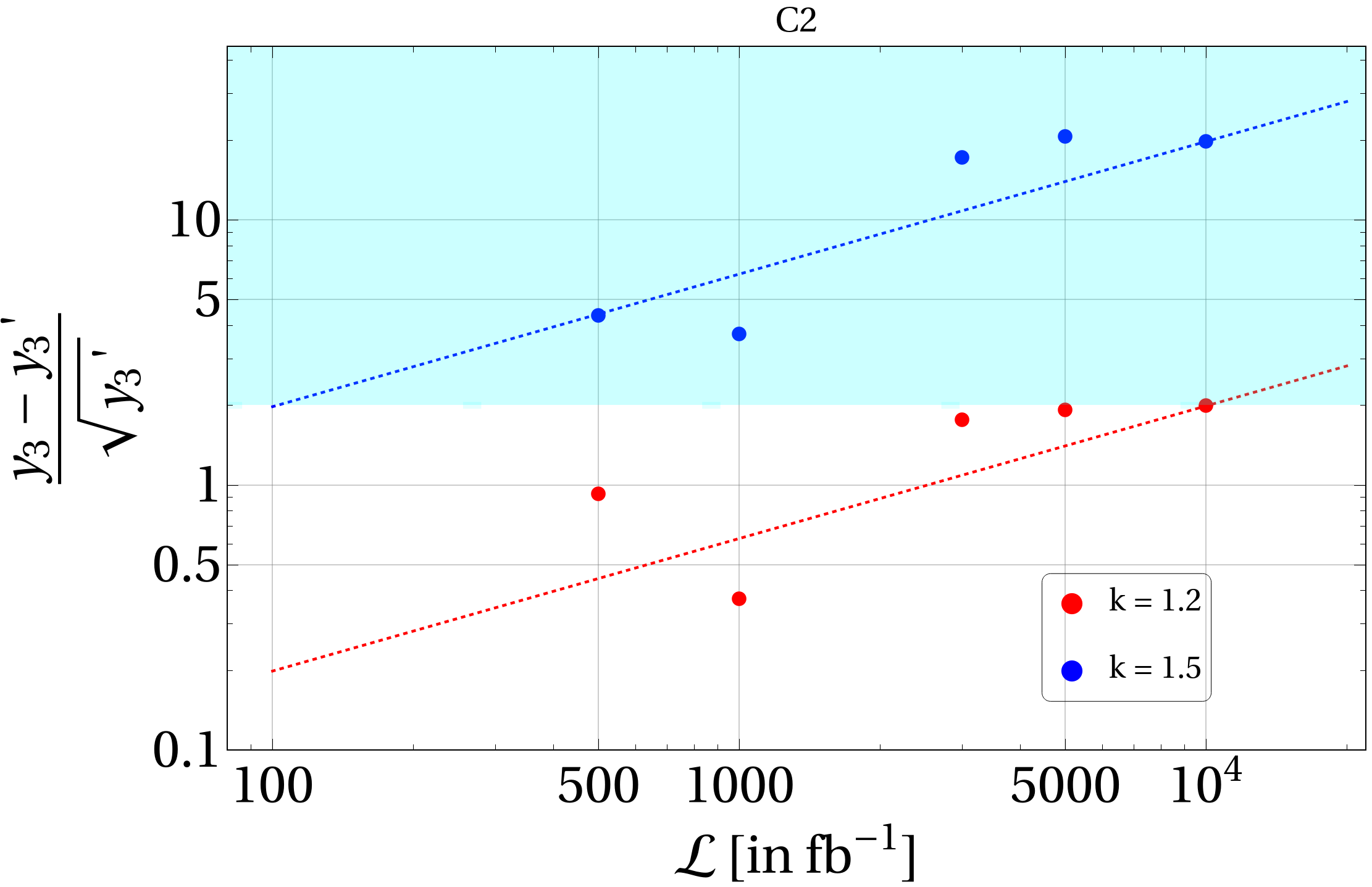}}\\
   \subfloat[]{\includegraphics[width=6.8cm,height=5.5cm]{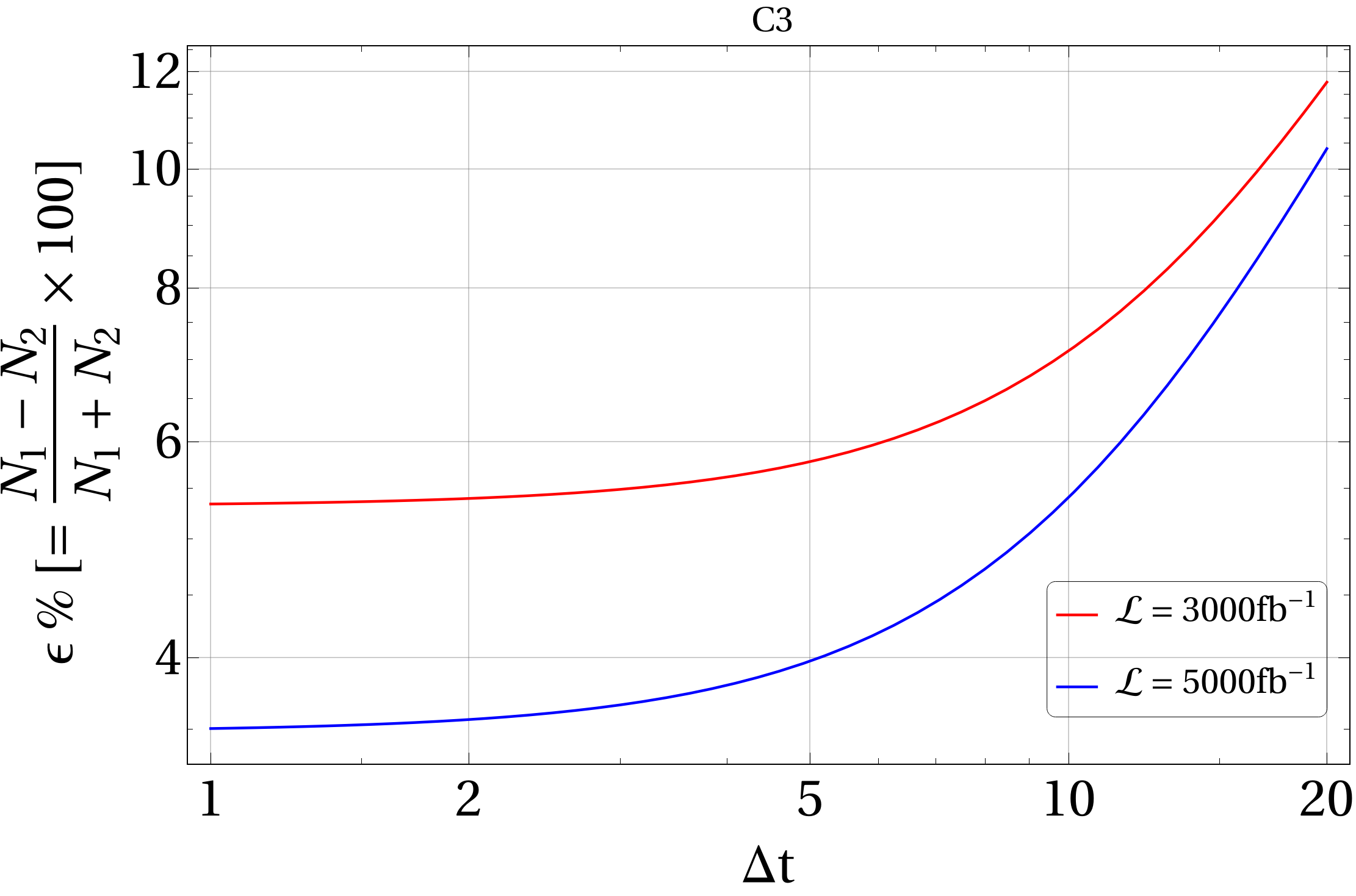}}~
 \subfloat[]{\includegraphics[width=6.8cm,height=5.5cm]{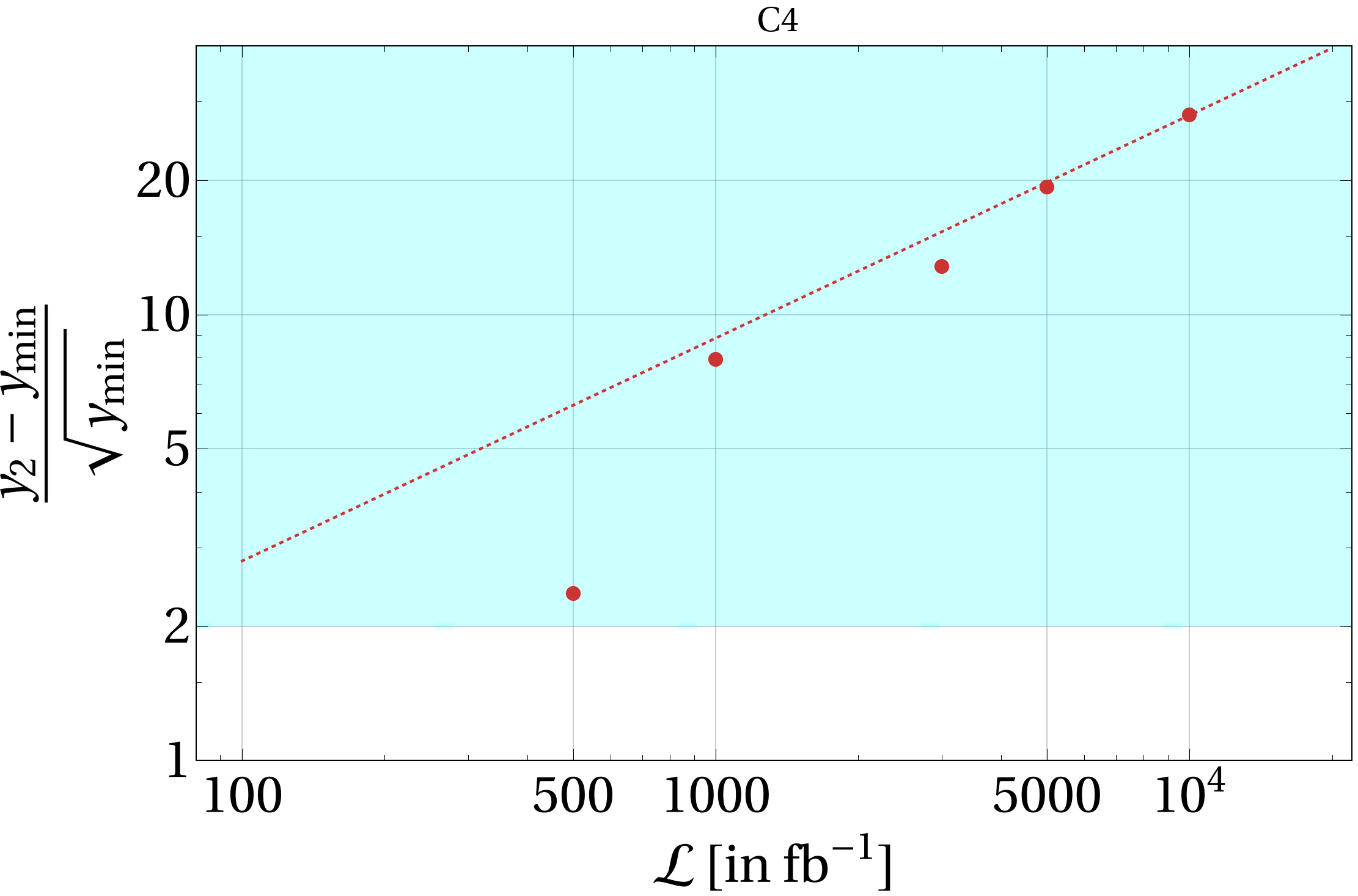}}
   
  \caption{Statistical conditions to distinguish the two peaks for BP1: (a) $R_{C1}$ as a function of ${\mathcal L}$ for different $n$ (defined in Equation~\ref{c1}), 
  (b) $R_{C2}$ vs $\mathcal{L}$ for different $k$ (defined in Equation~\eqref{c2}), (c) $R_{C3}$ as a function of $\Delta t$ (defined in Equation~\eqref{c3}) and (d) 
  $R_{C4}$ vs. $\mathcal{L}$ (defined in Equation~\eqref{c4}). Sky blue shaded region indicates $R_{C1,2,4}>2$. The dots indicate simulated points whereas the 
  dotted line indicates scaling with integrated luminosity as $\sim \sqrt{\mathcal{L}}$.}
  \label{sc1lum}
 \end{figure}
 
 
  \begin{figure}[htb!]
 
  \subfloat[]{\includegraphics[width=6.8cm,height=5.5cm]{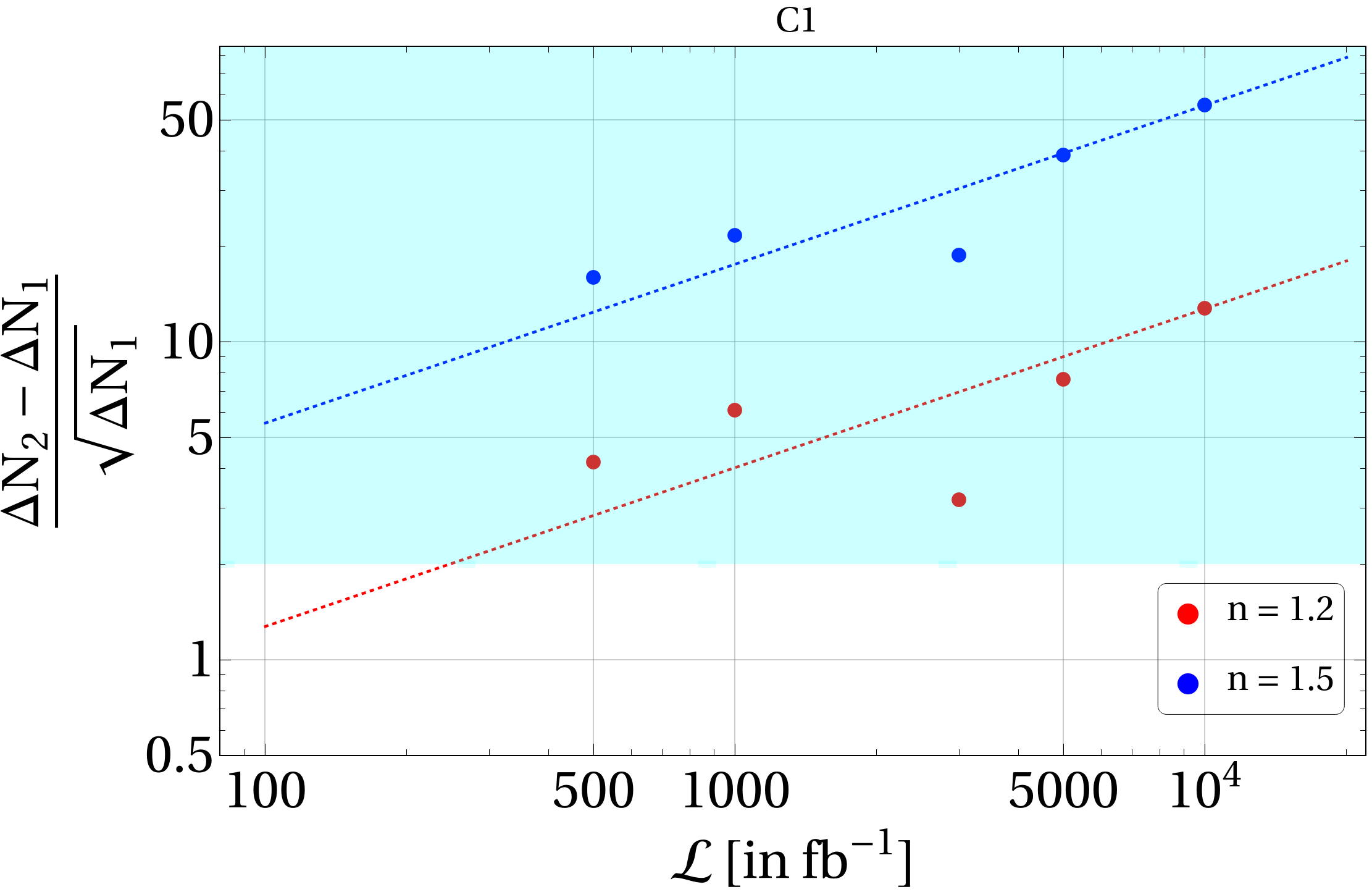}}~
    \subfloat[]{\includegraphics[width=6.8cm,height=5.5cm]{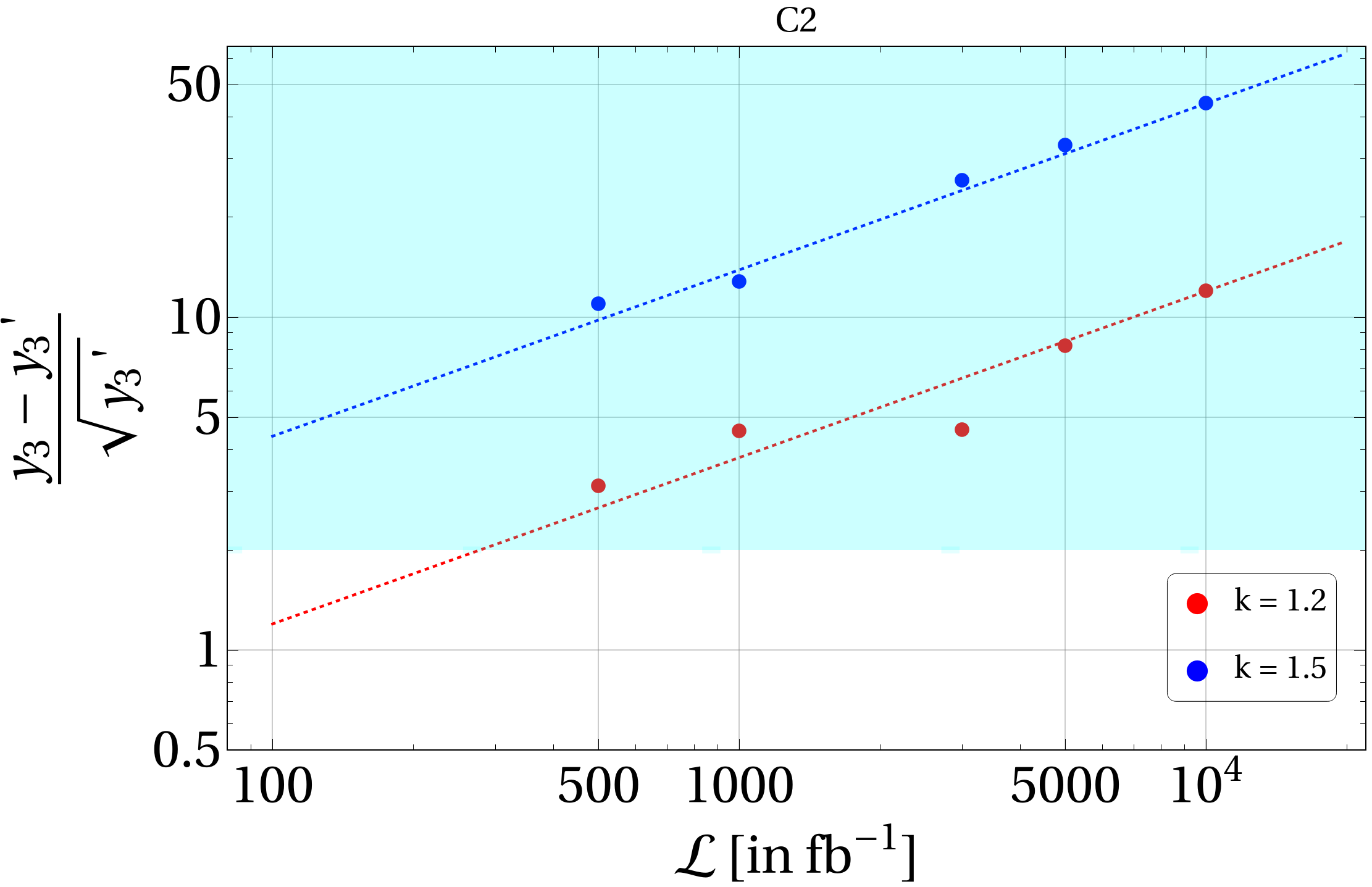}}\\
    \subfloat[]{\includegraphics[width=6.8cm,height=5.5cm]{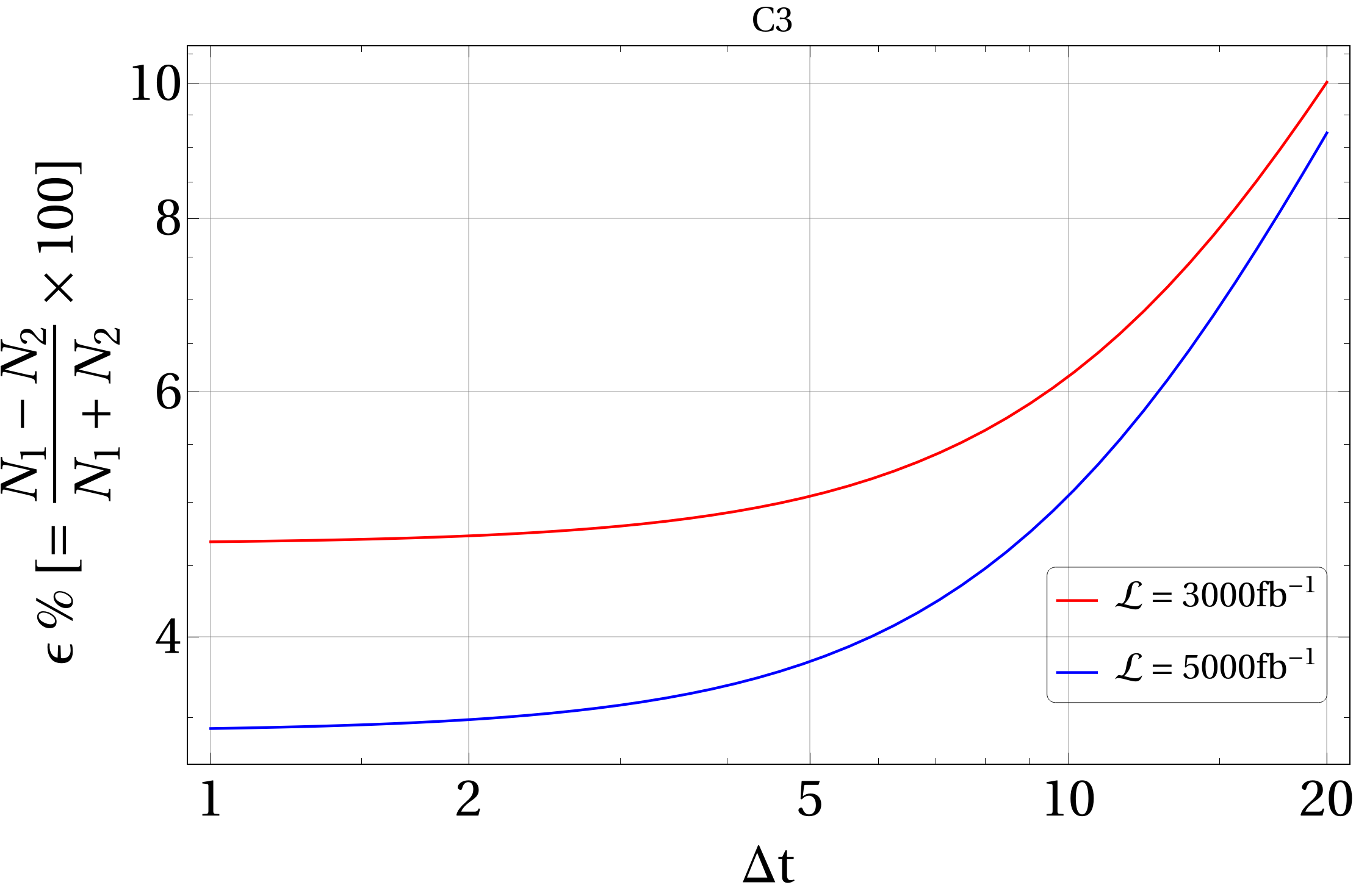}}~
 \subfloat[]{\includegraphics[width=6.8cm,height=5.5cm]{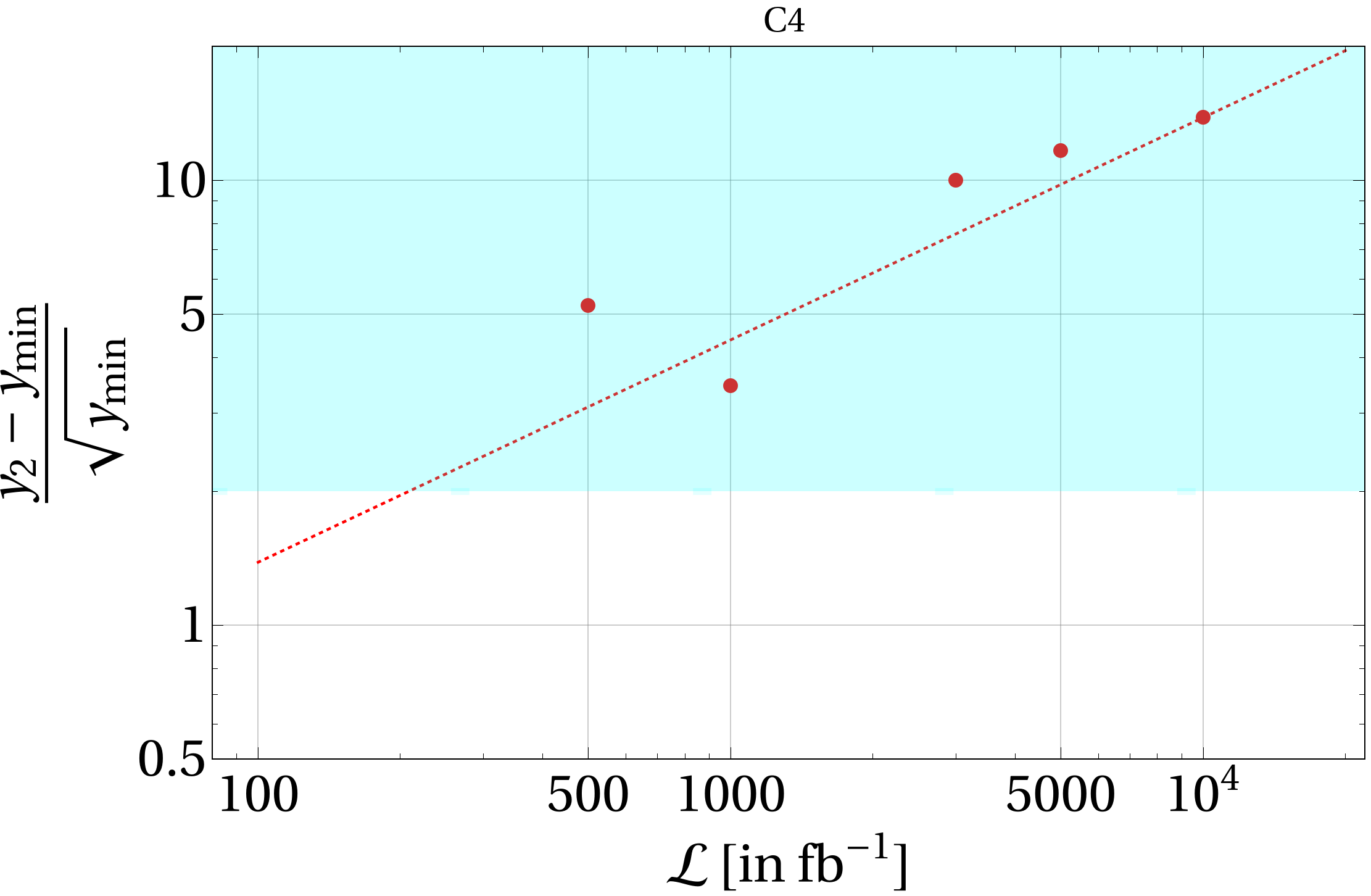}}~
  \caption{Same as Fig.~\ref{sc1lum} but for BP2.}
  \label{sc2lum}
 \end{figure}

 \begin{figure}[htb!]
 
  \subfloat[]{\includegraphics[width=6.8cm,height=5.5cm]{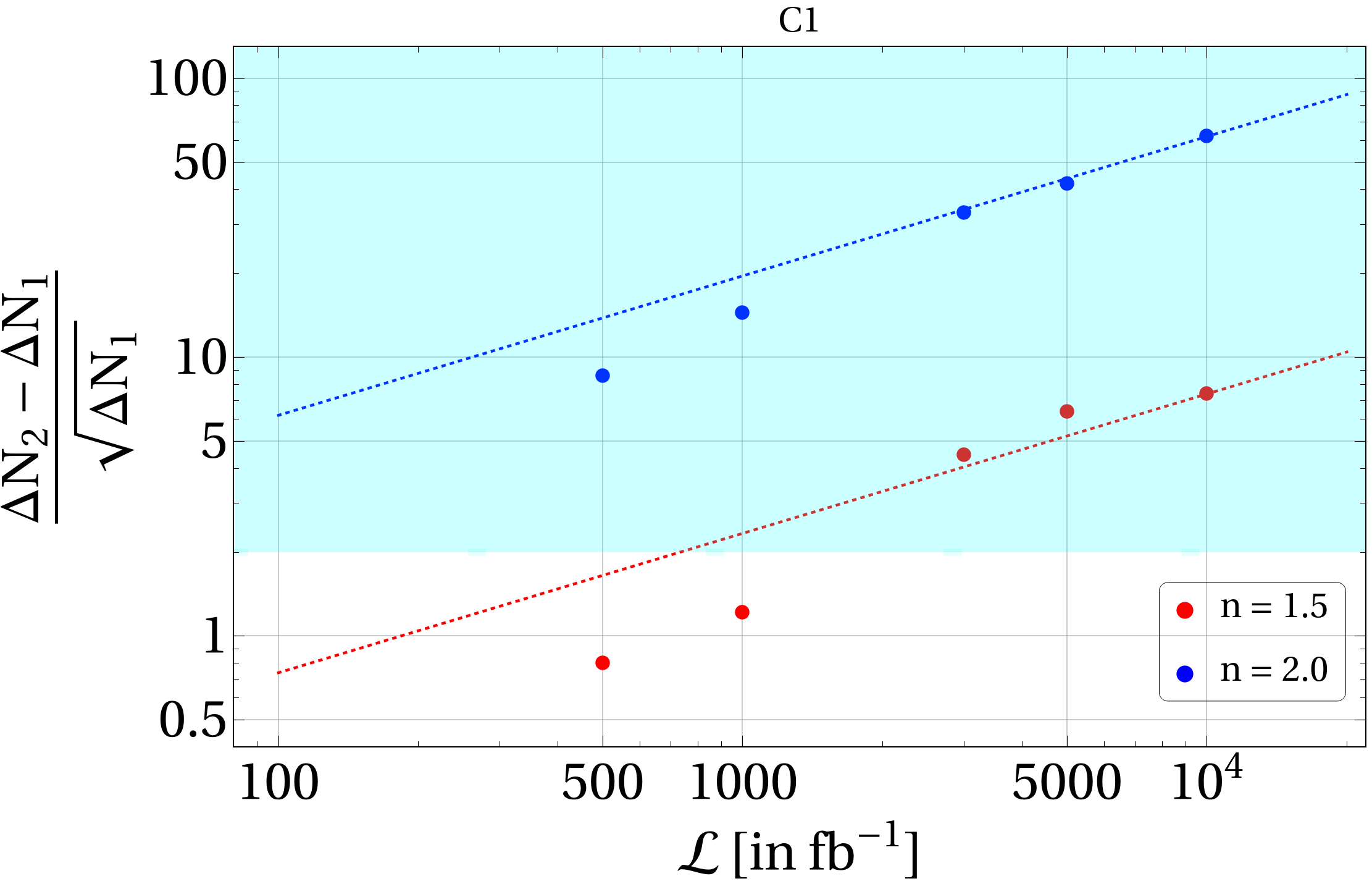}}~
    \subfloat[]{\includegraphics[width=6.8cm,height=5.5cm]{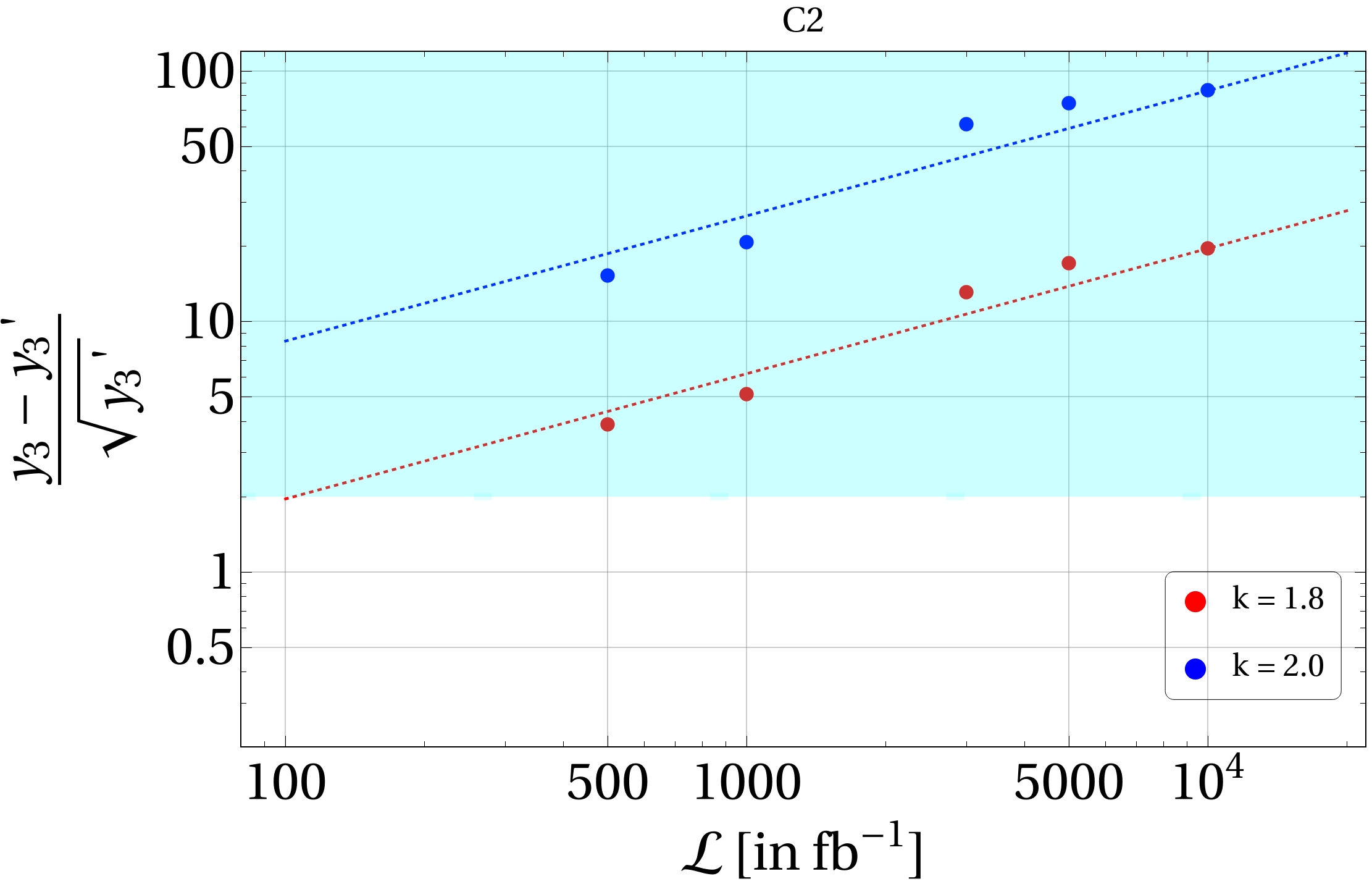}}\\

    \subfloat[]{\includegraphics[width=6.8cm,height=5.5cm]{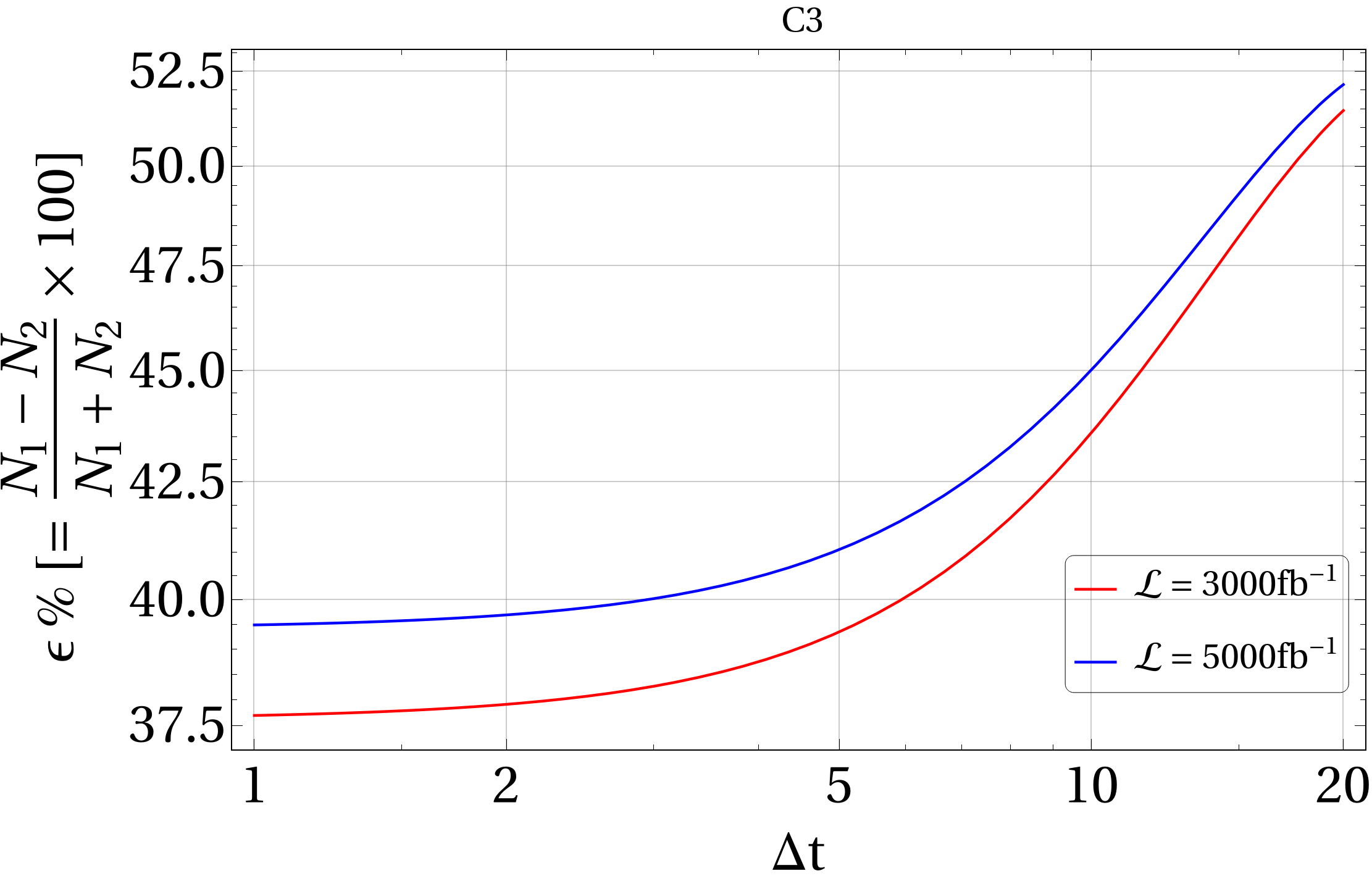}}~
 \subfloat[]{\includegraphics[width=6.8cm,height=5.5cm]{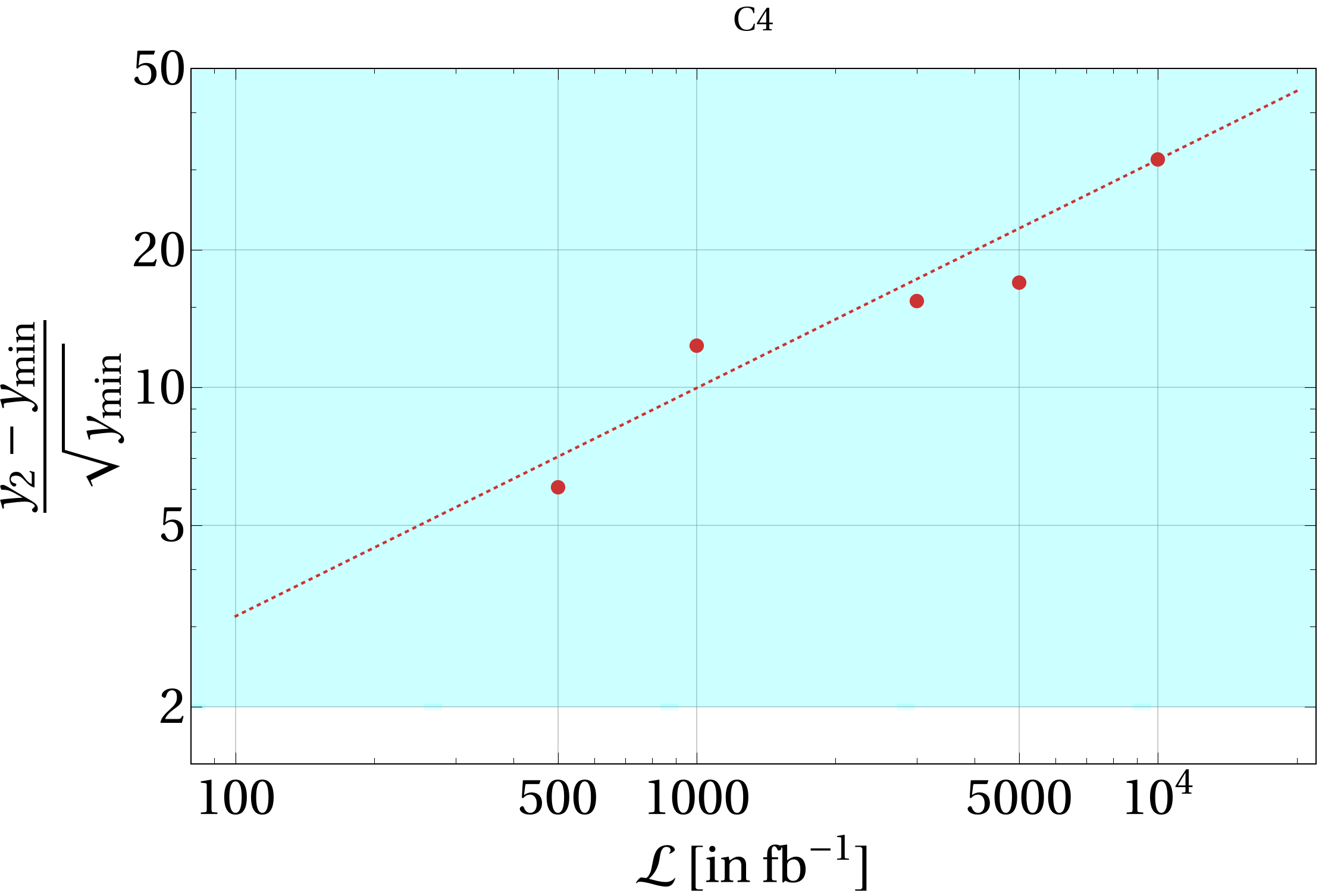}}
   
  \caption{Same as Fig.~\ref{sc1lum} but for BP3. }
  \label{sc3lum}
 \end{figure}
  \begin{figure}[htb!]
 
  \subfloat[]{\includegraphics[width=6.8cm,height=5.5cm]{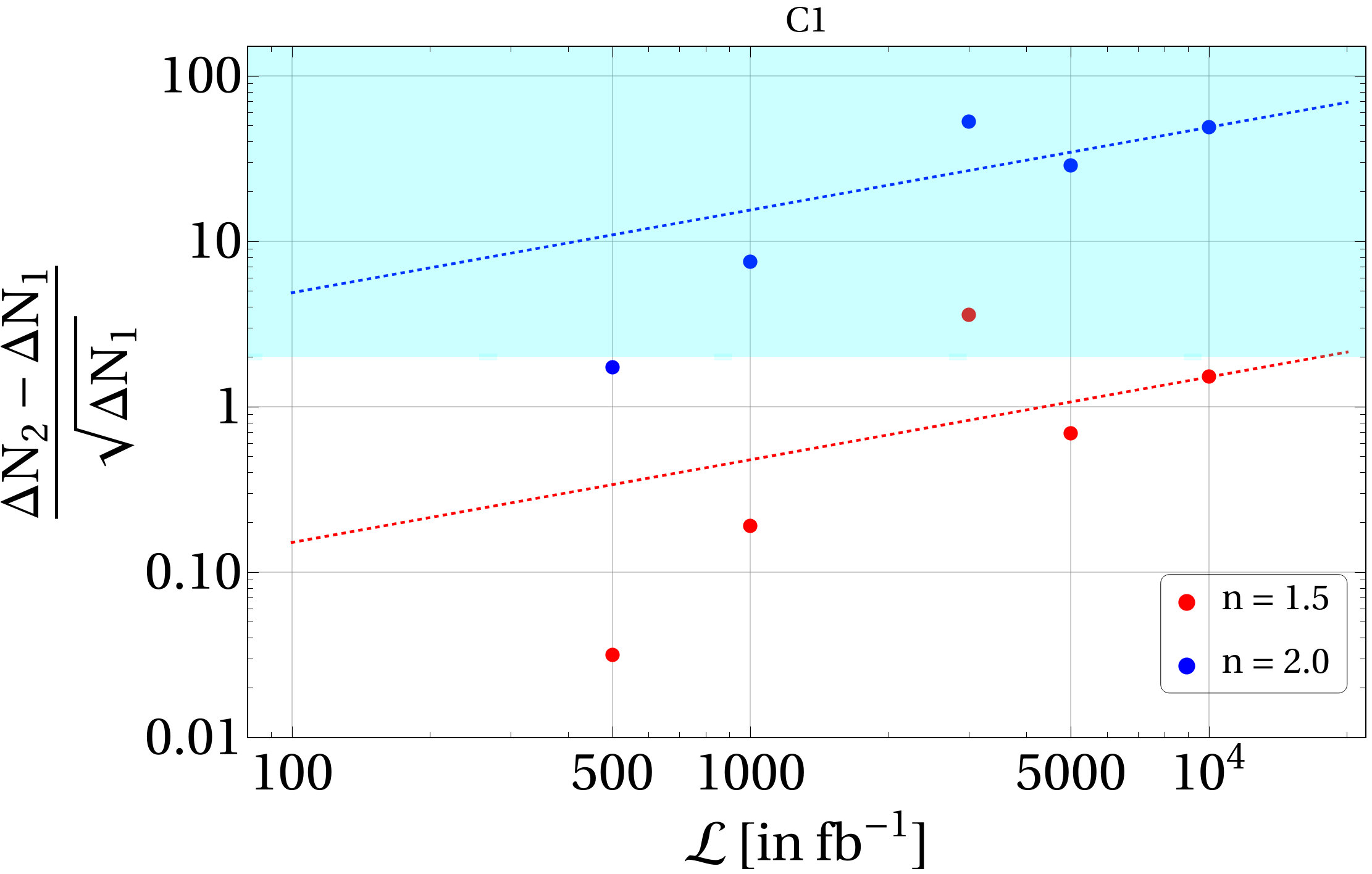}}~
    \subfloat[]{\includegraphics[width=6.8cm,height=5.5cm]{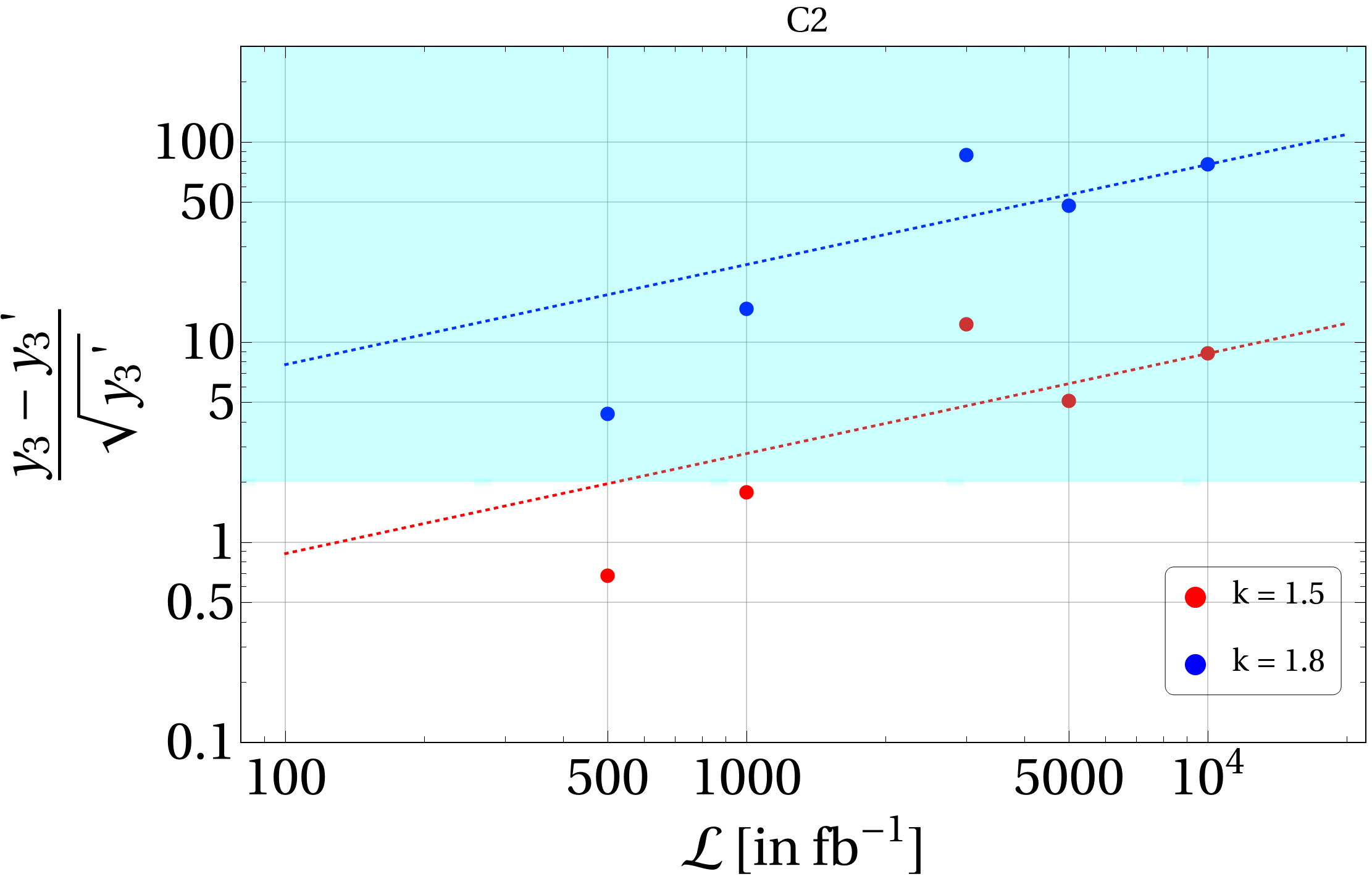}}\\

    \subfloat[]{\includegraphics[width=6.8cm,height=5.5cm]{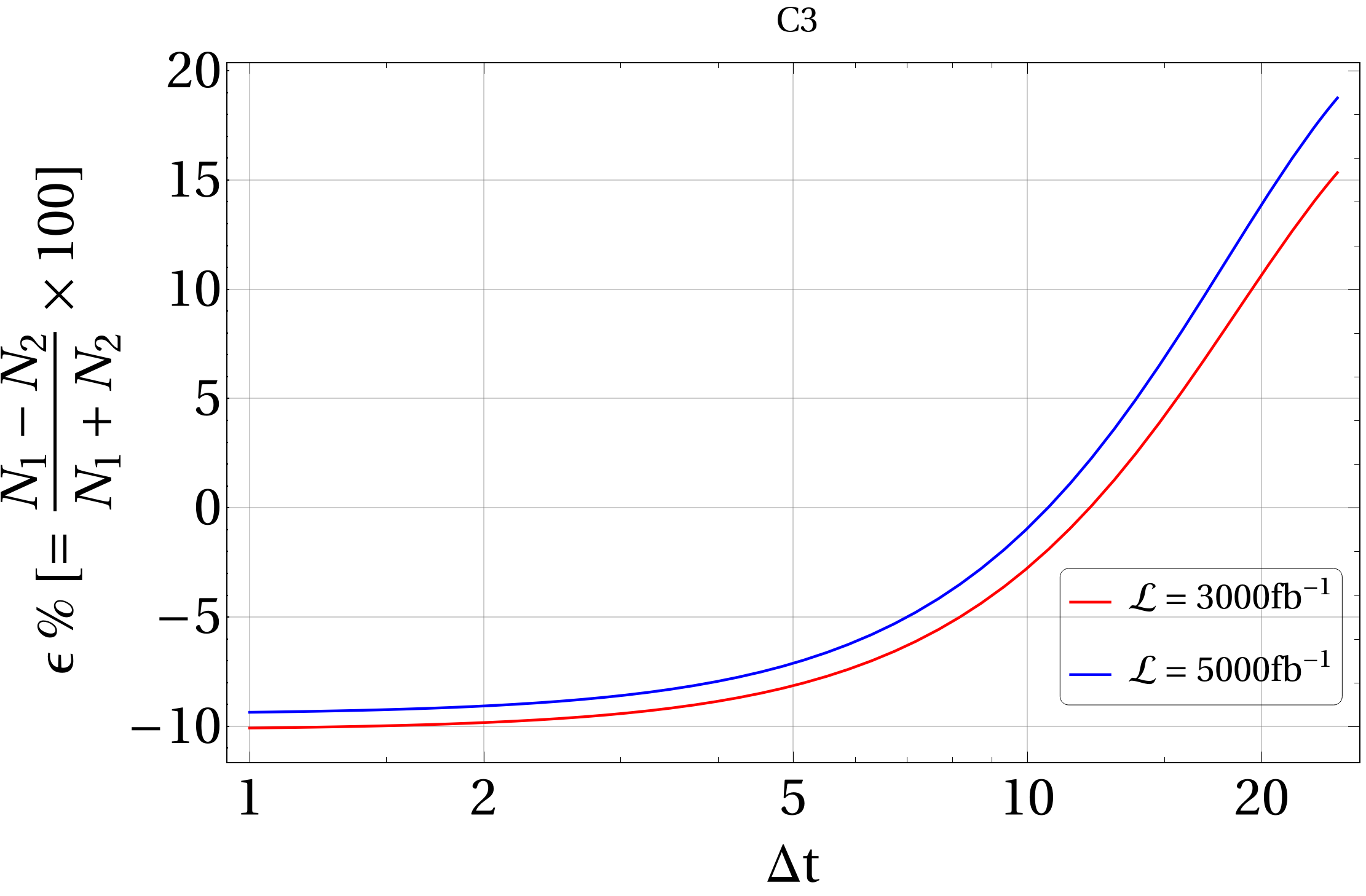}}~
 \subfloat[]{\includegraphics[width=6.8cm,height=5.5cm]{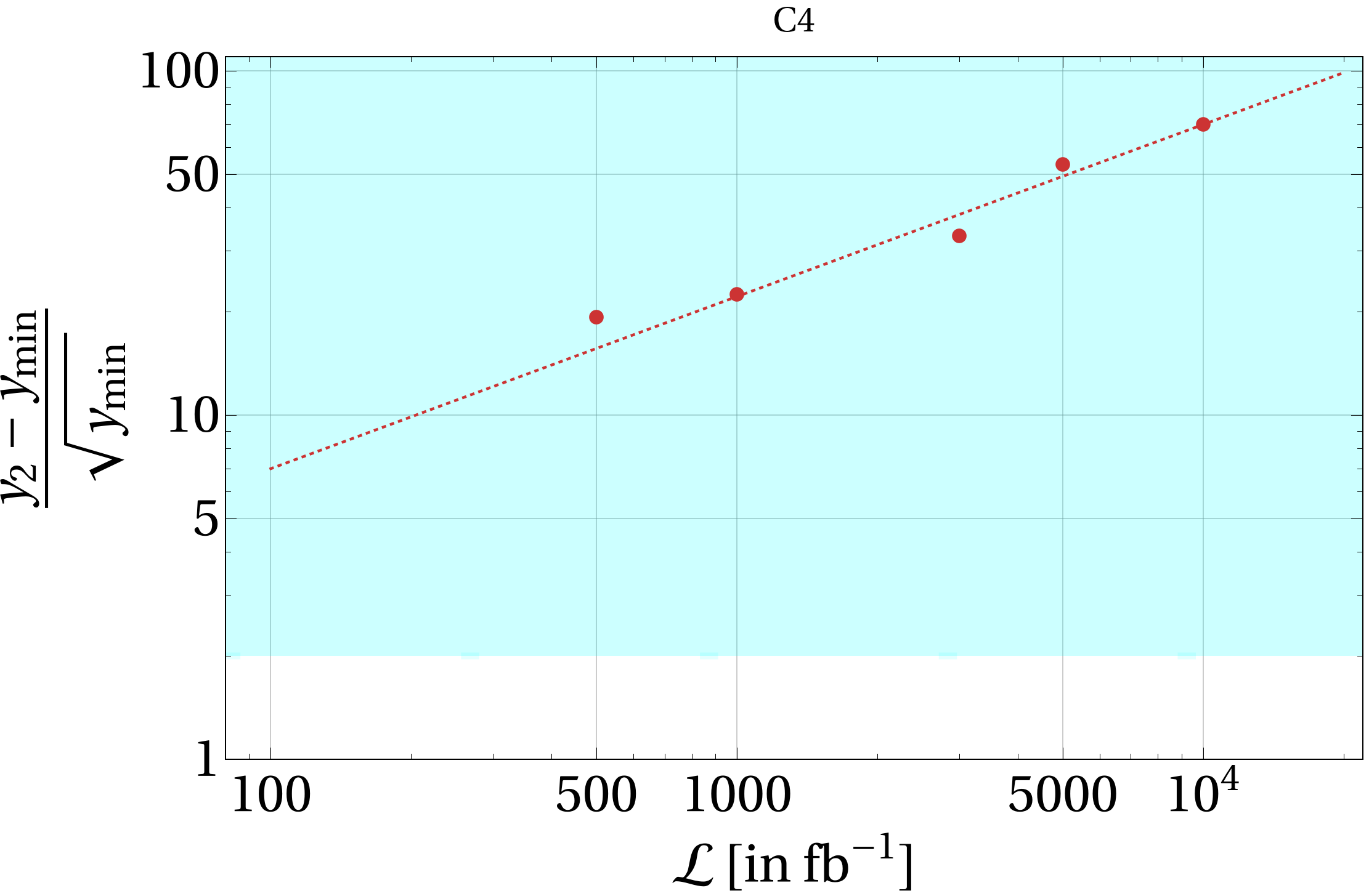}}~
   
  \caption{Same as Fig.~\ref{sc1lum} but for BP4. }
  \label{sc4lum}
 \end{figure}

 
The resulting $\slashed{E}$ distribution for BP1 along with Gaussian fits are shown in Fig.~\ref{scbp1} 
for various benchmark luminosities. Those for BP2, BP3 and BP4 are provided in 
Fig.~\ref{bp2bp3bp4} in the left, right and bottom panel figures for a fixed $\mathcal{L}=3000~{\rm fb}^{-1}$.
All the relevant parameters like $\mu_{1,2}$, $\sigma_{1,2}$ and $A_{1,2}$ for the Gaussian fit (following Eq.~\eqref{eq:gf}) 
are mentioned in the figure inset. The goodness of the fit is revealed from the parameter $\chi^2$ as well as $\chi^2$ per degrees of freedom, which are also 
mentioned in Fig.~\ref{scbp1}. Further details on the Gaussian fitting methodology along with $\chi^2$ evaluation can be found in Appendix \ref{app3}. 
One can see that at low luminosities, meagre statistics leads to larger fluctuations 
in the distributions, and in the resulting accuracy of Gaussian fitting. They get better with higher luminosities; 
for example, $\chi^2$ per degrees of freedom in Fig.~\ref{scbp1}(a) with $\mathcal{L}=500~{\rm fb}^{-1}$ is 1.91, while in Fig.~\ref{scbp1}(d), 
with $\mathcal{L}=5000~{\rm fb}^{-1}$, it turns out to be 0.76.

We are now all set to discuss the conditions for distinguishing two peaks in each of these cases. 
In Fig.~\ref{sc1lum}, we analyse BP1 in details. Fig.~\ref{sc1lum} (a) shows validation of C1 condition by plotting $R_{C1}$ 
as a function of integrated luminosity $\mathcal{L}$ for different values of $n$. The dots indicate the simulated points. 
Assuming that the point with highest luminosity provides the most accurate value of $R_{C1}$, 
we have scaled for other luminosities as $R_{C1} \sim \sqrt{\mathcal{L}}$ 
by the fitted line (which appears as a straight line in the log-log plot).
The sky-blue shaded region where $R_{C1}>2$, can be achieved for $n=1.5$ with a moderate
luminosity ($\mathcal{L} \sim 500 ~{\rm fb}^{-1}$). This indicates a very prominent presence of a second peak within 
$1.5 \sigma$ vicinity of the first one. It is obvious that $R_{C1}$ increases with $\mathcal{L}$ as statistics enhance 
(evident from Eqn.~\eqref{c1}); $R_{C1}$ also increases with $n$ as we approach the second peak. 
In Fig.~\ref{sc1lum}(b) we examine condition C2, where $R_{C2}$ is plotted as a function of $\mathcal{L}$ for various $k$ values. Again, we see that 
the sky-blue shaded region where $R_{C2}>2$, is achieved for $k=1.5$ with moderate luminosity. This means, number of events accumulated in
$k=1.5 \sigma$ apart from the first peak is larger than the number of events without the presence of the second peak by 2$\sigma$ or more. Again, as we go further 
away from the first peak, i.e. the larger the $k$ is, the larger $R_{C2}$ becomes\footnote{This is true {\it within} the range of the second peak. 
The value of $k$, where $R_{C2}$ becomes maximum indicate the presence of the second peak and marks the separation between the two peaks.}. 
The dependence of $R_{C2}$ on $\mathcal{L}$ is obvious, the larger is $\mathcal{L}$, the easier it is to sense the 
presence of a second peak. In Fig.~\ref{sc1lum}(c), we show the variation of $R_{C3}$ as a function $\Delta t ~(\equiv \Delta \slashed{E})$ 
for two representative luminosities. One can see that $R_{C3}$ remains almost constant for a small range of $\Delta t$, as indicated by Eqn.~\eqref{c3}, 
which actually marks the difference in height of the two peaks. But $R_{C3}$ starts increasing after a point, which mostly indicates the difference 
in the thickness of the Gaussian distributions around the peaks, instead of the difference in the heights of the peaks. Finally we verify condition 
C4 in Fig.~\ref{sc1lum}(d), where $R_{C4}$ is evaluated as a function of $\mathcal{L}$. We see that it is easy to satisfy condition C4 than others 
as the difference between the number of events at the second peak and that of the minima between them easily goes beyond 2$\sigma$ 
($R_{C4}>2$) even at small luminosities. The enhancement of $R_{C4}$ with $\mathcal{L}$ is also obvious. 
Figs.~\ref{sc2lum}, \ref{sc3lum}, \ref{sc4lum} present similar analysis for BP2, BP3 and BP4 respectively, where the features broadly remain the same.


A comparison between our chosen benchmarks in the light of the aforementioned distinction criteria is in order. $R_{C1}$ takes the largest value for BP2, 
since the relative height as well as the width of the second peak is large w.r.t the first peak in this case. 
On the other hand, $R_{C2}$ is highest for BP4, since the height of the second peak is largest there. 
$R_{C3}$ is largest for BP3, due to significant asymmetry in the heights of the two peaks. Consequently $R_{C1}$ and 
$R_{C2}$ take the lowest value for a specific $n$ and ${\cal L}$ in this case. $R_{C4}$ is maximum for BP4 due to large height and 
small width of the second peak. In general, BP1 performs best under C3, BP2 under C1 and BP4 under C2 as well as C4 criteria. 
BP3 does worse for all conditions, although C4 confirms the presence of a second peak clearly. This comparative analysis also exemplifies 
the qualitative distinction between the C1-C4 conditions and how each of them individually or together can be useful for distinguishing 
the two peaks.

 \begin{figure}[htb!]
 
  \subfloat[]{\includegraphics[width=6.8cm,height=5.5cm]{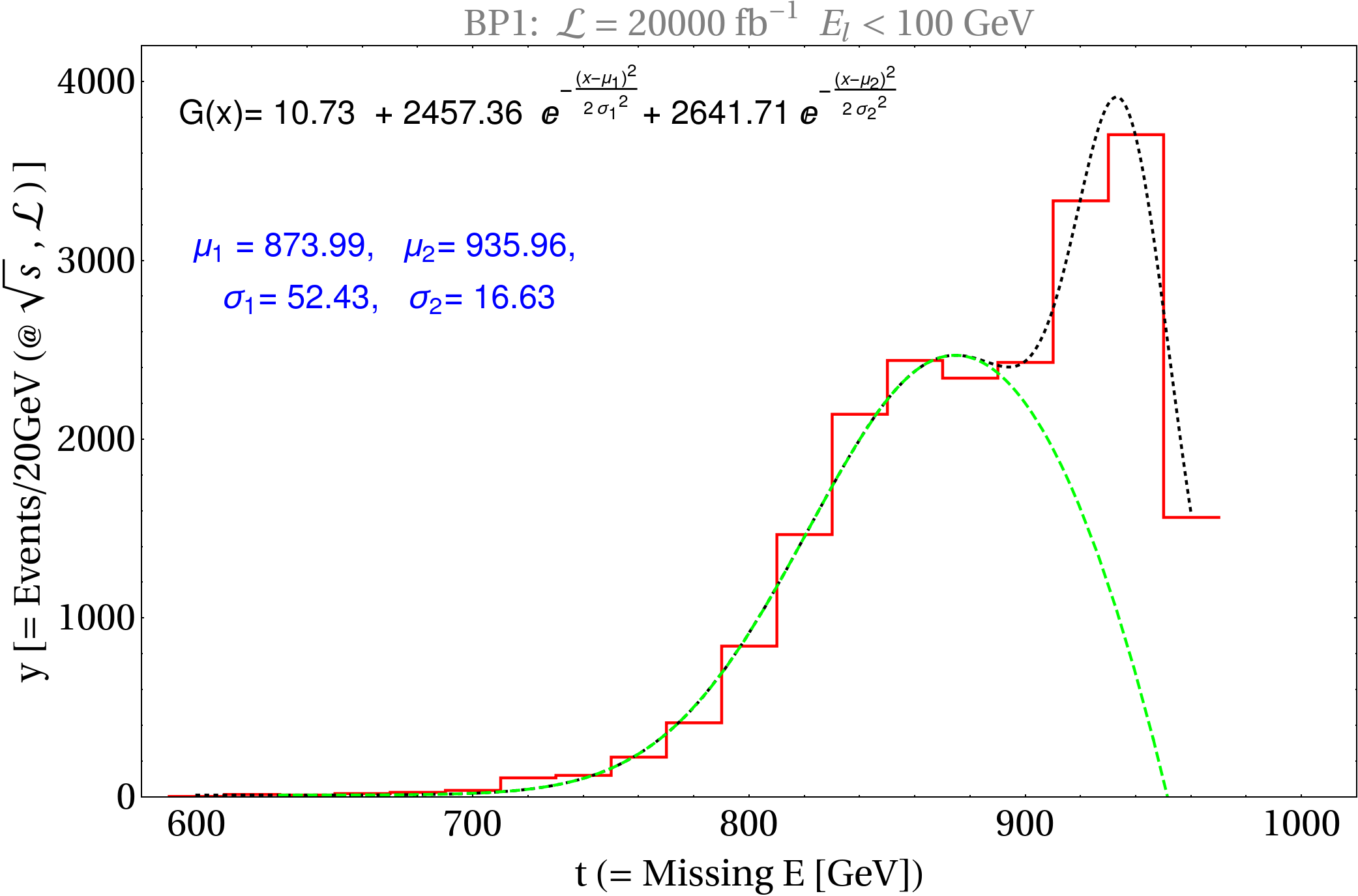}}~
    \subfloat[]{\includegraphics[width=6.8cm,height=5.5cm]{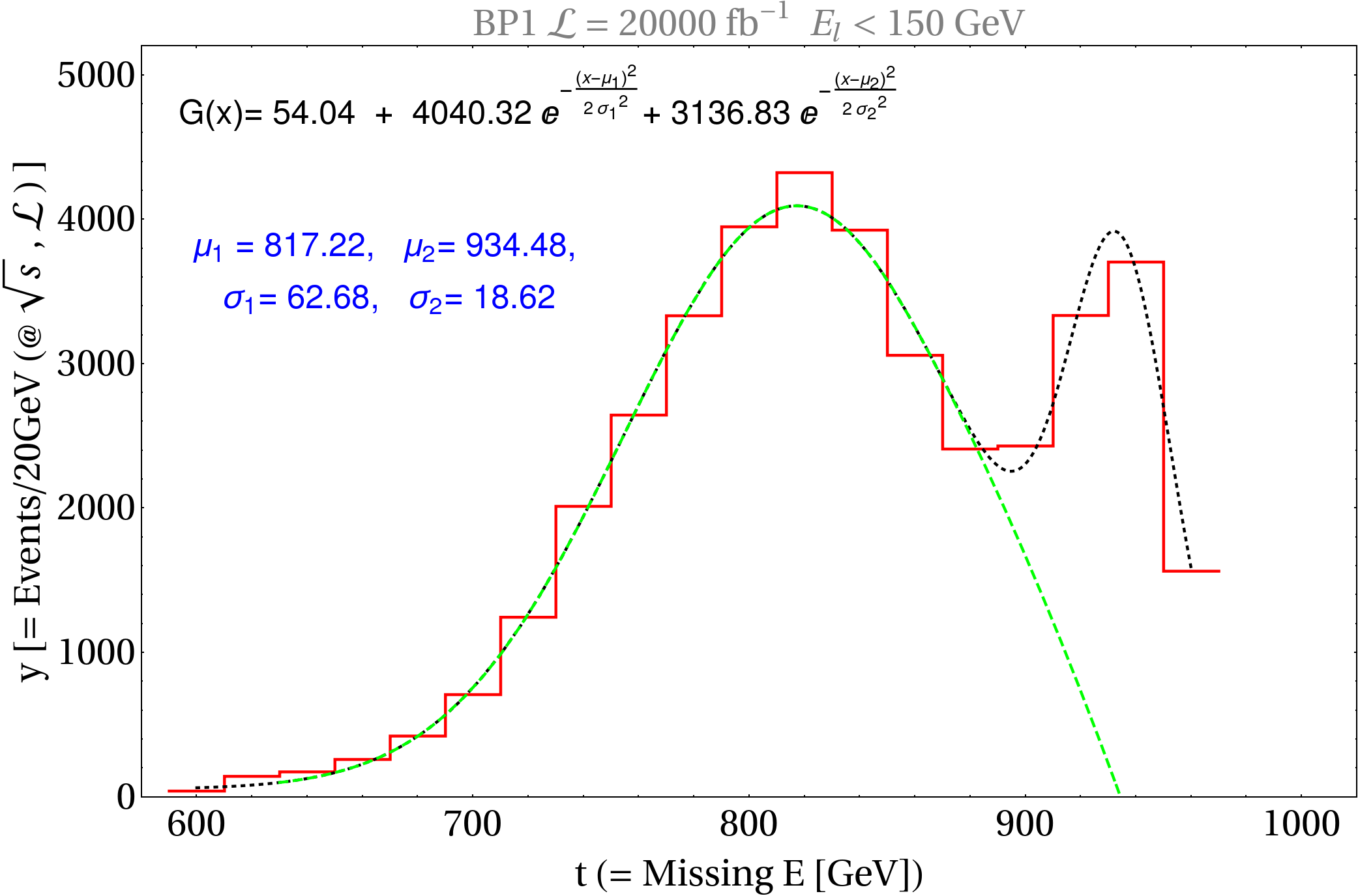}}~ \\
   
   \subfloat[]{\includegraphics[width=6.8cm,height=5.5cm]{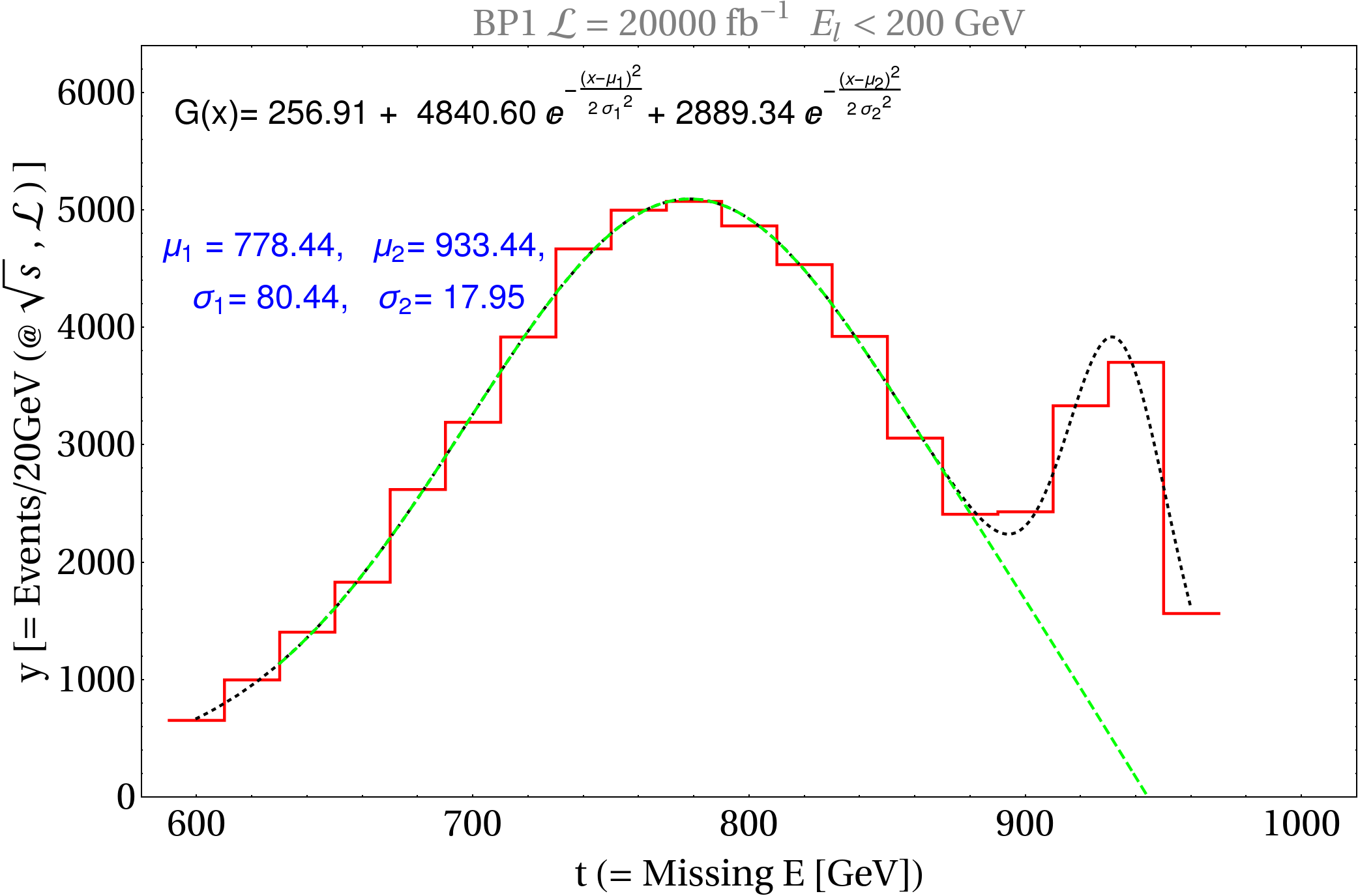}}
  
  \caption{$\slashed{E}$ distribution and Gaussian fitting for BP1 together with SM background with different cuts on leading lepton energy: (a) $E_{\ell_1}<$ 100 GeV, 
  (b) $E_{\ell_1}<$ 150 GeV and (c) $E_{\ell_1}<$ 200 GeV. The polarization of initial beams is chosen as P3 and $\mathcal{L} = 20 ab^{-1}$ has been used for illustration.}
  \label{sclepenergy}
 \end{figure}

\begin{figure}[htb!]
 
  \subfloat[]{\includegraphics[width=6.8cm,height=5.5cm]{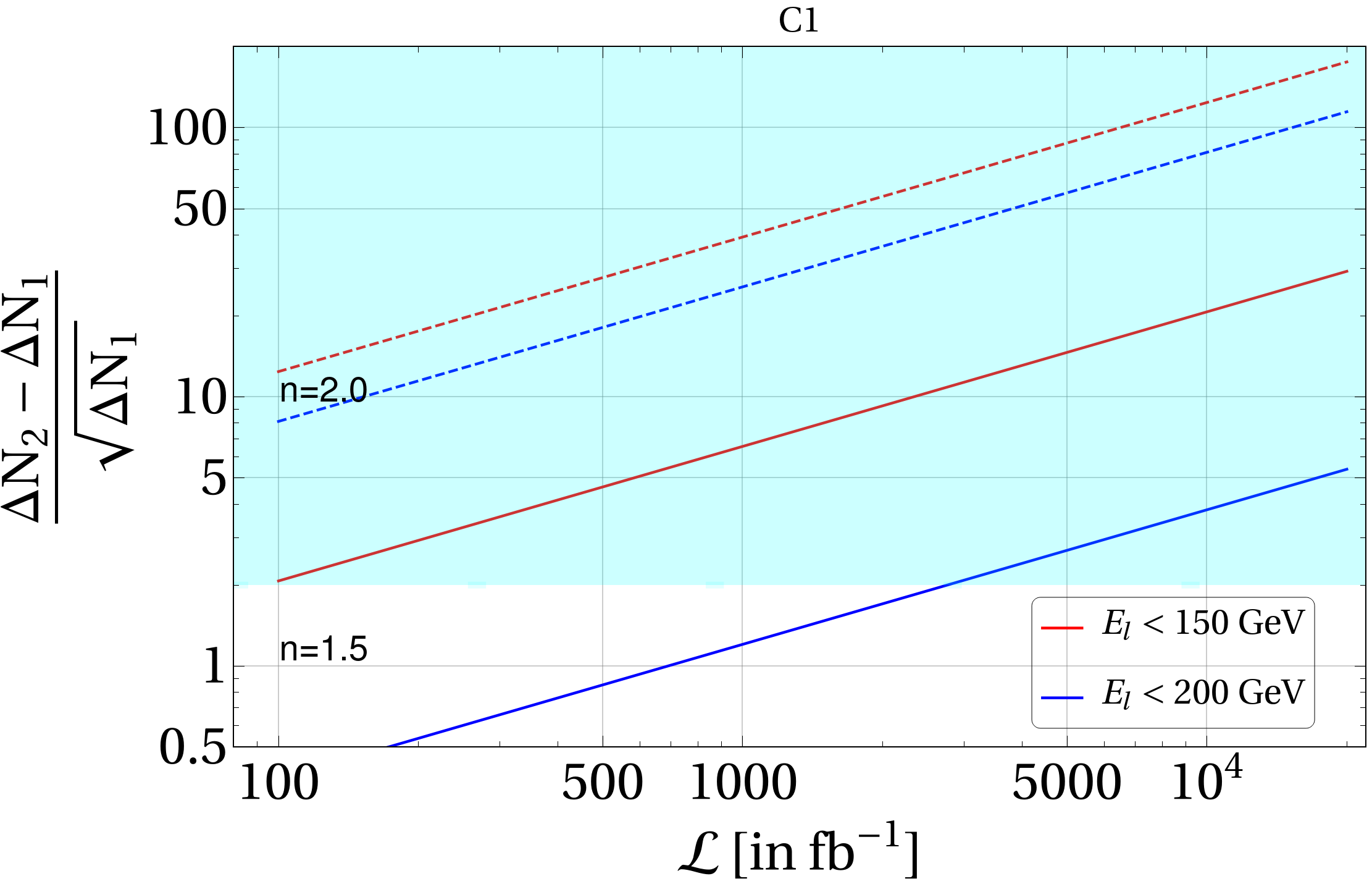}}~
    \subfloat[]{\includegraphics[width=6.8cm,height=5.5cm]{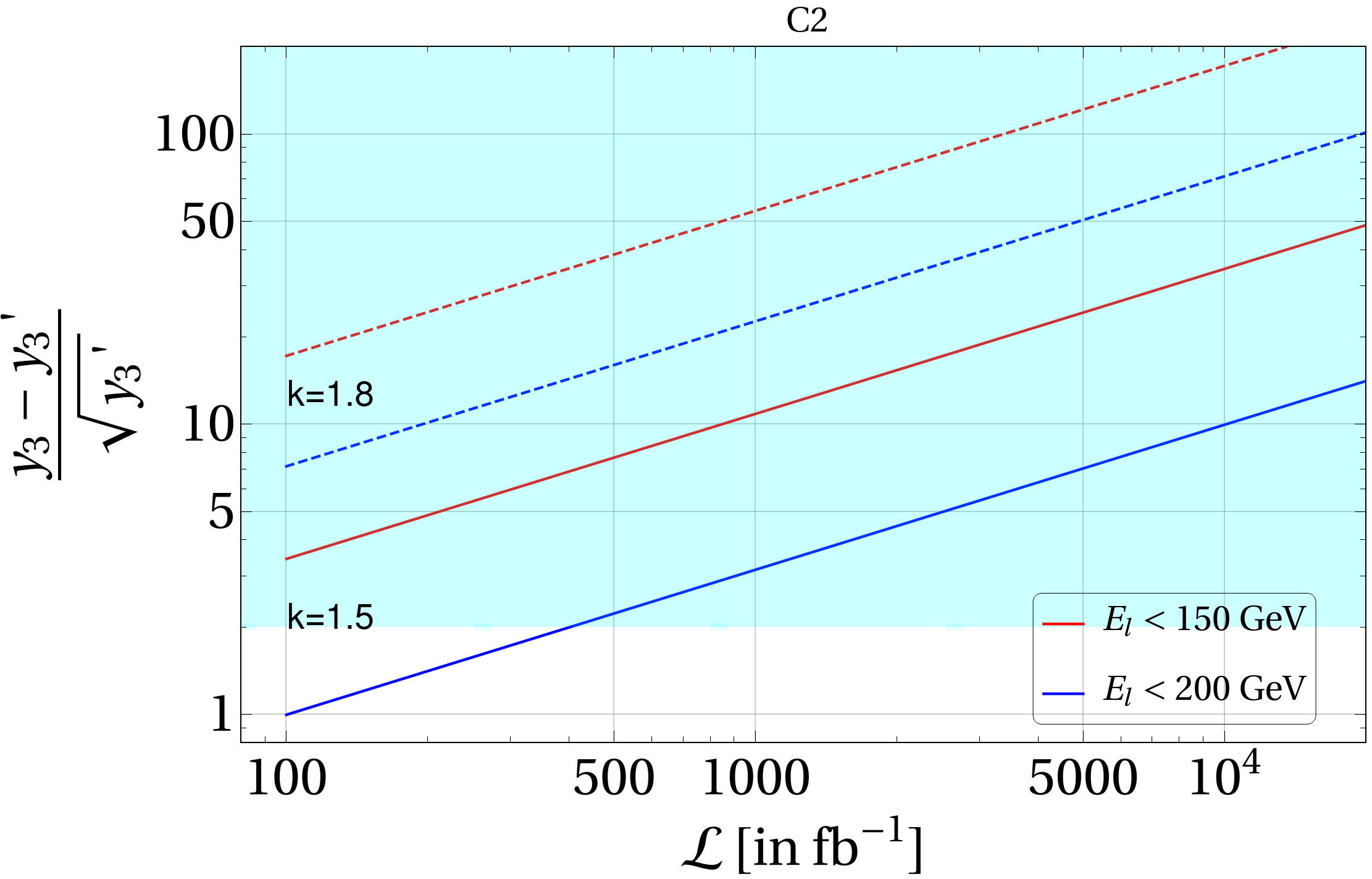}}~

    \subfloat[]{\includegraphics[width=6.8cm,height=5.5cm]{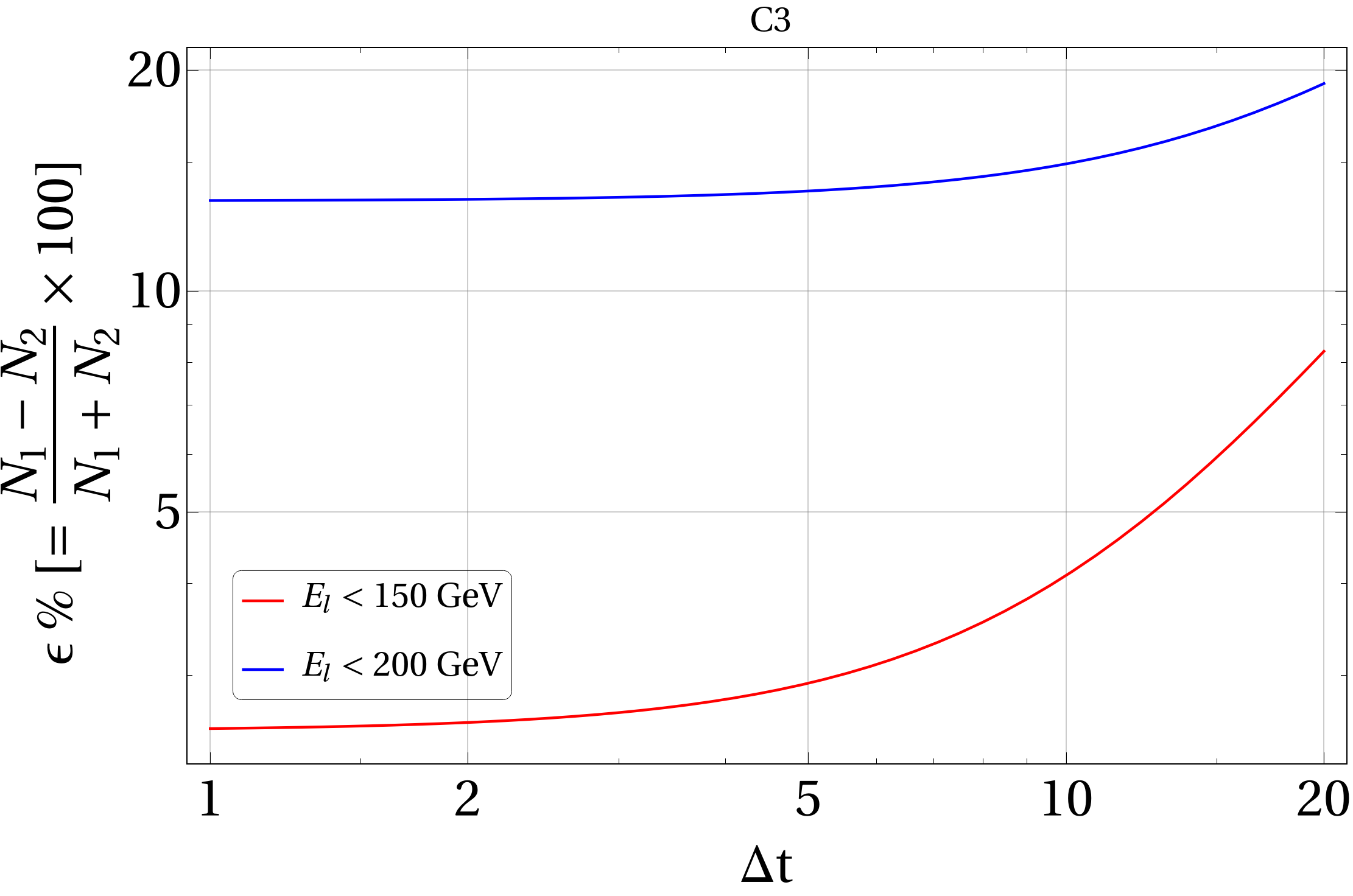}}~
    \subfloat[]{\includegraphics[width=6.8cm,height=5.5cm]{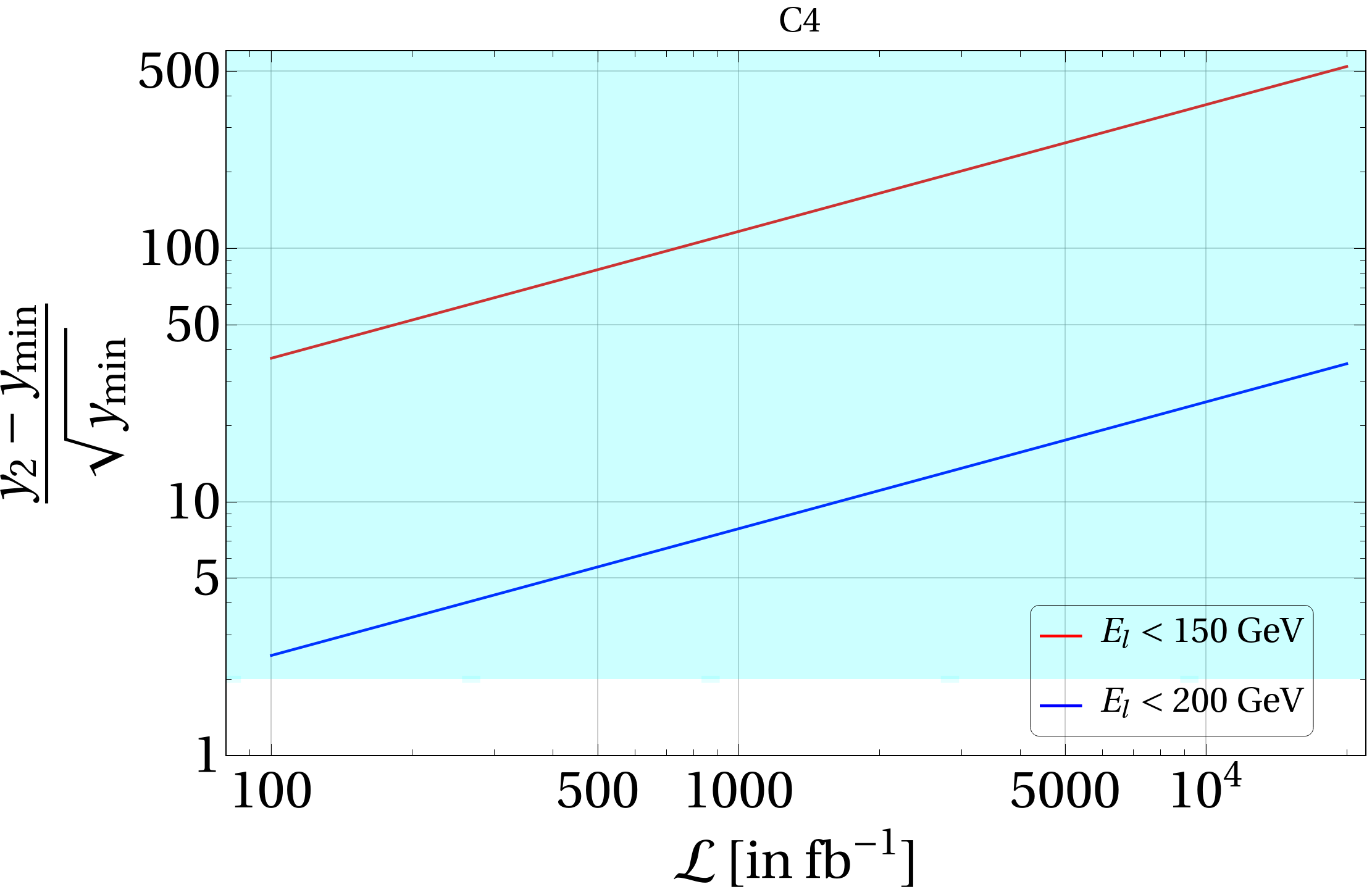}}~
  
  \caption{Validation of C1, C2, C3, C4 conditions for BP1 with two specified lepton energy cuts ($E_{\ell_1}<$ 150 GeV,  $E_{\ell_1}<$ 200 GeV): 
  (a) $R_{C1}$ as a function of $\mathcal{L}$ for different $n$ values, 
  (b) $R_{C2}$ as a function of $\mathcal{L}$ for different $k$ values, (c) $R_{C3}$ as function of $\Delta t$ and (d) $R_{C4}$ as a function of $\mathcal{L}$.}
  \label{lepenergyc1c2c3c4}
 \end{figure}

We analyse next the effect of lepton energy cut on the distinction criteria. In Fig.~\ref{sclepenergy}, we show $\slashed{E}$ distribution for BP1 together with SM 
background for different choices of energy cuts on the leading lepton; (a) $E_{\ell_1}<$ 100 GeV, (b) $E_{\ell_1}<$ 150 GeV and (c) $E_{\ell_1}<$ 200 GeV. 
For $E_{\ell_1}<$100 GeV, SM background gets reduced to a large extent. However, significant portion of the first peak of the signal also gets rejected. Consequently, 
the two-peak nature of the distribution disappears and only a small bump in the distribution remains. 
With both 150 GeV and 200 GeV cut, the reduction of background events is less but so is for the signal contribution, 
resulting a clear two-peak signal for both these cases. In Fig.~\ref{lepenergyc1c2c3c4}, we then quantify the distinguishability of the peaks for these cases
\footnote{We omit the case $E_{\ell_1}<$ 100 GeV as the two peak nature can barely be observed in this case.}.

%
%
%
%
%
%
 
The top panel of Fig.~\ref{lepenergyc1c2c3c4} investigates condition C1, middle panel evaluates C2, while the lower 
panel figures study conditions C3 and C4. 
One can see that at a given luminosity one has to choose higher $n(k)$ in order to get same $R_{C1}(R_{C2}$) values with $E_{\ell_1}<$ 200 GeV as compared to 
$E_{\ell_1}<$ 150 GeV. This is simply because the effect of second peak is subdued due to large background events contributing to first peak with the milder cut of 
$E_{\ell_1}<$ 200 GeV. This is also the reason for $R_{C3}$ to be larger with $E_{\ell_1}<$ 200 GeV compared to the case with $E_{\ell_1}<$ 150 GeV (see Fig.~\ref{lepenergyc1c2c3c4}(c)). $R_{C4}$ remains almost identical in both cases, as can be seen from Fig.~\ref{lepenergyc1c2c3c4}(d), 
as the lepton energy cut do not affect the second peak much. The effect of polarisation is studied in appendix \ref{app4}, where we also validate 
all the conditions C1-C4. We see that polarisation, although necessary for reducing SM background, doesn't alter the distinguishability of the two 
peaks significantly when the SM background is sufficiently suppressed. 

Let us finally summarize the key findings of this section.

\begin{itemize}
\item Conditions C1, C2, C3, C4 which involve $R_{C1}, R_{C2}$, $R_{C3}$ and $R_{C4}$ variables respectively,
can successfully distinguish double peak behaviour in the $\slashed{E}$ spectrum arising from two component DM signal. Among them, $R_{C4}$ 
turns out to be the variable with widest applicability.
\item All the conditions, if simultaneously satisfied, indicate a well separated and prominent second peak; although satisfying one condition is 
good enough to realise the presence of a second peak. 
\item $R_{C1}, R_{C2}$ (for specific $n(k)$) and $R_{C4}$ increase with integrated luminosity $\mathcal{L}$; therefore with larger luminosity, 
significant $R_{C1}$ and $R_{C2}$ values can be obtained at lower $n(k)$.
\item Low $R_{C3}$ is better for distinguishable second peak (unlike other variables), which increases with $\Delta t\equiv \Delta \slashed{E}$ 
and remains almost constant with integrated luminosity. 
\item At low luminosity, large statistical fluctuation becomes a roadblock while one tries to identify the two-peak signature in the $\slashed{E}$ 
distribution.
\item The distinction criteria are sensitive to the lepton energy cut in the chosen final state. If the cut is too stringent, 
the two-peak nature is lost, if the cut is too relaxed, the second peak becomes insignificant compared to the first, requiring an optimaisation. 
\item The distinction criteria are not too sensitive to the initial state polarization, given significant background reduction is already achieved.
\item If the contributions from both the DM components overlap significantly with each other or one contribution wins over the other 
completely, our proposed methods will not work, since in those cases, it will be similar to single-peak distribution.
\end{itemize}

\section{Summary and conclusion}
\label{summary}
We have suggested some methods of distinguishing two DM components, both of which
can be pair-produced in separate events at a collider. In particular, we study a scenario with 
two separate dark sectors, each capable of pair-producing HDSPs, which finally decay into DM 
pairs of either kinds via cascades. This results in double peaks in $\slashed{E}$ or $\slashed{E}_T$ distributions, 
whose identification and segregation constitute the quintessence of our investigation.

The key variables that play a role in producing distinguishable peaks are both of the DM masses ($\mdma,\mdmb$) 
and their mass-splitting with the corresponding HDSPs ($\Delta m_1, \Delta m_2$). 
We further demonstrate that, while $\slashed{E}_T$ is the canonical
label of invisible particles at hadron colliders, it is in $\slashed{E}$-distributions that the peaks are 
likely to be more prominent. This is because the DM masses do not play a role in $\slashed{E}_T$, 
while they show up in the $\slashed{E}$-distribution, thus making the peaks more distinct, when
the masses of the two DM-components are well-separated. Thus $e^-e^+$ colliders that have both the DM
components within their kinematic reach emerge as their best hunting grounds. In addition, the absence
of QCD backgrounds as well as the possibility of beam polarisation serves to reduce the background
to the DM signals. All these have been illustrated in the context of a two-component DM scenario, 
with one scalar and one spin-1/2 DM, each being the lightest state in a separate dark sector. 
Relic density and direct search constraints play an important role to shrink the allowed parameter space of the model for 
collider study, as we have demonstrated. It is further emphasized that, unless both the dark sectors lead to similar 
production rates for the corresponding DM pairs, the peak of one may get buried under the other. We have demonstrated,
with appropriate benchmarking, how this requirement carves out identifiable regions in the dual-DM
parameter space at an electon-positron collider with a given centre-of-mass energy.

We show further that the $WW$ background to $\ell^+\ell^- + \slashed{E}$ signal 
can either spoil or highlight the double-hump behaviour in the  $\slashed{E}$ distribution. 
We recommend the use of right-polarised electron beams and left-handed positron polarisation 
to reduce the $WW$ background contamination. A judicious cut on the lepton energy may help in keeping 
background $WW$ peak coincide nearly with one of the DM peaks.

Finally, we offer some prescriptions for distinguishing the two peaks in the $\slashed{E}$ distribution. 
For this purpose, we suggest a set of criteria which quantify the height, sharpness and separability
of one peak relative to the other. We also indicate the integrated  luminosities which make
these criteria useful,  keeping the SM background in consideration. Some of these criteria can be
useful even in cases where one goes beyond the cascading dark sector mode of
dual-DM production \cite{Bhattacharya:2022qck}. These distinguishability criteria are seen to be rather mildly sensitive 
to beam polarisation, once the SM background reduction has been achieved; on the other hand, 
they depend on the lepton energy cuts. On the whole, it is concluded that pushing the luminosity 
frontier to the level of several atobarns at an electron-positron
machine is a desideratum, if one aspires to distinguish a dual-DM scenario with the available energy reach. 

\section*{Acknowledgments}

SB and JL would like to acknowledge DST-SERB grant CRG/2019/004078 from Govt. of India.
PG would like to acknowledge the support from DAE, India for the Regional Centre
for Accelerator based Particle Physics (RECAPP), Harish Chandra Research Institute.

\section*{Appendix}
\begin{appendix}

\section{Some features of $\slashed{E_T}$, $\slashed{E}$ and $\slashed{M}$}
\label{app1}

In the limit of $\sqrt{s}\simeq 2\mhd$, HDSPs are produced almost at rest, decays further to SM fermion ($f$) and DM. Energy momentum conservation yields, 
\bea
|\vec{p}_{\rm DM}| = |\vec{p}_{f}|,  ~~~E_{\rm DM} + E_{f} = \mhd. 
\label{momenta}
\eea

%
%
%
 
Using equation of motion for DM (for both DMs at either end of the decay chain),

\begin{equation}
\sqrt{|\vec{p}_{DM}|^2 + \mdm^2} + |\vec{p}_{f}| = \mhd.
\end{equation}

Substituting $|\vec{p}_{f}|$ in terms of $|\vec{p}_{\rm DM}|$, from Equation~\ref{momenta}, 
\begin{eqnarray}
\sqrt{|\vec{p}_{\rm DM}|^2 + \mdm^2} + |\vec{p}_{\rm DM}| = \mhd\,, ~
\therefore |\vec{p}_{\rm DM}| = \frac{1}{2} \Delta m \left( 1+ \frac{\mdm}{\mdm + \Delta m} \right)\,.
\label{finaleq}
\end{eqnarray}

Denoting the momenta of the DM pair as $|\vec{p}^1_{\rm DM}|$ and $|\vec{p}^2_{\rm DM}|$, the angle between them as $\theta$, 
$\slashed{E_T}$ can be written as,

\begin{equation}
\slashed{E_T} = \sqrt{|\vec{p}^1_{\rm DM}|^2 + |\vec{p}^2_{\rm DM}|^2 + |\vec{p}^1_{\rm DM}||\vec{p}^2_{\rm DM}|\cos\theta}\,.
\end{equation}

It is maximum when $\theta = 0$; i.e. the two DM particles are colinear. Therefore,
\begin{equation}
\slashed{E_T}^{max} = |\vec{p}^1_{\rm DM}|+|\vec{p}^2_{\rm DM}| = \Delta m \left( 1+ \frac{\mdm}{\mdm + \Delta m} \right)= \Delta m \left( 1+r\right)\,;
\end{equation}
where $r=\frac{\mdm}{\mhd}$. On the other hand, following the energies of the two DM particles as,
\begin{eqnarray}
E^1_{\rm DM} = \sqrt{|\vec{p}^1_{\rm DM}|^2 + \mdm^2},  ~~
E^2_{\rm DM} = \sqrt{|\vec{p}^2_{\rm DM}|^2 + \mdm^2} \,;
\end{eqnarray}
we get,
\begin{equation}
\slashed{E} = E^1_{\rm DM} + E^2_{\rm DM} =\sqrt{|\vec{p}^1_{\rm DM}|^2 + \mdm^2} + \sqrt{|\vec{p}^2_{\rm DM}|^2 + \mdm^2}\,.
\label{eq:missenergytwo}
\end{equation}

\begin{figure}[!hptb]
	$$
	\includegraphics[width=7.5cm,height=6.5cm]{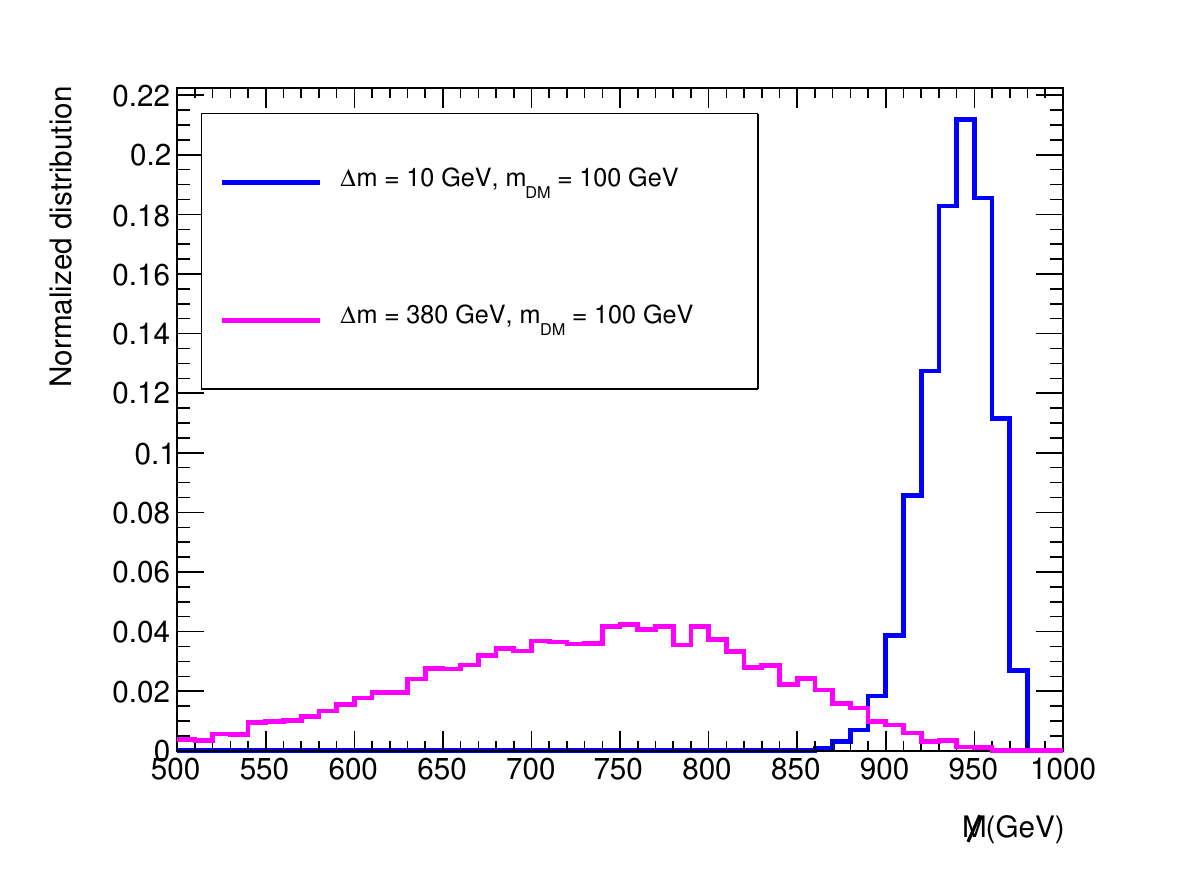}
        $$
	\caption{Normalised $\slashed{M}$ distribution for two cases with same DM mass and different $\Delta m$ (see figure insets for details). 
	The production and decay chain remains similar to that of Fig.~\ref{deltam_comparison}.}
	\label{plot:mmass}
\end{figure}

Let us now turn to $\slashed{M}$. We plot the normalised $\slashed{M}$ distribution in Fig.~\ref{plot:mmass} for the pair production 
of the charged component of the inert scalar doublet with fixed $\mdm$ and different $\Delta m$, 
where the peak shifts to the left with larger $\Delta m$. For hadronically quiet signals $\slashed{M}$ becomes,
\bea
\slashed{M} = \sqrt{\slashed{E}^2-|{\sum_i}~\vec{p}_{\ell_i}|^2}\,.
\label{missmass}
\eea
Evidently, $\slashed{M}$ distribution turns similar to $\slashed{E}$ distributions, and
does not offer much advantage in our context, unless $|{\sum}\vec{p}_{\ell_i}|$ is very large. 
However, such a situation can rarely occur for current planned $e^+e^-$ colliders. We consider a few situations to illustrate the same,

\begin{itemize}
\item ${\Delta m} = 10$ GeV and $\mdm = 50$ GeV

In this case, HDSP mass is low, so that the HDSPs are produced with significant boost. Therefore, the lepton and DM are almost collinear, 
while the two leptons are almost back-to-back for conservation of four-momenta. The effective visible momenta (second term in Eqn.~\ref{missmass}) 
is negligibly small, resulting almost overlapping $\slashed{M}$ and $\slashed{E}$ distributions as shown in left hand side of Fig.~\ref{missmass-missenergy}.  
A similar situation arises when HDSP mass is heavy with a reasonable large DM mass. Then the lepton momenta itself is negligible, producing similar 
$\slashed{E}$ and $\slashed{M}$ distributions.

\begin{figure}[!hptb]
	\centering
	\includegraphics[width=7.5cm,height=6cm]{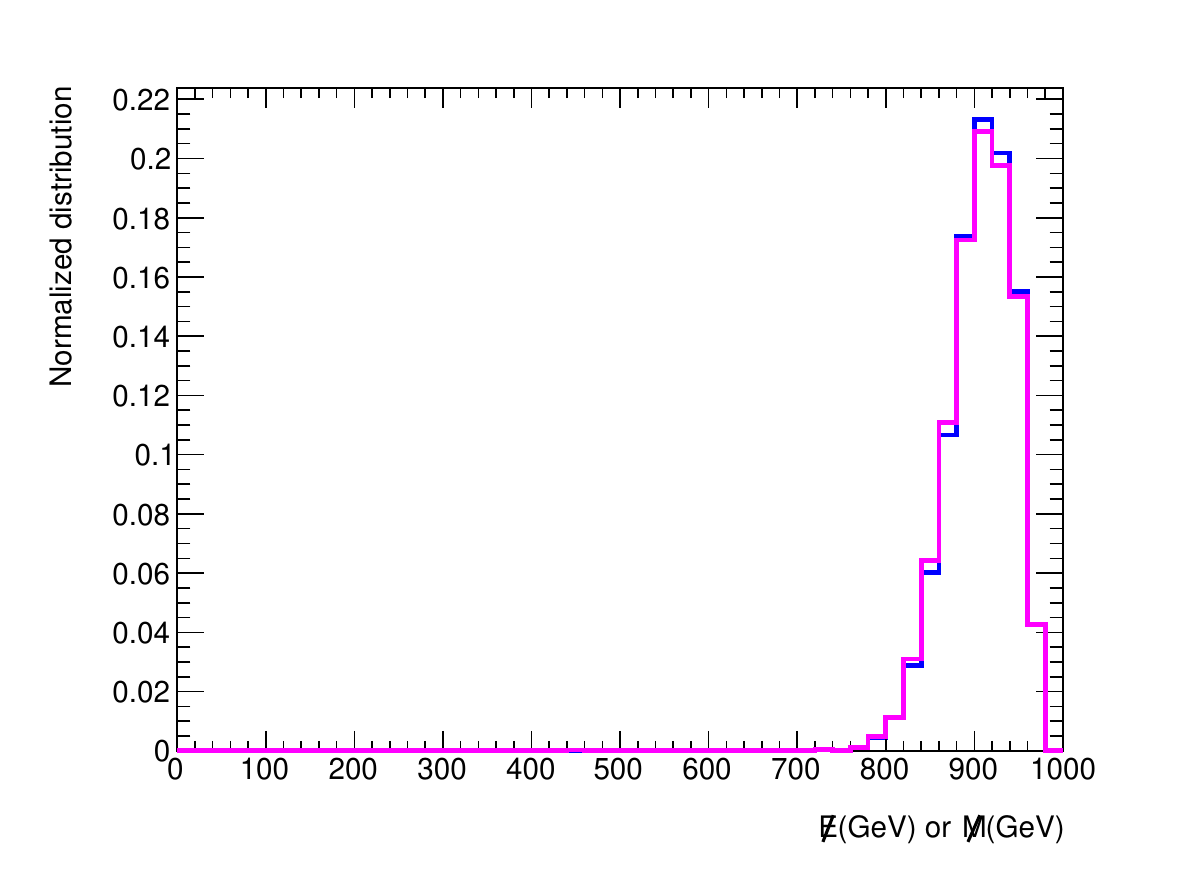}
	\includegraphics[width=7.5cm,height=6cm]{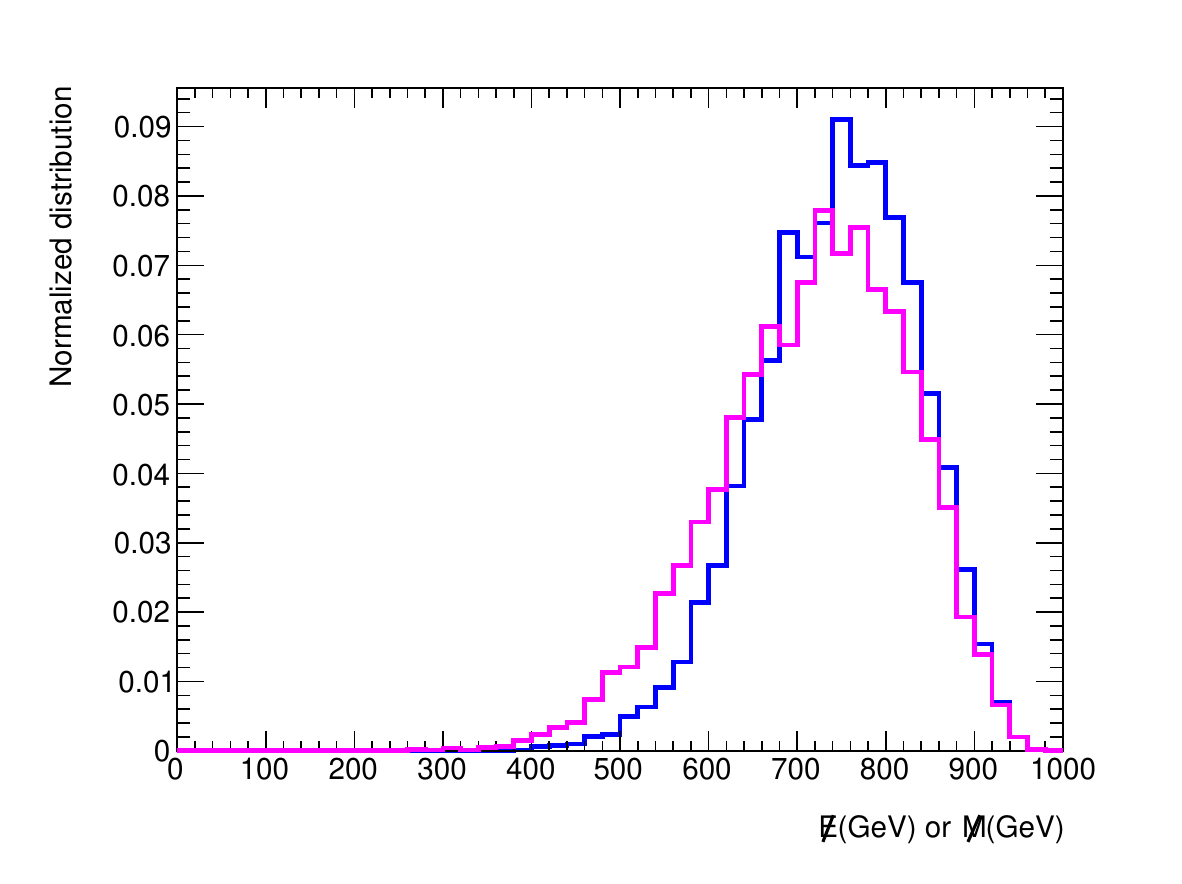}
	\caption{Comaparison between $\slashed{E}$ (blue) and $\slashed{M}$ (pink) distribution for 
	Left: $\{\Delta m, \mdm\} = \{10,50\}$ GeV, Right: $\{\Delta m, \mdm\} = \{450,20\}$ GeV.}
	\label{missmass-missenergy}
\end{figure}

\item ${\Delta m} = 450$ GeV and $\mdm = 20$ GeV

We consider next a scenario where, HDSP is massive and therefore produced almost at rest. In such cases, momenta of the lepton and DM produced at 
each end will almost fully cancel each other. Therefore, the leptons are not necessarily produced back to back. However, the DM being very light ensures 
the magnitudes of lepton momenta are substantial. Therefore, in this case $\slashed{M}$ distribution shows the maximum noticeable departure from the 
$\slashed{E}$ distribution. The deviation is however largest at the left tail-end (although $\lsim$ 10\%) as can be seen from the right hand plot of 
Fig.~\ref{missmass-missenergy}. 

\end{itemize}

It is possible to a have discernible difference between $\slashed{E}$ and $\slashed{M}$ distributions when leptons have considerable energy and are not back to back. 
At limited centre of energy of ILC such situations are rare to occur. But with high energy muon-colliders such a situation can arise and the two distributions can be significantly 
different.

\section{Annihilation, co-annihilation and elastic scattering of DM}
\label{app2}

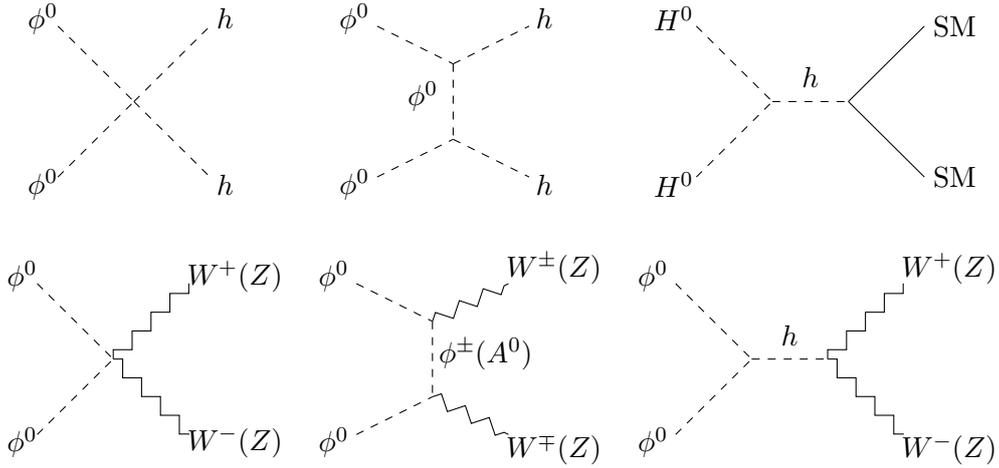
\begin{figure}[htb!]
 \begin{center}
    \begin{tikzpicture}[line width=0.4 pt, scale=1.0]
        \draw[dashed] (-6,1)--(-5,0);
\draw[dashed] (-6,-1)--(-5,0);
\draw[dashed] (-5,0)--(-4,1);
\draw[dashed] (-5,0)--(-4,-1);
\node at (-6.2,1.1) {$\phi^0$};
\node at (-6.2,-1.1) {$\phi^0$};
\node at (-3.8,1.1) {$h$};
\node at (-3.8,-1.1) {$h$};
\draw[dashed] (-1.8,1.0)--(-0.8,0.5);
\draw[dashed] (-1.8,-1.0)--(-0.8,-0.5);
\draw[dashed] (-0.8,0.5)--(-0.8,-0.5);
\draw[dashed] (-0.8,0.5)--(0.2,1.0);
\draw[dashed] (-0.8,-0.5)--(0.2,-1.0);
\node at (-2.1,1.1) {$\phi^0$};
\node at (-2.1,-1.1) {$\phi^0$};
\node at (-1.2,0.07) {$\phi^0$};
\node at (0.4,1.1) {$h$};
\node at (0.4,-1.1) {$h$};
        \draw[dashed] (2.4,1)--(3.4,0);
\draw[dashed] (2.4,-1)--(3.4,0);
\draw[dashed] (3.4,0)--(4.4,0);
\draw[solid] (4.4,0)--(5.4,1);
\draw[solid] (4.4,0)--(5.4,-1);
\node  at (2.1,-1.1) {$H^0$};
\node at (2.1,1.1) {$H^0$};
\node [above] at (3.9,0.05) {$h$};
\node at (5.8,1.0){{\rm SM}};
\node at (5.8,-1.0) {{\rm SM}};
     \end{tikzpicture}
 \end{center}
 \begin{center}
    \begin{tikzpicture}[line width=0.4 pt, scale=1.0]
        \draw[dashed] (-6,1)--(-5,0);
\draw[dashed] (-6,-1)--(-5,0);
\draw[snake] (-5,0)--(-4,1);
\draw[snake] (-5,0)--(-4,-1);
\node at (-6.2,1.1) {$\phi^0$};
\node at (-6.2,-1.1) {$\phi^0$};
\node at (-3.4,1.1) {$W^+(Z)$};
\node at (-3.4,-1.1) {$W^-(Z)$};
\draw[dashed] (-1.8,1.0)--(-0.8,0.5);
\draw[dashed] (-1.8,-1.0)--(-0.8,-0.5);
\draw[dashed] (-0.8,0.5)--(-0.8,-0.5);
\draw[snake] (-0.8,0.5)--(0.2,1.0);
\draw[snake] (-0.8,-0.5)--(0.2,-1.0);
\node at (-2.1,1.1) {$\phi^0$};
\node at (-2.1,-1.1) {$\phi^0$};
\node at (-0.1,0.07) {$\phi^{\pm}(A^0)$};
\node at (0.8,1.2) {$W^{\pm}(Z)$};
\node at (0.8,-1.2) {$W^{\mp}(Z)$};
        \draw[dashed] (2.4,1)--(3.4,0);
\draw[dashed] (2.4,-1)--(3.4,0);
\draw[dashed] (3.4,0)--(4.4,0);
\draw[snake] (4.4,0)--(5.4,1);
\draw[snake] (4.4,0)--(5.4,-1);
\node  at (2.1,-1.1) {$\phi^0$};
\node at (2.1,1.1) {$\phi^0$};
\node [above] at (3.9,0.05) {$h$};
\node at (6.0,1.2){$W^+(Z)$};
\node at (6.0,-1.2){$W^-(Z)$};
     \end{tikzpicture}
 \end{center}
\caption{Feynman diagrams for DM annihilation to SM particles for the Scalar DM ($\phi^0$). }
\label{Feyn-ann-SD}
 \end{figure}
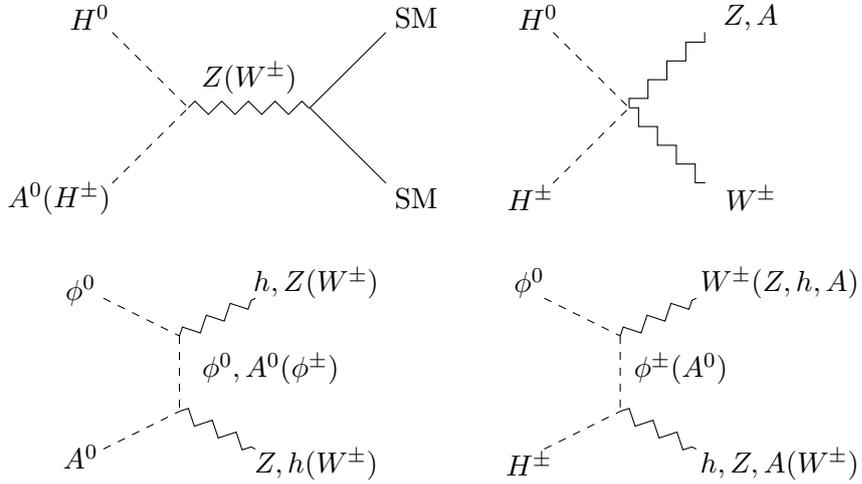
\begin{figure}[htb!]
 \begin{center}
    \begin{tikzpicture}[line width=0.4 pt, scale=1.0]
        \draw[dashed] (-5.0,1)--(-4.0,0);
\draw[dashed] (-5.0,-1)--(-4.0,0);
\draw[snake] (-4.0,0)--(-2.4,0);
\draw[solid] (-2.4,0)--(-1.4,1);
\draw[solid] (-2.4,0)--(-1.4,-1);
\node  at (-5.3,1.2) {$H^0$};
\node at (-5.7,-1.2) {$A^0(H^\pm)$};
\node [above] at (-3.2,0.05) {$Z(W^\pm)$};
\node at (-1.0,1.2){{\rm SM}};
\node at (-1.0,-1.2) {{\rm SM}};
         \draw[dashed] (0.8,1)--(1.8,0);
\draw[dashed] (0.8,-1)--(1.8,0);
\draw[snake] (1.8,0)--(2.8,1);
\draw[snake] (1.8,0)--(2.8,-1);
\node at (0.6,1.2) {$H^0$};
\node at (0.5,-1.2) {$H^\pm$};
\node at (3.4,1.2) {$Z,A$};
\node at (3.4,-1.2) {$W^{\pm}$};
     \end{tikzpicture}
 \end{center}
   \begin{center}
    ~~~~~~~~~~ ~~ \begin{tikzpicture}[line width=0.4 pt, scale=1.0]
       \draw[dashed] (-1.8,1.0)--(-0.8,0.5);
\draw[dashed] (-1.8,-1.0)--(-0.8,-0.5);
\draw[dashed] (-0.8,0.5)--(-0.8,-0.5);
\draw[snake] (-0.8,0.5)--(0.2,1.0);
\draw[snake] (-0.8,-0.5)--(0.2,-1.0);
\node at (-2.1,1.1) {$\phi^0$};
\node at (-2.1,-1.1) {$A^0$};
\node at (0.4,0.07) {$\phi^0,A^0 (\phi^\pm)$};
\node at (1.0,1.2) {$h,Z (W^\pm)$};
\node at (1.0,-1.2) {$Z,h(W^\pm)$};
        \draw[dashed] (4.0,1.0)--(5.0,0.5);
\draw[dashed] (4.0,-1.0)--(5.0,-0.5);
\draw[dashed] (5.0,0.5)--(5,-0.5);
\draw[snake] (5,0.5)--(6,1.0);
\draw[snake] (5,-0.5)--(6,-1.0);
\node at (3.8,1.2) {$\phi^0$};
\node at (3.8,-1.2) {$H^\pm$};
\node at (5.8,0.07) {$\phi^{\pm}(A^0)$};
\node at (7.1,1.2) {$W^{\pm}(Z,h,A)$};
\node at (7.1,-1.2) {$h,Z,A(W^\pm)$};
     \end{tikzpicture}
 \end{center}
\caption{Feynman diagrams for DM co-annihilation to SM particles for the Scalar DM ($\phi^0$) associated with heavy states ($A^0,\phi^\pm$). }
\label{Feyn-coann-SD}
 \end{figure}

 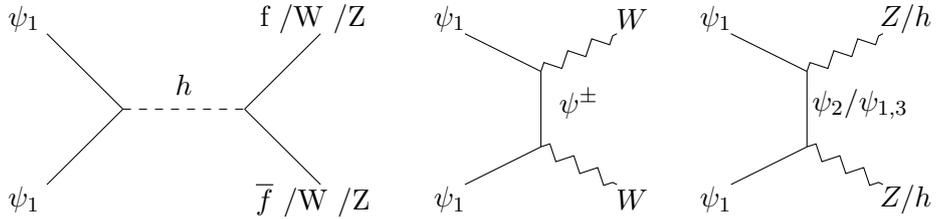
\begin{figure}[htb!]
   \begin{center}
     \begin{tikzpicture}[line width=0.4 pt, scale=1.0]
    \draw[solid] (-5.0,1)--(-4.0,0);
\draw[solid] (-5.0,-1)--(-4.0,0);
\draw[dashed] (-4.0,0)--(-2.4,0);
\draw[solid] (-2.4,0)--(-1.4,1);
\draw[solid] (-2.4,0)--(-1.4,-1);
\node  at (-5.3,1.2) {$\psi_1$};
\node at (-5.3,-1.2) {$\psi_1$};
\node [above] at (-3.2,0.05) {$h$};
\node at (-1.5,1.2){f /W /Z};
\node at (-1.5,-1.2) {$ \overline{f}$  /W /Z};
       \draw[solid] (0.5,1.0)--(1.5,0.5);
\draw[solid] (0.5,-1.0)--(1.5,-0.5);
\draw[solid] (1.5,0.5)--(1.5,-0.5);
\draw[snake] (1.5,0.5)--(2.5,1.0);
\draw[snake] (1.5,-0.5)--(2.5,-1.0);
\node at (0.3,1.2) {$\psi_1$};
\node at (0.3,-1.2) {$\psi_1$};
\node at (2.0,0.07) {$\psi^\pm$};
\node at (2.7,1.2) {$W$};
\node at (2.7,-1.2) {$W$};
        \draw[solid] (4.0,1.0)--(5.0,0.5);
\draw[solid] (4.0,-1.0)--(5.0,-0.5);
\draw[solid] (5.0,0.5)--(5,-0.5);
\draw[snake] (5,0.5)--(6,1.0);
\draw[snake] (5,-0.5)--(6,-1.0);
\node at (3.8,1.2) {$\psi_1$};
\node at (3.8,-1.2) {$\psi_1$};
\node at (5.7,0.07) {$\psi_2/\psi_{1,3}$};
\node at (6.3,1.2) {$Z /h$};
\node at (6.3,-1.2) {$Z /h$};
     \end{tikzpicture}
 \end{center}
\caption{Feynman diagrams for DM annihilation to SM particles for the Fermion DM ($\psi_1$). }
\label{Feyn-ann-FD}
 \end{figure}

 \begin{figure}[htb!]
   \begin{center}
     \begin{tikzpicture}[line width=0.4 pt, scale=1.0]
    \draw[solid] (-5.0,1)--(-4.0,0);
\draw[solid] (-5.0,-1)--(-4.0,0);
\draw[snake] (-4.0,0)--(-2.4,0);
\draw[solid] (-2.4,0)--(-1.4,1);
\draw[solid] (-2.4,0)--(-1.4,-1);
\node  at (-5.3,1.2) {$\psi_2$};
\node at (-5.3,-1.2) {$\psi_1$};
\node [above] at (-3.2,0.05) {$Z$};
\node at (-1.4,1.2){f/W /Z};
\node at (-1.4,-1.2) {$\overline{f}$/W/h};
       \draw[solid] (0.5,1.0)--(1.5,0.5);
\draw[solid] (0.5,-1.0)--(1.5,-0.5);
\draw[solid] (1.5,0.5)--(1.5,-0.5);
\draw[snake] (1.5,0.5)--(2.5,1.0);
\draw[snake] (1.5,-0.5)--(2.5,-1.0);
\node at (0.3,1.2) {$\psi_2$};
\node at (0.3,-1.2) {$\psi_1$};
\node at (2.0,0.07) {$\psi^\pm$};
\node at (2.7,1.2) {$W$};
\node at (2.7,-1.2) {$W$};
        \draw[solid] (4.0,1.0)--(5.0,0.5);
\draw[solid] (4.0,-1.0)--(5.0,-0.5);
\draw[solid] (5.0,0.5)--(5,-0.5);
\draw[snake] (5,0.5)--(6,1.0);
\draw[dashed] (5,-0.5)--(6,-1.0);
\node at (3.8,1.2) {$\psi_2$};
\node at (3.8,-1.2) {$\psi_1$};
\node at (5.6,0.07) {$\psi_{1,3}$};
\node at (6.3,1.2) {$Z$};
\node at (6.3,-1.2) {$h$};
     \end{tikzpicture}
 \end{center}
 \begin{center}
     \begin{tikzpicture}[line width=0.4 pt, scale=1.0]
    \draw[solid] (-5.0,1)--(-4.0,0);
\draw[solid] (-5.0,-1)--(-4.0,0);
\draw[dashed] (-4.0,0)--(-2.4,0);
\draw[solid] (-2.4,0)--(-1.4,1);
\draw[solid] (-2.4,0)--(-1.4,-1);
\node  at (-5.3,1.2) {$\psi_3$};
\node at (-5.3,-1.2) {$\psi_1$};
\node [above] at (-3.2,0.05) {$h$};
\node at (-1.4,1.2){f/ W /Z/h};
\node at (-1.4,-1.2) {$\overline{f}$/W/Z/h};
       \draw[solid] (0.5,1.0)--(1.5,0.5);
\draw[solid] (0.5,-1.0)--(1.5,-0.5);
\draw[solid] (1.5,0.5)--(1.5,-0.5);
\draw[snake] (1.5,0.5)--(2.5,1.0);
\draw[snake] (1.5,-0.5)--(2.5,-1.0);
\node at (0.3,1.2) {$\psi_3$};
\node at (0.3,-1.2) {$\psi_1$};
\node at (2.0,0.07) {$\psi^\pm$};
\node at (2.7,1.2) {$W$};
\node at (2.7,-1.2) {$W$};
    \draw[solid] (4.0,1.0)--(5.0,0.5);
\draw[solid] (4.0,-1.0)--(5.0,-0.5);
\draw[solid] (5.0,0.5)--(5,-0.5);
\draw[dashed] (5,0.5)--(6,1.0);
\draw[dashed] (5,-0.5)--(6,-1.0);
\node at (3.8,1.2) {$\psi_3$};
\node at (3.8,-1.2) {$\psi_1$};
\node at (5.6,0.07) {$\psi_{1,3}$};
\node at (6.3,1.2) {$h$};
\node at (6.3,-1.2) {$h$};
     \end{tikzpicture}
 \end{center}
 \begin{center}
     \begin{tikzpicture}[line width=0.4 pt, scale=1.0]
    \draw[solid] (-5.0,1)--(-4.0,0);
\draw[solid] (-5.0,-1)--(-4.0,0);
\draw[snake] (-4.0,0)--(-2.4,0);
\draw[solid] (-2.4,0)--(-1.4,1);
\draw[solid] (-2.4,0)--(-1.4,-1);
\node  at (-5.3,1.2) {$\psi^-$};
\node at (-5.3,-1.2) {$\psi_1$};
\node [above] at (-3.2,0.05) {$W^+$};
\node at (-1.4,1.2){f/W/W/W };
\node at (-1.4,-1.2) {$f^\prime$/h/Z/A};
       \draw[solid] (0.5,1.0)--(1.5,0.5);
\draw[solid] (0.5,-1.0)--(1.5,-0.5);
\draw[solid] (1.5,0.5)--(1.5,-0.5);
\draw[snake] (1.5,0.5)--(2.5,1.0);
\draw[snake] (1.5,-0.5)--(2.5,-1.0);
\node at (0.3,1.2) {$\psi^-$};
\node at (0.3,-1.2) {$\psi_1$};
\node at (2.0,0.07) {$\psi_2$};
\node at (2.7,1.2) {$W$};
\node at (2.7,-1.2) {$Z$};
        \draw[solid] (4.0,1.0)--(5.0,0.5);
\draw[solid] (4.0,-1.0)--(5.0,-0.5);
\draw[solid] (5.0,0.5)--(5,-0.5);
\draw[snake] (5,0.5)--(6,1.0);
\draw[dashed] (5,-0.5)--(6,-1.0);
\node at (3.8,1.2) {$\psi^-$};
\node at (3.8,-1.2) {$\psi_1$};
\node at (5.6,0.07) {$\psi_{1,3}$};
\node at (6.3,1.2) {$W^-$};
\node at (6.3,-1.2) {$h$};
     \end{tikzpicture}
 \end{center}
%
\caption{Feynman diagrams for DM co-annihilation to SM particles for the Fermion DM ($\psi_1$) associated with the heavy states, $\psi_2$, $\psi_3$ and $\psi^\pm$. }
\label{Feyn-coann-FD}
 \end{figure}
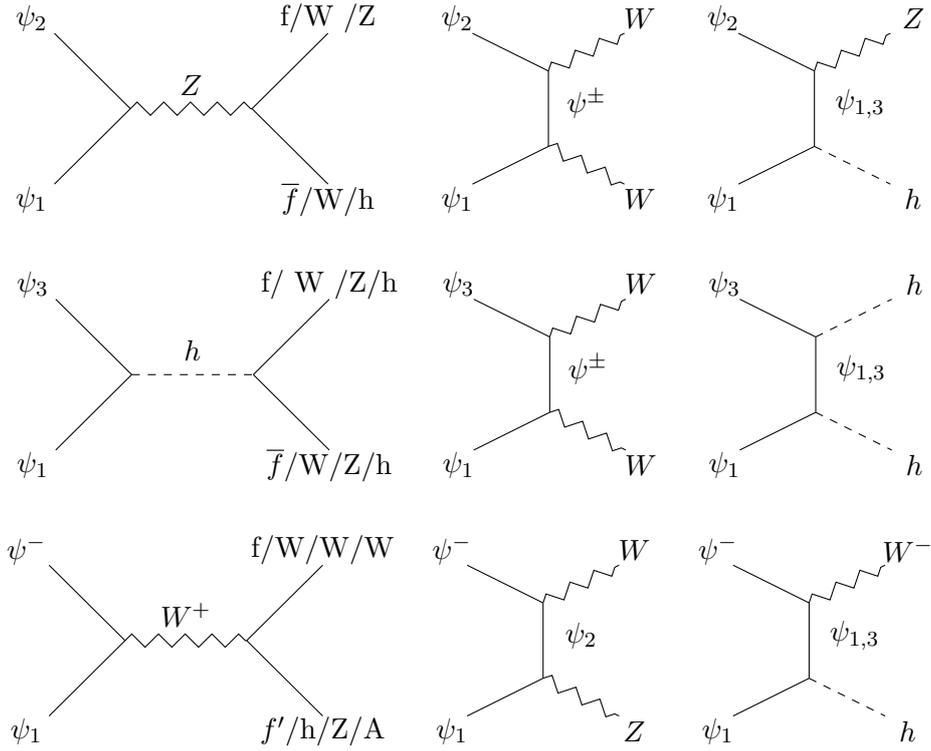

\begin{figure}[htb!]
\begin{center}
\begin{tikzpicture}[line width=0.6 pt, scale=1.2]
\draw[dashed] (8.4,1)--(9.4,0);
\draw[dashed] (8.4,-1)--(9.4,0);
\draw[snake] (9.4,0)--(10.4,0);
\draw[solid] (10.4,0)--(11.4,1);
\draw[solid] (10.4,0)--(11.4,-1);
\node  at (7.8,-1.2) {SDM};
\node at (7.8,1.2) {SDM};
\node [above] at (9.9,0.05) {$h/W/Z$};
\node at (12,1.2){FDM};
\node at (12,-1.2) {FDM};
\end{tikzpicture}
\end{center}
\caption{Feynman diagrams for DM conversion between scalar and fermion DM.}
\label{Feyn_conversion}
\end{figure}
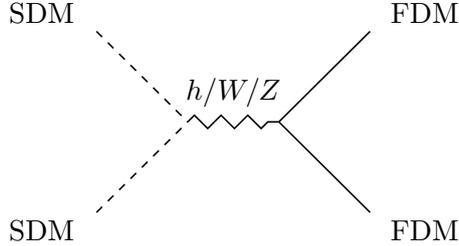

\noindent
The SDM ($\phi^0$) can annihilate and co-annihilate with other heavy states, $A^0$ and $\phi^\pm$ to SM particles via Higgs and gauge mediated 
interactions as shown in Fig.~\ref{Feyn-ann-SD} and Fig.~\ref{Feyn-coann-SD} respectively. Similarly, the FDM has Higgs and gauged mediated 
annihilation as well as co-annihilation processes to SM as shown in Fig.~\ref{Feyn-ann-FD} and Fig.~\ref{Feyn-coann-FD} respectively. 
Along with the standard annihilation and co-annihilation channels, the model also yield DM-DM conversion where 
one DM component can annihilate into the other as shown in Fig.~\ref{Feyn_conversion}. Both the dark sectors having SM gauge interaction, naturally 
allows them to be in thermal bath in the early universe and behave as WIMPs. The SDM in the mass range $m_W \lesssim m_{\phi^0} \lesssim 525$ GeV 
provides under abundance ($\Omega_{\rm DM} h^2 < 0.12-0.001 $) to form one component of the two DMs. For FDM, the gauge mediated annihilation is suppressed by the 
mixing angle $\sin\theta$. Relic under abundace for FDM is achieved both at very low $\Delta m_2$ via co-annihilation and at large $\Delta m_2$ ($\propto Y$) where 
Higgs-mediated annihilation provide required depletion. 

 \begin{figure}[htb!]
   \begin{center}
    \begin{tikzpicture}[line width=0.4 pt, scale=1.1]
\draw[dashed] (-1.8,1.0)--(-0.8,0.5);
\draw[solid] (-1.8,-1.0)--(-0.8,-0.5);
\draw[dashed] (-0.8,0.5)--(-0.8,-0.5);
\draw[dashed] (-0.8,0.5)--(0.2,1.0);
\draw[solid] (-0.8,-0.5)--(0.2,-1.0);
\node at (-2.1,1.1) {$\phi^0$};
\node at (-2.1,-1.1) {$n$};
\node at (-0.5,0.07) {$h$};
\node at (0.5,1.2) {$\phi^0$};
\node at (0.5,-1.2) {$n$};
\draw[solid] (4.0,1.0)--(5.0,0.5);
\draw[solid] (4.0,-1.0)--(5.0,-0.5);
\draw[dashed] (5.0,0.5)--(5,-0.5);
\draw[solid] (5,0.5)--(6,1.0);
\draw[solid] (5,-0.5)--(6,-1.0);
\node at (3.8,1.2) {$\psi_1$};
\node at (3.8,-1.2) {$n$};
\node at (5.3,0.07) {$h$};
\node at (6.5,1.2) {$\psi_1$};
\node at (6.5,-1.2) {$n$};
     \end{tikzpicture}
 \end{center}
\caption{Feynman diagrams for spin independent DM-neucleon scattering process for scalar and fermion DM. }
\label{DD}
 \end{figure}
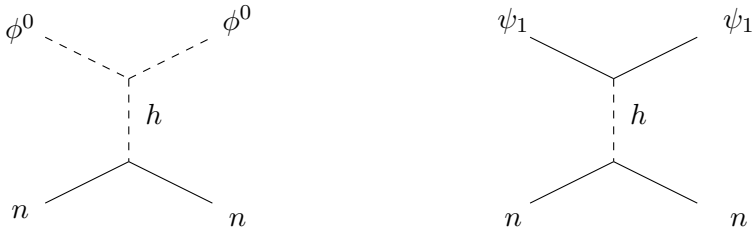

\noindent
Both the scalar ($\phi^0$) and fermion ($\psi_1$) DM can be detected through Higgs mediated $t-$ channel spin-independent (SI) DM-neucleon scattering events, 
as depicted in Fig.~\ref{DD}. At the tree-level, the DM-nucleon scattering cross-section for SDM
$\sigma_{\rm SI}(\phi^0) \propto \lambda_L^2/ m_{\phi^0}^2$, while for FDM, $\sigma_{\rm SI}(\psi_1) \propto Y^2/ m_{\psi_1}^2$ ($Y \propto \Delta m_2 \sin\theta $). 
It is clear that for FDM, large $\Delta m_2$ will result in large direct-detection cross-section, and therefore be disfavoured from the data, unless $\sin \theta$ is small. 
Such small values of $\sin \theta$ will of course lead to small annihilation and relic over-abundance. 
Therefore, $Z$ or Higgs resonance regions are only allowed for FDM.


\section{A sample benchmark from Region III}
\label{app3}

\begin{figure}[!hptb]
	\centering
	\subfloat[]{\includegraphics[width=7cm,height=6cm]{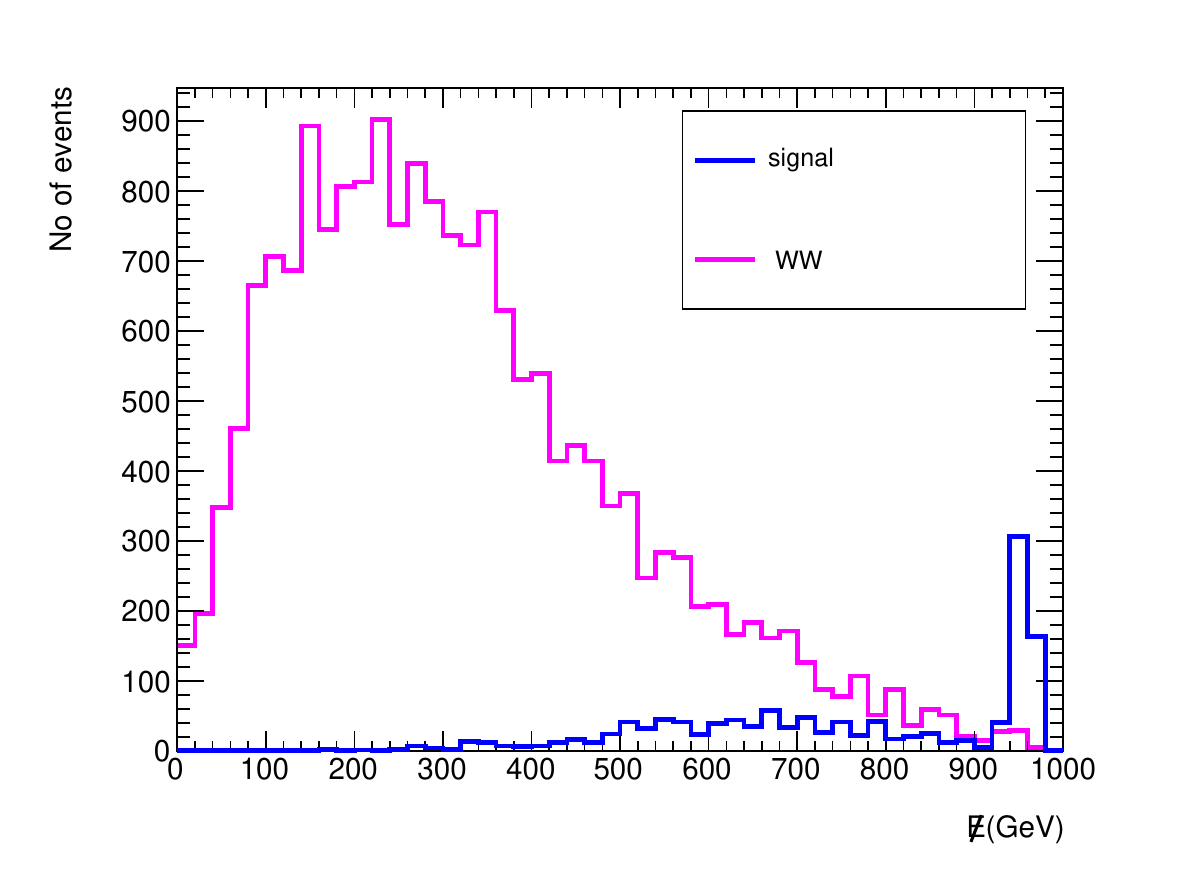}}
        \subfloat[]{\includegraphics[width=7cm,height=6cm]{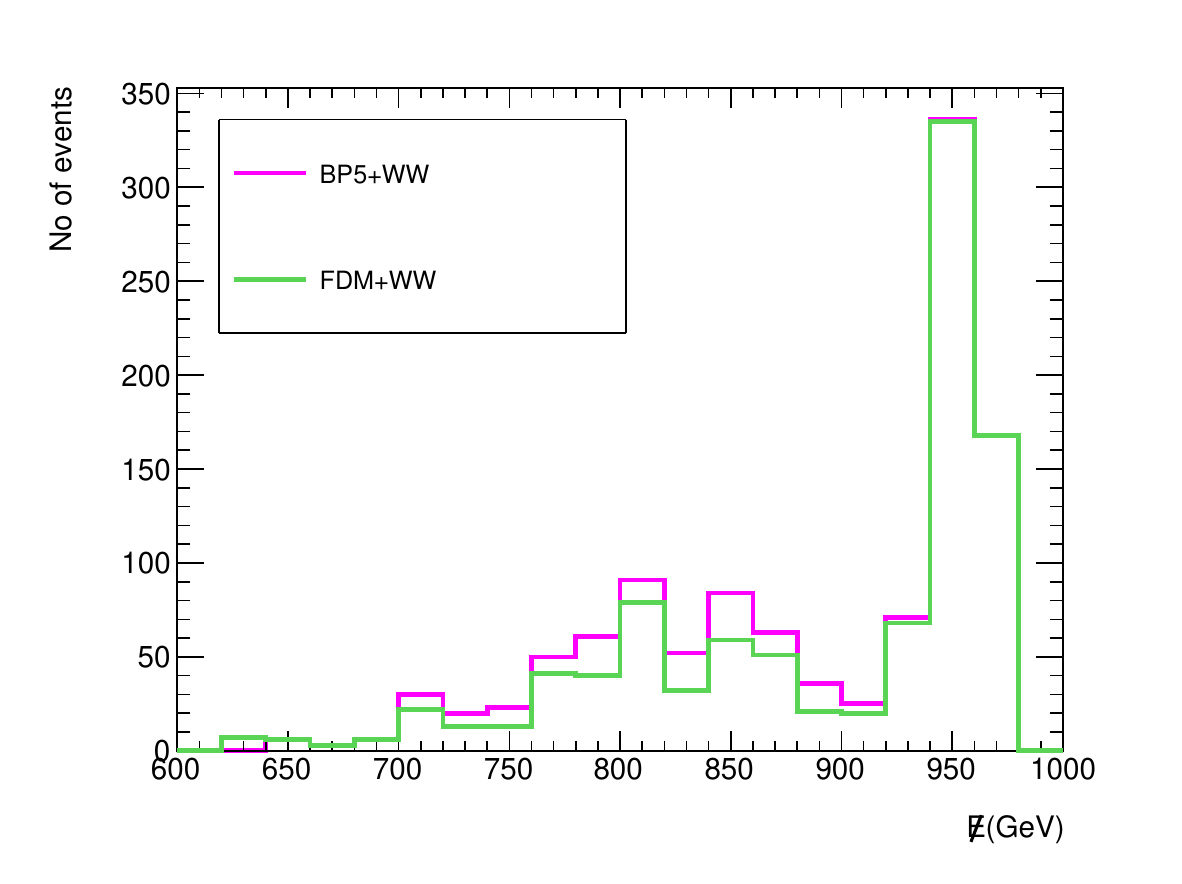}} 
	\caption{ $\slashed{E}$ distribution for (a) signal (BP5) and $WW$ background, (b) (pink) signal (BP5) + $WW$ background after applying cuts and (green) single component FDM + $WW$ background after applying cuts.}
	\label{bp5}
\end{figure}

Let us examine a benchmark point BP5 from Region III, given in Table \ref{table_crosssec}, 
where $m_{\phi^0} < m_{\psi_1}$ and $\Delta m_1 > \Delta m_2$. Since the scalar HDSP has larger production cross-section compared to the fermionic HDSP, we consider $m_{\phi^{\pm}} < m_{\psi^{\pm}}$, in order to have comparable cross-section for both DM sectors. 
 $m_{\phi^0} < m_{\psi_1}$ would imply that $\slashed{E}$ peak pertaining to SDM will appear 
on the left of the FDM peak. However the SDM peak will have smaller height and will be broader owing to large $\Delta m$. In such a case, the peak from scalar sector will be mostly 
buried under $WW$ tail(see Fig.~\ref{bp5}(a)).

\begin{table}[!hptb]
\begin{center}
\begin{tabular}{| c | c | c | c | c | c |}
\hline
Benchmark & $m_{\phi^0}$ and $\Delta m_1$ & $m_{\psi_1}$ and $\Delta m_2$ & $\sigma$(fb) & $\sigma$(scalar) & $\sigma$(fermion) \\
\hline
BP5 & 60 GeV, 60 GeV & 448 GeV, 40.0 GeV & 1.5 fb & 0.7 fb & 0.8 fb\\
\hline
\end{tabular}
\caption{Signal benchmark point and cross section for $P_{e^{-}} = 0.8, P_{e^{+}} = -0.3$ polarization at $\sqrt{s}=$ 1 TeV centre-of-mass energy.}
\label{table_crosssec}
\end{center}
\end{table}

\noindent
The $\slashed{E}$ distribution in this case not only fails to 
produce clearly separated peaks of comparable sizes, but yields very similar distribution to the scenario, when there exists only a single DM 
component and the background distribution contributes to a second peak-like behaviour. In Fig.~\ref{bp5}(b) we see, a single component FDM (pertaining to FDM sector of BP5) along with $WW$ background (green histogram), gives rise to $\slashed{E}$ distribution very similar to two-component DM scenario in BP5 along with $WW$ background (pink histogram). 
This is a rather general consequence of the lower peak from DM signal having a much flatter distribution. Although the two peaks are well-separated, 
the relative size of the two peaks makes it difficult to distinguish it from single-peak scenario. From the discussion above, one may thus 
conclude that Region III is by and large disfavored compared to Region IV, from the perspective of peak distinction. 

\section{Gaussian Fitting methodology}
\label{app4}
 \begin{figure}[!hptb]
	$$
	\includegraphics[width=7.5cm,height=6cm]{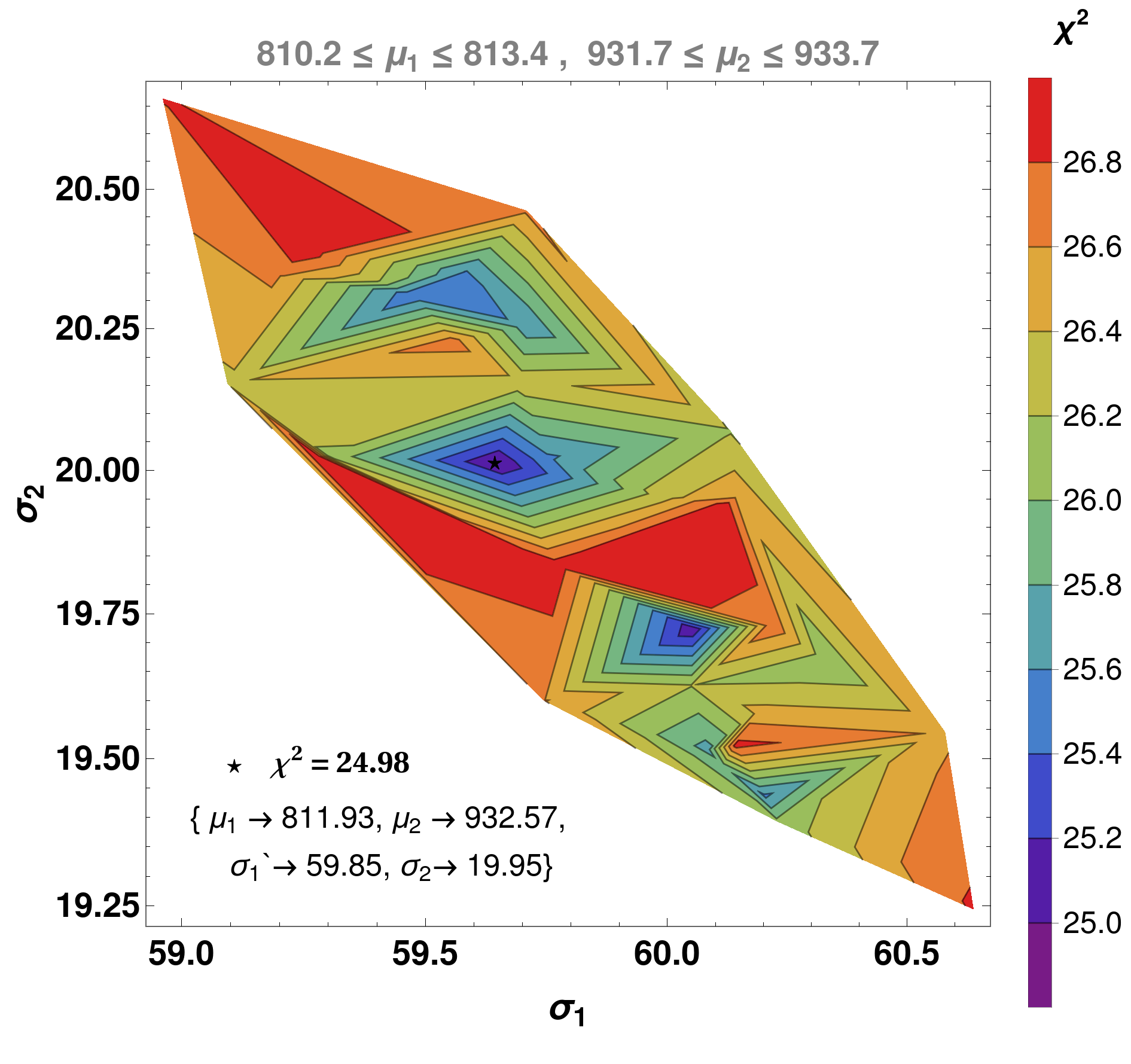}~~
	\includegraphics[width=7.5cm,height=6cm]{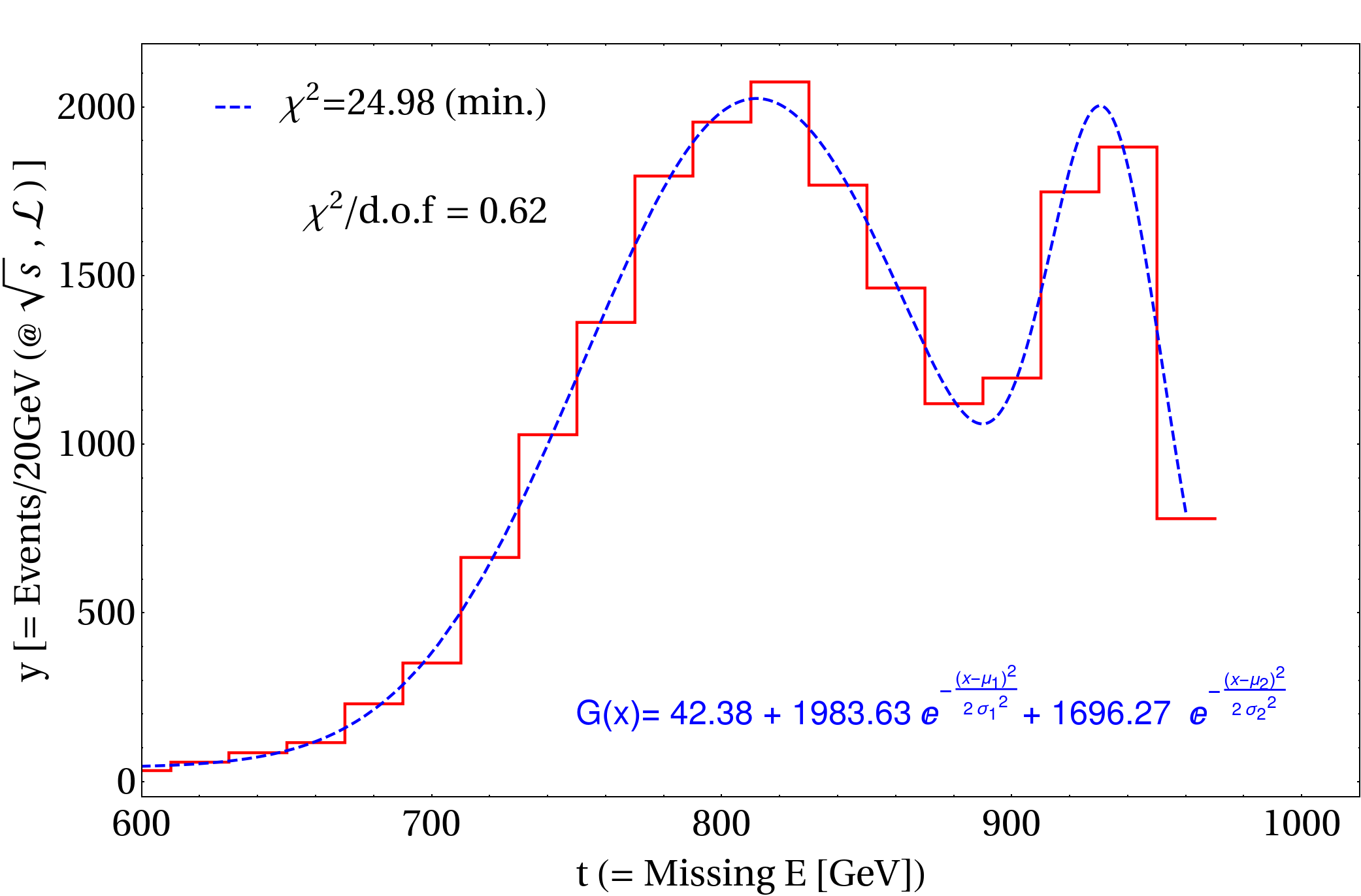}
	$$
	$$
    \includegraphics[width=7.5cm,height=6cm]{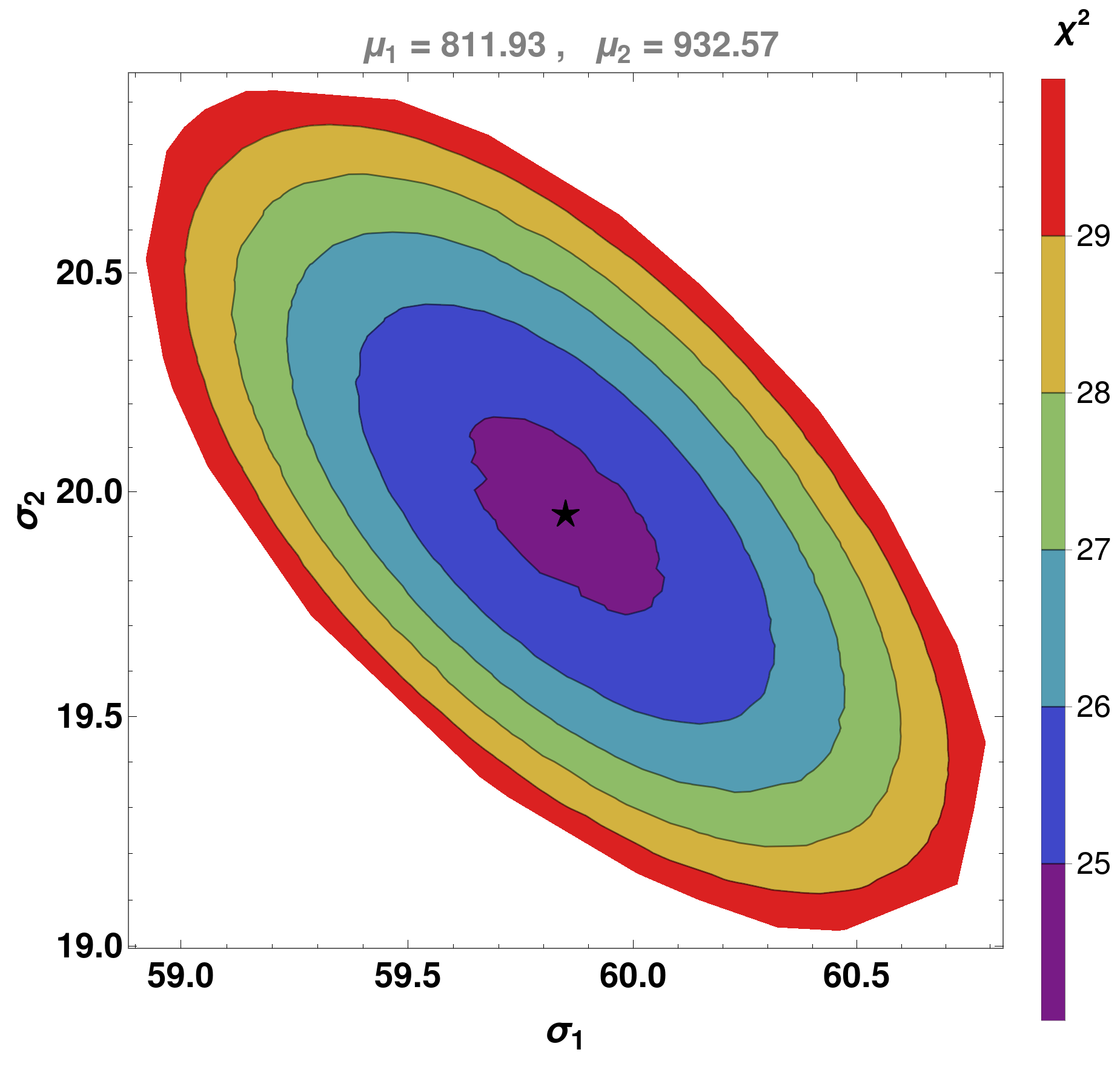}~
	$$
	\caption{[Top Left] $\chi^2$ defined by Eq.~\ref{eq:chisq} shown in different colored regions in the plane of $\sigma_1-\sigma_2$ for the random variation 
	of the parameters of the two peak Gaussian function $G$: $\{ \mu_1: \{810.2-813.4\}, \sigma_1: \{58-62\}; \mu_2: \{931.7-933.7\}, \sigma_2: \{18-22\} \}$. 
	[Top Right] The best fit two peak Gaussian function with minimum $\chi^2$ shown by blue dotted line, confronted with histogram data in solid red. 
	[Bottom] The variation of $\chi^2$ is again shown in the plane of $\sigma_1-\sigma_2$ keeping $\mu_1=811.93$ and $\mu_2=932.57$ fixed. 
	The $\star$ sign corresponds to the best fit parameters of the Gaussian function $G$, which also marks minimum $\chi^2$.}
	\label{fig:figA4}
\end{figure}
Consider that we have generated the histogram data by event simulation as,  
\bea
\mathcal{H}=\{\{x_H^1,y_H^1\},\{x_H^2,y_H^2\},....,\{x_H^n,y_H^n\}\} \,;
\eea
where $x_H \equiv \slashed{E},~y_H \equiv \frac{d\sigma}{d\slashed{E}}$ and $n$ refers to number of data points. 
We want to fit a two peak Gaussian function to this data as,
 \bea\label{eq:gfa}
 G(\mu_1,\sigma_1;\mu_2,\sigma_2)&=& A_1~ e^{-\frac{(x-\mu_1)^2}{2\sigma_1^2}}+A_2 ~ e^{-\frac{(x-\mu_2)^2}{2\sigma_2^2}} + \cal{B} ~ .
 \eea

Our goal is then to find out $\{\mu_1,\sigma_1\}$ of the first peak and $\{\mu_2,\sigma_2\}$ of the second peak of the above Gaussian function $G$ 
that best fit the histogram data $\mathcal{H}$. In order to do that we define $\chi^2$ function as:
 \bea\label{eq:chisq}
 \chi^2(\mu_1,\sigma_1;\mu_2,\sigma_2)=\sum_{i=1}^n {\frac{\Big(G(\mu_1,\sigma_1;\mu_2,\sigma_2)[x_H^i]-y_H^i \Big)^2}{y_H^i}}~~.
 \eea
The best fit function $G$ can be estimated by minimizing $\chi^2$; i.e. vary $(\mu_1,\sigma_1,\mu_2,\sigma_2)$, calculate $\chi^2$ using Eqn.~\ref{eq:chisq}, 
and choose the one that has minimum $\chi^2$. Note here that $A_1,~A_2$ and $\cal{B}$ 
of the function $G$ are automatically decided from the fitting (so the area under the curve remains the same) 
for a fixed set of $(\mu_1,\sigma_1,\mu_2,\sigma_2)$. For example, 
we randomly vary: $( \mu_1: \{810.2-813.4\}, \sigma_1: \{58-62\}, \mu_2: \{931.7-933.7\}, \sigma_2: \{18-22\} )$ 
for a particular data set, as shown in the top left panel of Fig.~\ref{fig:figA4} in the plane of $\sigma_1-\sigma_2$. The different color 
patches here correspond to different $\chi^2$ ranges. The minimum $\chi^2$ value, $\chi_{\rm min}^2=24.98$ provides the 
best fit parameters of $G$, marked by $\star$, where the values of the parameters turn out to be: 
$\{\mu_1^0,\sigma_1^0;\mu_2^0,\sigma_2^0\}=\{811.93,59.85;932.57,19.95\}$.  We further use this best fit Gaussian function, 
$G(\mu_1^0,\sigma_1^0;\mu_2^0,\sigma_2^0)$ and draw the distribution by blue dotted line on top of the Histogram data 
(red thick line) as shown in the right top panel of Fig.~\ref{fig:figA4}. The particular histogram data uses the simulated 
events for BP1 $\{\mdma,\mdmb,\Delta m_1,\Delta m_2\}=\{100,60.5,10,370\}$ GeV. This exercise has been repeated for all the cases
analysed in the text. We also estimate the $\chi^2/{\rm d.o.f}$ of the Gaussian fit where d.o.f refers to the 
number of histogram data sets. In this particular example, we find $\chi^2/{\rm d.o.f}=0.62$. Note here that any value for $\chi^2/{\rm d.o.f}<1$ is considered pretty accurate.  
In the bottom panel of Fig. \ref{fig:figA4}, we show the variation of $\chi^2$ in the plane of $\sigma_1-\sigma_2$ keeping $\mu_1=811.93$ and $\mu_2=932.57$ fixed. 
As expected, it shows a set of parabola having constant $\chi^2$ ranges. 

\section{Effect of polarisation in distinguishing two peaks}
\label{app5}
\begin{figure}[htb!]
 
  \subfloat[]{\includegraphics[width=6.8cm,height=5.5cm]{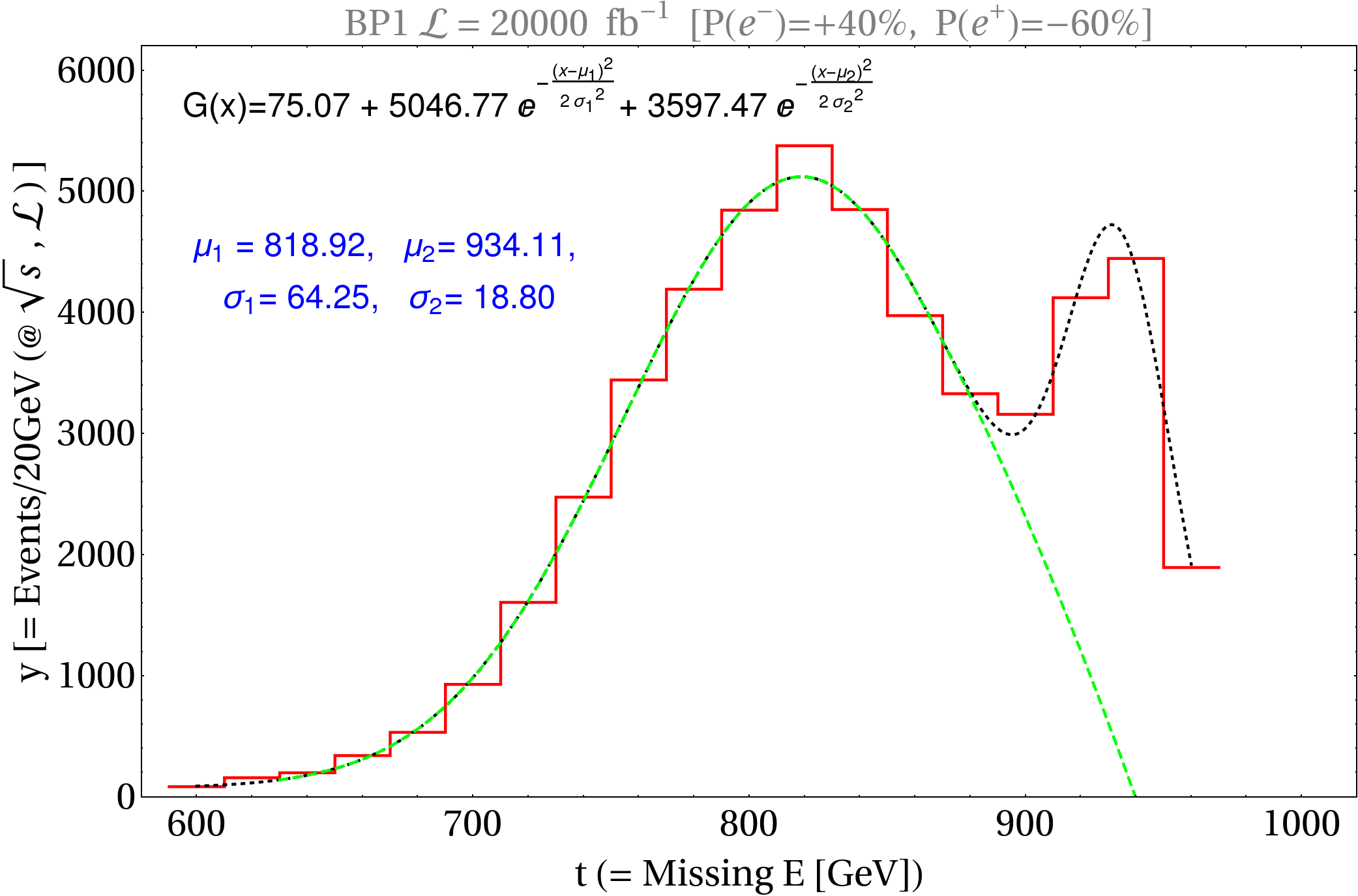}}~
   \subfloat[]{\includegraphics[width=6.8cm,height=5.5cm]{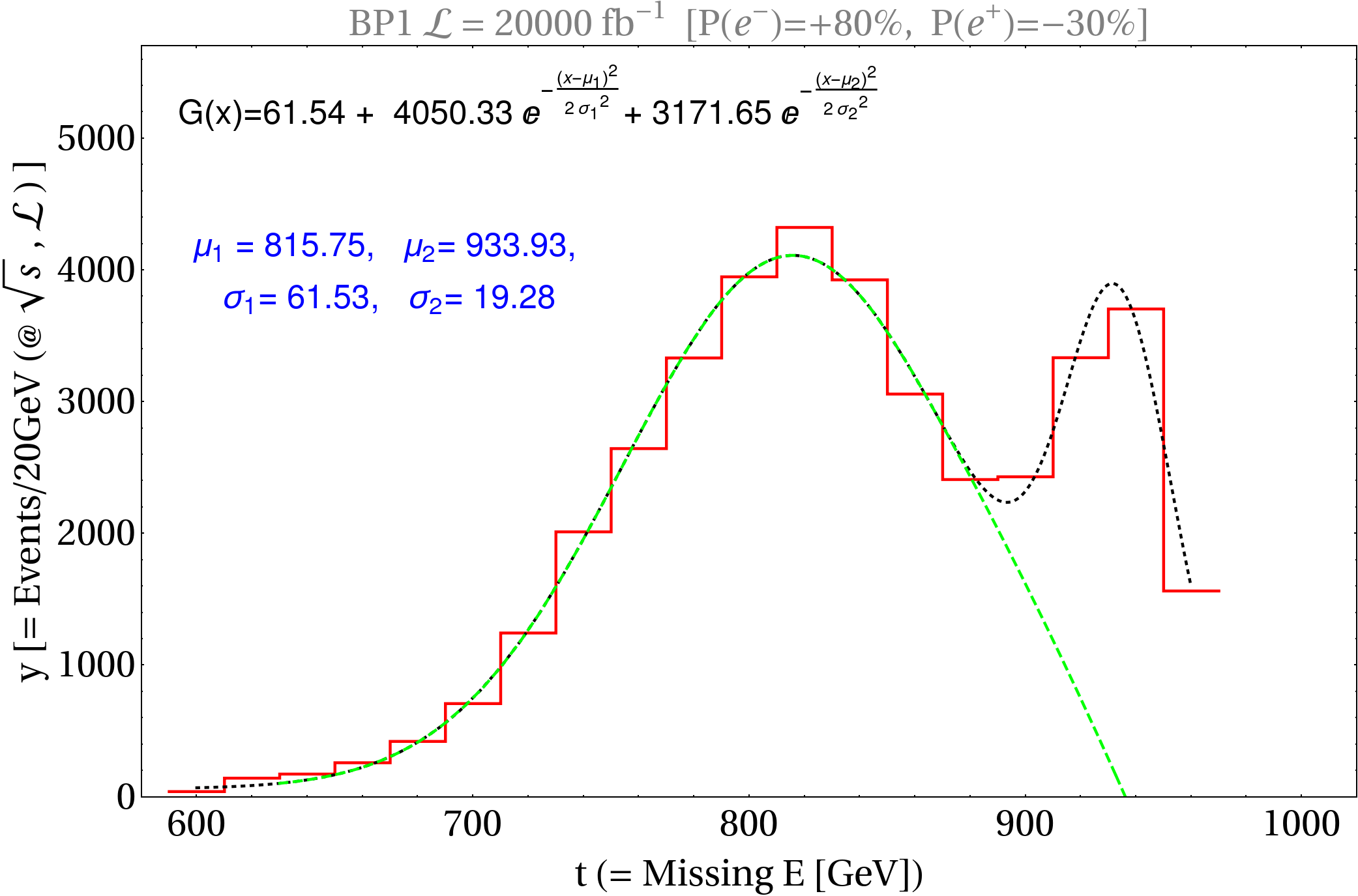}}\\
   \subfloat[]{\includegraphics[width=6.8cm,height=5.5cm]{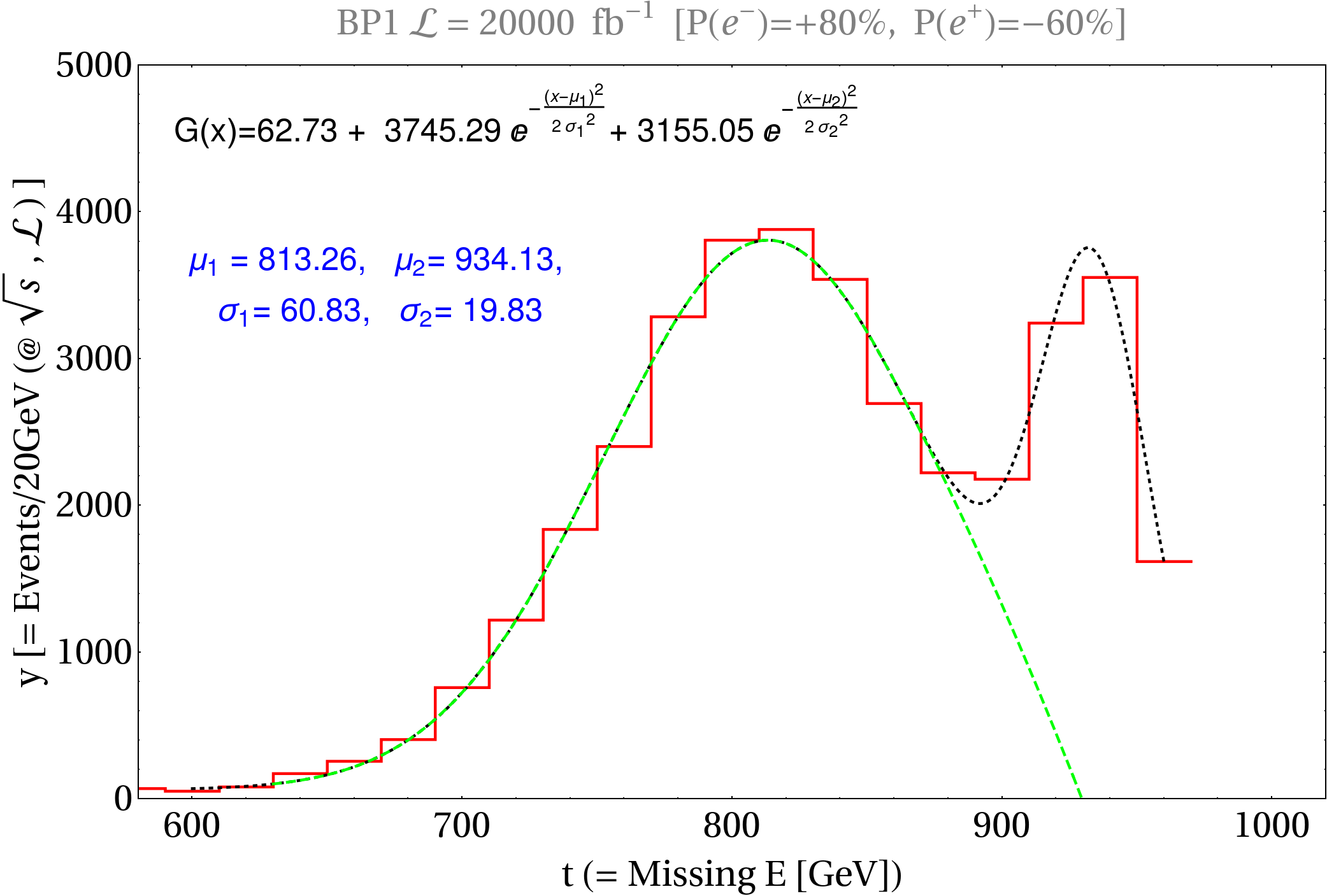}}~
   
  \caption{$\slashed{E}$ distribution for BP1 together with SM background and Gaussian fit for different beam polarizations: 
  (a) $P1^{'} \equiv \{P_{e^{-}}:+ 0.4, P_{e^{+}}: -0.6\};~(b)~P3 \equiv \{P_{e^{-}}:+ 0.8, P_{e^{+}}: -0.3\}$ and (c) $ P3^{'}\equiv \{P_{e^{-}}:+ 0.8, P_{e^{+}}: -0.6\}$.}
  \label{scpol}
 \end{figure}

We study the effect of polarization in two-peak identification in this section. In Fig.~\ref{scpol}, 
we show the Gaussian fit of the $\slashed{E}$ distribution for BP1 with right polarised electron and left polarised positron of three different degrees:  
$P1^{'} \equiv \{P_{e^{-}}:+ 0.4, P_{e^{+}}: -0.6\};~P3 \equiv \{P_{e^{-}}:+ 0.8, P_{e^{+}}: -0.3\}$ and $ P3^{'}\equiv \{P_{e^{-}}: 0.8, P_{e^{+}}: -0.6\}$. 
Following previous discussion, it is clear that SM background is largest for $P1^{'}$ and smallest for $ P3^{'}$. Consequently relative size of the second peak 
is highest for $ P3^{'}$ and smallest for $P1^{'}$. The luminosity is chosen high just to capture the effect of polarization.

\begin{figure}[htb!]
\subfloat[]{\includegraphics[width=5.8cm,height=4.5cm]{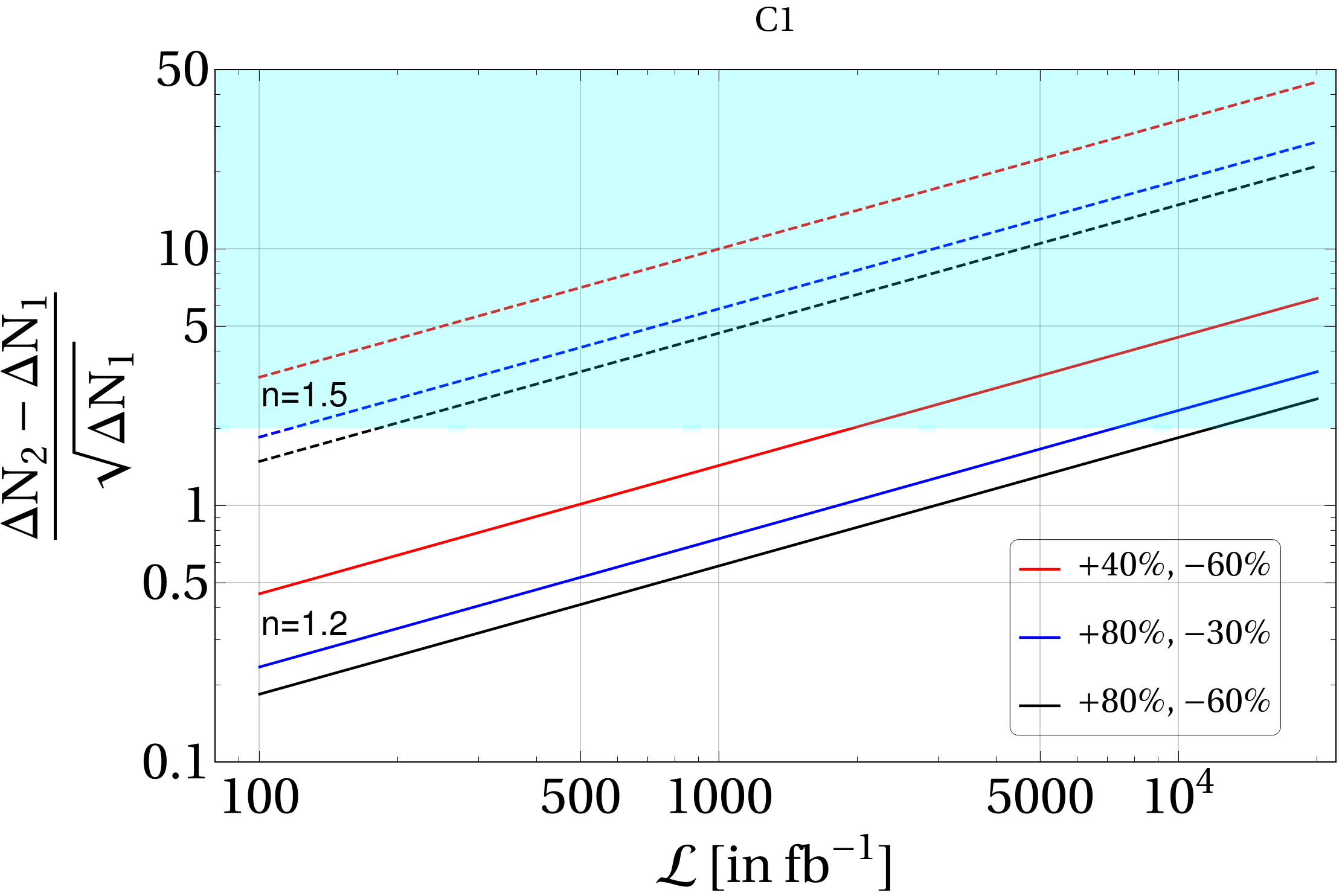}}~~
 \subfloat[]{\includegraphics[width=5.8cm,height=4.5cm]{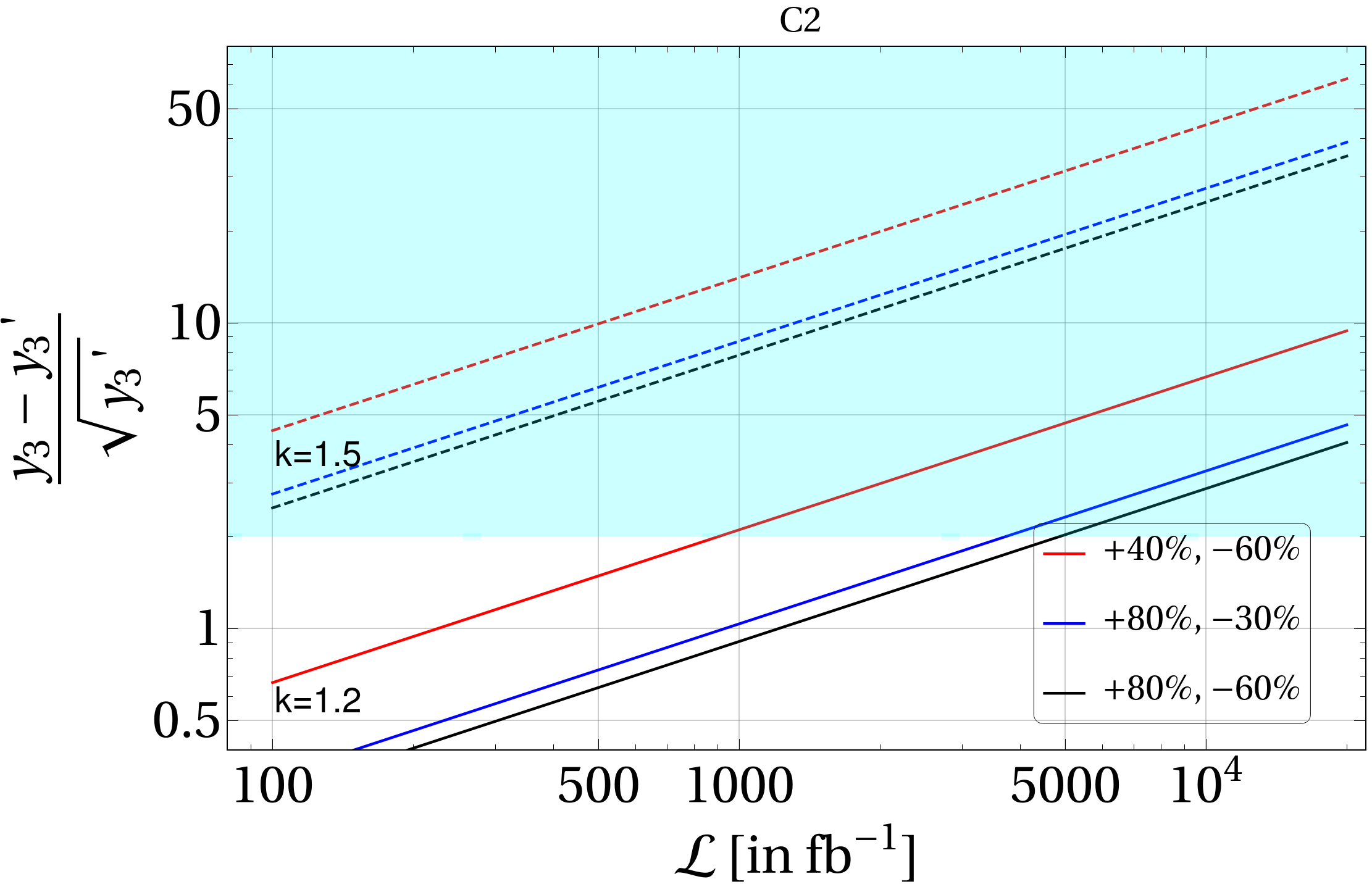}}\\
  \subfloat[]{\includegraphics[width=5.8cm,height=4.5cm]{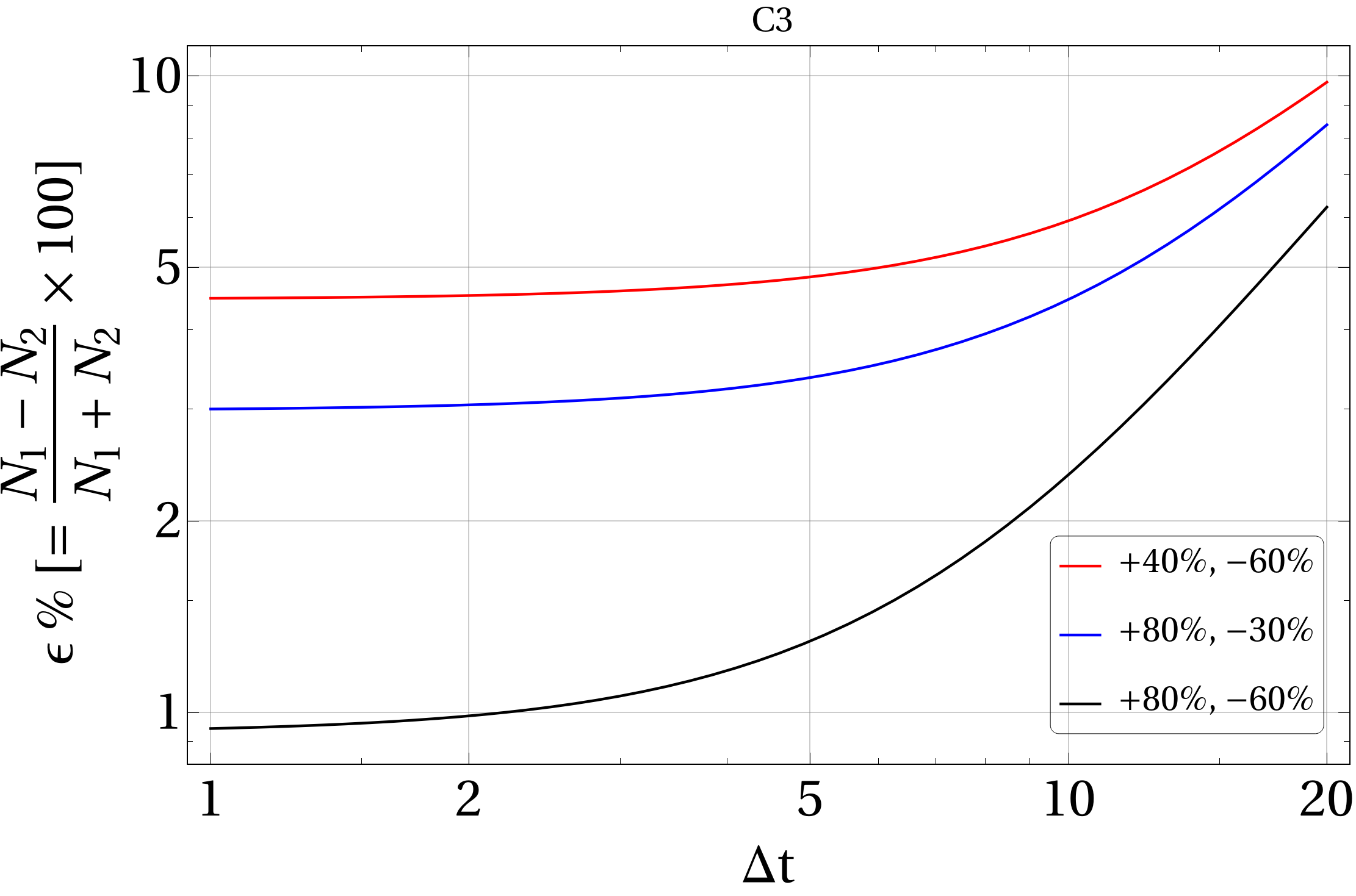}}~ ~   
  \subfloat[]{\includegraphics[width=5.8cm,height=4.5cm]{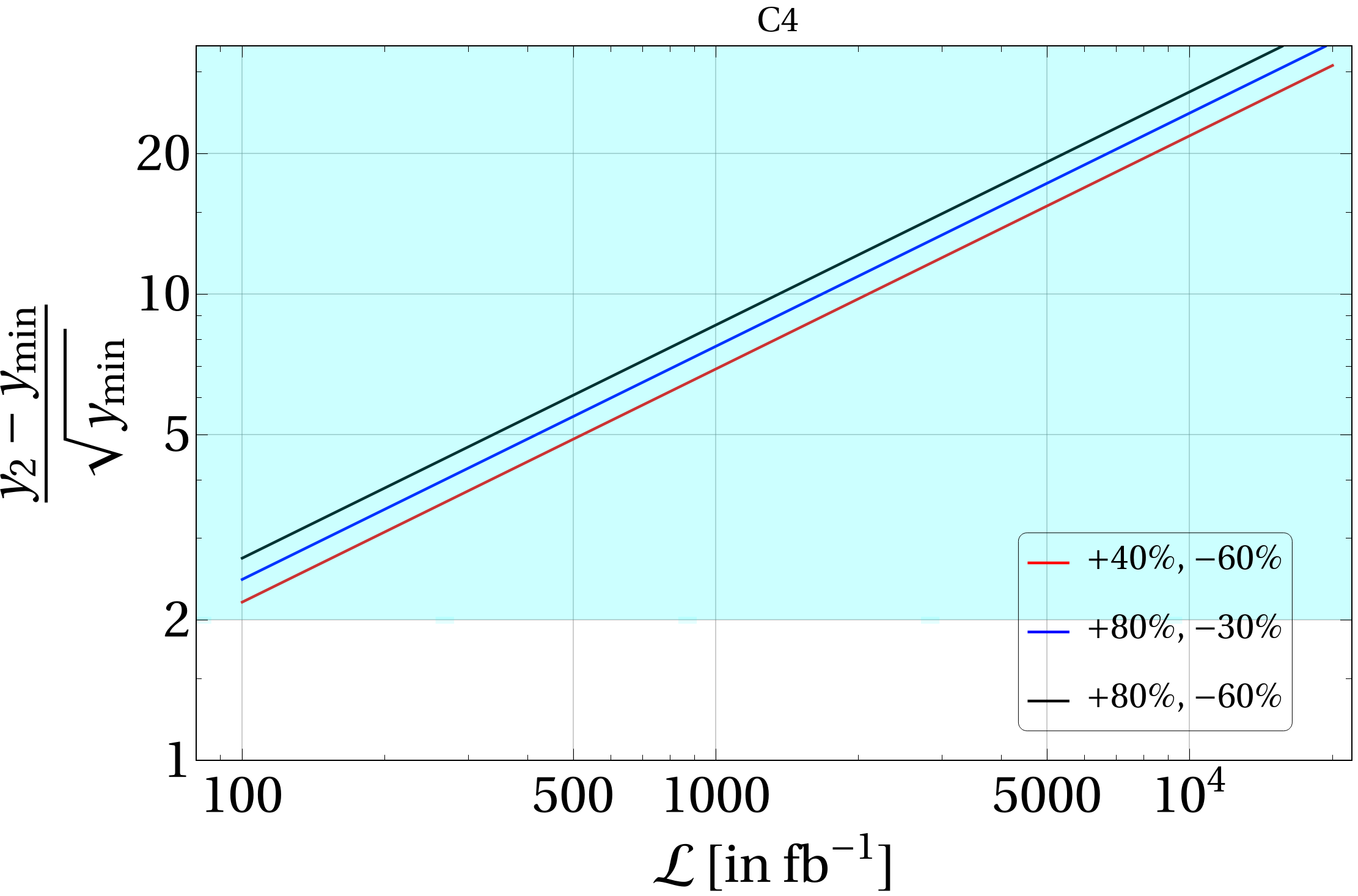}}
 \caption{Condition C1-C4 for BP1 for various polarization configurations, $P1^{'} \equiv \{P_{e^{-}}:+ 0.4, P_{e^{+}}: -0.6\};~P3 \equiv \{P_{e^{-}}:+ 0.8, P_{e^{+}}: -0.3\}$ and $ P3^{'}\equiv \{P_{e^{-}}: 0.8, P_{e^{+}}: -0.6\}$. (a) $R_{C1}$ as a function of $\mathcal{L}$ for different $n$ values (sky blue shaded region marks $R_{C1}\ge2$); (b) Same as (a) but for 
 $R_{C2}$; (c) $R_{C3}$ as function of $\Delta t$ and (d) Same as (a), but for $R_{C4}$.} 
 \label{polc1c4}
\end{figure}

We check conditions C1-C4 for all the aforementioned choices of polarisation for BP1. In Fig.~\ref{polc1c4}(a), we show the dependence of 
$R_{C1}$ as function of $\mathcal{L}$, for $n=1.2,1.5$. Here we see that $R_{C1}$ is largest for $P1^{'}$, 
simply because more number of events under the first peak due to background contamination. We check C2 next in Fig.~\ref{polc1c4}(b), 
where again $P1^{'}$ does best and larger $k$ is required to achieve $R_{C2}$ $\gsim 2$ at lower luminosities. The condition C3 is checked in 
Fig.~\ref{polc1c4} (c). Again, $R_{C3}$ is larger for $P1^{'}$ and lowest in $P3^{'}$ as it captures the height difference between the peaks. 
Fig.~\ref{polc1c4}(d) shows $R_{C4}$ as a function $\mathcal{L}$; here, $P3^{'}$ is maximum. 
We conclude that polarisation, although necessary for reducing SM background, 
doesn't alter the distinguishability of the two peaks significantly when varied within a range as 
$P1^{'}-P3^{'}$. 

%
%
%
%
%
%
%
%

\end{appendix}

\newpage
\bibliographystyle{jhep}
\bibliography{ref}

\providecommand{\href}[2]{#2}\begingroup\raggedright\begin{thebibliography}{10}

\bibitem{Rubin:1970zza}
V.~C. Rubin and W.~K. Ford, Jr., \emph{{Rotation of the Andromeda Nebula from a
  Spectroscopic Survey of Emission Regions}},
  \href{https://doi.org/10.1086/150317}{\emph{Astrophys. J.} {\bfseries 159}
  (1970) 379--403}.

\bibitem{Zwicky:1937zza}
F.~Zwicky, \emph{{On the Masses of Nebulae and of Clusters of Nebulae}},
  \href{https://doi.org/10.1086/143864}{\emph{Astrophys. J.} {\bfseries 86}
  (1937) 217--246}.

\bibitem{Hayashi:2006kw}
E.~Hayashi and S.~D.~M. White, \emph{{How Rare is the Bullet Cluster?}},
  \href{https://doi.org/10.1111/j.1745-3933.2006.00184.x}{\emph{Mon. Not. Roy.
  Astron. Soc.} {\bfseries 370} (2006) L38--L41},
  [\href{https://arxiv.org/abs/astro-ph/0604443}{{\ttfamily
  astro-ph/0604443}}].

\bibitem{Hu:2001bc}
W.~Hu and S.~Dodelson, \emph{{Cosmic microwave background anisotropies}},
  \href{https://doi.org/10.1146/annurev.astro.40.060401.093926}{\emph{Ann. Rev.
  Astron. Astrophys.} {\bfseries 40} (2002) 171--216},
  [\href{https://arxiv.org/abs/astro-ph/0110414}{{\ttfamily
  astro-ph/0110414}}].

\bibitem{Hinshaw:2012aka}
{\scshape WMAP} collaboration, G.~Hinshaw et~al., \emph{{Nine-Year Wilkinson
  Microwave Anisotropy Probe (WMAP) Observations: Cosmological Parameter
  Results}}, \href{https://doi.org/10.1088/0067-0049/208/2/19}{\emph{Astrophys.
  J. Suppl.} {\bfseries 208} (2013) 19},
  [\href{https://arxiv.org/abs/1212.5226}{{\ttfamily 1212.5226}}].

\bibitem{Spergel:2006hy}
{\scshape WMAP} collaboration, D.~N. Spergel et~al., \emph{{Wilkinson Microwave
  Anisotropy Probe (WMAP) three year results: implications for cosmology}},
  \href{https://doi.org/10.1086/513700}{\emph{Astrophys. J. Suppl.} {\bfseries
  170} (2007) 377}, [\href{https://arxiv.org/abs/astro-ph/0603449}{{\ttfamily
  astro-ph/0603449}}].

\bibitem{Planck:2018vyg}
{\scshape Planck} collaboration, N.~Aghanim et~al., \emph{{Planck 2018 results.
  VI. Cosmological parameters}},
  \href{https://doi.org/10.1051/0004-6361/201833910}{\emph{Astron. Astrophys.}
  {\bfseries 641} (2020) A6},
  [\href{https://arxiv.org/abs/1807.06209}{{\ttfamily 1807.06209}}].

\bibitem{Bertone:2004pz}
G.~Bertone, D.~Hooper and J.~Silk, \emph{{Particle dark matter: Evidence,
  candidates and constraints}},
  \href{https://doi.org/10.1016/j.physrep.2004.08.031}{\emph{Phys. Rept.}
  {\bfseries 405} (2005) 279--390},
  [\href{https://arxiv.org/abs/hep-ph/0404175}{{\ttfamily hep-ph/0404175}}].

\bibitem{Roszkowski:2017nbc}
L.~Roszkowski, E.~M. Sessolo and S.~Trojanowski, \emph{{WIMP dark matter
  candidates and searches—current status and future prospects}},
  \href{https://doi.org/10.1088/1361-6633/aab913}{\emph{Rept. Prog. Phys.}
  {\bfseries 81} (2018) 066201},
  [\href{https://arxiv.org/abs/1707.06277}{{\ttfamily 1707.06277}}].

\bibitem{Kolb:1990vq}
E.~W. Kolb and M.~S. Turner, \emph{{The Early Universe}}, {\emph{Front. Phys.}
  {\bfseries 69} (1990) 1--547}.

\bibitem{Hall:2009bx}
L.~J. Hall, K.~Jedamzik, J.~March-Russell and S.~M. West, \emph{{Freeze-In
  Production of FIMP Dark Matter}},
  \href{https://doi.org/10.1007/JHEP03(2010)080}{\emph{JHEP} {\bfseries 03}
  (2010) 080}, [\href{https://arxiv.org/abs/0911.1120}{{\ttfamily 0911.1120}}].

\bibitem{Aoki:2012ub}
M.~Aoki, M.~Duerr, J.~Kubo and H.~Takano, \emph{{Multi-Component Dark Matter
  Systems and Their Observation Prospects}},
  \href{https://doi.org/10.1103/PhysRevD.86.076015}{\emph{Phys. Rev.}
  {\bfseries D86} (2012) 076015},
  [\href{https://arxiv.org/abs/1207.3318}{{\ttfamily 1207.3318}}].

\bibitem{Liu:2011aa}
Z.-P. Liu, Y.-L. Wu and Y.-F. Zhou, \emph{{Enhancement of dark matter relic
  density from the late time dark matter conversions}},
  \href{https://doi.org/10.1140/epjc/s10052-011-1749-4}{\emph{Eur. Phys. J.}
  {\bfseries C71} (2011) 1749},
  [\href{https://arxiv.org/abs/1101.4148}{{\ttfamily 1101.4148}}].

\bibitem{Cao:2007fy}
Q.-H. Cao, E.~Ma, J.~Wudka and C.~P. Yuan, \emph{{Multipartite dark matter}},
  \href{https://arxiv.org/abs/0711.3881}{{\ttfamily 0711.3881}}.

\bibitem{Bhattacharya:2013hva}
S.~Bhattacharya, A.~Drozd, B.~Grzadkowski and J.~Wudka, \emph{{Two-Component
  Dark Matter}}, \href{https://doi.org/10.1007/JHEP10(2013)158}{\emph{JHEP}
  {\bfseries 10} (2013) 158},
  [\href{https://arxiv.org/abs/1309.2986}{{\ttfamily 1309.2986}}].

\bibitem{Esch:2014jpa}
S.~Esch, M.~Klasen and C.~E. Yaguna, \emph{{A minimal model for two-component
  dark matter}}, \href{https://doi.org/10.1007/JHEP09(2014)108}{\emph{JHEP}
  {\bfseries 09} (2014) 108},
  [\href{https://arxiv.org/abs/1406.0617}{{\ttfamily 1406.0617}}].

\bibitem{Karam:2016rsz}
A.~Karam and K.~Tamvakis, \emph{{Dark Matter from a Classically Scale-Invariant
  $SU(3)_X$}}, \href{https://doi.org/10.1103/PhysRevD.94.055004}{\emph{Phys.
  Rev.} {\bfseries D94} (2016) 055004},
  [\href{https://arxiv.org/abs/1607.01001}{{\ttfamily 1607.01001}}].

\bibitem{Ahmed:2017dbb}
A.~Ahmed, M.~Duch, B.~Grzadkowski and M.~Iglicki, \emph{{Multi-Component Dark
  Matter: the vector and fermion case}},
  \href{https://doi.org/10.1140/epjc/s10052-018-6371-2}{\emph{Eur. Phys. J.}
  {\bfseries C78} (2018) 905},
  [\href{https://arxiv.org/abs/1710.01853}{{\ttfamily 1710.01853}}].

\bibitem{Poulin:2018kap}
A.~Poulin and S.~Godfrey, \emph{{Multicomponent dark matter from a hidden
  gauged SU(3)}}, \href{https://doi.org/10.1103/PhysRevD.99.076008}{\emph{Phys.
  Rev.} {\bfseries D99} (2019) 076008},
  [\href{https://arxiv.org/abs/1808.04901}{{\ttfamily 1808.04901}}].

\bibitem{Aoki:2018gjf}
M.~Aoki and T.~Toma, \emph{{Boosted Self-interacting Dark Matter in a
  Multi-component Dark Matter Model}},
  \href{https://doi.org/10.1088/1475-7516/2018/10/020}{\emph{JCAP} {\bfseries
  1810} (2018) 020}, [\href{https://arxiv.org/abs/1806.09154}{{\ttfamily
  1806.09154}}].

\bibitem{YaserAyazi:2018lrv}
S.~Yaser~Ayazi and A.~Mohamadnejad, \emph{{Scale-Invariant Two Component Dark
  Matter}}, \href{https://doi.org/10.1140/epjc/s10052-019-6651-5}{\emph{Eur.
  Phys. J.} {\bfseries C79} (2019) 140},
  [\href{https://arxiv.org/abs/1808.08706}{{\ttfamily 1808.08706}}].

\bibitem{Aoki:2017eqn}
M.~Aoki, D.~Kaneko and J.~Kubo, \emph{{Multicomponent Dark Matter in Radiative
  Seesaw Models}},
  \href{https://doi.org/10.3389/fphy.2017.00053}{\emph{Front.in Phys.}
  {\bfseries 5} (2017) 53}, [\href{https://arxiv.org/abs/1711.03765}{{\ttfamily
  1711.03765}}].

\bibitem{Biswas:2013nn}
A.~Biswas, D.~Majumdar, A.~Sil and P.~Bhattacharjee, \emph{{Two Component Dark
  Matter : A Possible Explanation of 130 GeV $\gamma-$ Ray Line from the
  Galactic Centre}},
  \href{https://doi.org/10.1088/1475-7516/2013/12/049}{\emph{JCAP} {\bfseries
  1312} (2013) 049}, [\href{https://arxiv.org/abs/1301.3668}{{\ttfamily
  1301.3668}}].

\bibitem{Bhattacharya:2016ysw}
S.~Bhattacharya, P.~Poulose and P.~Ghosh, \emph{{Multipartite Interacting
  Scalar Dark Matter in the light of updated LUX data}},
  \href{https://doi.org/10.1088/1475-7516/2017/04/043}{\emph{JCAP} {\bfseries
  1704} (2017) 043}, [\href{https://arxiv.org/abs/1607.08461}{{\ttfamily
  1607.08461}}].

\bibitem{Bhattacharya:2017fid}
S.~Bhattacharya, P.~Ghosh, T.~N. Maity and T.~S. Ray, \emph{{Mitigating Direct
  Detection Bounds in Non-minimal Higgs Portal Scalar Dark Matter Models}},
  \href{https://doi.org/10.1007/JHEP10(2017)088}{\emph{JHEP} {\bfseries 10}
  (2017) 088}, [\href{https://arxiv.org/abs/1706.04699}{{\ttfamily
  1706.04699}}].

\bibitem{Barman:2018esi}
B.~Barman, S.~Bhattacharya and M.~Zakeri, \emph{{Multipartite Dark Matter in
  $SU(2)_N$ extension of Standard Model and signatures at the LHC}},
  \href{https://doi.org/10.1088/1475-7516/2018/09/023}{\emph{JCAP} {\bfseries
  1809} (2018) 023}, [\href{https://arxiv.org/abs/1806.01129}{{\ttfamily
  1806.01129}}].

\bibitem{Bhattacharya:2018cgx}
S.~Bhattacharya, P.~Ghosh and N.~Sahu, \emph{{Multipartite Dark Matter with
  Scalars, Fermions and signatures at LHC}},
  \href{https://doi.org/10.1007/JHEP02(2019)059}{\emph{JHEP} {\bfseries 02}
  (2019) 059}, [\href{https://arxiv.org/abs/1809.07474}{{\ttfamily
  1809.07474}}].

\bibitem{Bhattacharya:2019fgs}
S.~Bhattacharya, P.~Ghosh, A.~K. Saha and A.~Sil, \emph{{Two component dark
  matter with inert Higgs doublet: neutrino mass, high scale validity and
  collider searches}},
  \href{https://doi.org/10.1007/JHEP03(2020)090}{\emph{JHEP} {\bfseries 03}
  (2020) 090}, [\href{https://arxiv.org/abs/1905.12583}{{\ttfamily
  1905.12583}}].

\bibitem{Borah:2019aeq}
D.~Borah, R.~Roshan and A.~Sil, \emph{{Minimal Two-component Scalar Doublet
  Dark Matter with Radiative Neutrino Mass}},
  \href{https://arxiv.org/abs/1904.04837}{{\ttfamily 1904.04837}}.

\bibitem{Chakraborti:2018lso}
S.~Chakraborti and P.~Poulose, \emph{{Interplay of Scalar and Fermionic
  Components in a Multi-component Dark Matter Scenario}},
  \href{https://arxiv.org/abs/1808.01979}{{\ttfamily 1808.01979}}.

\bibitem{Chakraborti:2018aae}
S.~Chakraborti, A.~Dutta~Banik and R.~Islam, \emph{{Probing Multicomponent
  Extension of Inert Doublet Model with a Vector Dark Matter}},
  \href{https://arxiv.org/abs/1810.05595}{{\ttfamily 1810.05595}}.

\bibitem{Bhattacharya:2018cqu}
S.~Bhattacharya, A.~K. Saha, A.~Sil and J.~Wudka, \emph{{Dark Matter as a
  remnant of SQCD Inflation}},
  \href{https://doi.org/10.1007/JHEP10(2018)124}{\emph{JHEP} {\bfseries 10}
  (2018) 124}, [\href{https://arxiv.org/abs/1805.03621}{{\ttfamily
  1805.03621}}].

\bibitem{Yaguna:2021rds}
C.~E. Yaguna and O.~Zapata, \emph{{Fermion and scalar two-component dark matter
  from a $Z_4$ symmetry}},  \href{https://arxiv.org/abs/2112.07020}{{\ttfamily
  2112.07020}}.

\bibitem{Belanger:2021lwd}
G.~Belanger, A.~Mjallal and A.~Pukhov, \emph{{Two dark matter candidates: The
  case of inert doublet and singlet scalars}},
  \href{https://doi.org/10.1103/PhysRevD.105.035018}{\emph{Phys. Rev. D}
  {\bfseries 105} (2022) 035018},
  [\href{https://arxiv.org/abs/2108.08061}{{\ttfamily 2108.08061}}].

\bibitem{VanLoi:2021dzv}
D.~Van~Loi, N.~M. Duc and P.~V. Dong, \emph{{Dequantization of electric charge:
  Probing scenarios of cosmological multi-component dark matter}},
  \href{https://arxiv.org/abs/2106.12278}{{\ttfamily 2106.12278}}.

\bibitem{Yaguna:2021vhb}
C.~E. Yaguna and O.~Zapata, \emph{{Two-component scalar dark matter in Z$_{2n}$
  scenarios}}, \href{https://doi.org/10.1007/JHEP10(2021)185}{\emph{JHEP}
  {\bfseries 10} (2021) 185},
  [\href{https://arxiv.org/abs/2106.11889}{{\ttfamily 2106.11889}}].

\bibitem{DiazSaez:2021pfw}
B.~D\'\i{}az~S\'aez, K.~M\"ohling and D.~St\"ockinger, \emph{{Two real scalar
  WIMP model in the assisted freeze-out scenario}},
  \href{https://doi.org/10.1088/1475-7516/2021/10/027}{\emph{JCAP} {\bfseries
  10} (2021) 027}, [\href{https://arxiv.org/abs/2103.17064}{{\ttfamily
  2103.17064}}].

\bibitem{Chakrabarty:2021kmr}
N.~Chakrabarty, R.~Roshan and A.~Sil, \emph{{Two Component Doublet-Triplet
  Scalar Dark Matter stabilising the Electroweak vacuum}},
  \href{https://arxiv.org/abs/2102.06032}{{\ttfamily 2102.06032}}.

\bibitem{Nam:2020twn}
C.~H. Nam, D.~Van~Loi, L.~X. Thuy and P.~Van~Dong, \emph{{Multicomponent dark
  matter in noncommutative $B − L$ gauge theory}},
  \href{https://doi.org/10.1007/JHEP12(2020)029}{\emph{JHEP} {\bfseries 12}
  (2020) 029}, [\href{https://arxiv.org/abs/2006.00845}{{\ttfamily
  2006.00845}}].

\bibitem{Betancur:2020fdl}
A.~Betancur, G.~Palacio and A.~Rivera, \emph{{Inert doublet as multicomponent
  dark matter}},
  \href{https://doi.org/10.1016/j.nuclphysb.2020.115276}{\emph{Nucl. Phys. B}
  {\bfseries 962} (2021) 115276},
  [\href{https://arxiv.org/abs/2002.02036}{{\ttfamily 2002.02036}}].

\bibitem{Nanda:2019nqy}
D.~Nanda and D.~Borah, \emph{{Connecting Light Dirac Neutrinos to a
  Multi-component Dark Matter Scenario in Gauged $B-L$ Model}},
  \href{https://doi.org/10.1140/epjc/s10052-020-8122-4}{\emph{Eur. Phys. J. C}
  {\bfseries 80} (2020) 557},
  [\href{https://arxiv.org/abs/1911.04703}{{\ttfamily 1911.04703}}].

\bibitem{Bhattacharya:2019tqq}
S.~Bhattacharya, N.~Chakrabarty, R.~Roshan and A.~Sil, \emph{{Multicomponent
  dark matter in extended $U(1)_{B-L}$: neutrino mass and high scale
  validity}}, \href{https://doi.org/10.1088/1475-7516/2020/04/013}{\emph{JCAP}
  {\bfseries 04} (2020) 013},
  [\href{https://arxiv.org/abs/1910.00612}{{\ttfamily 1910.00612}}].

\bibitem{Elahi:2019jeo}
F.~Elahi and S.~Khatibi, \emph{{Multi-Component Dark Matter in a Non-Abelian
  Dark Sector}}, \href{https://doi.org/10.1103/PhysRevD.100.015019}{\emph{Phys.
  Rev. D} {\bfseries 100} (2019) 015019},
  [\href{https://arxiv.org/abs/1902.04384}{{\ttfamily 1902.04384}}].

\bibitem{Herrero-Garcia:2018lga}
J.~Herrero-Garcia, A.~Scaffidi, M.~White and A.~G. Williams,
  \emph{{Time-dependent rate of multicomponent dark matter: Reproducing the
  DAMA/LIBRA phase-2 results}},
  \href{https://doi.org/10.1103/PhysRevD.98.123007}{\emph{Phys. Rev. D}
  {\bfseries 98} (2018) 123007},
  [\href{https://arxiv.org/abs/1804.08437}{{\ttfamily 1804.08437}}].

\bibitem{Das:2022oyx}
A.~Das, S.~Gola, S.~Mandal and N.~Sinha, \emph{{Two-component scalar and
  fermionic dark matter candidates in a generic U$(1)_X$ model}},
  \href{https://arxiv.org/abs/2202.01443}{{\ttfamily 2202.01443}}.

\bibitem{Bhattacharya:2021rwh}
S.~Bhattacharya, S.~Chakraborti and D.~Pradhan, \emph{{Electroweak Symmetry
  Breaking and WIMP-FIMP Dark Matter}},
  \href{https://arxiv.org/abs/2110.06985}{{\ttfamily 2110.06985}}.

\bibitem{DuttaBanik:2016jzv}
A.~Dutta~Banik, M.~Pandey, D.~Majumdar and A.~Biswas, \emph{{Two component
  WIMP\textendash{}FImP dark matter model with singlet fermion, scalar and
  pseudo scalar}},
  \href{https://doi.org/10.1140/epjc/s10052-017-5221-y}{\emph{Eur. Phys. J. C}
  {\bfseries 77} (2017) 657},
  [\href{https://arxiv.org/abs/1612.08621}{{\ttfamily 1612.08621}}].

\bibitem{Choi:2021yps}
S.-M. Choi, J.~Kim, P.~Ko and J.~Li, \emph{{A multi-component SIMP model with
  $U(1)_{X} \to Z_{2} \to Z_{3}$}},
  \href{https://doi.org/10.1007/JHEP09(2021)028}{\emph{JHEP} {\bfseries 09}
  (2021) 028}, [\href{https://arxiv.org/abs/2103.05956}{{\ttfamily
  2103.05956}}].

\bibitem{XENON:2018voc}
{\scshape XENON} collaboration, E.~Aprile et~al., \emph{{Dark Matter Search
  Results from a One Ton-Year Exposure of XENON1T}},
  \href{https://doi.org/10.1103/PhysRevLett.121.111302}{\emph{Phys. Rev. Lett.}
  {\bfseries 121} (2018) 111302},
  [\href{https://arxiv.org/abs/1805.12562}{{\ttfamily 1805.12562}}].

\bibitem{PandaX-4T:2021bab}
{\scshape PandaX-4T} collaboration, Y.~Meng et~al., \emph{{Dark Matter Search
  Results from the PandaX-4T Commissioning Run}},
  \href{https://doi.org/10.1103/PhysRevLett.127.261802}{\emph{Phys. Rev. Lett.}
  {\bfseries 127} (2021) 261802},
  [\href{https://arxiv.org/abs/2107.13438}{{\ttfamily 2107.13438}}].

\bibitem{Herrero-Garcia:2017vrl}
J.~Herrero-Garcia, A.~Scaffidi, M.~White and A.~G. Williams, \emph{{On the
  direct detection of multi-component dark matter: sensitivity studies and
  parameter estimation}},
  \href{https://doi.org/10.1088/1475-7516/2017/11/021}{\emph{JCAP} {\bfseries
  1711} (2017) 021}, [\href{https://arxiv.org/abs/1709.01945}{{\ttfamily
  1709.01945}}].

\bibitem{Herrero-Garcia:2018qnz}
J.~Herrero-Garcia, A.~Scaffidi, M.~White and A.~G. Williams, \emph{{On the
  direct detection of multi-component dark matter: implications of the relic
  abundance}}, \href{https://doi.org/10.1088/1475-7516/2019/01/008}{\emph{JCAP}
  {\bfseries 1901} (2019) 008},
  [\href{https://arxiv.org/abs/1809.06881}{{\ttfamily 1809.06881}}].

\bibitem{Hernandez-Sanchez:2020aop}
J.~Hernandez-Sanchez, V.~Keus, S.~Moretti, D.~Rojas-Ciofalo and D.~Sokolowska,
  \emph{{Complementary Probes of Two-component Dark Matter}},
  \href{https://arxiv.org/abs/2012.11621}{{\ttfamily 2012.11621}}.

\bibitem{Konar:2009qr}
P.~Konar, K.~Kong, K.~T. Matchev and M.~Park, \emph{{Dark Matter Particle
  Spectroscopy at the LHC: Generalizing M(T2) to Asymmetric Event Topologies}},
  \href{https://doi.org/10.1007/JHEP04(2010)086}{\emph{JHEP} {\bfseries 04}
  (2010) 086}, [\href{https://arxiv.org/abs/0911.4126}{{\ttfamily 0911.4126}}].

\bibitem{Agashe:2010tu}
K.~Agashe, D.~Kim, D.~G.~E. Walker and L.~Zhu, \emph{{Using $M_{T2}$ to
  Distinguish Dark Matter Stabilization Symmetries}},
  \href{https://doi.org/10.1103/PhysRevD.84.055020}{\emph{Phys. Rev. D}
  {\bfseries 84} (2011) 055020},
  [\href{https://arxiv.org/abs/1012.4460}{{\ttfamily 1012.4460}}].

\bibitem{Giudice:2011ib}
G.~F. Giudice, B.~Gripaios and R.~Mahbubani, \emph{{Counting dark matter
  particles in LHC events}},
  \href{https://doi.org/10.1103/PhysRevD.85.075019}{\emph{Phys. Rev. D}
  {\bfseries 85} (2012) 075019},
  [\href{https://arxiv.org/abs/1108.1800}{{\ttfamily 1108.1800}}].

\bibitem{Belanger:2018sti}
G.~B\'elanger et~al., \emph{{LHC-friendly minimal freeze-in models}},
  \href{https://doi.org/10.1007/JHEP02(2019)186}{\emph{JHEP} {\bfseries 02}
  (2019) 186}, [\href{https://arxiv.org/abs/1811.05478}{{\ttfamily
  1811.05478}}].

\bibitem{Alimena:2019zri}
J.~Alimena et~al., \emph{{Searching for long-lived particles beyond the
  Standard Model at the Large Hadron Collider}},
  \href{https://doi.org/10.1088/1361-6471/ab4574}{\emph{J. Phys. G} {\bfseries
  47} (2020) 090501}, [\href{https://arxiv.org/abs/1903.04497}{{\ttfamily
  1903.04497}}].

\bibitem{Banerjee:2018uut}
S.~Banerjee, G.~B\'elanger, A.~Ghosh and B.~Mukhopadhyaya, \emph{{Long-lived
  stau, sneutrino dark matter and right-slepton spectrum}},
  \href{https://doi.org/10.1007/JHEP09(2018)143}{\emph{JHEP} {\bfseries 09}
  (2018) 143}, [\href{https://arxiv.org/abs/1806.04488}{{\ttfamily
  1806.04488}}].

\bibitem{Liew:2016oon}
S.~P. Liew, M.~Papucci, A.~Vichi and K.~M. Zurek, \emph{{Mono-X Versus Direct
  Searches: Simplified Models for Dark Matter at the LHC}},
  \href{https://doi.org/10.1007/JHEP06(2017)082}{\emph{JHEP} {\bfseries 06}
  (2017) 082}, [\href{https://arxiv.org/abs/1612.00219}{{\ttfamily
  1612.00219}}].

\bibitem{Kahlhoefer:2017dnp}
F.~Kahlhoefer, \emph{{Review of LHC Dark Matter Searches}},
  \href{https://doi.org/10.1142/S0217751X1730006X}{\emph{Int. J. Mod. Phys. A}
  {\bfseries 32} (2017) 1730006},
  [\href{https://arxiv.org/abs/1702.02430}{{\ttfamily 1702.02430}}].

\bibitem{Boveia:2018yeb}
A.~Boveia and C.~Doglioni, \emph{{Dark Matter Searches at Colliders}},
  \href{https://doi.org/10.1146/annurev-nucl-101917-021008}{\emph{Ann. Rev.
  Nucl. Part. Sci.} {\bfseries 68} (2018) 429--459},
  [\href{https://arxiv.org/abs/1810.12238}{{\ttfamily 1810.12238}}].

\bibitem{Abercrombie:2015wmb}
D.~Abercrombie et~al., \emph{{Dark Matter benchmark models for early LHC Run-2
  Searches: Report of the ATLAS/CMS Dark Matter Forum}},
  \href{https://doi.org/10.1016/j.dark.2019.100371}{\emph{Phys. Dark Univ.}
  {\bfseries 27} (2020) 100371},
  [\href{https://arxiv.org/abs/1507.00966}{{\ttfamily 1507.00966}}].

\bibitem{Abdallah:2015ter}
J.~Abdallah et~al., \emph{{Simplified Models for Dark Matter Searches at the
  LHC}}, \href{https://doi.org/10.1016/j.dark.2015.08.001}{\emph{Phys. Dark
  Univ.} {\bfseries 9-10} (2015) 8--23},
  [\href{https://arxiv.org/abs/1506.03116}{{\ttfamily 1506.03116}}].

\bibitem{Barman:2021hhg}
B.~Barman, S.~Bhattacharya, S.~Girmohanta and S.~Jahedi, \emph{{Catch 'em all:
  Effective Leptophilic WIMPs at the $e^+\,e^-$ Collider}},
  \href{https://arxiv.org/abs/2109.10936}{{\ttfamily 2109.10936}}.

\bibitem{Dolle:2009fn}
E.~M. Dolle and S.~Su, \emph{{The Inert Dark Matter}},
  \href{https://doi.org/10.1103/PhysRevD.80.055012}{\emph{Phys. Rev. D}
  {\bfseries 80} (2009) 055012},
  [\href{https://arxiv.org/abs/0906.1609}{{\ttfamily 0906.1609}}].

\bibitem{Dutta:2020xwn}
M.~Dutta, S.~Bhattacharya, P.~Ghosh and N.~Sahu, \emph{{Singlet-Doublet
  Majorana Dark Matter and Neutrino Mass in a minimal Type-I Seesaw Scenario}},
  \href{https://doi.org/10.1088/1475-7516/2021/03/008}{\emph{JCAP} {\bfseries
  03} (2021) 008}, [\href{https://arxiv.org/abs/2009.00885}{{\ttfamily
  2009.00885}}].

\bibitem{Kannike:2012pe}
K.~Kannike, \emph{{Vacuum Stability Conditions From Copositivity Criteria}},
  \href{https://doi.org/10.1140/epjc/s10052-012-2093-z}{\emph{Eur. Phys. J.}
  {\bfseries C72} (2012) 2093},
  [\href{https://arxiv.org/abs/1205.3781}{{\ttfamily 1205.3781}}].

\bibitem{Chakrabortty:2013mha}
J.~Chakrabortty, P.~Konar and T.~Mondal, \emph{{Copositive Criteria and
  Boundedness of the Scalar Potential}},
  \href{https://doi.org/10.1103/PhysRevD.89.095008}{\emph{Phys. Rev.}
  {\bfseries D89} (2014) 095008},
  [\href{https://arxiv.org/abs/1311.5666}{{\ttfamily 1311.5666}}].

\bibitem{LEP}
\emph{Searches for supersymmetric particles in e + e- collisions up to 208 gev
  and interpretation of the results within the mssm},
  \href{https://doi.org/10.1140/epjc/s2003-01355-5}{\emph{The European Physical
  Journal C} {\bfseries 31} (Dec, 2003) 421–479}.

\bibitem{Pierce:2007ut}
A.~Pierce and J.~Thaler, \emph{{Natural Dark Matter from an Unnatural Higgs
  Boson and New Colored Particles at the TeV Scale}},
  \href{https://doi.org/10.1088/1126-6708/2007/08/026}{\emph{JHEP} {\bfseries
  08} (2007) 026}, [\href{https://arxiv.org/abs/hep-ph/0703056}{{\ttfamily
  hep-ph/0703056}}].

\bibitem{Degrande:2011ua}
C.~Degrande, C.~Duhr, B.~Fuks, D.~Grellscheid, O.~Mattelaer and T.~Reiter,
  \emph{{UFO - The Universal FeynRules Output}},
  \href{https://doi.org/10.1016/j.cpc.2012.01.022}{\emph{Comput. Phys. Commun.}
  {\bfseries 183} (2012) 1201--1214},
  [\href{https://arxiv.org/abs/1108.2040}{{\ttfamily 1108.2040}}].

\bibitem{Belanger:2018ccd}
G.~B\'elanger, F.~Boudjema, A.~Goudelis, A.~Pukhov and B.~Zaldivar,
  \emph{{micrOMEGAs5.0 : Freeze-in}},
  \href{https://doi.org/10.1016/j.cpc.2018.04.027}{\emph{Comput. Phys. Commun.}
  {\bfseries 231} (2018) 173--186},
  [\href{https://arxiv.org/abs/1801.03509}{{\ttfamily 1801.03509}}].

\bibitem{MAGIC:2016xys}
{\scshape MAGIC, Fermi-LAT} collaboration, M.~L. Ahnen et~al., \emph{{Limits to
  Dark Matter Annihilation Cross-Section from a Combined Analysis of MAGIC and
  Fermi-LAT Observations of Dwarf Satellite Galaxies}},
  \href{https://doi.org/10.1088/1475-7516/2016/02/039}{\emph{JCAP} {\bfseries
  02} (2016) 039}, [\href{https://arxiv.org/abs/1601.06590}{{\ttfamily
  1601.06590}}].

\bibitem{Fermi-LAT:2015att}
{\scshape Fermi-LAT} collaboration, M.~Ackermann et~al., \emph{{Searching for
  Dark Matter Annihilation from Milky Way Dwarf Spheroidal Galaxies with Six
  Years of Fermi Large Area Telescope Data}},
  \href{https://doi.org/10.1103/PhysRevLett.115.231301}{\emph{Phys. Rev. Lett.}
  {\bfseries 115} (2015) 231301},
  [\href{https://arxiv.org/abs/1503.02641}{{\ttfamily 1503.02641}}].

\bibitem{Adolphsen:2013kya}
\emph{{The International Linear Collider Technical Design Report - Volume 3.II:
  Accelerator Baseline Design}},
  \href{https://arxiv.org/abs/1306.6328}{{\ttfamily 1306.6328}}.

\bibitem{Bambade:2019fyw}
P.~Bambade et~al., \emph{{The International Linear Collider: A Global
  Project}},  \href{https://arxiv.org/abs/1903.01629}{{\ttfamily 1903.01629}}.

\bibitem{Zarnecki:2020ics}
{\scshape CLICdp, ILD concept group} collaboration, A.~F. Zarnecki, \emph{{On
  the physics potential of ILC and CLIC}},
  \href{https://doi.org/10.22323/1.376.0037}{\emph{PoS} {\bfseries CORFU2019}
  (2020) 037}, [\href{https://arxiv.org/abs/2004.14628}{{\ttfamily
  2004.14628}}].

\bibitem{Alwall:2014hca}
J.~Alwall, R.~Frederix, S.~Frixione, V.~Hirschi, F.~Maltoni, O.~Mattelaer
  et~al., \emph{{The automated computation of tree-level and next-to-leading
  order differential cross sections, and their matching to parton shower
  simulations}}, \href{https://doi.org/10.1007/JHEP07(2014)079}{\emph{JHEP}
  {\bfseries 07} (2014) 079},
  [\href{https://arxiv.org/abs/1405.0301}{{\ttfamily 1405.0301}}].

\bibitem{deFavereau:2013fsa}
{\scshape DELPHES 3} collaboration, J.~de~Favereau, C.~Delaere, P.~Demin,
  A.~Giammanco, V.~Lema\^\i{}tre, A.~Mertens et~al., \emph{{DELPHES 3, A
  modular framework for fast simulation of a generic collider experiment}},
  \href{https://doi.org/10.1007/JHEP02(2014)057}{\emph{JHEP} {\bfseries 02}
  (2014) 057}, [\href{https://arxiv.org/abs/1307.6346}{{\ttfamily 1307.6346}}].

\bibitem{Behnke:2013lya}
H.~Abramowicz et~al., \emph{{The International Linear Collider Technical Design
  Report - Volume 4: Detectors}},
  \href{https://arxiv.org/abs/1306.6329}{{\ttfamily 1306.6329}}.

\bibitem{Bhattacharya:2022qck}
S.~Bhattacharya, P.~Ghosh, J.~Lahiri and B.~Mukhopadhyaya, \emph{{Mono-X signal
  and two component dark matter: new distinction criteria}},
  \href{https://arxiv.org/abs/2211.10749}{{\ttfamily 2211.10749}}.

\end{thebibliography}\endgroup

\end{document}